\documentclass[a4paper,11pt]{article}

\usepackage{pdfpages}
\usepackage[T1]{fontenc}
\usepackage[utf8]{inputenc}
\usepackage{feynmp-auto}
\usepackage{ragged2e} 
\usepackage{geometry}
\usepackage{a4wide}
\usepackage{slashed}
\usepackage{graphicx}
\usepackage{tikz}
\usepackage{afterpage}
\usepackage{amsmath}
\usetikzlibrary{decorations.pathmorphing}
\usepackage{epsf}
\usepackage{tikz-feynman}
\usetikzlibrary{patterns} 
\usepackage{amssymb}
\usepackage[normalem]{ulem}

\usepackage{cite}
\usepackage{multirow}
\usepackage{subfigure}
\usepackage{appendix}
\usepackage{tikz}
\usepackage{amsfonts,amsthm,euscript,braket,xcolor}
\usepackage{breqn}
\newcommand{\be}{\begin{equation}}
	\newcommand{\ee}{\end{equation}}

\newcommand{\Rmnum}[1]{\expandafter\@slowromancap\romannumeral #1@}
\newcommand{\bea}{\begin{eqnarray}}
	\newcommand{\eea}{\end{eqnarray}}

\usepackage{adjustbox}         
\newcommand{\sech}{\operatorname{sech}} 

\newcommand\scalemath[2]{\scalebox{#1}{\mbox{\ensuremath{\displaystyle #2}}}}
\numberwithin{equation}{section}

\usepackage{ulem}
\usepackage[colorlinks=true, linkcolor=blue, citecolor=blue, urlcolor=blue]{hyperref}

\begin{document}
\begin{titlepage}
    \begin{center}

        {\renewcommand{\rmdefault}{qcs} 
        \normalfont

        {\Large UNIVERSIDADE FEDERAL DO ABC \\ PROGRAMA DE PÓS-GRADUAÇÃO EM FÍSICA}

        \vspace{3cm}

        {\large Bruno Pinheiro Toniato}
        
        \vspace{3cm}
        
        \textbf{\Large Holographic QCD and quarkonium melting: Finite temperature, density, and external field effects in self-consistent dynamical models}

        \vspace{3cm}

        {\large Master Dissertation}

        \vfill

        {\large Santo André, SP - Brazil \\ 
        2026}

        }

    \end{center}
\end{titlepage}

\newpage
\thispagestyle{empty}   
\null                   
\newpage

\begin{center}
    {\large Bruno Pinheiro Toniato}
    
    \vspace{2cm}
    
    {\Large Holographic QCD and quarkonium melting: Finite temperature, density, and external field effects in self-consistent dynamical models}

\end{center}

\vspace{2cm}

\begin{flushright}
  \begin{minipage}{.45\textwidth}
    Dissertação apresentada ao Programa de
    Pós-Graduação em Física da Universidade
    Federal do ABC como requisito parcial à
    obtenção do título de Mestre em Física.

    \vspace{1cm}
    
    Orientador: Prof. Dr. Roldão da Rocha.

  \end{minipage}
\end{flushright}

\begin{center}
    \vfill

    {\large Santo André, SP - Brazil \\ 
    2026}
    \thispagestyle{empty}
\end{center}

\includepdf[pages=-]{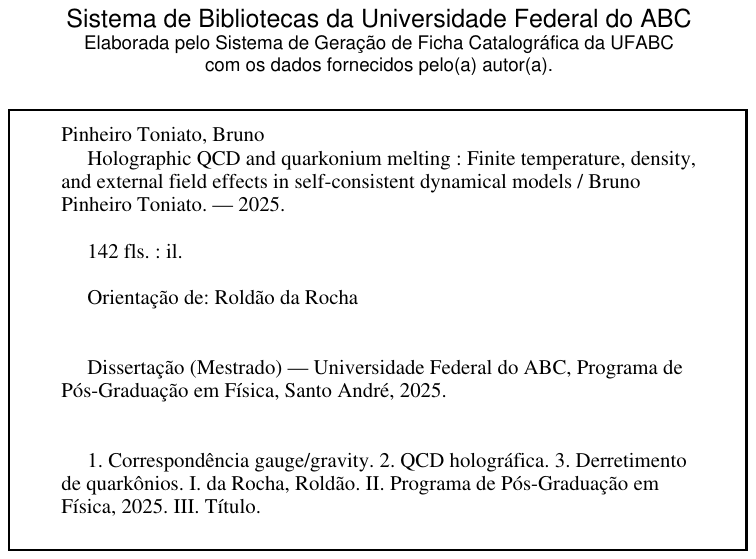}

\newpage

\null\vfill

\thispagestyle{empty}

\noindent Este exemplar foi revisado e alterado em relação à versão original, de acordo com as
observações levantadas pela banca examinadora no dia da defesa, sob responsabilidade única do(a) autor(a) e com a anuência do(a) (co)orientador(a).

\newpage

\newpage
\thispagestyle{empty}   
\null                   
\includepdf[pages=-]{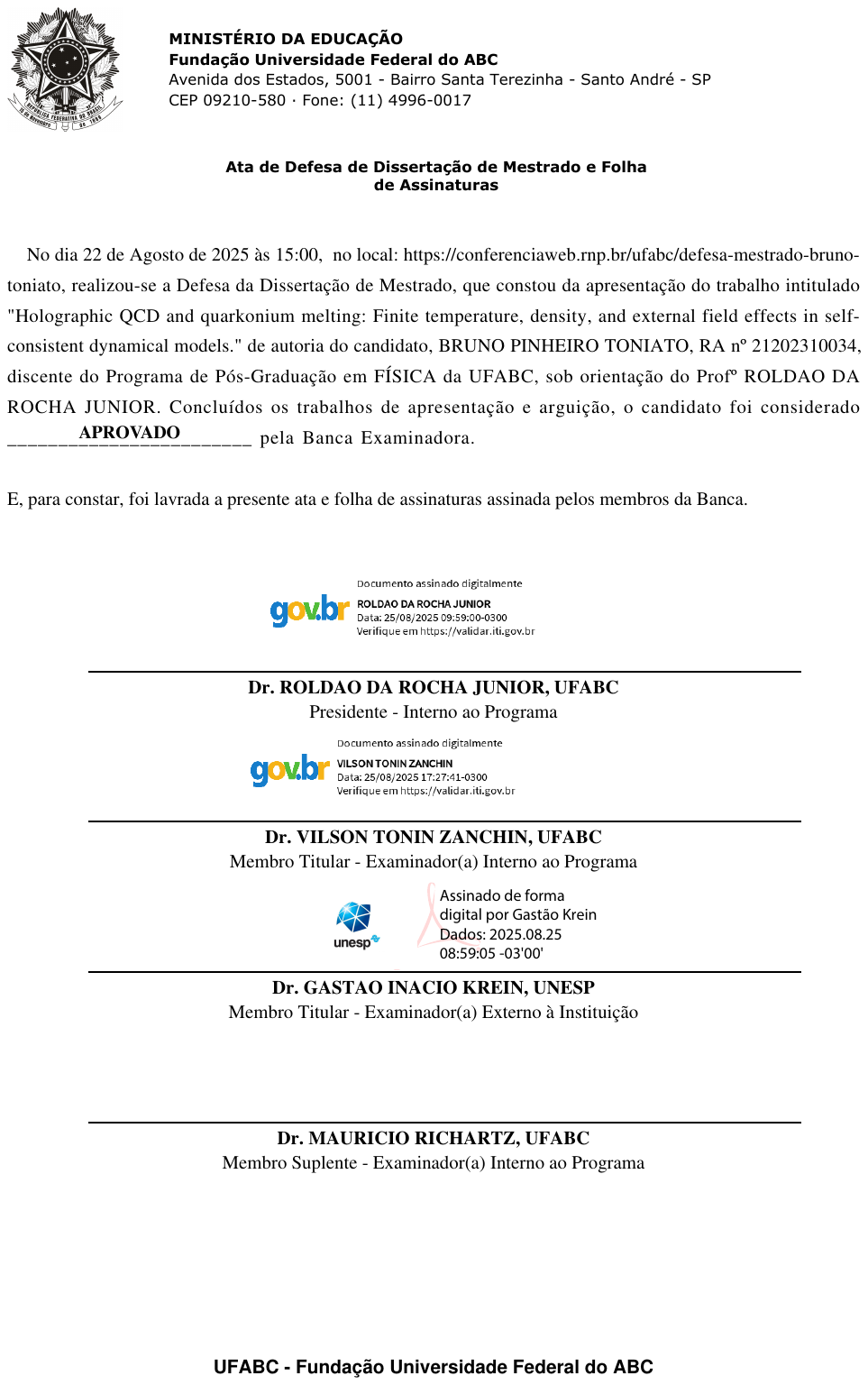}
\newpage

\newpage
\thispagestyle{empty}   
\null                   
\newpage

\null\vfill

\thispagestyle{empty}

\noindent This study was financed in part by the Coordenação de Aperfeiçoamento de Pessoal de Nível Superior – Brasil (CAPES) – Finance Code 001 and Fundação de Amparo à Pesquisa do Estado de São Paulo – Brasil (FAPESP) – Grants 2022/15325-1, 2023/09097-9.

\newpage
\thispagestyle{empty}   
\null                   
\newpage

\newpage

\begin{center}

        {\renewcommand{\rmdefault}{qcs} 
        \normalfont

        {\Large ACKNOWLEDGEMENTS}

        }

    \end{center}

\vspace{1cm}

{\renewcommand{\rmdefault}{qcs} 
        
        All my accomplishments rest on the support of my parents, Carlos Toniato and Lucimara Pinheiro Toniato, who have always put my education first. Their sacrifices and unwavering belief sustained me through every setback and milestone. The education they prioritized carried me to heights I never dreamed possible. From my first astronomy book at age six to their encouragement throughout university, they have always been there for me. My love and gratitude for them are infinite and eternal. For her hard work in helping raise me, I must also thank my grandmother, Maria da Graça, whose unmatched kindness and compassion have deeply shaped the course of my life. 

        The people at HRAC/Centrinho-USP are among the most remarkable professionals I have known. Their importance in my life is tremendous, as I was born with a craniofacial anomaly. After six major surgeries, one undertaken while I was writing this dissertation, my quality of life has improved greatly. I therefore extend my sincere thanks to the doctors, nurses, and staff at HRAC/Centrinho-USP who have cared for me over the years.

        I am deeply grateful to my advisor, Prof. Roldão da Rocha, for his understanding and patience. His accomplishments in gauge/gravity duality have continually inspired me, and I have learned a great deal from his example and guidance. In addition to being a world-class researcher, Prof. Roldão has a profound grasp of human nature; when a student is in need, he does everything in his power to help. 

        My heartfelt thanks also go to Prof. David Dudal, whose excellence as a researcher is matched by his kindness. During my six-month stay at KU Leuven (Kortrijk), he was exceptionally welcoming and made me feel at ease in Belgium. Our collaboration was fruitful, and I have learned a great deal from him. He served as a true co-advisor for the work presented here.

        I owe a debt of gratitude to everyone at KU Leuven, especially the international community, for their kindness and warm welcome. Spending time with everyone at De Blauwe Frituur and Shooters was a highlight of the past two years. I must single out Felipe Kenji Nakano, who helped me with many bureaucratic matters and everyday tasks. From obtaining my residence card to setting up a somewhat complex washing machine with instructions in Dutch, he was there for me with unfailing kindness.

        Lastly, the emotional support of my friends in the Mathematical Physics Laboratory ($\Phi$RMa Lab to the initiated) at UFABC was crucial to the completion of this work. Bruno Silva, Deborah Fabri, Eduardo Albacete, Flame Souza, Lucas Tobias, Pedro Croti, Pedro Zottolo and Vitor Fernandes Guimarães were so nice to me that I cannot even express my gratitude properly. With Lucas Tobias and Pedro Croti, I spent an unforgettable week in Copenhagen. I shared lunch and dinner with Eduardo Albacete almost every day at UFABC, and I cherish our conversations to this day. Pedro Zottolo and Bruno Silva were invaluable classmates in the courses we took together. Vitor Fernandes Guimarães and Deborah Fabri were the very best older “research siblings” I could have asked for. I also deeply admire Flame Souza for the courage with which she fights for her right to exist and be herself as she is meant to be. I love you all.

}

\thispagestyle{empty}

\newpage
\thispagestyle{empty}   
\null                   
\newpage

\begin{center}
    \section*{Resumo}
\end{center}
A correspondência AdS/CFT fornece uma poderosa ferramenta para modelar teorias de gauge fortemente acopladas e, portanto, investigar fenômenos não perturbativos da QCD. Neste trabalho, após uma revisão das ideias fundamentais que encapsulam a correspondência AdS/CFT, apresentamos um modelo holográfico dinâmico e autoconsistente de QCD no arcabouço de Einstein-Maxwell-dilaton, obtido a partir das equações acopladas dos campos, para estudar o espectro de massas e o comportamento de derretimento de mésons pesados e exóticos em temperatura e densidade finitas. Análises em temperatura finita revelam uma transição confinamento-desconfinamento e o derretimento sequencial de quarkônios. Em densidade finita, o aumento do potencial químico acelera o derretimento dos mésons, com funções espectrais evoluindo suavemente através da linha de transição de fase. Finalmente, utilizando um modelo não linear de Einstein-Born-Infeld-dilaton, efeitos de campo magnético demonstram uma mudança da catálise magnética inversa para catálise magnética, destacando o impacto da anisotropia espacial na estabilidade dos quarkônios.

\vspace{1cm}

\noindent Palavras-chave: Correspondência gauge/gravity, QCD holográfica, derretimento de quarkônios
\thispagestyle{empty}

\newpage
\thispagestyle{empty}   
\null                   
\newpage

\begin{center}
    \section*{Abstract}
\end{center}
\label{sec1}
The AdS/CFT correspondence provides a powerful framework for modeling strongly coupled gauge theories and, as a consequence, investigating non-perturbative phenomena in QCD. In this work, following an overview of the ideas that encapsulate the AdS/CFT correspondence, we present a self-consistent dynamical holographic QCD model within the Einstein-Maxwell-dilaton framework, derived from the coupled field equations, to study the mass spectra and melting behavior of heavy and exotic mesons at finite temperature and density. Finite temperature analyses reveal a confinement-deconfinement transition and sequential quarkonia melting. At finite density, an increase in chemical potential accelerates meson melting, with spectral functions evolving smoothly across the phase transition line. Finally, using a nonlinear Einstein-Born-Infeld-dilaton model, magnetic field effects demonstrate a shift from inverse magnetic catalysis to magnetic catalysis, highlighting the impact of spatial anisotropy on quarkonium stability.

\vspace{1cm}

\noindent Keywords: Gauge/gravity correspondence, holographic QCD, quarkonium melting
\thispagestyle{empty}

\newpage
\thispagestyle{empty}
\newpage

\begin{center}
    \section*{List of Papers}
\end{center}

\noindent This dissertation is based on the following published papers:

\vspace{.2cm}

\begin{itemize}
    \item[(I)] Paper \cite{Toniato:2025gts} by Bruno Toniato, David Dudal, Subhash Mahapatra, Roldao da Rocha, and Siddhi Swarupa Jena. Holographic QCD model for heavy and exotic mesons at finite density: A self-consistent dynamical approach. Phys. Rev. D, {\bf 111}, 126021 (2025)                   [arXiv:2502.12694[hep-th]]. 
    \item[(II)] Paper \cite{Jena:2024cqs} by Siddhi Swarupa Jena, Jyotirmoy Barman, Bruno Toniato, David Dudal, and Subhash Mahapatra. A dynamical Einstein-Born-Infeld-dilaton model and holographic quarkonium melting in a magnetic field. JHEP {\bf 12} 096, (2024) [arXiv:2408.14813[hep-th]]. 
\end{itemize}

\newpage

\tableofcontents

\newpage

\section{Introduction}\label{intro}

One of the great triumphs of Quantum Field Theory (QFT) is the remarkable sense of inevitability and rigidity embedded within its formal structure. Such properties of the theory emerge from its two foundational principles: Quantum Mechanics and Special Relativity. Individually revolutionary, their combination into QFT imposes stringent and far-reaching constraints on the imaginable theories describing fundamental interactions at large distances. Unlike Classical Mechanics, which allows for significant theoretical flexibility, the internal consistency of QFT is exceptionally delicate and, as such, small modifications can lead to catastrophic breakdowns of the theory.

This rigidity and sense of inevitability become explicit when examining the allowed helicities of interacting massless particles. The consistency of interactions, soft IR behavior and locality/unitarity drastically limits these helicities to a discrete and restricted set: $h = 0, \frac{1}{2}, 1, \frac{3}{2}, 2$. Not only is this enumeration surprisingly finite, but each helicity value corresponds precisely to fundamental fields observed, or predicted, to occur in nature: scalars ($h=0$), fermions ($h=1/2$), gauge bosons ($h=1$), gravitinos ($h=3/2$), and, notably, the graviton ($h=2$). Weinberg’s seminal soft theorems \cite{Weinberg:1965nx} show that any consistent interacting massless helicity-$2$ particle must couple universally (with the same strength) to all forms of energy--momentum (the field theory version of the equivalence principle). Thus, in that sense, gravity itself appears not as an independent choice but rather as an unavoidable prediction arising from QFT’s structural consistency: a consequence of combining Special Relativity with Quantum Mechanics.

It is indeed astonishing that the combination of these two fundamental principles places such profound restrictions on the structure of physical theories. Spin-$1$ particles, constrained by gauge symmetry, lead directly to Yang–Mills theories, which form the backbone of our understanding of the Standard Model interactions. Massless spin-$3/2$ particles are intimately connected to supersymmetric theories of gravity \cite{vanNieuwenhuizen:1981ae}. Most remarkably, spin-$2$ particles are forced into the unique role of mediating gravitational interactions, inevitably leading to General Relativity (GR). Therefore, GR emerges as the universal low-energy effective theory of an interacting massless helicity-$2$ field. This remarkable rigidity emphasizes how deeply intertwined these foundational physical principles are and how little freedom exists in constructing consistent fundamental theories.

Since QFT imposes strict conditions on particle interactions and severely restricts the types of theories that can exist, its extension into realms beyond the Standard Model remains an immense theoretical challenge. Developing a consistent framework capable of describing gravity at the quantum level has been a persistent difficulty in theoretical physics, primarily due to inherent incompatibilities between the conventional formulations of General Relativity and Quantum Mechanics. String Theory is often regarded as a consistent extension of QFT that incorporates both gravity and Quantum Mechanics without internal contradictions. Unlike standard QFTs, which describe elementary particles as point-like objects, String Theory postulates fundamental constituents as one-dimensional strings. The extended nature of these fundamental objects allows String Theory to naturally resolve divergences and inconsistencies that plague attempts to quantize gravity directly within traditional QFT methods. Remarkably, gravity emerges naturally within String Theory as a necessary consequence, embodied by a massless spin-$2$ state \cite{Becker:2006dvp}. 

Among the many theoretical insights derived from String Theory, the AdS/CFT correspondence has stood out prominently. Originally proposed by Maldacena \cite{Maldacena:1997re}, this duality conjecture has profoundly impacted our modeling of strongly coupled systems. The AdS/CFT correspondence establishes a remarkable equivalence between gravitational theories in higher dimensional Anti-de Sitter (AdS) spacetimes and Conformal Field Theories (CFTs) residing on the AdS boundary. Inspired by this duality, a myriad of holographic models has been developed to investigate strongly coupled phenomena, particularly in Quantum Chromodynamics (QCD), where traditional perturbative methods falter at low energies. By translating challenging QCD calculations into more tractable gravitational computations in AdS, holography has provided new perspectives and powerful computational tools to tackle non-perturbative QCD effects such as confinement \cite{Dudal:2015kza} and the quark–gluon plasma \cite{Braga:2023qee}. Indeed, extracting mesonic properties such as the mass spectra and spectral functions directly from first principles in QCD remains an immense challenge due to the strongly coupled, non-perturbative nature of low-energy QCD. Traditional perturbative methods, valid only at high energies, fail in the regime where bound states like mesons dominate. Although lattice QCD provides a powerful non-perturbative approach, it faces significant computational limitations, especially at finite density or in real-time dynamics. In contrast, holographic models inspired by the AdS/CFT correspondence offer a complementary approach, circumventing these difficulties by mapping intricate QCD computations onto more accessible gravitational problems. This dissertation leverages the AdS/CFT-inspired holographic framework to explore finite temperature, finite density, and external field effects on heavy and exotic mesons through self-consistent dynamical holographic QCD models. Here, the term ``self-consistent'' means that the background solves coupled Einstein–matter equations rather than inserting fields by hand, some other AdS/CFT-inspired models.

In this setting, heavy quarkonia play a particularly privileged role: the heavy-quark mass introduces a hard scale that tends to produce compact bound states, making quarkonium observables comparatively clean probes of the surrounding medium and especially valuable in the phenomenology of relativistic heavy-ion collisions, where in-medium modification and suppression have long been associated with deconfinement and color screening \cite{Matsui:1986dk,Brambilla:2010cs,Rapp:2008tf}. This connection naturally motivates a focus on quarkonium (and, more broadly, heavy and exotic mesons) at finite temperature and density, because one expects a characteristic pattern of sequential melting in which less tightly bound states dissolve earlier as the medium is heated or densified. A quantitatively sharp way to diagnose these effects is through spectral functions, which encode the real-time response of the system: peaks identify quasibound states, while thermal broadening and peak disappearance provide direct measures of in-medium widths and dissociation \cite{Aarts:2011sm,Laine:2006ns}. However, precisely these real-time and finite density observables are among the most difficult to extract from first-principles QCD, since lattice simulations are formulated in Euclidean time (making spectral reconstruction an ill-posed analytic continuation problem) and finite baryon density is hindered by severe computational obstacles. Holography is therefore especially well-suited to the quarkonium problem: within AdS/CFT-inspired models one can compute retarded correlators—and hence spectral functions—directly from classical bulk dynamics with horizon boundary conditions, providing a complementary non-perturbative framework to study how heavy and exotic mesons respond to hot and dense QCD matter and to external fields.

It is important to stress at the outset what can, and cannot, be claimed in a holographic approach to QCD. Unlike the original AdS/CFT setting, where the boundary theory is conformal and the dual bulk description is known in a precise string-theoretic construction, QCD is neither conformal nor supersymmetric, and no exact weakly curved gravitational dual of real-world QCD is currently known. The holographic QCD models employed here should therefore be understood as AdS/CFT-inspired, bottom-up effective descriptions: they are engineered to capture key qualitative and, when appropriately calibrated, semi-quantitative features of the strongly coupled regime (such as confinement scales, thermodynamics, and spectral behavior), while inevitably relying on modeling choices and assumptions (e.g.\ large-$N_c$ and strong-coupling limits, simplified matter content, and a restricted set of operators). With this perspective, the goal is not to replace first-principles approaches, but to provide a complementary framework that offers direct access to dynamical observables (notably real-time correlators and spectral functions at finite $T$, $\mu$, and external fields), and to test the robustness of physical mechanisms across a controlled, self-consistent gravitational setting.

This work is organized as follows: In Sec.~\ref{qcdover}, we present an overview of Quantum Field Theory fundamentals, covering essential concepts both at zero and finite temperature, including symmetries, renormalization, and thermal field theory methods. Sec.~\ref{adsover} introduces the AdS/CFT correspondence, detailing its foundational principles and significance in modeling strongly coupled gauge theories. Sec.~\ref{spectra} discusses the holographic computation of the mesonic mass spectra, focusing on the historical hard-wall and soft-wall models. In Secs.~\ref{EMDspec}–\ref{metingmu}, we systematically investigate the properties of heavy and exotic mesonic states using the Einstein-Maxwell-dilaton (EMD) framework, a mathematically self-consistent holographic model. We explore the mass spectra and melting process at finite temperature and density of these heavy and exotic mesons. In Secs.~\ref{modEBID}–\ref{memb}, we examine the influence of external magnetic fields on the quarkonium dissociation through a nonlinear self-consistent Einstein-Born-Infeld-dilaton (EBID), highlighting the shift from inverse magnetic catalysis to magnetic catalysis and the spatial anisotropy in the spectral functions introduced by a background magnetic field. Finally, in Sec.~\ref{conc}, we conclude with a comprehensive summary of our key findings, alongside an outlook on future research avenues and potential implications.

\newpage

\section{Quantum Field Theory at zero and finite temperature}\label{qcdover}

A holographic description of mesons begins with the language of Quantum Field Theory, as all the boundary observables that one later matches to bulk dynamics: Propagators, correlation functions, spectral densities are defined in that framework. This Section revisits Quantum Field Theory both at zero temperature, where one studies fields in the vacuum, and at finite temperature, where the same fields form a thermal ensemble. The zero temperature part recalls how actions, path integrals, and generating functionals encode the dynamics of the quantum system and yield the Feynman, retarded, and advanced propagators that will become boundary conditions for their dual bulk fields. The finite temperature part then extends the formalism to systems in thermal equilibrium by compactifying time, imposing Kubo–Martin–Schwinger periodicity, introducing Matsubara modes, and performing the analytic continuation that converts Euclidean correlators into real-time Green’s functions. Together, these tools supply the spectral functions, masses, widths, and transport coefficients that are essential for holographic computations for both vacuum mesons and the hot quark–gluon plasma created in heavy-ion collisions. 

Throughout, we follow the pedagogical treatments of Zee’s Quantum Field Theory in a Nutshell \cite{Zee:2003mt} and Peskin and Schroeder’s An Introduction to Quantum Field Theory \cite{Peskin:1995ev} for the zero temperature formalism, and Kapusta and Gale’s Finite-Temperature Field Theory: Principles and Applications \cite{Kapusta:2006pm} for the finite temperature case.

\subsection{Correlation functions}
The exploration of Quantum Field Theory here begins by considering a real scalar field, \(\phi(x)\), defined in a four-dimensional Minkowski spacetime characterized by coordinates \(x = (t, \vec{x})\)\footnote{We work in natural units in this work, so $c=\hbar=1$ unless stated otherwise.} and a mostly-minus metric signature $\eta_{\mu \nu}= \text{diag} (1,-1,\ldots,-1)$ for this Section to align with most of the particle physics literature \cite{Peskin:1995ev, Zee:2003mt}. Although we work primarily in four-dimensional Minkowski spacetime, we denote the spacetime dimension by $d$ when writing formulas in a general form. The dynamics of this scalar field are encapsulated by an action of the form
\begin{equation}\label{la}
S = \int d^4x \, \mathcal{L}(\phi, \partial_\mu \phi) \,,
\end{equation}
where \(\mathcal{L}(\phi, \partial_\mu \phi)\) is the Lagrangian density, which, in general, specifies interactions and potential terms for the scalar field.

An essential concept in analyzing quantum fields is the set of two-point correlation functions. A particularly basic one is the Wightman (``greater'') two-point function, defined as the vacuum expectation value of two field operators at distinct spacetime points:
\begin{equation}\label{greenF}
G_{W}(x,y)\equiv \langle 0|\phi(x)\phi(y)|0 \rangle \equiv \langle \phi(x)\phi(y) \rangle \,.
\end{equation}

Green's functions (or propagators) are special two-point functions that invert the differential operator appearing in the equations of motion, with a choice of boundary (or causality) prescription. For a free scalar field of mass \(m\), the equation of motion is the Klein--Gordon equation,
\begin{equation}
(\Box + m^2)\phi(x)=0\,,
\end{equation}
and the corresponding propagators \(G_{s}(x,y)\) satisfy the inhomogeneous equation
\begin{equation}\label{KGgreen}
(\Box_x + m^2)\,G_{s}(x,y)= -i\,\delta^{(4)}(x-y)\,,
\end{equation}
where the subscript \(s\in\{F,R,A\}\) specifies the choice of time-ordering/causality prescription.

To enforce causality explicitly, we define the retarded and advanced Green's functions in terms of commutators:
\begin{equation}
G_R(x,y)\equiv -i\,\theta(x^0-y^0)\,\langle[\phi(x),\phi(y)]\rangle\,,
\end{equation}
\begin{equation}
G_A(x,y)\equiv -i\,\theta(y^0-x^0)\,\langle[\phi(y),\phi(x)]\rangle\,,
\end{equation}
so that \(G_R\) (\(G_A\)) has support only for \(x^0>y^0\) (\(x^0<y^0\)). In particular, the commutator vanishes at spacelike separation, ensuring causal propagation.

Another crucial propagator is the Feynman (time-ordered) Green's function, which is the object that appears directly in perturbation theory. The time-ordering symbol \(\mathcal{T}\) arranges operators so that later times appear to the left:
\begin{equation}\label{FeynmanProp}
G_F(x,y)\equiv \langle \mathcal{T}\,\phi(x)\phi(y)\rangle
= \theta(x^0-y^0)\langle \phi(x)\phi(y)\rangle
+ \theta(y^0-x^0)\langle \phi(y)\phi(x)\rangle\,.
\end{equation}

With these conventions, \(G_F\), \(G_R\), and \(G_A\) all satisfy Eq.~\eqref{KGgreen}, differing only by their boundary/causality prescriptions.

To systematically compute correlation functions in QFT, it is useful to introduce a generating functional \(Z[J]\), defined as
\begin{equation}\label{ge}
Z[J] = \int \mathcal{D}\phi \exp \left\{i\int d^4x[\mathcal{L}(\phi, \partial_\mu \phi) + J(x)\phi(x)]\right\}\,,
\end{equation}
where \(\mathcal{D}\phi\) denotes a functional integral over all field configurations, and \(J(x)\) is an auxiliary external source field introduced to probe the system.

Correlation functions, which are related to measurable physical quantities such as scattering amplitudes and response functions, are computed as functional derivatives of the generating functional \eqref{ge}. For example, the vacuum expectation value of the field operator is obtained as
\begin{equation}
\langle \phi(x) \rangle = \frac{1}{Z_0} \left( -i \frac{\delta}{\delta J(x)} \right) Z[J] \bigg|_{J=0}\,,
\end{equation}
where \( Z_0 = Z[J=0] \) is the vacuum generating functional.

More generally, the two-point time-ordered correlation function (Feynman propagator) is given by
\begin{equation}
\langle \mathcal{T}\phi(x)\phi(y) \rangle = \frac{1}{Z_0} \left( -i \frac{\delta}{\delta J(x)} \right) \left( -i \frac{\delta}{\delta J(y)} \right) Z[J] \bigg|_{J=0}\,.
\end{equation}

Higher-order \( n \)-point Green’s functions are obtained by taking additional functional derivatives with respect to the source field \( J(x) \).

These \( n \)-point Green’s functions are not directly physical observables but are crucial intermediates. To obtain physically observable scattering amplitudes (such as cross-sections), one uses the Lehmann–Symanzik–Zimmermann (LSZ) reduction formula. This procedure relates S-matrix elements (scattering amplitudes) to time-ordered correlation functions by amputating external propagators and taking the fields on-shell. For scalar fields, the LSZ formula schematically takes the form \cite{Peskin:1995ev,Zee:2003mt}
\begin{equation}
\langle p_1, \ldots, p_n \text{ out} | q_1, \ldots, q_m \text{ in} \rangle = 
\int \prod_i d^4x_i \, e^{i p_i \cdot x_i} (\Box_{x_i} + m^2) \cdots \langle \mathcal{T} \phi(x_1) \cdots \phi(x_n) \cdots \rangle\,.
\end{equation}
where the differential operators \( (\Box + m^2) \) act on each external leg to isolate the interacting part of the Green’s function, and the on-shell condition is imposed for the external momenta.

Thus, functional derivatives of the generating functional not only encode quantum fluctuations and field correlations but also provide the foundation for computing measurable particle scattering processes in QFT.

\subsection{Renormalization, running coupling and the $\phi^4$ theory $\beta$-function}

When the interacting Lagrangian of the theory~(\ref{la}) is sufficiently complicated, it is often useful to work within perturbation theory, expanding the integrand of the generating functional $Z[J]$~(\ref{ge}) as a power series in the coupling. This approach provides a controlled (typically asymptotic) approximation when the renormalized coupling is small at the relevant physical scale.

Quantum Field Theories frequently exhibit ultraviolet (UV) divergences in perturbative loop calculations. To extract finite, physically meaningful predictions, these divergences must be treated systematically through \emph{renormalization}, in which bare quantities are expressed in terms of finite, measurable (renormalized) quantities plus counterterms.

As an example, consider a real scalar field theory with quartic interaction, defined by the bare Lagrangian
\begin{equation}
\mathcal{L}_B = \frac{1}{2} (\partial_\mu \phi_B)(\partial^\mu \phi_B) - \frac{1}{2} m_B^2 \phi_B^2 - \frac{g_B}{4!} \phi_B^4\,,
\end{equation}
where $\phi_B$, $m_B$, and $g_B$ denote the bare field, mass, and coupling constant, respectively.

To renormalize the theory, we express bare quantities in terms of renormalized quantities and counterterms:
\begin{equation}
\phi_B = Z^{1/2}_\phi \, \phi\,, \qquad
m_B^2 = m^2 + \delta m^2\,, \qquad
g_B = \mu^{\epsilon} (g + \delta g)\,,
\end{equation}
where $Z_\phi$ is the field-strength renormalization constant, $\delta m^2$ and $\delta g$ are counterterms, and $\mu$ is the renormalization scale introduced in dimensional regularization ($d = 4 - \epsilon$).

The renormalized Lagrangian then becomes
\begin{equation}\label{bare}
\mathcal{L} =
\frac{1}{2} Z_\phi (\partial_\mu \phi)(\partial^\mu \phi)
- \frac{1}{2} Z_\phi (m^2 + \delta m^2) \phi^2
- \frac{Z_\phi^2}{4!} (g + \delta g)\, \mu^{\epsilon}\, \phi^4 \,.
\end{equation}

The beta function $\beta(g)$ describes how the renormalized coupling changes with the renormalization scale $\mu$. It is defined (holding the bare coupling fixed) by
\begin{equation}\label{betaf}
\beta(g) \equiv \mu \left.\frac{d g}{d \mu}\right|_{g_B}\,.
\end{equation}

To determine the running of the coupling $g$ for~(\ref{bare}), we analyze loop corrections to the four-point vertex function. At second order in the coupling, the relevant contribution to the four-point function is~\cite{Peskin:1995ev,Zee:2003mt}\footnote{The field-strength renormalization first contributes at two loops, so $Z_\phi$ can be set to unity at one loop for the present purpose. The mass counterterm $\delta m^2$ is already fixed at one loop from the two-point function; here we focus on the coupling renormalization from the four-point function.}
\begin{equation}
\frac{1}{2!}\left(\frac{-ig}{4!}\right)^2
\int d^4x\, d^4y \;
\left\langle
\mathcal{T}\left\{
\phi^4(x)\, \phi^4(y)\, \phi(x_1)\phi(x_2)\phi(x_3)\phi(x_4)
\right\}
\right\rangle \,,
\end{equation}

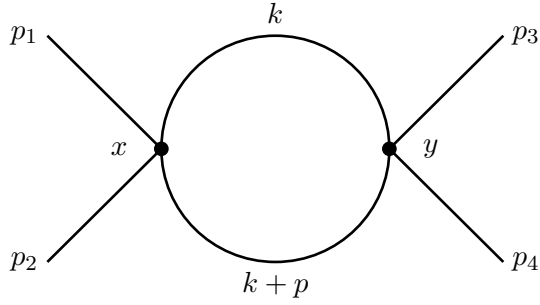
\begin{figure}[h!]
\begin{center}
\begin{tikzpicture}[line width=1 pt, scale=1.5]

  \coordinate (L) at (-1,0);  
  \coordinate (R) at (1,0);   

  \draw (-2,1) -- (L);
  \draw (-2,-1) -- (L);
  \draw (R) -- (2,1);
  \draw (R) -- (2,-1);

  \draw (L) arc (180:0:1);
  \draw (R) arc (0:-180:1);

  \filldraw (L) circle (1.5pt);
  \filldraw (R) circle (1.5pt);

  \node[left] at (-1.2,0) {\(x\)};
  \node[right] at (1.2,0) {\(y\)};
  \node at (-2.2,1) {\(p_1\)};
  \node at (-2.2,-1) {\(p_2\)};
  \node at (2.2,1) {\(p_3\)};
  \node at (2.2,-1) {\(p_4\)};
  \node at (0,1.2) {\(k\)};
  \node at (0,-1.2) {\(k + p\)};

\end{tikzpicture}
\end{center}
\caption{$\phi^4$ theory fish, or bubble, diagram.}
\label{feyn}
\end{figure}

At one-loop order, the correction arises from the ``fish'' (bubble) diagram, shown in Fig.~\ref{feyn}. The $s$-channel contribution to the one-loop correction is
\begin{equation}
\mathcal{M}^{(1)}_s =
-\frac{g^2}{2}
\int \frac{d^d k}{(2\pi)^d}
\frac{i}{k^2 - m^2 + i\epsilon}\,
\frac{i}{(k + p)^2 - m^2 + i\epsilon}\,.
\end{equation}
where $p=p_1+p_2$ is the total incoming momentum.

Using a Feynman parameter, we combine denominators as
\begin{equation}
\frac{1}{AB} = \int_0^1 dx \, \frac{1}{\big[A x + B(1 - x)\big]^2}\,,
\end{equation}

This yields
\begin{equation}
\mathcal{M}^{(1)}_s =
\frac{g^2}{2}\int_0^1 dx
\int \frac{d^d k}{(2\pi)^d}
\frac{1}{\big[(k + x p)^2 - \Delta + i\epsilon\big]^2}\,,
\end{equation}
with
\begin{equation}
\Delta = m^2 - x(1-x)p^2\,,
\end{equation}

Shifting the integration variable to $q = k + x p$ (equivalently, up to $x\to 1-x$), the integral becomes
\begin{equation}\label{intdim}
\mathcal{M}^{(1)}_s =
\frac{g^2}{2}\int_0^1 dx
\int \frac{d^d q}{(2\pi)^d}
\frac{1}{(q^2 - \Delta + i\epsilon)^2}\,.
\end{equation}

To evaluate~(\ref{intdim}) in dimensional regularization, we perform a Wick rotation $q^0 \mapsto i q_E^0$, so that $q^2 \mapsto -q_E^2$ and $d^d q \mapsto i\, d^d q_E$. The loop integral becomes
\begin{equation}
I \equiv i \int \frac{d^d q_E}{(2\pi)^d}\,\frac{1}{(q_E^2 + \Delta)^2}\,,
\end{equation}

The angular integration produces the surface area $\Omega_d$ of the unit $(d-1)$-sphere:
\begin{equation}
\Omega_d = \frac{2 \pi^{d/2}}{\Gamma(d/2)}\,,
\end{equation}

Thus,
\begin{align}
I &= \frac{i}{(2\pi)^d}\,\Omega_d \int_0^\infty dq_E \, \frac{q_E^{d-1}}{(q_E^2 + \Delta)^2}\,, \nonumber \\
  &= \frac{i}{2}\,\frac{\Omega_d}{(2\pi)^d} \int_0^\infty dt \, \frac{t^{\frac{d}{2} - 1}}{(t + \Delta)^2}\,,
\end{align}
where we used $t=q_E^2$ so that $dt = 2 q_E\,dq_E$. Using the standard identity~\cite{Gradshteyn:1943cpj}
\begin{equation}
\int_{0}^{\infty} dx\,\frac{x^{\mu-1}}{(x+\beta)^{\nu}}
=\beta^{\,\mu-\nu}\,B(\mu,\nu-\mu)\,,
\end{equation}
with $\mu=\tfrac d2$, $\nu=2$, and $\beta=\Delta$, we find
\begin{align}
\int_{0}^{\infty}dt\,\frac{t^{\frac d2-1}}{(t+\Delta)^2}
&=\Delta^{\,\frac d2-2}\,
B\!\left(\tfrac d2,\,2-\tfrac d2\right)
=\Delta^{\,\frac d2-2}\,
\Gamma\!\left(\tfrac d2\right)\,
\Gamma\!\left(2-\tfrac d2\right)\,.
\end{align}

Therefore,
\begin{align}
I
&=\frac{i}{2}\,\frac{\Omega_d}{(2\pi)^d}\,
\Delta^{\frac d2-2}\,
\Gamma\!\left(\tfrac d2\right)\,
\Gamma\!\left(2-\tfrac d2\right) \nonumber\\
&= i\,(4\pi)^{-d/2}\,
\Gamma\!\left(2-\tfrac d2\right)\,
\Delta^{\,\frac d2-2}\,.
\end{align}

In $d=4-\epsilon$ dimensions,
\begin{equation}
\Gamma\!\left(2 - \frac{d}{2}\right)
=\Gamma\!\left(\frac{\epsilon}{2}\right)
\simeq \frac{2}{\epsilon} - \gamma_E + \mathcal{O}(\epsilon)\,,
\end{equation}
so the divergent part of the $s$-channel contribution is
\begin{equation}
\mathcal{M}^{(1)}_s\Big|_{\text{div}}
=
\frac{i g^2}{(4\pi)^2}\,\frac{1}{\epsilon}\,.
\end{equation}

Including the three channels ($s$, $t$, and $u$), the total one-loop divergence is
\begin{equation}
\mathcal{M}^{(1)}_{\text{div}}
=
\frac{3i g^2}{(4\pi)^2}\,\frac{1}{\epsilon}\,.
\end{equation}

In a minimal-subtraction (MS) scheme, the coupling counterterm is chosen to cancel the pole:
\begin{equation}
\delta g = \frac{3g^2}{16\pi^2 \epsilon} + \mathcal{O}(g^3)\,.
\end{equation}

Requiring that the bare coupling $g_B=\mu^\epsilon(g+\delta g)$ be independent of $\mu$ then gives the one-loop beta function (in the $4$D limit $\epsilon\to 0$):
\begin{equation}
\boxed{\beta(g) = \mu \left.\frac{d g}{d \mu}\right|_{g_B}
= \frac{3}{16\pi^2} g^2 + \mathcal{O}(g^3)\,.}
\end{equation}

Since $\beta(g)>0$, the coupling grows with the renormalization scale $\mu$: $\phi^4$ theory is not asymptotically free. In particular, the running coupling exhibits Landau-pole (``triviality'') behavior in the ultraviolet, so sufficiently high scales lie outside the domain of validity of perturbation theory.

\subsection{Yang-Mills theory}

Gauge theories, which model all fundamental forces except (quantum) gravity, describe interactions through local symmetry principles.
We first review the abelian example, Quantum Electrodynamics (QED), and then generalize to non-abelian Yang--Mills theory, which underpins the Standard Model's strong and electroweak sectors.

The QED Lagrangian for a Dirac fermion is
\begin{equation}
\mathcal{L}_{\text{QED}}
= -\frac14 F_{\mu\nu}F^{\mu\nu}
+ \bar{\psi}\bigl(i\gamma^\mu D_\mu - m\bigr)\psi \,,
\label{eq:QEDLagrangian}
\end{equation}
with
\begin{align}
F_{\mu\nu} &\;=\;\partial_\mu A_\nu - \partial_\nu A_\mu
&\text{(field strength)}\,,\\[4pt]
D_\mu &\;=\;\partial_\mu + i e A_\mu
&\text{(covariant derivative)}\,.
\end{align}

Local $U(1)$ gauge invariance,
\begin{equation}
\psi(x) \mapsto e^{i\alpha(x)} \psi(x)\,,
\qquad
A_\mu(x)\mapsto A_\mu(x)-\frac{1}{e}\,\partial_\mu\alpha(x)\,,
\end{equation}
forces the appearance of $A_\mu$ in \eqref{eq:QEDLagrangian}. The interaction term $-e\,\bar\psi\gamma^\mu A_\mu\psi$ couples the conserved Noether current
\(
j^\mu=\bar{\psi}\gamma^\mu\psi
\)
to the photon. Varying the action gives
\begin{align}
\partial^\mu F_{\mu\nu} &= e\,\bar{\psi}\gamma_\nu\psi
&\text{(Maxwell--Amp\`ere)}\,,\\
(i\gamma^\mu D_\mu-m)\psi &= 0
&\text{(Dirac in a background $A_\mu$)}\,.
\end{align}

As in the $\phi^4$ example of the previous Section, QED is renormalizable \cite{Zee:2003mt,Peskin:1995ev}. For a single Dirac fermion of unit charge, the one-loop $\beta$-function is positive,
\(
\beta(e)\equiv\mu\frac{de}{d\mu}= \tfrac{e^{3}}{12\pi^{2}}+\cdots,
\)
so the effective electric charge grows logarithmically with energy (Landau-pole behavior), much like $\phi^4$ theory.

If a theory contains \emph{several} gauge fields, the key question is the structure of the underlying gauge \emph{algebra}.
When all generators commute, the algebra is abelian (e.g.\ a product of $U(1)$ factors), and the dynamics reduces to independent copies of QED.
To obtain intrinsically new dynamics, notably self-interactions among gauge bosons, we require generators that \emph{fail} to commute.
Promoting the single photon field \(A_\mu\) to a multiplet \(A_\mu^{a}\) labeled by an index \(a\) in a compact Lie group \(G\) therefore leads to a \emph{non-abelian} gauge theory, traditionally called \emph{Yang--Mills theory}.
The group is spanned by generators \(T^{a}\) (in some representation) satisfying the Lie-algebra relations
\begin{equation} \label{gene}
    [T^{a},T^{b}] =  i f^{abc}\,T^{c}\,,
\end{equation}
with real, totally antisymmetric structure constants \(f^{abc}\).
In what follows, we construct the covariant derivative, field-strength tensor, and gauge-invariant Lagrangian that arise from this generalization.

If we replace $U(1)$ by a compact Lie group $G$ (e.g.\ $SU(N_c)$),
the gauge field is matrix-valued,
\(
A_\mu = A_\mu^a T^a\,,
\)
with generators $T^a$ obeying \eqref{gene} and normalization $\mathrm{tr}(T^aT^b)=\tfrac12\delta^{ab}$.
Importantly, $SU(N_c=3)$ is the gauge group of Quantum Chromodynamics.

For a field $\Phi$ in a representation $R$, the covariant derivative is
\begin{equation}
D_\mu \Phi = \bigl(\partial_\mu - i g_{YM}\,A_\mu^a T_R^a\bigr)\Phi\,,
\end{equation}
and the field strength is most transparently defined by the commutator
\begin{equation}
[D_\mu,D_\nu] = - i g_{YM}\,F_{\mu\nu}^a T_R^a\,,
\end{equation}
which yields, in components,
\begin{equation}
F_{\mu\nu}^a = \partial_\mu A_\nu^a - \partial_\nu A_\mu^a
              + g_{YM}\, f^{abc} A_\mu^b A_\nu^c\,.
\end{equation}

Under a local gauge transformation $U(x)=e^{i\alpha^a(x)T^a}$, matter fields transform as $\Phi\mapsto U\Phi$, while the covariant derivative transforms as $D_\mu \mapsto U D_\mu U^{-1}$.
Equivalently, the gauge potential transforms as
\begin{equation}
A_\mu \mapsto U A_\mu U^{-1} + \frac{i}{g_{YM}}\,U\,\partial_\mu U^{-1}\,.
\end{equation}

To first order in $\alpha^a(x)$ this gives, in components,
\begin{equation}
A_\mu^a(x)\mapsto A_\mu^a(x) + \frac{1}{g_{YM}}\,\partial_\mu\alpha^a(x) - f^{abc}\,\alpha^b(x)\,A_\mu^c(x)\,.
\end{equation}

The gauge field is therefore \emph{not} gauge invariant; instead it transforms inhomogeneously so that $D_\mu$ transforms covariantly.
Correspondingly, the field strength transforms \emph{covariantly} (in the adjoint representation),
\begin{equation}
F_{\mu\nu} \mapsto U F_{\mu\nu} U^{-1}
\qquad\Longrightarrow\qquad
F_{\mu\nu}^a \mapsto F_{\mu\nu}^a - f^{abc}\,\alpha^b\,F_{\mu\nu}^c\,.
\end{equation}

This covariance is precisely what makes the Yang--Mills action gauge invariant. The pure-gauge Lagrangian is
\begin{equation}
  \mathcal{L}_{\text{YM}}
  \;=\;
  -\frac14 F_{\mu\nu}^a F^{\mu\nu a} \,,
\label{eq:YMLagrangian}
\end{equation}
and is invariant under $G$. Varying \eqref{eq:YMLagrangian} yields the Yang--Mills equations of motion,
\begin{equation}
    D^\mu F_{\mu\nu}^a =0\,,
\end{equation}
with the covariant derivative in the adjoint representation, together with the Bianchi identity $D_{[\mu}F_{\nu\rho]}=0$.

The one-loop $\beta$-function for pure $SU(N_c)$ (no matter fields) is \cite{Peskin:1995ev,Zee:2003mt}
\begin{equation}
\beta(g)
= -\frac{11\,N_c}{48\pi^{2}}\,g^{3}+\mathcal{O}(g^{5})\,.
\end{equation}

The negative sign shows that the interaction becomes weaker at high energies, with $g(\mu)\to 0$ as $\mu\to\infty$; pure Yang--Mills theory is therefore asymptotically free.
With $N_f$ massless Dirac fermions in the fundamental representation, the coefficient becomes
\(
\beta(g)= -\frac{1}{16\pi^{2}}\Bigl(\tfrac{11}{3}N_c-\tfrac{2}{3}N_f\Bigr)g^{3}+\cdots,
\)
so asymptotic freedom persists for $N_f<\tfrac{11}{2}N_c$ (e.g.\ for QCD with $N_c=3$, one requires $N_f\le 16$).

Adding Dirac fermions $\psi_f^i$ (color index $i=1,\dots,N_c$) yields
\begin{equation}\label{YMDirac}
\mathcal{L}_{\text{YM+Dirac}}
=
-\frac14 F_{\mu\nu}^a F^{\mu\nu a}
+\sum_{f=1}^{N_f}\bar{\psi}_f\bigl(i\gamma^\mu D_\mu - m_f\bigr)\psi_f\,.
\end{equation}
with $D_\mu=\partial_\mu - i g\,A_\mu^a T^a$ (here $T^a$ is in the fundamental representation).

In summary, QED illustrates the gauge principle in its simplest form, while Yang--Mills theory reveals the richer dynamics of non-abelian gauge fields: self-interactions, asymptotic freedom, and (in QCD) confinement. These structures form the backbone of the Standard Model.

\subsection{Dynamics of quarks and gluons}

Quantum Chromodynamics is the $SU(3)$ gauge theory that governs the strong interaction.
Its basic dynamical variables are an eight-component gluon field $A_\mu = A_\mu^{a} T^{a}$ and six flavors of Dirac quarks $\psi_f^i$, each quark carrying a color index $i = 1,2,3$.
The matrices $T^{a} = \tfrac12 \lambda^{a}$ are one-half of the Gell--Mann matrices, satisfying the relations introduced in the previous Section.

Varying the action \eqref{YMDirac} gives the Yang--Mills equation of motion
\begin{equation}
    D^\mu F_{\mu\nu}^{a} = g_s\,\bar{\psi}_f \gamma_\nu T^{a} \psi_f\,,
\end{equation}
alongside the Dirac equations $\bigl(i\gamma^\mu D_\mu - m_f\bigr)\psi_f = 0$ for each flavor.
Because gluons carry color charge, the $g_s f^{abc} A_\mu^{b} A_\nu^{c}$ term generates trilinear and quadrilinear gluon interactions that have no analogue in the abelian theory of electromagnetism.
Perturbative renormalization of these self-interactions drives the strong coupling to weaker values at short distances. Indeed, in dimensional regularization one finds the one-loop beta function \cite{Peskin:1995ev,Zee:2003mt}
\begin{equation}
    \beta(g_s) \equiv \mu \frac{d g_s}{d\mu}
  = -\frac{g_s^{3}}{16\pi^{2}}
      \left(\frac{11}{3}C_A - \frac{4}{3}T_F N_f\right)
      + \mathcal{O}(g_s^{5})\,,
\end{equation}
with $C_A = 3$ and $T_F = \tfrac12$.
For the physical case $N_f = 6$ (at scales above the heavy-quark thresholds), the coefficient remains negative, so $g_s(\mu)$ decreases logarithmically as the renormalization scale $\mu$ grows; this is the celebrated property of asymptotic freedom.
Integrating the renormalization--group equation gives
\begin{equation}
    \alpha_s(\mu) = \frac{g_s^2(\mu)}{4\pi}
  \simeq
  \frac{4\pi}{\beta_0 \ln(\mu^2/\Lambda_{\mathrm{QCD}}^2)}\,,
  \qquad
  \beta_0 = 11 - \frac{2}{3} N_f\,.
\end{equation}

This introduces the intrinsic QCD scale $\Lambda_{\mathrm{QCD}}\!\sim\!200\,\mathrm{MeV}$, which sets the scale where the perturbative running coupling becomes $\mathcal{O}(1)$; below it, perturbation theory is no longer controlled and non-perturbative dynamics such as confinement dominate. Consequently, many zero temperature phenomena involving protons, neutrons, mesons, and so on require genuinely non-perturbative methods.

At the classical level and in the limit $m_f\to0$, the Lagrangian \eqref{YMDirac} is invariant under an $SU(N_f)_L\times SU(N_f)_R$ chiral flavor symmetry.
Quantum effects spoil the conservation of the axial singlet current through the Adler--Bell--Jackiw anomaly \cite{Peskin:1995ev,Zee:2003mt}
\begin{equation}
    \partial_\mu j_5^\mu
  = \frac{g_s^{2}N_f}{16\pi^{2}} F_{\mu\nu}^{a}\,\tilde F^{\mu\nu a}\,.
\end{equation}
but the non-singlet axial currents remain conserved in perturbation theory and are spontaneously broken in the QCD vacuum, giving rise to nearly massless pseudoscalar pseudo-Goldstone bosons.

Finally, the non-trivial topology of the $SU(3)$ gauge group permits instanton solutions classified by an integer winding number $Q$.
The true vacuum is therefore a $\theta$-parametrized superposition of topological sectors, characterized by the term \cite{Peccei:1977hh}
\begin{equation}
    \mathcal{L}_{\theta}
  = \theta\,\frac{g_s^{2}}{32\pi^{2}}
      F_{\mu\nu}^{a}\,\tilde F^{\mu\nu a}\,.
\end{equation}

Experimental bounds on strong CP violation, as seen in \cite{PhysRevLett.97.131801}, require $|\theta|\lesssim10^{-10}$, a smallness that motivates the Peccei--Quinn mechanism and the axion.

In summary, QCD exhibits asymptotically free quarks and gluons at short distances, but confines them into hadrons at long distances, while simultaneously displaying an intricate interplay of chiral symmetry breaking, axial anomalies, topological structure, and running couplings. 

\subsection{Spectral functions in QCD}

The dynamical information contained in a Quantum Field Theory is elegantly encoded in its
\emph{spectral functions}. For a Hermitian operator\footnote{%
Typical choices are local, gauge-invariant currents such as the vector current
$j_\mu(x)=\bar\psi(x)\gamma_\mu\psi(x)$, the axial current
$j_\mu^{5}(x)=\bar\psi(x)\gamma_\mu\gamma_5\psi(x)$, or composite operators like $G_{\mu\nu}^2$.}
$\mathcal{O}(x)$, we define the spectral function $\rho(\omega,p)$ by
\begin{equation}
\rho(\omega,p)
=
-\operatorname{Im}\,G^{R}(\omega,p)\,,
\qquad
G^{R}(\omega,p)
=
-i\!\int d^4x\,e^{i\omega t-ip\cdot x}\,
        \theta(t)\,\langle[\mathcal{O}(x),\mathcal{O}(0)]\rangle \,.
\label{eq:spectral-definition}
\end{equation}
where $G^{R}$ is the retarded two-point function. The Källén--Lehmann representation implies
that $\rho(\omega,p)$ is odd in $\omega$ and non-negative for $\omega>0$, measuring the
distribution of physical states created by $\mathcal{O}$ carrying four-momentum $(\omega,p)$.

Spectral functions enter observable quantities through dispersion relations.
For example, the time-ordered correlator $\Pi(q^2)$ of a conserved current obeys a dispersion relation of the form
\begin{equation}
\Pi(q^2) = \frac{1}{\pi}\int_{0}^{\infty}\!ds\,\frac{\rho(s)}{s-q^2-i\epsilon}\,,
\label{eq:dispersion}
\end{equation}
so knowledge of $\rho$ fixes the entire correlator and, by the optical theorem, cross-sections such as
$e^+e^-\!\to\,\text{hadrons}$. In thermal QCD, $\rho(\omega,0)$ at small $\omega$ determines transport
coefficients through Kubo relations, e.g.\ the shear viscosity or heavy-quark diffusion constant.

At $T=0$ and for $q^{0}>0$, the discontinuity (imaginary part) of the time-ordered correlator across the physical cut coincides with that of the retarded correlator (up to the sign convention adopted here), so using $\rho(s)=\Im\,\Pi(s+i0^+)$ in the dispersion relation is consistent with our earlier definition of the spectral function from $G^{R}$.

Now we consider how one would go about computing spectral functions in QCD. Analytically, perturbation theory is
reliable for $q^2\gg \Lambda_{\text{QCD}}^2$. As an explicit one-loop example, let us compute the spectral density
carried by the flavor-non-singlet vector current \(j_\mu(x)=\bar\psi(x)\gamma_\mu\psi(x)\) in perturbative QCD.
The momentum-space correlator is\footnote{Here we take a single quark flavor of mass \(m\).}
\begin{equation} \label{momcor}
    \Pi_{\mu\nu}(q)
   = i\!\int d^4x\,e^{iqx}\,
     \langle 0|\,\mathcal{T}\,\bigl\{j_\mu(x)\,j_\nu(0)\bigr\}\!|0\rangle \,,
\end{equation}

By Lorentz invariance and current conservation, \eqref{momcor} decomposes as
\[
\Pi_{\mu\nu}(q)=
 (q_\mu q_\nu - q^2 \eta_{\mu\nu})\,\Pi(q^2)\,,
\]
so the task reduces to evaluating the scalar function $\Pi(q^2)$.
For $q^0>0$, we define the (scalar) vector-channel spectral density by the discontinuity across the physical cut,
\begin{equation}
\rho_{\mu\nu}(q) \equiv (q_\mu q_\nu - q^2\eta_{\mu\nu})\,\rho_V(s)\,,
\qquad
\rho_V(s) \equiv \operatorname{Im}\,\Pi(s+i0)\,,
\qquad s\equiv q^2\,.
\end{equation}

With these conventions, the optical theorem gives $R(s)=12\pi\,\rho_{\rm em}(s)$ for the electromagnetic current,
and at leading order in perturbation theory this reproduces the standard partonic result
$R^{(0)}(s)=N_c\sum_f Q_f^2\,v\,(3-v^2)/2$ with $v=\sqrt{1-4m^2/s}$ \cite{Chetyrkin:1996ela}.

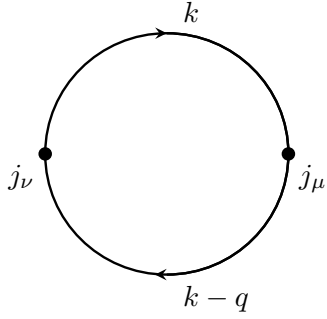
\begin{figure}[h!]
\begin{center}

\begin{tikzpicture}[line width=0.9 pt, scale=1.6,
  midarrowA/.style={
    postaction={decorate},
    decoration={markings, mark=at position 0.0001 with {\arrow{stealth}}}
  },
  midarrowB/.style={
    postaction={decorate},
    decoration={markings, mark=at position 0.9999 with {\arrow{stealth reversed}}}
  }]

  \draw[midarrowA] (0, 1) arc (  90: -90:1);

  \draw[midarrowB] (0,-1) arc (-90: 270:1);

  \filldraw (-1,0) circle (1.3pt);
  \filldraw ( 1,0) circle (1.3pt);

  \node[below left]  at (-1,0) {$j_\nu$};
  \node[below right] at ( 1,0) {$j_\mu$};

  \node[above right,xshift=2pt] at (0, 1) {$k$};
  \node[below right,xshift=2pt] at (0,-1) {$k-q$};

\end{tikzpicture}
\caption{Amputated quark loop diagram.} \label{vacdiag}
\end{center}
\end{figure}

In the relevant diagram, because we are interested in the correlator itself (not in an S-matrix element),
we do not attach external propagators. In that sense, the diagram is ``amputated'': the dots stand for local
operator insertions that inject and remove four-momentum $q$ but do not correspond to external on-shell states.
This is shown in Fig.~\ref{vacdiag}, a closed quark loop with two current insertions \cite{Muta:1998vi}:
\begin{equation}\label{ampcor}
    \Pi_{\mu\nu}(q)
   = -iN_c
       \int\!\frac{d^d k}{(2\pi)^d}\,
        \frac{\mathrm{Tr}\,\bigl[\gamma_\mu\bigl(\slashed{k}+m\bigr)
                              \gamma_\nu\bigl(\slashed{k}-\slashed{q}+m\bigr)\bigr]}
             {\bigl(k^2-m^2+i\epsilon\bigr)
              \bigl[(k-q)^2-m^2+i\epsilon\bigr]}  \,,
\end{equation}
with $d=4-\epsilon$.

Taking the trace in \eqref{ampcor}, using
$\operatorname{Tr}(\gamma_\mu\gamma_\alpha\gamma_\nu\gamma_\beta)
  =4\bigl(\eta_{\mu\alpha}\eta_{\nu\beta}
          -\eta_{\mu\nu}\eta_{\alpha\beta}
          +\eta_{\mu\beta}\eta_{\nu\alpha}\bigr)$, gives
\begin{equation}
\mathrm{Tr}\,\bigl[\gamma_\mu(\slashed{k}+m)\gamma_\nu(\slashed{k}-\slashed{q}+m)\bigr]
=
4\Bigl[\,k_\mu(k-q)_\nu + k_\nu(k-q)_\mu
             - \eta_{\mu\nu}\bigl(k\!\cdot\!(k-q)-m^2\bigr)\Bigr] \,,
\end{equation}

To extract the scalar function $\Pi(q^2)$, contract the decomposition
$\Pi_{\mu\nu}=(q_\mu q_\nu-q^2\eta_{\mu\nu})\Pi(q^2)$ with
$P^{\mu\nu}\equiv(q^\mu q^\nu-q^2\eta^{\mu\nu})$ and use
$P^{\mu\nu}P_{\mu\nu}=(d-1)q^4$. This yields the projector formula
\begin{equation}
\Pi(q^2)=\frac{1}{(d-1)\,q^4}\,P^{\mu\nu}\,\Pi_{\mu\nu}(q)\,,
\end{equation}
and, after a short algebra (using symmetry under $k\mapsto k-q$),
\begin{equation}
\Pi(q^{2})
  =
  \frac{4N_c}{(d-1)\,q^{2}}
  \int\!\frac{d^{d}k}{(2\pi)^{d}}\,
  \frac{k\!\cdot\!(k-q)-m^{2}}
       {(k^{2}-m^{2}+i\epsilon)\bigl[(k-q)^{2}-m^{2}+i\epsilon\bigr]}\,.
\label{eq:scalarPi}
\end{equation}

The crucial factor $1/q^2$ in front ensures that $\Pi(q^2)$ (and hence $\rho_V$) is dimensionless.

We proceed as in the renormalization Section, introducing a Feynman parameter, combining denominators, shifting the loop momentum, and evaluating the resulting integral in dimensional regularization. One finds the standard representation
\begin{equation}
\Pi(q^2)
=
\frac{2N_c\,\mu^\epsilon}{(4\pi)^{d/2}}
\int_0^1\!dx\,x(1-x)\,
\Gamma\,\Bigl(2-\tfrac{d}{2}\Bigr)\,
\bigl[m^2-x(1-x)q^2-i0\bigr]^{\frac{d}{2}-2}\,,
\end{equation}
where the factor $x(1-x)$ originates from the projector/numerator algebra in \eqref{eq:scalarPi}.
Expanding at $d=4-\epsilon$ gives
\begin{equation}
\Pi(s)
=
\frac{N_c}{2\pi^{2}}
\int_0^1\!dx\,x(1-x)\,
\Bigl[
\frac{2}{\epsilon}-\gamma_E+\log(4\pi)
-\log\,\Bigl(\frac{m^2-x(1-x)s-i0}{\mu^2}\Bigr)
\Bigr]
+\mathcal{O}(\epsilon)\,,
\end{equation}

The above $\mu$ is the renormalization scale introduced in dimensional regularization (from the $\mu^\epsilon$ factor) to keep $\Pi$ dimensionless in $d=4-\epsilon$. The spectral density is the discontinuity across the branch cut on the real $s$ axis. Only the logarithm contributes,
since $\operatorname{Im}\log(a-i0)= -\pi\,\theta(-a)$. Therefore,
\begin{equation}
\rho_V(s)\equiv \operatorname{Im}\Pi(s+i0)
=
\frac{N_c}{2\pi^{2}}
\int_0^1\!dx\,x(1-x)\,
\pi\,\theta\,\bigl(x(1-x)s-m^2\bigr)\,,
\end{equation}

For $s>4m^2$ the inequality $x(1-x)s>m^2$ holds for $x_-\le x\le x_+$ with
$x_{\pm}=\tfrac12\bigl(1\pm\sqrt{1-4m^{2}/s}\bigr)$. Carrying out the $x$-integration gives the
standard leading-order result \cite{Chetyrkin:1996ela}:
\begin{equation}
\boxed{%
\rho_{V}(s)
  =
  \frac{N_c}{12\pi}\,
  \Bigl(1+\frac{2m^{2}}{s}\Bigr)
    \sqrt{1-\frac{4m^{2}}{s}}\;
  \theta\,\bigl(s-4m^{2}\bigr)\,.}
\label{eq:spectralFinal}
\end{equation}
with tensor structure
\(
\rho_{\mu\nu}(q)
   = \bigl(q_\mu q_\nu - q^2 \eta_{\mu\nu}\bigr)\,\rho_V(s)
\).
In the massless limit $m\to0$,
\begin{equation}
\rho_V(s)\xrightarrow[m\to0]{}\frac{N_c}{12\pi}\,\theta(s)\,,
\end{equation}
and, using $R(s)=12\pi\sum_f Q_f^2\,\rho_V(s)$, one recovers the well-known partonic result
$R(s)=N_c\sum_f Q_f^{2}$ \cite{Chetyrkin:1996ela}.

The formula \eqref{eq:spectralFinal} is obtained by treating quarks as free, on-shell partons. It therefore describes
the \emph{quark--antiquark continuum}: the probability that the current produces a free $q\bar q$ pair of invariant mass
$\sqrt{s}$. No bound-state structure can appear at this order; hence, the result is smooth above the threshold $s=4m^2$
and vanishes below it.

\begin{figure}[h!]
  \centering
  \includegraphics[width=0.75\textwidth]{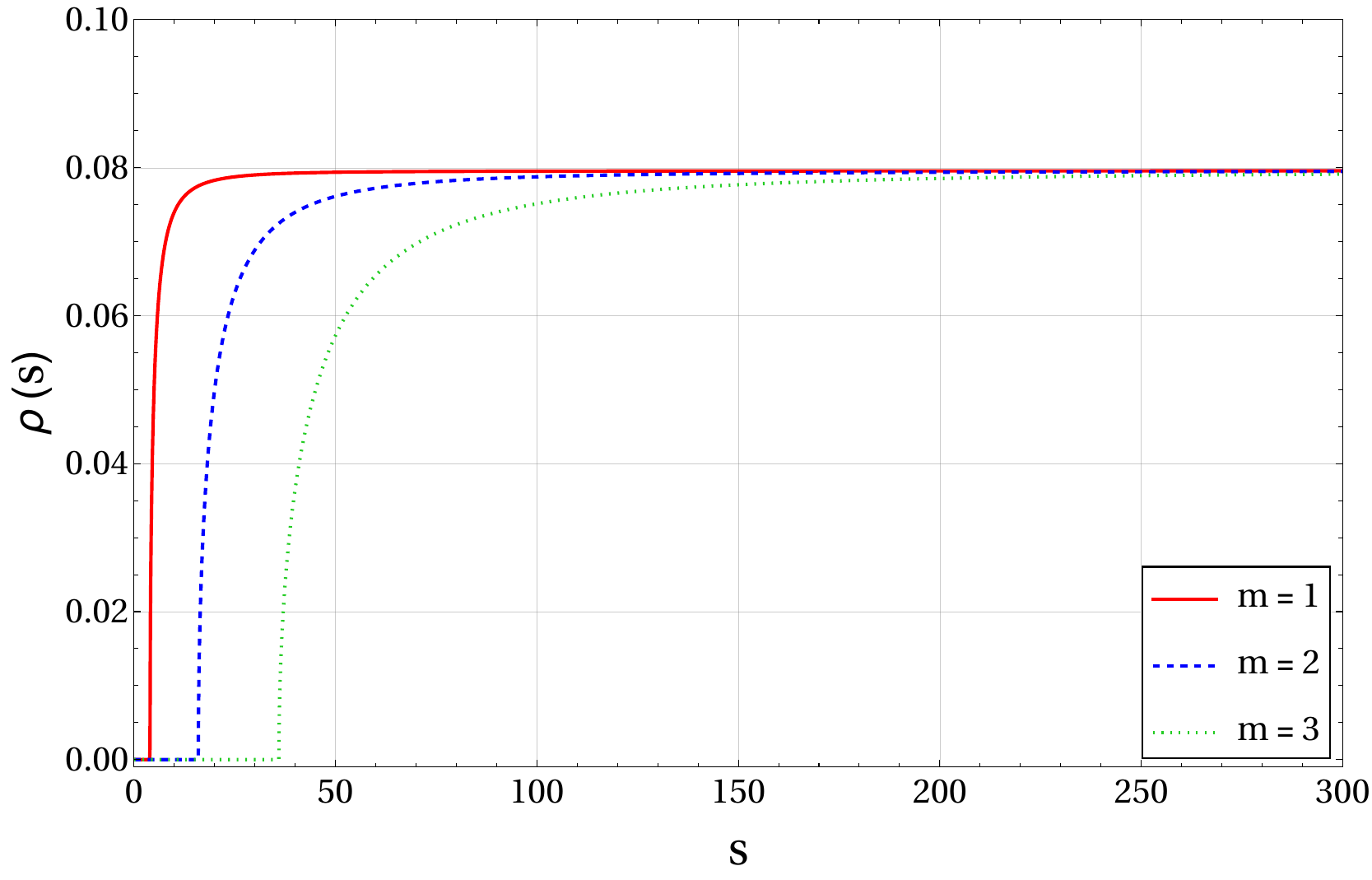}
  \caption{Leading-order continuum spectral density $\rho_V(s)$ as a function of $s$ for three representative quark
  masses. The threshold occurs at $s=4m^2$, and for $s\gg m^2$ the curves approach the universal plateau
  $\rho_V(s)\to N_c/(12\pi)=1/(4\pi)$ (for $N_c=3$).}
  \label{fig:rho_full_rhoV}
\end{figure}

A useful way to visualize the quark--antiquark continuum is to plot the full leading-order spectral density
$\rho_V(s)$ as a function of the invariant mass $s=q^2$. As shown in Fig.~\ref{fig:rho_full_rhoV}, the spectral
density vanishes below the physical threshold $s=4m^2$ and turns on continuously above it, reflecting the phase-space
factor $\sqrt{1-4m^2/s}$ for producing an on-shell $q\bar q$ pair. Increasing the quark mass shifts the threshold to
larger $s$ and delays the approach to the ultraviolet regime, but at sufficiently large $s$ all curves converge to the
same constant plateau: the mass dependence becomes negligible and the spectral density approaches its universal
high-energy limit. Indeed, from \eqref{eq:spectralFinal} one finds
\begin{equation}
\lim_{s\to\infty}\rho_V(s)
=\frac{N_c}{12\pi}
\qquad\left(\text{in particular, for }N_c=3:\ \lim_{s\to\infty}\rho_V(s)=\frac{1}{4\pi}\right),
\end{equation}
which is independent of the quark mass.

Once the leading quark bubble is under control, nothing in principle prevents us from adding gluon exchange,
quark-mass counterterms, and vertex corrections. After renormalization in the $\overline{\text{MS}}$ scheme the
vector-current spectral density acquires the well-known power series
\begin{equation}
\rho_{V}(s,\mu)
   =
   \rho_{V}^{(0)}(s)\,
   \Bigl[1
         + \frac{\alpha_{s}(\mu)}{\pi}
         + c_{2}\,\left(\frac{\alpha_{s}(\mu)}{\pi}\right)^{\!2}
         + c_{3}\,\left(\frac{\alpha_{s}(\mu)}{\pi}\right)^{\!3}
         + \ldots
   \Bigr]\,.
\label{eq:rhoPertSeries}
\end{equation}
where $\rho_{V}^{(0)}(s)=N_{c}/(12\pi)$ for $m\!=\!0$, and the coefficients $c_n$ have been worked out up to $n=4$
in \cite{Baikov:2008jh}. The renormalization scale $\mu$ is arbitrary, but choosing $\mu^{2}\simeq s$ minimizes the
large logarithms produced by loop integrals. Higher loops remain under perturbative control so long as
$\alpha_s(s)\ll 1$.

However, when $s$ approaches a few~$\mathrm{GeV}^{2}$ the coupling becomes $\alpha_{s}(s)=\mathcal{O}(1)$ and the series
\eqref{eq:rhoPertSeries} ceases to be reliable. Physically, quarks and gluons turn into hadrons as soon as the hard scale
drops to a few hundred $\mathrm{MeV}$; the true spectral function develops a tower of resonant poles
(\(\rho,\omega,\phi,\dots\)) and multiparticle thresholds that cannot be captured by any finite order in $\alpha_s$.

To interpolate between the ultraviolet partonic expansion and the infrared regime of bound states, one typically resorts to genuinely non-perturbative methods. In the operator--product expansion (OPE), short-distance physics is separated from long-distance dynamics by organizing correlators into Wilson coefficients multiplying local operators, so that, for instance, power corrections such as $\langle\,:\!G^{2}\!:\,\rangle/s^{2}$ supplement the perturbative series~\eqref{eq:rhoPertSeries} and encode confinement-scale physics through condensates \cite{Wilson:1969zs}. QCD sum rules then relate appropriately weighted moments of $\rho_V$ to these condensates, often introducing an effective continuum threshold $s_0$ to model the transition to the perturbative continuum \cite{Shifman:1978bx}. Lattice QCD provides a first-principles non-perturbative approach by computing Euclidean correlators and reconstructing $\rho_V$ through the (numerically ill-posed) inverse problem connecting $G_E(\tau)$ to the spectral density, thereby capturing resonances and thresholds without expanding in $\alpha_s$ \cite{Aarts:2015cua}. Finally, holographic QCD replaces strongly coupled gauge dynamics by a dual classical gravity description in five dimensions; at large $N_c$ and $T=0$ this typically yields a spectral density dominated by an infinite tower of narrow meson poles (schematically $\rho_V(s)\propto \sum_n f_n^2\,\delta(s-M_n^2)$), while at finite temperature the replacement of AdS by an AdS--Schwarzschild geometry generically broadens these delta-function peaks into smooth, Breit--Wigner-like structures characterized by in-medium masses and widths \cite{Son:2002sd}.

All these frameworks agree that perturbation theory is quantitatively reliable only when the hard scale is sufficiently large (typically several~$\mathrm{GeV}$), whereas at lower scales the growth of the running coupling and confinement necessitate non-perturbative descriptions of the spectral function. In this work, we adopt the holographic approach.

When a single, relatively narrow resonance dominates a given channel, the retarded two-point function of the corresponding current is governed by a simple pole displaced from the real axis by the resonance's finite lifetime. For a vector meson of physical mass $M$ and decay width $\Gamma$, one may write, in momentum space and at vanishing spatial momentum,
\begin{equation}
G^{R}(\omega,0)
   \;\simeq\;
   \frac{|f|^{2}}
        {(\omega+i0)^{2}-M^{2}+iM\Gamma}\,,
\label{eq:GR_BW}
\end{equation}
where $f$ denotes the current--meson coupling. Consistent with the convention \eqref{eq:spectral-definition}, the corresponding spectral function is
\begin{equation}
\rho(\omega)
   \;=\;
   -\,\mathrm{Im}\,G^{R}(\omega+i0)
   \;\simeq\;
   \frac{|f|^{2}\,M\Gamma}{
         \bigl(\omega^{2}-M^{2}\bigr)^{2}+M^{2}\Gamma^{2}}\,.
\label{eq:BW_rho}
\end{equation}

This is a Lorentzian (Breit--Wigner) shape centered near $\omega=M$ with a width controlled by $\Gamma$ (for a narrow resonance, the full width at half maximum is approximately $\Gamma$). In vacuum QCD, this parametrization describes, for example, the $\rho$ meson, with empirical fits giving $M_\rho\simeq770\,\text{MeV}$ and $\Gamma_\rho\simeq150\,\text{MeV}$. In a medium, the pole migrates deeper into the complex plane as interactions broaden the state, so $\Gamma=\Gamma(T,\mu)$ increases with temperature or density and the Breit--Wigner peak widens, eventually dissolving into a smooth continuum. Holographic QCD captures the same qualitative mechanism: replacing thermal AdS by an AdS--Schwarzschild black hole endows the dual bulk gauge field with quasinormal frequencies $\Omega_n=\omega_n-i\Gamma_n/2$, and the boundary spectral function becomes a sum of overlapping Breit--Wigner-like peaks whose widths $\Gamma_n$ grow with the Hawking temperature, smoothly interpolating between narrow meson poles at $T=0$ and a nearly featureless continuum in the quark--gluon plasma.

Finally, transport coefficients follow from the low-frequency behavior of spectral functions through Kubo relations; for example, they are typically determined by the $\omega\to0$ limit of $\rho(\omega,0)/\omega$ in the appropriate channel. Spectral functions therefore constitute the bridge between formal correlation functions of the QCD Lagrangian and physical observables ranging from hadronic cross-sections and decay widths to the transport properties of the quark--gluon plasma.

\subsection{Planar limit of Yang-Mills theory}\label{secYM}

The discovery by 't~Hooft in \cite{tHooft:1973alw} that non-abelian gauge theories simplify dramatically
when the number of colors $N_c$ is sent to infinity has become a cornerstone of modern field theory.
The key observation is that if one lets
\begin{equation}
N_c \to \infty\,,
\qquad\text{while keeping}\qquad
\lambda \equiv g_{\text{YM}}^{2}\,N_c
\;\;\text{fixed}\,,
\label{eq:tHooftLimit}
\end{equation}
then Feynman diagrams reorganize into a topological expansion classified by the genus of the two-dimensional
surface on which they can be drawn without line crossings. Here $\lambda$ is the 't~Hooft coupling: it remains the
genuine dynamical parameter of the planar theory, whereas $1/N_c$ plays the role of a topological loop-counting
variable.

We write the gluon field as $A_\mu = A_\mu^{a}T^{a}$ with $\operatorname{tr}(T^{a}T^{b})=\tfrac12\delta^{ab}$ and
introduce the double-line (ribbon) representation, in which each adjoint propagator carries two fundamental color
indices $(i,j)$, one flowing forward and the other backward. It is convenient to make the $N_c$-counting manifest by
rescaling the gauge field as $A_\mu \mapsto g_{\text{YM}} A_\mu$, so that the Yang--Mills action takes the schematic form
\begin{equation}
    S_{\text{YM}} \sim \frac{N_c}{\lambda}\int d^4x\,\operatorname{tr}\Bigl[(\partial A)^2 + A^2(\partial A) + A^4\Bigr]\,.
\end{equation}

With this normalization, each propagator contributes a factor $\lambda/N_c$, each interaction vertex contributes a
factor $N_c/\lambda$, and each closed color index loop contributes a factor $N_c$. Therefore, a connected vacuum
diagram with $V$ vertices, $E$ propagators, and $F$ closed color loops scales as
\begin{equation}
N_c^{F}\left(\frac{\lambda}{N_c}\right)^{E}\left(\frac{N_c}{\lambda}\right)^{V}
= N_c^{\,F-E+V}\,\lambda^{\,E-V}\,,
\end{equation}

Using Euler's relation $F-E+V = 2-2h$ for a surface of genus $h$, the amplitude can be written as
\begin{equation}
\mathcal{A}_{h}(\lambda,N_c)
=
N_c^{\,2-2h}\,F_{h}(\lambda)\,,
\end{equation}
where $F_h(\lambda)$ denotes the sum of all genus-$h$ diagrams and their dependence on the 't~Hooft coupling.
Planar graphs ($h=0$), therefore, scale as $N_c^{2}$, torus graphs ($h=1$) as $N_c^{0}$, and so on. Summing all
connected vacuum diagrams, one obtains for the logarithm of the partition function
\begin{equation}
\log Z
=
\sum_{h=0}^{\infty} N_c^{\,2-2h}\,F_{h}(\lambda)
=
N_c^{2}\,F_{0}(\lambda)
+ F_{1}(\lambda)
+ \mathcal{O}\,\bigl(1/N_c^{2}\bigr)\,.
\label{eq:stringGenus}
\end{equation}

Eq.~\eqref{eq:stringGenus} is formally identical to the genus expansion of a closed String Theory with string
coupling $g_{s}\sim 1/N_c$; this analogy underlies gauge/string duality and provides a heuristic bridge to AdS/CFT.

Introducing $N_f$ fundamental quark flavors adds single-line propagators, so a quark loop carries only one free color
index rather than two. In the 't~Hooft limit with fixed $N_f$, a connected diagram with $L_q$ quark loops scales as
$N_c^{2-2h-L_q}$, and quark vacuum bubbles are suppressed by $1/N_c$ relative to purely gluonic graphs. (In the
Veneziano limit, where $N_f/N_c$ is held fixed, quark loops need not be suppressed in the same way.)

Connected $n$-point functions of single-trace operators
$\mathcal{O}(x)=\operatorname{tr}F^{2}(x)$, $\operatorname{tr}\bar\psi\gamma_\mu\psi(x)$, ..., scale as
\begin{equation}
\langle\!\langle\,
  \mathcal{O}_{1}\mathcal{O}_{2}\cdots\mathcal{O}_{n}
\,\rangle\!\rangle_{\text{conn}}
\sim
N_c^{\,2-n}\,.
\end{equation}
so at leading order correlation functions factorize into products of two-point functions. Dynamically, this implies
that color-singlet hadrons become narrow at large $N_c$: meson widths scale as $\Gamma_M\sim 1/N_c$, while glueball
widths scale as $\Gamma_G\sim 1/N_c^{2}$ \cite{Manohar:1998xv}. In the planar limit, the gauge theory therefore behaves like a weakly
interacting gas of stable color-singlet states, even though the underlying dynamics can remain strongly coupled for
arbitrary~$\lambda$.

In AdS/CFT language, the identifications $g_s\sim 1/N_c$ and $\alpha'/R^2\sim 1/\sqrt{\lambda}$ map the topological
expansion \eqref{eq:stringGenus} onto the string genus expansion and organize finite-$N_c$ and finite-$\lambda$
corrections. Thus the large-$N_c$ limit not only systematizes Yang--Mills perturbation theory but also provides the
indispensable starting point for its stringy holographic dual.

\subsection{QFT at finite temperature}\label{sec27}

Starting from a Minkowskian path integral, one performs the Wick rotation
\(
(t,x)\mapsto(-i\tau,x)
\)
so that the metric becomes Euclidean and the oscillatory phase $e^{iS}$ is replaced by
the damped weight $e^{-S_E}$, improving convergence.\footnote{Whereas $e^{iS}$ is highly oscillatory and
numerically challenging, $e^{-S_E}$ is exponentially suppressed. We denote Euclidean quantities with a subscript $E$.}

The generating functional transforms to
\begin{equation}
  Z_E[J]=\int\!\mathcal{D}\phi\,
  \exp\!\Bigl[-\!\int_0^\beta\!d\tau\!\int\!d^3x\,
  \bigl\{\mathcal{L}_E(\tau,x)-J(\tau,x)\,\phi(\tau,x)\bigr\}\Bigr]\,,
  \label{eq:ZE}
\end{equation}
which, after restricting the Euclidean time interval to $0\le\tau<\beta$, yields the thermal partition function.
At vanishing chemical potential, one has
\begin{equation}
  Z_E[0]=\int\!\mathcal{D}\phi\,e^{-S_E[\phi]}=\mathrm{Tr}\,e^{-\beta H},
\qquad
S_E[\phi]=\int_0^\beta\!d\tau\!\int\!d^3x\,\mathcal{L}_E,
\qquad
\beta=\frac1T\,.
\label{eq:TrForm}
\end{equation}
which, for finite density, one replaces $H\to H-\mu N$.

The Kubo--Martin--Schwinger (KMS) condition enforces periodicity for bosons and antiperiodicity for fermions
\cite{Kapusta:2006pm}, $\phi(0,x)=\pm\phi(\beta,x)$.
Consequently, every field admits a discrete Fourier expansion
\(
\phi(\tau,x)=\sum_{n\in\mathbb Z}\phi(\omega_n,x)\,e^{i\omega_n\tau}
\)
with Matsubara frequencies
\begin{equation}
\omega^{\text{B}}_n=\frac{2\pi n}{\beta}\,,\qquad
\omega^{\text{F}}_n=\frac{(2n+1)\pi}{\beta}\,.
\end{equation}

In an interacting theory the functional integral \eqref{eq:ZE} can rarely be evaluated in closed form,
but for a free real scalar one finds \cite{Kapusta:2006pm} up to an additive constant and vacuum subtraction
\begin{equation}
\ln Z_E
= -\frac12\,V\sum_{n\in\mathbb Z}\,\int\!\frac{d^3k}{(2\pi)^3}\,
\ln \bigl[\omega_n^2+E_k^2\bigr],
\qquad
E_k=\sqrt{k^2+m^2}\,.
\end{equation}

Thermodynamic observables follow from derivatives of $\ln Z_E$; for instance
$p=(T/V)\ln Z_E$ and $S=\partial_T\!\left(T\ln Z_E\right)$ at fixed $V$.

In the imaginary-time formalism, the fundamental object is the Euclidean time-ordered two-point function
\begin{equation}
  G_E(\omega_n,k)
  =\int_0^\beta\!d\tau\!\int\!d^3x\;
  e^{-i\omega_n\tau-ik \cdot x}\,
  \langle T_E\,\mathcal{O}(\tau,x)\mathcal{O}(0)\rangle\,,
  \label{eq:GE_def}
\end{equation}
with $T_E$ denoting imaginary-time ordering.

Real-time correlators are obtained by analytic continuation in the complex frequency plane. For bosonic Matsubara
frequencies, one has
\begin{equation}
  G_E(\omega_n,k) = G_R(i\omega_n,k)\,,
  \qquad \omega_n = 2\pi nT,
\label{eq:Ret_Euc}
\end{equation}
and $G_R(\omega,k)$ follows by $i\omega_n\to\omega+i0^+$.

A useful way to display the temperature dependence is through the spectral density,
\(
\rho(\omega,k) \equiv -\,\mathrm{Im}\,G_R(\omega,k),
\)
which is an odd function of $\omega$ and non-negative for $\omega>0$.
For vanishing spatial momentum, the Euclidean correlator in real space reads
\begin{equation}
  G_E(\tau,0)
  =\int_{0}^{\infty}\!\frac{d\omega}{\pi}\;
   \rho(\omega)\;
   \frac{\cosh\bigl[(\beta/2-\tau)\omega\bigr]}
        {\sinh(\beta\omega/2)}\,.
  \label{eq:spectral}
\end{equation}

Inverting \eqref{eq:spectral} to recover $\rho(\omega)$ from noisy lattice data is a numerically ill-posed problem,
which explains the intrinsic difficulty of accessing real-time information directly from Euclidean simulations
\cite{Asakawa:2000tr}.

This limitation is particularly acute for dynamical observables, such as spectral functions and transport
coefficients, which are defined in real time and are directly related to retarded correlators. Since the lattice
formulation is intrinsically Euclidean, extracting $\rho(\omega)$ requires an analytic continuation from a discrete set
of Matsubara frequencies or from $G_E(\tau)$, an inverse problem that is exponentially ill-conditioned in the presence
of statistical noise. This motivates complementary non-perturbative approaches that access real-time response functions
more directly. In particular, within gauge/gravity duality one can compute retarded correlators in Minkowski signature
by solving classical bulk equations with in-falling boundary conditions at the black hole horizon, thereby obtaining
$\rho(\omega,k)$ without performing a numerical analytic continuation \cite{Son:2002sd}.

\subsection{Kubo's linear response theory}

Sufficiently close to local thermal equilibrium, the long-wavelength dynamics of a
Quantum Field Theory are governed by relativistic hydrodynamics. The regime of
validity is characterized by external frequencies and momenta small compared to a
microscopic scale set by the temperature,
\begin{equation}
  \omega,k\;\ll\;T\,,
\end{equation}
so that observables may be organized in a derivative expansion.

At the level needed here, hydrodynamics follows from the conservation of the
energy--momentum tensor,
\begin{equation}
  \partial_\mu T^{\mu\nu}=0\,,
  \label{eq:conservation}
\end{equation}
together with constitutive relations expressing $T^{\mu\nu}$ in terms of local
thermodynamic variables.

With mostly-minus signature $\eta_{\mu\nu}=\mathrm{diag}(+,-,-,-)$, we introduce the local rest-frame
four-velocity $u^\mu$ normalized by $u^\mu u_\mu=1$ and the orthogonal projector
$\Delta^{\mu\nu}=\eta^{\mu\nu}-u^\mu u^\nu$. The most general stress tensor consistent with
rotational symmetry reads, up to first order in gradients \cite{Kapusta:2006pm},
\begin{equation}
  T^{\mu\nu}
  = (\varepsilon+p)\,u^\mu u^\nu - p\,\eta^{\mu\nu}
    + \pi^{\mu\nu}\,,
  \qquad
  \pi^{\mu\nu}
  = -\eta\,\sigma^{\mu\nu}-\zeta\,\Delta^{\mu\nu}\,\partial_\alpha u^\alpha\,,
  \label{eq:Tmunu}
\end{equation}
where the shear tensor
\begin{equation}
      \sigma^{\mu\nu}
  =\Delta^{\mu\alpha}\Delta^{\nu\beta}
    \Bigl(\partial_\alpha u_\beta+\partial_\beta u_\alpha
    -\tfrac23 \Delta_{\alpha\beta}\,\partial_\lambda u^\lambda\Bigr) \,.
\end{equation}
is symmetric and traceless, and the transport coefficients $\eta$ (shear) and
$\zeta$ (bulk) parametrize dissipation.

Linearizing Eqs.~\eqref{eq:conservation}--\eqref{eq:Tmunu} around a static homogeneous state, one finds two
universal hydrodynamic excitations \cite{Kapusta:2006pm}:
\begin{align}
  \text{Shear diffusion: } &\omega(k)=-\,i\,D_\eta\,k^{2}
  +\mathcal O(k^{4})\,,\qquad
  D_\eta=\frac{\eta}{\varepsilon+p}\,,                                          \\
  \text{Sound: } &\omega(k)=c_s\,k
      -\frac{i}{2}\,\Gamma_s\,k^{2}+\mathcal O(k^{3})\,,\qquad
  \Gamma_s=\frac{4\eta/3+\zeta}{\varepsilon+p}\,,
\end{align}
with $c_s^{\,2}=\partial p/\partial\varepsilon$ the adiabatic sound speed. The attenuation constants $D_\eta$
and $\Gamma_s$ are fixed entirely by the viscosities and the equation of state.

Transport coefficients admit a microscopic definition through real-time thermal Green functions. Coupling an external
metric perturbation $h_{\mu\nu}$ to $T^{\mu\nu}$ and applying linear response yields the Kubo relations
\begin{equation}
  \eta
  =\lim_{\omega\to0}\frac{1}{\omega}\,
    \rho_s(\omega,0)\,,
  \qquad
  \zeta
  =\lim_{\omega\to0}\frac{1}{\omega}\,
    \rho_b(\omega,0)\,,
  \label{eq:Kubo}
\end{equation}
where we define spectral densities (consistent with Sec.~\ref{sec27}) by
\begin{align}
  \rho_s(\omega,0)
    &\equiv -\,\mathrm{Im}\,G^{R}_{12,12}(\omega,0)\,, \\
  G^{R}_{12,12}(\omega,0)
    &=-i\!\int d^4x\,e^{i\omega t}\theta(t)\,
      \langle[T_{12}(t,\mathbf{x}),T_{12}(0)]\rangle\,, \\[4pt]
  \rho_b(\omega,0)
    &\equiv -\,\frac{1}{9}\,\mathrm{Im}\,G^{R}_{\,T^{i}{}_{i},\,T^{j}{}_{j}}(\omega,0)\,, \\
  G^{R}_{\,T^{i}{}_{i},\,T^{j}{}_{j}}(\omega,0)
    &=-i\!\int d^4x\,e^{i\omega t}\theta(t)\,
      \langle[T^{i}{}_{i}(t,\mathbf{x}),T^{j}{}_{j}(0)]\rangle\,.
\end{align}

Eq.~\eqref{eq:Kubo} implies, for instance, that the retarded correlator in the shear channel behaves as
\(
  G^{R}_{12,12}(\omega,0)
  =-\,i\,\eta\,\omega+\mathcal O(\omega^{2}),
\)
so that the low-frequency slope of its imaginary part determines the shear viscosity.

To make the connection with Sec.~\ref{sec27} explicit, we use the same spectral-function convention
\begin{equation}
\rho(\omega,k) \equiv -\,\mathrm{Im}\,G_R(\omega,k)\,,
\label{eq:rho_convention_hydro}
\end{equation}
so that $\rho(\omega,k)$ is odd in $\omega$ and non-negative for $\omega>0$ for Hermitian operators.
With this definition, the Euclidean correlators are related to $\rho$ by the same spectral representation as in
Eq.~\eqref{eq:spectral}:
\begin{equation}
  G_E(\tau,0)
  =\int_{0}^{\infty}\!\frac{d\omega}{\pi}\;
   \rho(\omega,0)\;
   \frac{\cosh\bigl[(\beta/2-\tau)\omega\bigr]}
        {\sinh(\beta\omega/2)}\,,
\label{eq:GE_from_rho_hydro}
\end{equation}
which highlights why transport coefficients are difficult to extract from lattice simulations: they are controlled by
the $\omega\to0$ behavior of $\rho(\omega,0)$, precisely where the Euclidean kernel is least sensitive.

This is also the natural entry point for holography. In strongly coupled theories admitting a gravity dual, the
retarded correlators appearing in \eqref{eq:Kubo} can be obtained directly from classical bulk perturbations by imposing
in-falling boundary conditions at the horizon, avoiding numerical analytic continuation. In this way, holography gives
a controlled computation of transport and real-time response in the same framework used later in this dissertation to
compute mesonic spectral functions at finite temperature, density, and external fields.

\newpage

\section{An overview of the AdS/CFT correspondence}\label{adsover}

In this Section, following the pedagogical treatment in \cite{Ammon:2015wua}, we review core principles of
gauge/gravity duality. We begin with the geometry of Anti-de Sitter (AdS) space and then motivate the
Maldacena conjecture \cite{Maldacena:1997re} and its usefulness as a tool for strongly coupled dynamics.
Although QCD is not conformal and no exact weakly curved dual of real-world QCD is known, the AdS/CFT framework
inspires effective holographic models that can capture qualitative, and often semi-quantitative, features of
non-perturbative QCD.

\subsection{AdS geometry}

Throughout this subsection, we employ the mostly-minus convention
$\eta_{MN}=\mathrm{diag}(1,-1,\ldots,-1)$ and denote the bulk spacetime dimension by $d$ (so the conformal boundary
has dimension $d-1$).

Anti-de Sitter space is a maximally symmetric spacetime of constant negative curvature. A convenient way to
construct $\mathrm{AdS}_d$ is via its embedding in a flat ambient space $\mathbb{R}^{2,d-1}$ with two timelike
directions. Concretely, introduce coordinates
\begin{equation}
\left(x_0,\, x_1, \ldots,\, x_{d-1}, \, x_d\right)\,,
\end{equation}
in signature $(2,d-1)$, with ambient-space line element
\begin{equation}
ds^2 = dx_0^2 - \sum_{i=1}^{d-1} dx_i^2 + dx_d^2\,,
\end{equation}

Imposing the hyperboloid constraint
\begin{equation}
x_0^2 - \sum_{i=1}^{d-1} x_i^2 + x_d^2 = -\,L^2\,.
\end{equation}
defines an $\mathrm{AdS}_d$ hypersurface whose isometry group is $\mathrm{SO}(2,d-1)$.

From the viewpoint of General Relativity, AdS arises as a vacuum solution of the Einstein equations with a negative
cosmological constant. Starting from the Einstein--Hilbert action
\begin{equation}
S = \frac{1}{16\pi G}\int d^d x\,\sqrt{-g}\,\bigl(\mathcal{R}-2\Lambda\bigr)\,,
\end{equation}
variation with respect to $g_{MN}$ yields
\begin{equation}
\mathcal{R}_{MN}-\frac12\,g_{MN}\,\mathcal{R} = -\,\Lambda\,g_{MN}\,.
\end{equation}

For $\mathrm{AdS}_d$, one has $\Lambda<0$, and the constant-curvature solution reads \cite{Ammon:2015wua}
\begin{equation}
\mathcal{R}_{MN}=-\,\frac{d-1}{L^2}\,g_{MN}\,,
\qquad
\mathcal{R} = -\,\frac{d(d-1)}{L^2}\,,
\qquad
\Lambda = -\,\frac{(d-1)(d-2)}{2L^2}\,.
\end{equation}
where $L$ is the AdS radius and sets the overall curvature scale.

A common coordinate system in holography is the Poincar\'e patch, which makes the conformal boundary at $z=0$
manifest. In these coordinates, the metric takes the form
\begin{equation} \label{adspatch}
ds^2 = \frac{L^2}{z^2}\Bigl(dx_0^2-d\vec{x}^{\,2}-dz^2\Bigr)\,,
\end{equation}
where $\vec{x}\in\mathbb{R}^{d-2}$ and $z>0$ ranges from the boundary at $z\to 0$ to the deep interior at
$z\to\infty$. The boundary at $z=0$ lies at infinite proper distance, but it is a conformal boundary on which one
can naturally define the dual $(d-1)$-dimensional field theory.

A central geometric intuition for holography is the UV/IR connection: in the Poincar\'e patch, the radial coordinate
$z$ sets an energy scale in the boundary theory. Small $z$ corresponds to ultraviolet physics (short distances),
whereas large $z$ corresponds to infrared physics (long distances). This identification underlies the holographic
interpretation of renormalization-group flow and will be important later when dynamical backgrounds are used to model
confinement scales, finite temperature horizons, and in-medium modifications of spectral observables.

Global coordinates cover the entire spacetime rather than a patch. A standard form of the global AdS$_d$ metric is
\cite{Nastase:2015wjb}
\begin{equation}
ds^2=\frac{L^2}{\cos^2\rho}\Bigl(d\tau^2-d\rho^2-\sin^2\rho\,d\Omega_{d-2}^2\Bigr)\,,
\qquad 0\le \rho<\frac{\pi}{2}\,.
\end{equation}

\begin{figure}[h!]
\centering
\includegraphics[width=0.5\textwidth]{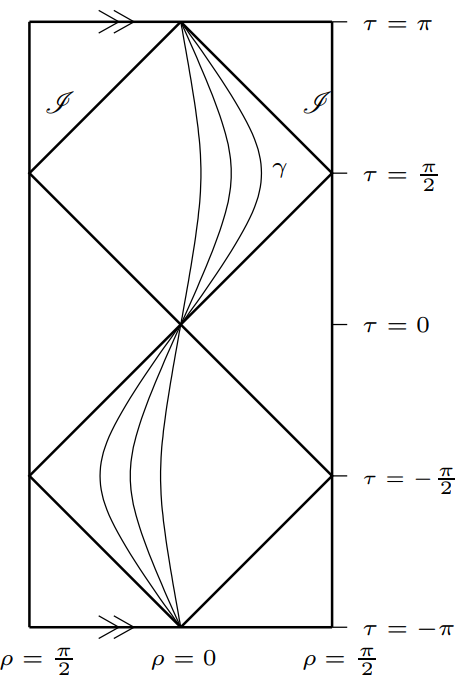}
\caption{Penrose diagram of $AdS_5$ spacetime. Figure taken, and adapted, from \cite{Gruppuso:2004db}.}
\label{adspen}
\end{figure}

A useful way to capture the global causal structure is to analyze the corresponding Penrose diagram. Since the
overall conformal factor $L^2/\cos^2\rho$ does not affect null trajectories, one can conformally rescale the metric
to a finite ``cylinder'', in which the AdS boundary at $\rho=\pi/2$ appears at a finite coordinate location. In the
Penrose diagram, the boundary is timelike, shown as the vertical edge of the diagram, and light rays travel at
$45^\circ$ angles; thus massless signals can reach the boundary in finite coordinate time. Fig.~\ref{adspen}
illustrates this structure for $AdS_5$. The arrows indicate the periodic identification of global time in AdS; in
applications to AdS/CFT one typically considers the universal cover, in which $\tau\in\mathbb{R}$ and the same strip
extends infinitely in the vertical direction.

Consequently, the Penrose diagram of $\mathrm{AdS}$ is a rectangular strip in which the vertical boundaries represent
the timelike conformal boundary of the spacetime. It encodes the causal relationships between bulk and boundary
regions and highlights a key feature relevant for holography: bulk excitations can influence the boundary (and vice
versa) within finite coordinate time, once appropriate boundary conditions are specified.

\subsection{Black hole thermodynamics}\label{sec:bh_thermo}

The Schwarzschild solution of the vacuum Einstein equations $\mathcal{R}_{\mu\nu}=0$ has line element%
\footnote{We set $c=\hbar=k_B=1$ and keep $G$ explicit throughout.}
\begin{equation}
    ds^{2}= f(r)\,dt^{2}-f(r)^{-1}dr^{2}-r^{2}d\Omega_{2}^{2}\,,
    \qquad
    f(r)=1-\frac{r_s}{r}\,,
    \qquad r_s=2GM\,.
\end{equation}
which contains a \emph{coordinate} singularity at the horizon $r_h=r_s$.

Introducing the tortoise coordinate
\begin{equation}
  r_{*}=r+r_s\ln\bigl|r/r_s-1\bigr|\,,
  \qquad \frac{dr_*}{dr}=f(r)^{-1}\,,
\end{equation}
and advanced/retarded null variables $v=t+r_*$ and $u=t-r_*$, one defines Kruskal--Szekeres coordinates (for $r>r_s$) by
\begin{equation}\label{kruscalc}
  V=\exp\,\left(\frac{v}{2r_s}\right)\,,
  \qquad
  U=-\exp\,\left(-\frac{u}{2r_s}\right)\,,
\end{equation}
for which the metric becomes
\begin{equation}\label{kruscalmetric}
    ds^{2}= \frac{4r_s^{3}}{r}\;
          e^{-\,r/r_s}\,
          dU\,dV
          -r^{2}d\Omega_{2}^{2}\,.
\end{equation}
with $r$ understood as an implicit function of $UV$ via $UV=-(1-r_s/r)\,e^{r/r_s}$. Eq.~\eqref{kruscalmetric} is
manifestly regular at $r=r_s$ and shows that the event horizon is a null hypersurface rather than a physical singularity.

We now Wick rotate the Schwarzschild time, $t\mapsto -i\tau$. A direct way to see the thermal periodicity is to work
with the Euclidean Schwarzschild metric,
\begin{equation}
    ds_E^{2}= f(r)\,d\tau^{2}+f(r)^{-1}dr^{2}+r^{2}d\Omega_{2}^{2}\,,
\end{equation}

Close to the horizon, set $r=r_s+\epsilon$ with $\epsilon\ll r_s$. Then $f(r)\simeq \epsilon/r_s$, and the $(\tau,r)$
submanifold becomes
\begin{equation}
ds_{E}^{2}\simeq \frac{\epsilon}{r_s}\,d\tau^{2}+\frac{r_s}{\epsilon}\,d\epsilon^{2}+r_s^{2}d\Omega_2^{2}\,,
\end{equation}

Defining a radial coordinate $\rho$ by $\epsilon=\rho^{2}/(4r_s)$, one finds
\begin{equation}
ds_{E}^{2}\simeq d\rho^{2}+\rho^{2}d\theta^{2}+r_s^{2}d\Omega_2^{2},
\qquad
\theta\equiv \frac{\tau}{2r_s}\,,
\end{equation}

This is flat $\mathbb{R}^{2}$ in polar coordinates $(\rho,\theta)$ times $S^{2}$. Regularity at $\rho=0$ (the horizon)
requires $\theta\sim\theta+2\pi$, hence
\begin{equation}
\tau\sim\tau+\beta,
\qquad
\beta=4\pi r_s = 8\pi GM\,,
\end{equation}

In Euclidean finite temperature field theory one identifies $\beta=1/T$, so the Hawking temperature is
\begin{equation}\label{hawking}
\boxed{T_{H}=\frac{1}{8\pi GM}\,.}
\end{equation}

Thus, the Hawking temperature emerges as the condition for avoiding a conical defect at the Euclidean horizon.

Treating the black hole energy as $E=M$, the first law gives $dS = dM/T_H$, and using \eqref{hawking} yields
\begin{equation}
S(M)=\int \frac{dM}{T_H}= \int 8\pi GM\,dM = 4\pi G M^{2}
= \frac{A}{4G}\,,
\qquad A=4\pi r_s^2\,.
\end{equation}
which is the Bekenstein--Hawking entropy.

A complementary derivation uses the (regulated) Euclidean gravitational action \cite{Gibbons:1976ue,Wald:1984rg}
\begin{equation}
  I_E =
  -\frac{1}{16\pi G}
     \int_{\mathcal M}\! d^4x\;\sqrt{g}\;\mathcal{R}
  -\frac{1}{8\pi G}
     \int_{\partial\mathcal M}\!\! d^3x\;\sqrt{h}\;
     \bigl(K-K_0\bigr)\,,
\label{eq:IE-def}
\end{equation}
where $h_{ab}$ is the induced metric on the boundary $\partial\mathcal M$, $K=h^{ab}\nabla_a n_b$ is the trace of the
extrinsic curvature (with outward unit normal $n^a$), and $K_0$ is the corresponding quantity for an isometric
embedding of the boundary into flat space. For Euclidean Schwarzschild, $\mathcal{R}=0$, so the bulk term vanishes and
the on-shell action comes entirely from the boundary. Evaluating the GHY term on a large-$r$ hypersurface $r=r_0$ and
subtracting the flat-space counterterm, one finds the finite result
\begin{equation}
I_{\mathrm{ren}}=\frac{\beta M}{2}
=\frac{\beta r_s}{4G}
=\frac{A}{4G}\,,
\end{equation}
with the period fixed by regularity, $\beta=T_H^{-1}=8\pi GM$. Interpreting the semiclassical partition function as
$Z(\beta)\simeq e^{-I_{\mathrm{ren}}}$, one has
\begin{equation}\label{logbeta}
\log Z(\beta)=-\,\frac{\beta^{2}}{16\pi G}\,,
\end{equation}
and standard thermodynamic relations gives
\begin{equation}
E=-\frac{\partial}{\partial\beta}\log Z
=\frac{\beta}{8\pi G}=M\,,
\qquad
S=\log Z+\beta E
=\frac{\beta^{2}}{16\pi G}
=\frac{A}{4G}\,.
\end{equation}

The corresponding Helmholtz free energy is
\begin{equation}
F=-\frac{1}{\beta}\log Z=\frac{\beta}{16\pi G}=\frac{M}{2}\,,
\end{equation}
and the heat capacity is
\begin{equation}
C\equiv \frac{dE}{dT}=-\,8\pi G M^{2}<0\,.
\end{equation}

Thus, an asymptotically flat Schwarzschild black hole has negative heat capacity: as it radiates, its temperature rises
and the canonical ensemble is unstable. In contrast, sufficiently large AdS black holes have $C>0$ and participate in
the Hawking--Page transition.

In AdS/CFT, the Euclidean time period $\beta$ of a black hole (or black brane) geometry is identified with the inverse
temperature of the dual field theory. The gravitational on-shell action computes the thermodynamic potential, and the
horizon entropy corresponds to the entropy of the thermal state. This relation between horizon regularity, temperature,
and entropy is the basic reason AdS--Schwarzschild backgrounds are the natural holographic laboratory for
finite temperature QCD-like physics studied in later chapters.

In a more general case, let $M,J,Q$ be the mass, angular momentum, and electric charge of a stationary black hole. With
surface gravity $\kappa$, electric potential $\Phi$, and angular velocity $\Omega$, the identifications
\(T=\kappa/(2\pi)\) and \(S=A/(4G)\) map the following geometric statements to the ordinary laws of thermodynamics
\cite{Wald:1984rg}:
\begin{itemize}
    \item[(0th)] Uniform temperature: for any stationary black hole, the surface gravity is constant over the horizon,
    $\kappa=\mathrm{const}$ on $\mathcal{H}$.
    \item[(1st)] First law:
    \(
    dM = \frac{\kappa}{8\pi G}\,dA + \Omega\,dJ + \Phi\,dQ
    \)
    (and additional work terms for other charges).
    \item[(2nd)] Area theorem: for any classically allowed process, $\delta A\ge0$ and hence $\delta S_{\mathrm{BH}}\ge0$.
    Including matter entropy gives the generalized second law
    \(
    \delta(S_{\mathrm{BH}}+S_{\mathrm{m}})\ge0
    \).
    \item[(3rd)] It is impossible, by any finite sequence of manipulations, to reduce $\kappa$ to zero; hence $T=0$ is
    unattainable in finite steps.
\end{itemize}

Finally, one may define the thermodynamic ensembles:
\begin{itemize}
  \item[(I)] Microcanonical: fixed $(M,J,Q)$; the entropy $S(M,J,Q)$ is the fundamental potential.
  \item[(II)] Canonical: fixed $T$; the Helmholtz free energy $F=-T\log Z$ is the fundamental potential.
  \item[(III)] Grand-canonical: fixed $(T,\Omega,\Phi)$; the Gibbs free energy $G=F-\Omega J-\Phi Q$ is the fundamental potential.
\end{itemize}

The area law also motivates the covariant (Bousso) entropy bound $S\le A/(4G)$, suggesting that the maximal
information content in a region scales with its boundary area and underpinning the holographic principle.

\subsection{Conformal Field Theory}

Conformal Field Theories (CFTs) are Quantum Field Theories invariant under conformal transformations, i.e.\ coordinate
transformations that preserve the metric up to a local Weyl factor. Equivalently, a conformal transformation acts as
\begin{equation}
  dx^{\prime\mu}\,dx'_{\mu}=\Omega(x)^{-2}\,dx^{\mu}\,dx_{\mu}\,,
\end{equation}
for some positive function $\Omega(x)$.

In a renormalized Quantum Field Theory, couplings run with the renormalization scale $\mu$ and the flow is governed by
the beta function, shown in Eq.~\eqref{betaf}. If $\beta(g)=0$ for all couplings (or at a fixed point), the theory is
scale invariant. In $d>2$ dimensions, it is widely expected that under mild assumptions (unitarity, Poincar\'e
invariance, and a well-defined stress tensor) scale invariance implies full conformal invariance, so that such fixed
points define CFTs.

For Minkowski signature in $d$ spacetime dimensions, the conformal group is isomorphic to $\mathrm{SO}(2,d)$.
Its generators include translations $P_\mu$,
\begin{equation}
  P_\mu = i\,\partial_\mu\,,
\end{equation}
Lorentz transformations $M_{\mu\nu}$,
\begin{equation}
  M_{\mu\nu}= i\bigl(x_\mu\,\partial_\nu-x_\nu\,\partial_\mu\bigr)\,,
\end{equation}
dilatations $D$,
\begin{equation}
  D=i\,x^\mu\partial_\mu\,,
\end{equation}
and special conformal transformations $K_\mu$,
\begin{equation}
  K_\mu
  = i\Bigl(2x_\mu\,x^\nu\partial_\nu-x^2\partial_\mu\Bigr)\,.
\end{equation}

Together, these satisfy the conformal algebra associated with $\mathrm{SO}(2,d)$ \cite{Natsuume:2014sfa}.

A local operator $\mathcal{O}(x)$ is assigned a scaling dimension $\Delta$ if under a uniform rescaling
$x^\mu\mapsto x^{\prime\mu}=\alpha x^\mu$ it transforms as
\begin{equation}\label{scale}
  \mathcal{O}'(x')=\alpha^{-\Delta}\,\mathcal{O}(x)\,.
\end{equation}

In a CFT, operators are organized into representations of the conformal algebra. In particular, a (scalar) primary
operator $\mathcal{O}_\Delta$ is defined by the condition that it is annihilated by the generators of special
conformal transformations at the origin,
\begin{equation}
  [K_\mu,\mathcal{O}_\Delta(0)]=0\,,
\end{equation}
or equivalently (in radial quantization) that the corresponding state $|\mathcal{O}_\Delta\rangle$ satisfies
$K_\mu|\mathcal{O}_\Delta\rangle=0$. Descendants are generated by the action of translations $P_\mu$ on primaries, so
each primary operator defines a conformal family obtained by repeated derivatives.

Conformal symmetry strongly constrains correlation functions. For example, the two-point function of identical scalar
primaries in Euclidean signature is fixed up to normalization:
\begin{equation}
  \langle\mathcal{O}_\Delta(x)\,\mathcal{O}_\Delta(0)\rangle
  =\frac{C_{\mathcal{O}}}{(x^2)^\Delta}\,,
\end{equation}
while three-point functions are determined up to a set of structure constants, and higher-point functions are
constrained by conformal symmetry and crossing symmetry \cite{Ammon:2015wua}.

A crucial link to AdS/CFT is that $\mathrm{AdS}_{d+1}$ has isometry group $\mathrm{SO}(2,d)$, matching the conformal
group of a CFT in $d$ dimensions. This symmetry matching is one of the key kinematical motivations for expecting a
duality between gravity on $\mathrm{AdS}_{d+1}$ and a Conformal Field Theory living on its $d$-dimensional boundary.

Conformal invariance also fixes how primary operators transform under a general conformal map $x\mapsto x'$ with local
Weyl factor $\Omega(x)$. For a scalar primary of dimension $\Delta$ one has
\begin{equation}
\mathcal{O}'(x')=\Omega(x)^{\Delta}\,\mathcal{O}(x)\,,
\end{equation}
which reduces to the dilatation law \eqref{scale} for $\Omega=\alpha^{-1}$ (constant rescaling). As a consequence,
conformal symmetry strongly constrains correlation functions. For example, the two-point function of scalar primaries
is fixed up to an overall normalization,
\begin{equation}
\langle \mathcal{O}_i(x)\,\mathcal{O}_j(0)\rangle
=\frac{C_i\,\delta_{ij}}{(x^2)^{\Delta_i}}\,,
\end{equation}
and the three-point function is determined up to a single structure constant,
\begin{equation}
\langle \mathcal{O}_1(x_1)\,\mathcal{O}_2(x_2)\,\mathcal{O}_3(x_3)\rangle
=\frac{C_{123}}{
x_{12}^{\Delta_1+\Delta_2-\Delta_3}\,
x_{23}^{\Delta_2+\Delta_3-\Delta_1}\,
x_{31}^{\Delta_3+\Delta_1-\Delta_2}}\,,
\qquad x_{ij}\equiv |x_i-x_j|\,.
\end{equation}

More generally, a CFT is largely characterized by its ``CFT data'': the spectrum of primary scaling dimensions
$\{\Delta_i\}$ and the OPE coefficients (structure constants) $\{C_{ijk}\}$ \cite{Ammon:2015wua,Natsuume:2014sfa}.
In the holographic setting, these symmetries and data control the ultraviolet behavior of boundary correlators, while
real-time response functions (retarded correlators and spectral densities) are obtained by imposing appropriate
boundary conditions on the dual bulk fields.

\subsection{Partition functions in CFTs}

A convenient way to organize correlation functions in a CFT is through a generating functional in the presence of
classical sources coupled to local operators. Introducing sources $\phi_{\Delta_j}(x)$ for a set of operators
$\mathcal{O}_{\Delta_j}(x)$ of scaling dimension $\Delta_j$, one defines the CFT partition function \cite{Ammon:2015wua}
\begin{equation}
Z_{\mathrm{CFT}}[\phi_{\Delta_j}]
=
\Biggl\langle
\exp\!\Bigl[
 i \int d^d x \sum_j \phi_{\Delta_j}(x)\,\mathcal{O}_{\Delta_j}(x)
\Bigr]
\Biggr\rangle_{\mathrm{CFT}}\,,
\label{eq:ZCFTsources}
\end{equation}
where $\langle\cdots\rangle_{\mathrm{CFT}}$ denotes the path integral over the CFT fields (with the appropriate
normalization).

It is often useful to work with the generating functional of connected correlators,
\begin{equation}
W_{\mathrm{CFT}}[\phi]\equiv -i\ln Z_{\mathrm{CFT}}[\phi]\,.
\end{equation}

Then connected $n$-point functions follow from functional differentiation:
\begin{equation}
\bigl\langle \mathcal{O}_{\Delta_{1}}(x_{1})\cdots \mathcal{O}_{\Delta_{n}}(x_{n})\bigr\rangle_{\!c}
=
\frac{1}{i^{n}}
\frac{\delta^{n} W_{\mathrm{CFT}}[\phi]}
     {\delta \phi_{\Delta_{1}}(x_{1})\cdots \delta \phi_{\Delta_{n}}(x_{n})}
\Biggr|_{\phi=0}\,.
\label{eq:conncorr}
\end{equation}

In this case, without the logarithm, derivatives of $Z_{\mathrm{CFT}}$ generate time-ordered correlators but also include disconnected pieces.

Conformal invariance fixes the scaling of sources. For a scalar primary $\mathcal{O}_\Delta$ with
$\mathcal{O}'_\Delta(x')=\alpha^{-\Delta}\mathcal{O}_\Delta(x)$ under $x'=\alpha x$, invariance of the source
deformation $\int d^d x\,\phi_\Delta(x)\mathcal{O}_\Delta(x)$ implies that the source transforms as
\begin{equation}
\phi'_\Delta(x')=\alpha^{-(d-\Delta)}\,\phi_\Delta(x)
\Longleftrightarrow
\phi'_\Delta(x)=\alpha^{d-\Delta}\,\phi_\Delta(\alpha x)\,.
\label{eq:source_scaling}
\end{equation}

Therefore, $\phi_\Delta$ carries scaling dimension $d-\Delta$, as expected for a coupling to an operator of dimension
$\Delta$.

This formalism is precisely the entry point for AdS/CFT: sources $\phi_{\Delta_j}(x)$ on the boundary are identified
with boundary values of bulk fields in $\mathrm{AdS}_{d+1}$, and the gravity on-shell action provides the generating
functional for connected correlators. In later chapters, we will use the real-time version of this statement, where
retarded correlators (and hence spectral functions) are obtained by imposing in-falling boundary conditions at a
black hole horizon in the bulk.

\subsection{$\mathcal{N}=4$ SYM theory}

In ordinary Quantum Field Theory, the Coleman--Mandula theorem forbids any non-trivial unification of spacetime and
internal symmetries. The obstruction is bypassed by enlarging the symmetry algebra to a \emph{graded} Lie algebra,
as in the Haag--\L{}opusza\'nski--Sohnius extension, which allows fermionic generators $Q_\alpha^{\,I}$ (supercharges).
In particular, the supercharges close into translations \cite{Weinberg:2000cr}:
\begin{equation}
  \{Q_\alpha^{\,I},\bar Q_{\dot\beta J}\}
  =2(\sigma^\mu)_{\alpha\dot\beta}P_\mu\,\delta^{I}_{\;J}\,,
  \qquad
  I,J=1,\dots,\,\mathcal N\,,
\end{equation}
while the bosonic sector contains the Lorentz algebra together with internal symmetries. In $3+1$ dimensions, an
interacting local gauge theory without gravity can have at most $\mathcal N=4$ supersymmetry, corresponding to
16 real supercharges.

The paradigmatic example is $\mathcal N=4$ supersymmetric Yang--Mills theory with gauge group $SU(N_c)$ (the overall
$U(1)$ factor in $U(N_c)$ decouples). Its field content consists of a gauge field $A_\mu$, four gaugini $\lambda^I$
($I=1,\dots,4$), and six real scalars $\phi^i$ ($i=1,\dots,6$), all in the adjoint representation. The action can be
written compactly as the dimensional reduction of ten-dimensional $\mathcal N=1$ SYM \cite{Weinberg:2000cr}:
\begin{equation}
  S
  =\frac{1}{g_{\mathrm{YM}}^{2}}
    \int d^{4}x\;
    \operatorname{Tr}\,\Bigl[
      -\tfrac14 F_{\mu\nu}F^{\mu\nu}
      +\tfrac12 D_\mu\phi^i D^{\mu}\phi^i
      -\tfrac14[\phi^i,\phi^j]^2
      + i\,\bar\lambda_I\slashed{D}\lambda^I
      -\bar\lambda_I\Gamma^{\,i}_{IJ}[\phi^i,\lambda^J]
    \Bigr] .
\end{equation}

A remarkable property of $\mathcal N=4$ SYM is that quantum corrections to the gauge coupling cancel exactly:
$\beta(g_{\mathrm{YM}})=0$. The theory is therefore conformal at all scales. Its bosonic global symmetry is
\begin{equation}
  SO(4,2)\times SU(4)_R \simeq SO(4,2)\times SO(6)_R\,.
\end{equation}
the direct product of the four-dimensional conformal group and the $R$-symmetry. Including supersymmetries, the full
superconformal symmetry group is $PSU(2,2|4)$.

The complexified coupling
\(
\tau=\frac{\theta}{2\pi}+\frac{4\pi i}{g_{\mathrm{YM}}^{2}}
\)
enjoys an $SL(2,\mathbb{Z})$ S-duality, which acts on $\tau$ by fractional linear transformations and exchanges
electric and magnetic degrees of freedom. Taking the 't~Hooft limit,
\(
\lambda=g_{\mathrm{YM}}^{2}N_c
\)
fixed as $N_c\to\infty$, reorganizes perturbation theory into a genus expansion dominated by planar diagrams, the same
large-$N_c$ structure discussed in Sec.~\ref{secYM}. These properties make $\mathcal N=4$ SYM the canonical arena for the
AdS/CFT correspondence and a useful benchmark for real-time observables (such as transport coefficients) in strongly
coupled gauge theories.

Moreover, the bosonic symmetry $SO(4,2)\times SO(6)$ matches the isometry group of
$\mathrm{AdS}_5\times S^5$, anticipating the structure of the gravitational dual that appears in the Maldacena
conjecture.

\subsection{Supergravity as the low-energy limit of String Theory}

String Theory provides a consistent framework for quantum gravity by positing that fundamental particles are not
point-like, but rather excitations of one-dimensional extended objects (strings). A string is characterized by its
tension
\begin{equation}
  T_s=\frac{1}{2\pi\alpha'}\,,
\end{equation}
where $\alpha'$ is the Regge slope and sets the fundamental length scale $\ell_s$ via $\alpha'=\ell_s^{\,2}$. The
classical dynamics of a string are governed by a worldsheet action that extremizes the area swept out by the string
in spacetime.

Quantum consistency requires the bosonic closed string to live in $d=26$ dimensions; however, the bosonic theory
contains a tachyon and is therefore unstable around the perturbative vacuum. Introducing spacetime supersymmetry leads
to superstring theories, which remove the tachyonic instability and provide a consistent perturbative framework. In
ten dimensions there are five consistent superstring theories \cite{Becker:2006dvp}: Type~I (open and closed strings
with $SO(32)$ gauge symmetry), Type~IIA and Type~IIB (closed strings with non-chiral and chiral supersymmetry,
respectively), and the heterotic theories with gauge groups $SO(32)$ and $E_8\times E_8$.

Quantization of the string yields an infinite tower of states, including massless modes and massive excitations with
masses set by the string scale $m_s\sim 1/\ell_s\sim 1/\sqrt{\alpha'}$. At energies well below this scale,
$E\ll m_s$, the massive string modes decouple, and the dynamics are captured by an effective field theory of the
massless sector. This low-energy effective theory is a ten-dimensional supergravity theory: each of the five
superstring theories reduces to a corresponding ten-dimensional supergravity in the $\alpha'\to 0$ limit.

For example, in Type~IIB String Theory the massless spectrum includes the graviton (metric fluctuations), the dilaton
(which controls the string coupling), the Kalb--Ramond two-form $B_{MN}$, and Ramond--Ramond (RR) gauge fields.
Accordingly, the low-energy action takes the schematic form
\begin{equation}
S_{\text{SUGRA}}
=\frac{1}{16\pi G_{10}}
\int d^{10}x\,\sqrt{-g}\,\Bigl(\mathcal{R}+\cdots\Bigr)\,,
\end{equation}
where the ellipsis denotes the kinetic terms and interactions of the remaining supergravity fields, and $G_{10}$ is
the ten-dimensional Newton constant.

Supergravity is a reliable approximation when two independent expansions are under control. First, higher-derivative
corrections from integrating out massive string modes are suppressed when curvatures are small in string units, so
that $\alpha' \mathcal{R}\ll 1$ (equivalently $\ell_s/L\ll 1$ for a background with curvature scale $L$). Second,
quantum (string loop) corrections are controlled by the string coupling $g_s$, which is set by the dilaton vacuum
expectation value, $g_s=e^{\langle\phi\rangle}$; the classical (tree-level) limit corresponds to $g_s\ll 1$. In
perturbation theory, string interactions are described by worldsheet topologies of different genus $h$, leading to the
topological expansion \cite{Becker:2006dvp}
\begin{equation}
\mathcal{A}=\sum_{h=0}^{\infty} g_s^{\,2h-2}\,F_h(\alpha')\,.
\end{equation}

This mirrors the genus expansion in the large-$N_c$ gauge theory discussed in \eqref{eq:stringGenus}, with the
identification $g_s\sim 1/N_c$ in AdS/CFT.

In backgrounds relevant for holography, it is common to express the $\alpha'$ expansion in terms of the ratio
$\ell_s/L$. Taking $\ell_s/L\ll 1$ suppresses higher-derivative terms and makes the Einstein--Hilbert action a good
approximation. Moreover, the gravitational coupling is set by $G_{10}\propto g_s^{2}\ell_s^{8}$ in ten dimensions, so
$g_s\ll 1$ suppresses quantum gravity corrections.

Since String Theory reduces to classical gravity in this regime, the bulk partition function can be approximated by a
saddle-point expansion around classical solutions $\phi_{\mathrm{cl}}$ (solutions of the classical equations of motion
$\delta S/\delta\phi=0$). To leading order,
\begin{equation}
Z_{\text{bulk}}[\phi_0]
\;\sim\;
\sum_{\{\phi_{\mathrm{cl}}\}}
\exp\,\left(i\,S_{\text{on-shell}}[\phi_{\mathrm{cl}}]\right)\,.
\end{equation}
where $\phi_0$ denotes boundary data for the bulk fields. This saddle-point approximation is the starting point for
the AdS/CFT dictionary: it allows one to relate the on-shell gravitational action (evaluated on solutions with fixed
boundary conditions) to the generating functional of the dual field theory, as developed in the next subsection.

\subsection{Equivalence between Type IIB String Theory and $\mathcal{N}=4$ SYM}

Type IIB String Theory is a ten-dimensional theory of closed strings whose spectrum contains a massless sector
(described at low energies by Type IIB supergravity) together with an infinite tower of massive excitations at the
string scale $m_s\sim 1/\ell_s$, where $\ell_s^2=\alpha'$. The perturbative expansion is controlled by the string
coupling $g_s$.

A crucial non-perturbative ingredient is the existence of D$p$-branes, $(p+1)$-dimensional hypersurfaces on which open
strings can end. Their tension is
\begin{equation}
T_{Dp}=\frac{1}{(2\pi)^p\,g_s\,\ell_s^{\,p+1}}\,,
\end{equation}
and they carry Ramond--Ramond charge, so they act as sources for the corresponding supergravity fields. A stack of
$N_c$ coincident D3-branes ($p=3$) admits two complementary low-energy descriptions \cite{Ammon:2015wua}. In the open-string
picture, the light open-string modes localized on the branes give a four-dimensional $\mathcal N=4$ supersymmetric
Yang--Mills theory with gauge group $SU(N_c)$ (the overall $U(1)$ decouples): it contains a gauge field $A_\mu$, six
real scalars $\Phi^i$, and four Weyl fermions, all in the adjoint representation. The Yang--Mills coupling is related
to the string coupling by
\begin{equation}
g_{\mathrm{YM}}^{2}=4\pi g_s\,,
\end{equation}
and since $\beta(g_{\mathrm{YM}})=0$, the theory is conformal at all energy scales.

In the closed-string picture, the same D3-branes gravitate and backreact on the geometry. In the near-horizon limit,
their supergravity solution approaches $\mathrm{AdS}_5\times S^5$, with metric
\begin{equation}
ds^2=\frac{L^2}{z^2}\Bigl(dt^2-d\vec{x}^{\,2}-dz^2\Bigr)+L^2\,d\Omega_5^2\,,
\end{equation}
where $\vec{x}\in\mathbb{R}^3$, and the common curvature radius $L$ of $\mathrm{AdS}_5$ and $S^5$ is fixed by the
D3-brane charge as
\begin{equation}\label{rpa}
L^4 = 4\pi g_s N_c\,\ell_s^{\,4}\,.
\end{equation}

This background realizes the bosonic symmetry $SO(2,4)\times SO(6)$, matching the conformal and $R$-symmetry groups of
$\mathcal N=4$ SYM.

Maldacena's conjecture \cite{Maldacena:1997re} is that the common low-energy, near-horizon sector of this single
brane system can be decoupled, yielding an exact equivalence between Type IIB String Theory on
$\mathrm{AdS}_5\times S^5$ and $\mathcal N=4$ SYM in four dimensions. A useful way to organize the parameter matching
is through the 't~Hooft coupling $\lambda=g_{\mathrm{YM}}^{2}N_c$. Combining \eqref{rpa} with $g_{\mathrm{YM}}^{2}=4\pi
g_s$ gives
\begin{equation}
\left(\frac{L}{\ell_s}\right)^4=\lambda\,,
\end{equation}
so the supergravity limit (suppression of $\alpha'$ corrections) corresponds to $\lambda\gg 1$.
String loop corrections are controlled by $g_s$, and can be expressed in gauge-theory variables as
$g_s=g_{\mathrm{YM}}^{2}/(4\pi)=\lambda/(4\pi N_c)$. Equivalently, in ten dimensions one has
$G_{10}\propto g_s^2\,\ell_s^{\,8}$, so quantum-gravity corrections are suppressed at large $N_c$. Parametrically,
this implies
\begin{equation}
\left(\frac{\ell_P}{L}\right)^8 \sim \frac{1}{N_c^2}\,,
\end{equation}
up to numerical factors, where $\ell_P$ is the ten-dimensional Planck length. Thus the classical supergravity regime
corresponds to
\begin{equation}
N_c\gg 1,\qquad \lambda\gg 1\,.
\end{equation}
confirming that AdS/CFT is a strong/weak duality: strongly coupled $\mathcal N=4$ SYM is captured by weakly coupled
gravity on $\mathrm{AdS}_5\times S^5$.

Beyond providing a concrete realization of the holographic principle, the correspondence has several far-reaching
implications \cite{Ammon:2015wua}. It relates black hole thermodynamics in AdS to thermal physics in the dual field
theory and enables direct computations of real-time response (e.g.\ transport coefficients such as $\eta/s$) at strong
coupling. Moreover, although $\mathcal N=4$ SYM is conformal and does not confine, deformations of the AdS/CFT setup
motivate holographic models closer to QCD; this perspective will be developed in the next Section.

\subsection{The holographic dictionary}

In AdS/CFT, one relates a $d$-dimensional Conformal Field Theory to a gravitational (string) theory on
$\mathrm{AdS}_{d+1}$. The bulk theory is described by a partition function with fixed boundary conditions
for the bulk fields \cite{Nastase:2015wjb}:
\begin{equation}
Z_{\mathrm{AdS}}[\phi_0]
=
\int \mathcal{D}\Phi\;
\exp\!\Bigl\{ i\,S_{\mathrm{bulk}}[\Phi] \Bigr\}
\Biggr|_{\Phi|_{\partial \mathrm{AdS}}=\phi_0(x)}\,,
\end{equation}
where $\Phi$ collectively denotes bulk fields and $\phi_0(x)$ their boundary values. The AdS/CFT correspondence asserts
that this is equal to the generating functional of the dual CFT in the presence of sources,
\begin{equation}
Z_{\mathrm{CFT}}[\phi_0]=Z_{\mathrm{AdS}}[\phi_0]\,,
\end{equation}
with the identification that $\phi_0(x)$ is the source coupled to the dual operator. In practice, it is often more
useful to work with the connected generating functional $W_{\mathrm{CFT}}[\phi_0]\equiv -i\ln Z_{\mathrm{CFT}}[\phi_0]$,
since functional derivatives of $W_{\mathrm{CFT}}$ generate connected correlators. In the classical gravity limit,
the bulk path integral is dominated by saddle points and one has schematically
\begin{equation}
W_{\mathrm{CFT}}[\phi_0]\;\simeq\;S_{\mathrm{bulk,on\text{-}shell}}[\Phi]\Big|_{\Phi|_{\partial \mathrm{AdS}}=\phi_0}\,.
\end{equation}
which is the practical starting point for computing CFT correlation functions from bulk dynamics.

To illustrate the dictionary, consider a scalar field $\phi(x,z)$ in the Poincar\'e patch of $\mathrm{AdS}_{d+1}$
(with boundary coordinates $x^\mu$ and radial coordinate $z$) with action
\begin{equation}
S=\frac{1}{2}\int d^{d}x\,dz\;\sqrt{-g}\,
\left[g^{MN}\partial_M\phi\,\partial_N\phi-m^2\phi^2\right]\,,
\end{equation}

The resulting equation of motion is
\begin{equation}
z^{d+1}\partial_z\!\left[z^{1-d}\partial_z\phi(x,z)\right]
-z^2\Box\,\phi(x,z)-m^2L^2\,\phi(x,z)=0\,,
\end{equation}
where $\Box=\eta^{\mu\nu}\partial_\mu\partial_\nu$ is the $d$-dimensional boundary d'Alembertian. Near the boundary
($z\to 0$), solutions behave as \cite{Ammon:2015wua}
\begin{equation}
\phi(x,z)\;\approx\;\phi_0(x)\,z^{\Delta_-}+\sigma(x)\,z^{\Delta_+}\,,
\qquad
\Delta_\pm=\frac{d}{2}\pm\sqrt{\frac{d^2}{4}+m^2R^2}\,.
\end{equation}

In the standard quantization, the coefficient of the leading term is identified with the source $\phi_0(x)$ for a
scalar operator $\mathcal{O}_\Delta$ of dimension $\Delta=\Delta_+$, while the subleading coefficient $\sigma(x)$ is
proportional to the expectation value $\langle\mathcal{O}_\Delta\rangle$. The mass--dimension relation becomes
\begin{equation}
m^2L^2=\Delta(\Delta-d)\,,
\end{equation}

This relation allows negative $m^2$ without instability provided the Breitenlohner--Freedman bound is satisfied,
\begin{equation}
m^2L^2\ge -\frac{d^2}{4}\,,
\end{equation}

When $-d^2/4\le m^2L^2<-d^2/4+1$, an alternative quantization exchanging $\Delta_+$ and $\Delta_-$ is also possible.

For higher-spin fields, the same logic applies: the falloff near the boundary determines the operator dimension. For a
$p$-form field in $\mathrm{AdS}_{d+1}$, one has the standard generalization \cite{Ammon:2015wua}
\begin{equation}\label{bulkmass}
\boxed{m^2L^2=(\Delta-p)(\Delta+p-d)\,.}
\end{equation}

In particular, a conserved CFT current $J^\mu$ has $\Delta=d-1$, and inserting $p=1$ into \eqref{bulkmass} gives
$m^2=0$, consistent with a massless bulk gauge field. This is the precise sense in which global symmetries of the
boundary theory correspond to gauge symmetries in the bulk: coupling a boundary source $A_\mu^{(0)}$ to $J^\mu$ via
$\int d^d x\,A_\mu^{(0)}J^\mu$ is invariant under $A_\mu^{(0)}\to A_\mu^{(0)}+\partial_\mu\alpha$ only if the current is
conserved, $\partial_\mu J^\mu=0$ (or $\nabla_\mu J^\mu=0$ in curved space).

Putting these ingredients together, the holographic dictionary states that: the CFT generating functional in the
presence of sources is reproduced by the bulk path integral with boundary conditions fixed by those sources; in the
regime relevant for classical gravity (large $N_c$ and strong 't~Hooft coupling), the bulk partition function is
well-approximated by the exponential of the on-shell action; and operator quantum numbers in the CFT (dimension and
spin) are encoded in the near-boundary asymptotics and bulk masses of the dual fields. Finally, all known CFTs of
interest in AdS/CFT contain a conserved stress-energy tensor $T_{\mu\nu}$ of dimension $\Delta=d$, and the dictionary
identifies its dual bulk field as a massless spin-2 excitation, the graviton, which is why the correspondence is
fundamentally a gauge/gravity duality.

\subsection{Potential application of holography to QCD}

The AdS/CFT correspondence provides a useful organizing principle for modeling aspects of strongly coupled QCD, even
though no exact weakly curved string dual of real-world QCD is known. The basic strategy is therefore to construct
\emph{effective} holographic descriptions that reproduce selected QCD phenomena. Two complementary philosophies are
commonly pursued: \emph{top-down} models, in which one engineers a brane construction in String Theory whose low-energy
gauge dynamics shares qualitative features with QCD, and \emph{bottom-up} models, in which one builds a five-dimensional
gravity theory by phenomenological input, aiming to capture key QCD observables without requiring a complete UV
completion in String Theory.

A central motivation for expecting a stringy description at all comes from the large-$N_c$ limit introduced by
't~Hooft. One generalizes $SU(3)$ to $SU(N_c)$ and takes $N_c\to\infty$ while keeping the 't~Hooft coupling
$\lambda=g_{\mathrm{YM}}^{2}N_c$ fixed, so that perturbation theory reorganizes into a topological (genus) expansion
dominated by planar diagrams. This diagrammatics resembles a string worldsheet expansion and suggests that confining
gauge theories may admit an effective higher-dimensional description, although important differences remain between QCD
and the best-understood holographic example, $\mathcal{N}=4$ SYM.

In particular, QCD is not conformal and develops a dynamically generated scale, exhibits confinement and a mass gap at
low energies, and contains quarks in the fundamental representation with approximate chiral symmetry and its
spontaneous breaking. Holographic QCD models must therefore incorporate these ingredients beyond the AdS$_5\times S^5$
setup \cite{Ammon:2015wua}. 

In practice, this is achieved by explicitly breaking conformal invariance through a
non-trivial background (for example, a warped geometry and/or dilaton profile) that introduces an infrared scale
associated with confinement; by implementing a discrete hadron spectrum via an effective IR modification such as a
hard-wall or soft-wall; and by adding fields (bottom-up) or flavor branes (top-down) that realize chiral symmetry and
its breaking in the mesonic sector. Finite temperature physics is incorporated by replacing the vacuum geometry with a
black hole (or black brane) background, whose Hawking temperature is identified with the gauge-theory temperature; this
provides a tractable framework for studying the quark--gluon plasma and real-time observables such as spectral
functions and transport coefficients. A well-known example is the universal strong-coupling result $\eta/s=1/4\pi$ in
simple holographic models, which is often used as a benchmark for strongly coupled plasma dynamics.

Despite the absence of an exact dual for QCD, these AdS/CFT-inspired constructions offer a controlled way to model
non-perturbative phenomena and to compute real-time quantities that are difficult to access by other means. In the
remainder of this dissertation, we will develop and apply such holographic QCD models to explore mesonic spectral
properties at finite temperature, density, and in external fields.

A final caveat is worth stating explicitly. Most practical holographic calculations are performed in the regime of
large $N_c$ and strong 't~Hooft coupling, where the bulk description reduces to classical gravity, while real QCD has
$N_c=3$ and a running coupling that is not parametrically large in all regimes. Bottom-up holographic QCD models are
therefore not derived from first principles and necessarily involve modeling choices (e.g.\ for the warp factor,
dilaton potential, and flavor sector). Their predictions should be viewed as controlled results within a given
effective framework: they are particularly valuable for identifying robust qualitative trends and for computing
real-time observables in strongly coupled regimes, but they are not a substitute for lattice QCD or perturbative QCD
in the domains where those methods apply.

\subsection{Chemical potential in QCD and holography}

Many QCD phenomena of phenomenological interest, such as dense nuclear matter, heavy-ion collisions at finite baryon density,
and the low-temperature phase structure, require introducing a chemical potential $\mu$ for a conserved charge,
typically baryon number. The grand-canonical ensemble is defined by
\begin{equation}
Z(T,\mu)=\mathrm{Tr}\,\exp\!\big[-\beta\,(H-\mu N)\big]\,,
\qquad
\Omega(T,\mu)\equiv -T\ln Z\,,
\qquad
\beta=\frac{1}{T}\,,
\end{equation}
and the first law takes the form
\begin{equation}
dE = T\,dS - P\,dV + \mu\, dN\,.
\end{equation}
so that $\mu$ is conjugate to the conserved charge $N$. In field theory, a chemical potential can be viewed as a
constant source for the time component of a conserved current: adding $\int d^4x\, \mathcal{A}_\mu J^\mu$ with
$\mathcal{A}_0=\mu$ implements the grand-canonical deformation and modifies thermodynamics and correlation functions.

It is useful to keep in mind the qualitative QCD phase diagram in the $(T,\mu_B)$ plane. At $\mu_B=0$, lattice
calculations indicate a rapid but analytic crossover between a hadronic resonance gas and a deconfined quark--gluon
plasma at a pseudo-critical temperature $T_c(0)\approx 155~\mathrm{MeV}$ \cite{HotQCD:2019TcKappa}, while at small
$\mu_B$ the continuation of this crossover line can be parameterized by a mild curvature,
\begin{equation}
T_c(\mu_B)\simeq T_c(0)\left[1-\kappa\left(\frac{\mu_B}{T_c(0)}\right)^2+\mathcal{O}(\mu_B^4)\right]\,.
\end{equation}
with $\kappa=\mathcal{O}(10^{-2})$ in current lattice determinations \cite{HotQCD:2019TcKappa,Ding:2024Curvature}.
Direct lattice access to large $\mu_B$ is obstructed by the fermion sign problem \cite{Nagata:2022SignProblem}, and
information at finite density is often inferred from Taylor expansions around $\mu_B=0$, analytic continuation from
imaginary $\mu_B$, and effective models constrained by lattice and experiment. A long-standing possibility is the
existence of a critical end point (CEP) at $(T_{\rm CEP},\mu_{B,{\rm CEP}})$ where the crossover turns into a
first-order transition toward lower $T$ and larger $\mu_B$; if present, the first-order line would exhibit latent heat
and satisfy a Clausius--Clapeyron relation
\begin{equation}
\frac{dT}{d\mu_B}\bigg|_{\text{1st order}}=-\frac{\Delta n_B}{\Delta s}\,,
\end{equation}
where $\Delta n_B$ and $\Delta s$ are discontinuities of baryon density and entropy density across the coexistence
line. At very low temperature and sufficiently high density, asymptotic freedom and attractive quark--quark channels
suggest color-superconducting phases (e.g.\ 2SC and CFL), potentially relevant for the cores of neutron stars.

Thermodynamics and fluctuations at finite $\mu_B$ are encoded in the pressure $p(T,\mu_B)=T\,V^{-1}\ln Z$ and its
derivatives. The baryon density and generalized susceptibilities,
\begin{equation}
n_B(T,\mu_B)=\frac{\partial p}{\partial \mu_B}\,,
\qquad
\chi_n^B(T,\mu_B)=\frac{\partial^n}{\partial (\mu_B/T)^n}\left(\frac{p}{T^4}\right)\,,
\end{equation}
control conserved-charge cumulants via fluctuation--dissipation \cite{Asakawa:2016FluctuationsReview}, with commonly
used ratios
\begin{equation}
\frac{M}{\sigma^2}=\frac{\chi_1^B}{\chi_2^B}\,,\qquad
S\,\sigma=\frac{\chi_3^B}{\chi_2^B}\,,\qquad
\kappa\,\sigma^2=\frac{\chi_4^B}{\chi_2^B}\,.
\end{equation}
which serve as sensitive probes of criticality. In heavy-ion collisions, the effective $\mu_B$ at chemical freeze-out
is small at top LHC energies and increases along the RHIC Beam Energy Scan \cite{Chen:2024BESReview}; this motivates
searches for nonmonotonic behavior in fluctuation observables (noting that experimentally one often measures net-proton
cumulants as a proxy for net-baryon number \cite{Luo:2017CPFluctuationsOverview}, and acceptance/non-equilibrium effects
require care). Recent high-order net-proton fluctuation measurements by STAR provide a prominent example of such
observables \cite{STAR:2023C56NetProton,STAR:2025C56Erratum}. These QCD-side considerations set targets for effective
descriptions, including holographic models, which should reproduce the small-$\mu_B$ crossover behavior, allow for
first-order structure and a CEP if present, and capture the pattern of susceptibilities that govern fluctuations.

Holography provides a complementary description in which a global $U(1)$ symmetry of the boundary theory is dual to a
bulk gauge field $A_M$. The central statement of the dictionary is that the CFT generating functional with sources is
equal to the bulk gravitational partition function with matching boundary data, and at large $N_c$ and strong coupling
the latter is well approximated by the on-shell bulk action. For a conserved current $J^\mu$, the corresponding source
is the boundary value of the bulk gauge potential, so turning on a chemical potential amounts to fixing the asymptotic
value of the time component,
\begin{equation}
\mu \;\longleftrightarrow\; A_t^{(0)}\equiv \lim_{z\to 0} A_t(z)\,,
\end{equation}
in a static, homogeneous state. Gauge invariance implies that the physical chemical potential is the potential
difference between boundary and horizon. At finite temperature the relevant backgrounds contain a horizon at $z=z_h$,
and regularity of the Euclidean gauge potential fixes a convenient gauge choice
\begin{equation}
A_t(z_h)=0\,,
\end{equation}
so that $\mu=A_t(0)-A_t(z_h)=A_t(0)$. The charge density is read off from the conserved radial electric flux. For a
bulk Maxwell sector with gauge-kinetic function $f(\phi)$,
\begin{equation}
\nabla_M\,\big(f(\phi)F^{MN}\big)=0\,,
\end{equation}
and in static, translationally invariant backgrounds this implies
\begin{equation}
\partial_z\!\Big(\sqrt{-g}\,f(\phi)\,F^{zt}\Big)=0
\quad\Longrightarrow\quad
\sqrt{-g}\,f(\phi)\,F^{zt}=\rho\,,
\end{equation}
where $\rho$ is constant along the radial direction and equals the boundary charge density
$\rho=\langle J^t\rangle$. Using $F^{zt}=g^{zz}g^{tt}\partial_z A_t$, one obtains the first integral
\begin{equation}
\boxed{\partial_z A_t(z)=\frac{\rho}{\sqrt{-g}\,f(\phi)\,g^{zz}g^{tt}}\,,
\qquad
\mu=\int_{z_h}^{0}\!dz\;\frac{\rho}{\sqrt{-g}\,f(\phi)\,g^{zz}g^{tt}}\,.}
\end{equation}
which makes explicit that $\mu$ is fixed by the line integral of the radial electric field from the horizon to the
boundary once $\rho$ is specified. In the grand-canonical ensemble, one prescribes $\mu$ and solves the coupled bulk
equations with $A_t(z_h)=0$, extracting $\rho$ from the asymptotic flux; in the canonical ensemble one instead fixes
$\rho$ and allows $A_t^{(0)}$ to adjust, with the thermodynamic potential obtained by a Legendre transform of the
on-shell action.

In phenomenological applications to QCD, the relevant $U(1)$ is typically baryon number, so one identifies
$\mu=\mu_B$. Holographic QCD models often represent this sector by a bulk gauge field dual to the baryon current; more
detailed flavor structure can be incorporated by enlarging the bulk gauge sector, but the elementary identifications
remain universal: the chemical potential is the boundary value of $A_t$, the density is the conserved radial flux, the
choice of ensemble is encoded in boundary conditions for $A_t$ (Dirichlet) or its normal derivative (Neumann), and the
thermodynamic potential is the Euclidean on-shell action with the appropriate counterterms. This setup, together with
horizon regularity, is the basis for computing phase diagrams, susceptibilities, and real-time response at finite
density in the holographic models developed in the Sections that follow.

\newpage

\section{Holographic computation of the mesonic mass spectra}\label{spectra}

The AdS/CFT correspondence provides a useful framework for studying strongly coupled gauge theories by mapping them to
higher-dimensional gravitational models. In QCD-motivated holography, one of the most direct applications is the
computation of mesonic mass spectra: one introduces appropriate bulk fields propagating in a five-dimensional geometry
and interprets their normalizable fluctuations as hadron-like excitations in an effective four-dimensional dual
description.

We begin with the simplest setting, a free scalar field in pure $\mathrm{AdS}_5$ in the Poincar\'e patch. Because this
background is maximally symmetric and contains no intrinsic infrared scale, the resulting fluctuations do not form a
discrete set of bound states; instead, one finds a continuum of modes. This illustrates why additional IR structure is
required to mimic confinement. In later subsections, we will introduce such structure through standard deformations,
most notably hard-wall and soft-wall models, which break conformal invariance by imposing an IR cutoff or by turning on
a non-trivial background profile (e.g.\ a dilaton). These modifications yield normalizable solutions with discrete
eigenvalues, providing a simple holographic mechanism for generating mesonic resonance spectra.

\subsection{Free scalar field in $\mathrm{AdS}_5$: absence of a discrete spectrum}\label{sec41}

Following \cite{Zou:2018eam}, we analyze a free scalar field in the $\mathrm{AdS}_5$ background. The action is
\begin{equation}\label{Acfree}
S=\frac{1}{2}\int d^{4}x \, dz \,\sqrt{-g}\,
\Bigl(
g^{MN}\,\partial_{M} \Phi \,\partial_{N} \Phi -\mu^{2} \,\Phi^{2}
\Bigr)\,,
\end{equation}
where $\mu$ is the five-dimensional mass. Varying with respect to $\Phi$ gives the curved-space Euler--Lagrange
equation
\begin{equation}\label{KGcurved}
\frac{1}{\sqrt{-g}}\,\partial_{M}\!\left(\sqrt{-g}\,g^{MN}\,\partial_{N}\Phi\right)+\mu^{2}\Phi=0\,,
\end{equation}
or, equivalently, $\nabla^{M}\nabla_{M}\Phi+\mu^{2}\Phi=0$.

In the Poincar\'e patch, we write the $\mathrm{AdS}_5$ metric as
\begin{equation}\label{AdSmetricPoincare}
ds^{2}=e^{2A(z)}\bigl(\eta_{\alpha\beta}dx^{\alpha}dx^{\beta}-dz^{2}\bigr)\,,
\qquad
e^{2A(z)}=\frac{L^{2}}{z^{2}}\,,
\qquad
A(z)=\ln\!\left(\frac{L}{z}\right)\,,
\end{equation}
with mostly-minus $\eta_{\alpha\beta}=\mathrm{diag}(1,-1,-1,-1)$. Then
\begin{equation}
g^{MN}=e^{-2A(z)}\eta^{MN}\,,
\qquad
\sqrt{-g}=e^{5A(z)}\,.
\end{equation}
where $\eta^{MN}$ is the 5D flat space metric. 

Substituting into \eqref{KGcurved} yields
\begin{equation}\label{free1}
\eta^{\alpha\beta}\partial_{\alpha}\partial_{\beta}\Phi
-\partial_{z}^{2}\Phi
-3A'(z)\,\partial_{z}\Phi
+\mu^{2}e^{2A(z)}\Phi
=0\,.
\end{equation}

Using $A'(z)=-1/z$ and $e^{2A(z)}=L^{2}/z^{2}$, this becomes
\begin{equation}\label{free1b}
\left[\eta^{\alpha\beta}\partial_{\alpha}\partial_{\beta}
-\partial_z^2+\frac{3}{z}\partial_z+\frac{(\mu L)^2}{z^2}\right]\Phi(x,z)=0\,,
\end{equation}

We Fourier transform along the boundary direction,
\begin{equation}\label{FourierPhi}
\Phi(x,z)=\int\frac{d^{4}q}{(2\pi)^{4}}\,e^{iq\cdot x}\,\psi(q,z)\,,
\end{equation}
so that $\eta^{\alpha\beta}\partial_{\alpha}\partial_{\beta}\to -q^{2}$ with
$q^{2}\equiv\eta^{\alpha\beta}q_\alpha q_\beta$. The radial equation becomes
\begin{equation}\label{momentumfree}
\left(
-\partial_{z}^{2}
+\frac{3}{z}\,\partial_{z}
+\frac{(\mu L)^{2}}{z^{2}}
-q^{2}
\right)\psi(q,z)=0\,.
\end{equation}

For timelike momentum one identifies $q^{2}=m^{2}$ with the four-dimensional mass parameter. We can write Eq.~\eqref{momentumfree} in a standard Bessel equation form, introducing the dimensionless variable $u=qz$ and the Ans\"atze
\begin{equation}\label{ansatzBessel}
\psi(q,z)=z^{\alpha}\,\xi(u)\,,
\qquad u=qz\,,
\end{equation}
where primes on $\xi$ denote derivatives with respect to $u$. Since $\partial_z = q\,\partial_u$, one finds
\begin{align}
\partial_z \psi
&= \alpha z^{\alpha-1}\xi(u) + q z^{\alpha}\xi'(u)\,, \label{dzpsi}\\
\partial_z^2 \psi
&= \alpha(\alpha-1)z^{\alpha-2}\xi(u)
+2\alpha q z^{\alpha-1}\xi'(u)
+q^2 z^{\alpha}\xi''(u)\,, \label{dz2psi}
\end{align}

Substituting \eqref{dzpsi}--\eqref{dz2psi} into \eqref{momentumfree} gives
\begin{align}
0
={}&
-\alpha(\alpha-1)z^{\alpha-2}\xi
-2\alpha q z^{\alpha-1}\xi'
-q^2 z^{\alpha}\xi''
+3\alpha z^{\alpha-2}\xi
+3q z^{\alpha-1}\xi'
+(\mu L)^2 z^{\alpha-2}\xi
-q^2 z^{\alpha}\xi \nonumber\\
={}&
z^{\alpha-2}\Bigl[\!-\alpha(\alpha-1)+3\alpha+(\mu L)^2\Bigr]\xi
+q z^{\alpha-1}\bigl(-2\alpha+3\bigr)\xi'
-q^2 z^\alpha\bigl(\xi''+\xi\bigr)\,, \label{afterSub}
\end{align}

Divide by $q^2 z^\alpha$ and use $z^{-1}=q/u$ and $z^{-2}=q^2/u^2$ to obtain
\begin{equation}
\xi''(u)+\frac{2\alpha-3}{u}\,\xi'(u)
+\left[1-\frac{-\alpha(\alpha-1)+3\alpha+(\mu L)^2}{u^2}\right]\xi(u)=0\,,
\end{equation}

Choosing $\alpha$ so that the coefficient of $\xi'(u)$ becomes $1/u$ fixes
\begin{equation}
2\alpha-3=1 \Longrightarrow \alpha=2\,.
\end{equation}

With $\alpha=2$, the equation reduces to the Bessel equation
\begin{equation}\label{besseleq}
u^{2}\,\xi''(u)+u\,\xi'(u)+\bigl(u^{2}-\nu^{2}\bigr)\xi(u)=0\,,
\qquad
\nu^{2}=(\mu L)^{2}+4\,.
\end{equation}

Therefore, the general solution is
\begin{equation}\label{generalBessel}
\psi(q,z)=z^{2}\Bigl[A\,J_{\nu}(qz)+B\,Y_{\nu}(qz)\Bigr]\,.
\end{equation}
with $J_{\nu}$ and $Y_{\nu}$ the Bessel functions of the first and second kind and $A,B$ constants determined by
boundary conditions. Near the AdS boundary $z\to 0$, the $Y_\nu$ branch is more singular; in the standard quantization
one typically keeps the less singular/normalizable behavior, consistent with the AdS/CFT identification
$\Delta=2+\nu$ for $\mathrm{AdS}_5/\mathrm{CFT}_4$.

It is also useful to cast \eqref{momentumfree} into Schr\"odinger form. Defining
\begin{equation}
\psi(q,z)=z^{3/2}\,\chi(q,z)\,,
\end{equation}
eliminates the first-derivative term and yields
\begin{equation}
\boxed{-\partial_z^2\chi(q,z)+V(z)\chi(q,z)=q^{2}\chi(q,z)\,,
\qquad
V(z)=\frac{4(\mu L)^2+15}{4z^{2}}\,.}
\end{equation}

Crucially, in the Poincar\'e patch $z\in(0,\infty)$ this potential does not generate a discrete set of normal modes:
there is no infrared boundary condition that would quantize $q^{2}$. As a result, solutions exist for continuous
values of $q^{2}$, reflecting the scale invariance of pure $\mathrm{AdS}_5$ and the absence of an intrinsic confinement
scale. To obtain a discrete spectrum reminiscent of mesonic resonances in QCD-like theories, one must break conformal
invariance by introducing additional IR structure, such as an explicit cutoff in $z$ (hard-wall) or a background
profile (soft-wall), which restricts normalizable solutions and leads to quantized masses.

\subsection{The hard-wall model}

A simple way to obtain a discrete spectrum from the pure $\mathrm{AdS}_5$ wave equation is to impose a ``hard-wall''
in the radial direction at a finite position $z=z_0$. Physically, following the seminal hard-wall construction
\cite{Erlich:2005qh}, this introduces an infrared scale by restricting the bulk coordinate to the finite interval
\begin{equation}
0< z \le z_0\,,
\end{equation}
thereby mimicking confinement by excluding the deep-interior region $z>z_0$.

In practice, one supplements the ultraviolet condition at $z\to 0$ (regularity/normalizability, which typically
eliminates the more singular $Y_\nu$ branch) with an infrared boundary condition at $z=z_0$. For the scalar example of
Sec.~\ref{sec41}, a convenient choice is a Dirichlet hard-wall condition,
\begin{equation}
\psi(q,z)\Big|_{z=z_0}=0\,,
\end{equation}
which quantizes the allowed values of the four-dimensional invariant momentum $q^2$.

Using the general solution \eqref{generalBessel}, regularity near $z\to 0$ selects
\begin{equation}
\psi(q,z)=A\,z^{2}\,J_{\nu}(qz)\,,
\end{equation}
so the hard-wall condition gives
\begin{equation}
z_0^{2}J_{\nu}(qz_0)=0
\Longrightarrow\
J_{\nu}(qz_0)=0\,.
\end{equation}

Hence $q$ must coincide with a zero of $J_\nu$. Denoting the $s$-th positive zero by $j_{\nu,s}$, the discrete mass
tower is
\begin{equation}
\boxed{q_s=\frac{j_{\nu,s}}{z_0}\,,
\qquad
M_s^2=q_s^2\,.}
\label{eq:hardwall_spectrum}
\end{equation}

This discretization is a direct consequence of imposing an IR boundary condition at finite $z_0$, which turns the
radial problem into a Sturm--Liouville eigenvalue problem.

\begin{figure}[h!]
\centering
\includegraphics[width=0.75\textwidth]{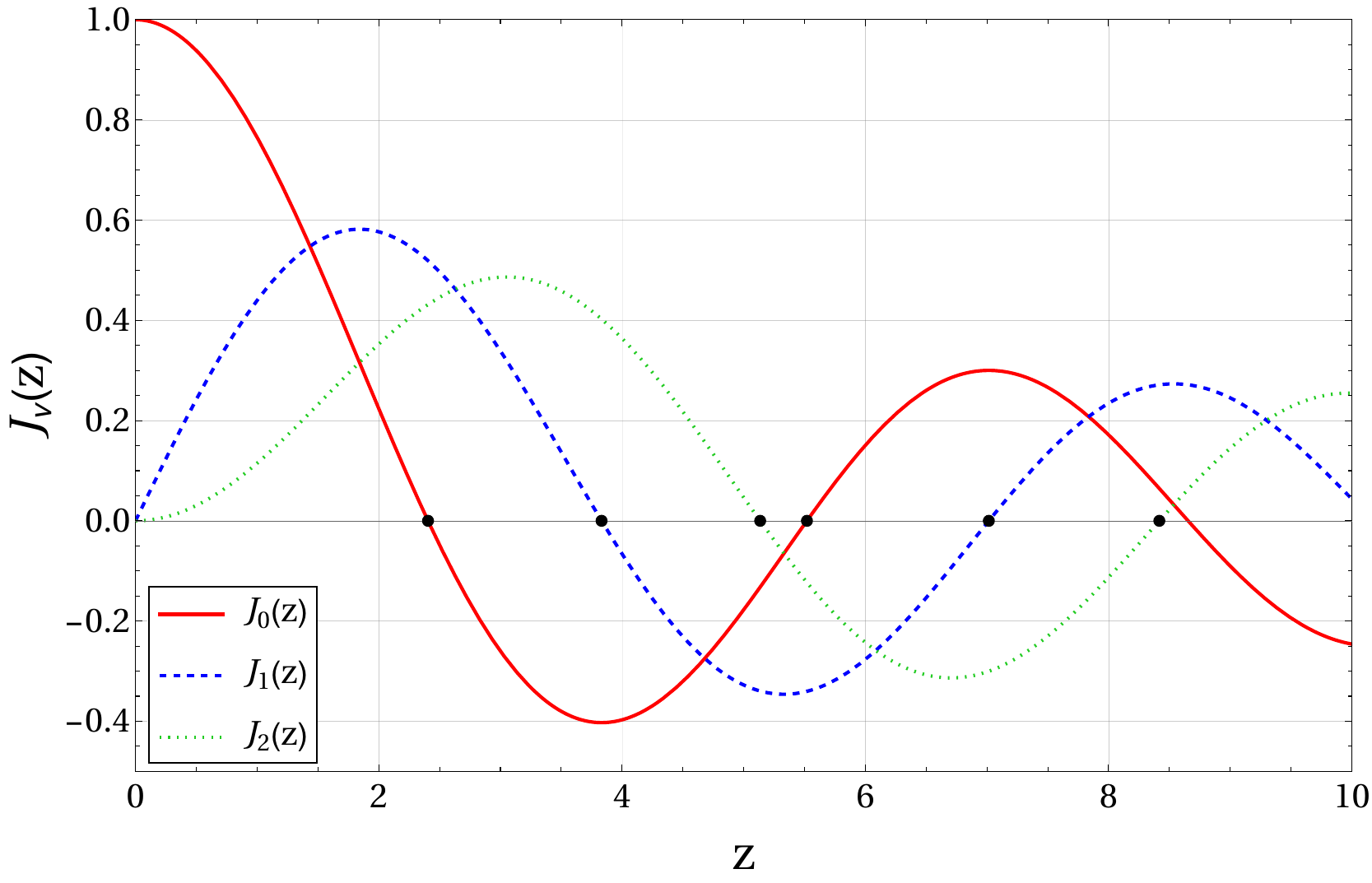}
\caption{The first three Bessel functions $J_{\nu}(z)$ (for representative orders) and their zeros.}
\label{figbessel}
\end{figure}

For illustration, Fig.~\ref{figbessel} shows the first few Bessel functions and their zeros; the quantization
condition simply picks out those zeros as the allowed values of $qz_0$.

The zeros $j_{\nu,s}$ can be obtained numerically. For $s\gg 1$ (and moderate $\nu$), they admit the asymptotic
expansion
\begin{equation}
j_{\nu,s}
\approx
\left(s+\frac{\nu}{2}-\frac{1}{4}\right)\pi
-\frac{4\nu^2-1}{8\pi\left(s+\tfrac{\nu}{2}-\tfrac{1}{4}\right)}
+\cdots\,.
\end{equation}
which makes explicit that $j_{\nu,s}\sim s\pi$ at large $s$, so the hard-wall model yields $M_s\sim s/z_0$ and
therefore $M_s^2\sim s^2$ (in contrast with linear Regge behavior $M_s^2\sim s$).

A standard phenomenological choice (in the original hard-wall AdS/QCD model \cite{Erlich:2005qh}) is to fix $z_0$ from
the lightest vector meson mass. In that channel one finds a Bessel quantization condition involving $J_0$, and fitting
the lightest root $j_{0,1}\simeq 2.405$ to $m_\rho\simeq 776\,\mathrm{MeV}$ gives
\begin{equation}
\frac{1}{z_0}\simeq \frac{m_\rho}{j_{0,1}}\simeq 323\,\mathrm{MeV}\,.
\end{equation}

With such value, Eq.~\eqref{eq:hardwall_spectrum} predicts, for $\nu=0$,
\begin{align}
m_{1} &= \frac{j_{0,1}}{z_{0}}
\approx 2.405 \times 323\,\mathrm{MeV}
\approx 777\,\mathrm{MeV}\,, \nonumber\\
m_{2} &= \frac{j_{0,2}}{z_{0}}
\approx 5.520 \times 323\,\mathrm{MeV}
\approx 1783\,\mathrm{MeV}\,, \nonumber\\
m_{3} &= \frac{j_{0,3}}{z_{0}}
\approx 8.654 \times 323\,\mathrm{MeV}
\approx 2795\,\mathrm{MeV}\,.
\end{align}
The ground state is fixed by construction, while the higher excitations illustrate both the utility and the
limitations of the hard-wall: it produces discrete states and a mass gap, but tends to overestimate higher radial
excitations and does not reproduce linear Regge trajectories, motivating the soft-wall improvement discussed next.

In summary, the hard-wall model introduces an explicit IR scale through the cutoff at $z=z_0$, converting the
continuous spectrum of pure $\mathrm{AdS}_5$ into a discrete tower of modes. Although it is a simplified picture of
confinement, implemented by a sharp boundary condition rather than a dynamical mechanism, it provides a transparent
holographic realization of mass quantization and serves as a baseline for more refined constructions.

\subsection{The soft-wall model}

A smooth way to generate a discrete spectrum in a holographic setting is to introduce a background dilaton profile,
as first proposed in the soft-wall model of \cite{Karch:2006pv}. For a scalar field $\Phi(x,z)$ in $\mathrm{AdS}_5$,
one modifies the action by a dilaton factor,
\begin{equation}\label{eq:softwall_action}
S=\frac{1}{2}\int d^{4}x\,dz\,\sqrt{-g}\;e^{-\phi(z)}
\Bigl(g^{MN}\partial_M\Phi\,\partial_N\Phi-\mu^{2}\Phi^{2}\Bigr)\,,
\end{equation}
where $A(z)=\ln(L/z)$ so that $e^{2A}=L^2/z^2$ as usual. The minus sign in $e^{-\phi}$ is essential: for
$\phi(z)=\kappa z^2$ it suppresses the deep-interior region and effectively provides an infrared ``soft'' cutoff.

Varying \eqref{eq:softwall_action} with respect to $\Phi$ yields
\begin{equation}\label{eq:softwall_eom_full}
\frac{1}{\sqrt{-g}}\,\partial_M\!\left(\sqrt{-g}\,e^{-\phi}\,g^{MN}\partial_N\Phi\right)+e^{-\phi}\mu^2\Phi=0\,.
\end{equation}

In the $\mathrm{AdS}_5$ metric \eqref{AdSmetricPoincare}, this becomes
\begin{equation}\label{eq:softwall_eom}
\eta^{\alpha\beta}\partial_\alpha\partial_\beta\Phi
-\partial_z^2\Phi
+\left[\phi'(z)-3A'(z)\right]\,\partial_z\Phi
+\mu^2 e^{2A(z)}\Phi
=0\,,
\end{equation}
where primes denote derivatives with respect to $z$.

Performing the 4D Fourier transform, $\Phi(x,z)=\int\frac{d^4q}{(2\pi)^4}e^{iq\cdot x}\,\psi(q,z)$, gives
\begin{equation}\label{eq:softwall_radial}
-\psi''(z)+\left[\phi'(z)-3A'(z)\right]\psi'(z)
+\left[\frac{(\mu L)^2}{z^2}-q^2\right]\psi(z)=0\,.
\end{equation}

For timelike momentum we identify $q^2=m^2$. It is convenient to remove the first-derivative term by the standard redefinition
\begin{equation}
\psi(z)=e^{\frac{1}{2}B(z)}\,\chi(z)\,,
\qquad
B(z)\equiv \phi(z)-3A(z)\,,
\end{equation}
which casts \eqref{eq:softwall_radial} into Schr\"odinger form
\begin{equation}\label{eq:softwall_sch}
-\chi''(z)+V_{\mathrm{sw}}(z)\,\chi(z)=q^2\,\chi(z)\,,
\end{equation}
with effective potential
\begin{equation}\label{eq:Vsw}
\boxed{V_{\mathrm{sw}}(z)
=\frac{4(\mu L)^2+15}{4z^2}
+\frac{1}{4}\phi'(z)^2-\frac{1}{2}\phi''(z)
+\frac{3}{2}\frac{\phi'(z)}{z}\,.}
\end{equation}
where we used $A'(z)=-1/z$ for $\mathrm{AdS}_5$.

For the canonical soft-wall choice $\phi(z)=\kappa z^2$ with $\kappa>0$, one has $\phi'(z)=2\kappa z$ and
$\phi''(z)=2\kappa$, and the potential becomes
\begin{equation}\label{eq:Vsw_kz2}
V_{\mathrm{sw}}(z)
=\kappa^2 z^2
+\frac{4(\mu L)^2+15}{4z^2}
+2\kappa\,.
\end{equation}

Eq.~\eqref{eq:softwall_sch} is then the radial equation of a one-dimensional harmonic oscillator with an
additional $1/z^2$ term. Normalizable solutions exist only for a discrete set of eigenvalues, yielding a linear
trajectory in the radial quantum number. Writing $\nu\equiv\sqrt{(\mu L)^2+4}$ (as in Sec.~\ref{sec41}), the spectrum takes
the form
\begin{equation}\label{eq:softwall_spectrum}
q_n^2 = 4\kappa\left(n+1+\frac{\nu}{2}\right)\,,
\qquad n=0,1,2,\ldots\,,
\end{equation}
and the corresponding normalizable wavefunctions may be written in terms of associated Laguerre polynomials as
\begin{equation}\label{eq:softwall_wf}
\chi_n(z)\propto z^{\nu+1/2}\,e^{-\frac{1}{2}\kappa z^2}\,
L_n^{(\nu)}(\kappa z^2)\,,
\qquad
\psi_n(z)=e^{\frac{1}{2}B(z)}\chi_n(z)\,.
\end{equation}

The key qualitative point is that the soft-wall background yields $q_n^2\propto n$, i.e.\ linear Regge-like behavior
in the squared masses.

A limitation of the original soft-wall model is that the chosen dilaton profile is imposed by hand and does not arise
from solving five-dimensional Einstein equations coupled consistently to matter. In the next part of this work, we
address this by adopting a dynamical Einstein--(Maxwell)--dilaton setup in which the background is determined from the
coupled field equations, providing a self-consistent holographic framework that simultaneously breaks conformal
invariance, generates an IR scale, and supports realistic mesonic spectra.

\subsection{Matrix-Numerov method and the scalar meson \( f_0 \)}

The Schr\"odinger-type eigenvalue problem
\(\,[-\partial_z^2+V(z)]\psi(z)=M^2\psi(z)\), for several backgrounds of interest, does not admit a simple closed-form solution, and one must determine the
eigenvalues and eigenfunctions numerically. To make our computation reproducible, we briefly summarize the numerical
scheme we employ. For the potentials encountered in this chapter, the resulting boundary-value problems are not
particularly stiff, and a high-accuracy finite-difference method such as Numerov provides an efficient approach.

The Numerov formula applies to second-order linear equations of the form
\begin{equation}
\psi''(x)=f(x)\,\psi(x)\,,
\end{equation}
discretized on an evenly spaced grid \(x_i=x_{\min}+id\) with step size \(d\). The Numerov update for the next point
takes the form
\begin{equation}\label{numerov}
\psi_{i+1}=
\frac{\psi_{i-1}\,(12-d^2 f_{i-1})-2\psi_i\,(5d^2 f_i+12)}
{d^2 f_{i+1}-12}
+\mathcal{O}(d^6)\,.
\end{equation}

When applied to the Schr\"odinger equation \(\psi''=(V-M^2)\psi\), this yields a discretization that is accurate to
\(\mathcal{O}(d^6)\) for smooth potentials.

Applying Numerov to the Schr\"odinger-like equation \eqref{eq:softwall_sch} gives the standard three-point stencil
\begin{equation}\label{numericf0}
-\frac{\psi_{i-1}-2\psi_i+\psi_{i+1}}{d^2}
+\frac{V_{i-1}\psi_{i-1}+10V_i\psi_i+V_{i+1}\psi_{i+1}}{12}
=
M^2\,\frac{\psi_{i-1}+10\psi_i+\psi_{i+1}}{12}\,,
\end{equation}
where \(V_i\equiv V(z_i)\) and \(M^2\) is the eigenvalue.

Representing \(\psi=(\psi_1,\ldots,\psi_N)^{T}\) as a vector on the interior grid points and introducing the matrices
\begin{equation}
A=\frac{I_{-1}-2I_0+I_{1}}{d^2}\,,
\qquad
B=\frac{I_{-1}+10I_0+I_{1}}{12}\,,
\qquad
C=\mathrm{diag}(V_1,\ldots,V_N)\,,
\end{equation}
with \(I_p\) denoting the matrix with ones on the \(p\)-th diagonal and zeros elsewhere, Eq.~(\ref{numericf0}) may be
written compactly as the generalized eigenvalue problem
\begin{equation}
\bigl[A + B\,C\bigr]\psi = M^2\,B\,\psi\,,
\end{equation}

Multiplying by \(B^{-1}\) from the left yields the standard matrix-Numerov eigenvalue equation
\begin{equation}\label{finalnum}
\boxed{\left[B^{-1}A + C\right]\psi = M^2\,\psi\,.}
\end{equation}

Boundary conditions are implemented by restricting to the interior grid and imposing Dirichlet conditions at the
endpoints, \(\psi(z_{\min})=\psi(z_{\max})=0\). Numerically, this corresponds to solving on an interval
\([z_{\min},z_{\max}]\) with a sufficiently large \(z_{\max}\) so that the low-lying eigenvalues are insensitive to
the IR cutoff.

\begin{figure}[htb!]
    \centering
    \begin{minipage}[t]{0.49\textwidth}
        \centering
        \includegraphics[width=\textwidth]{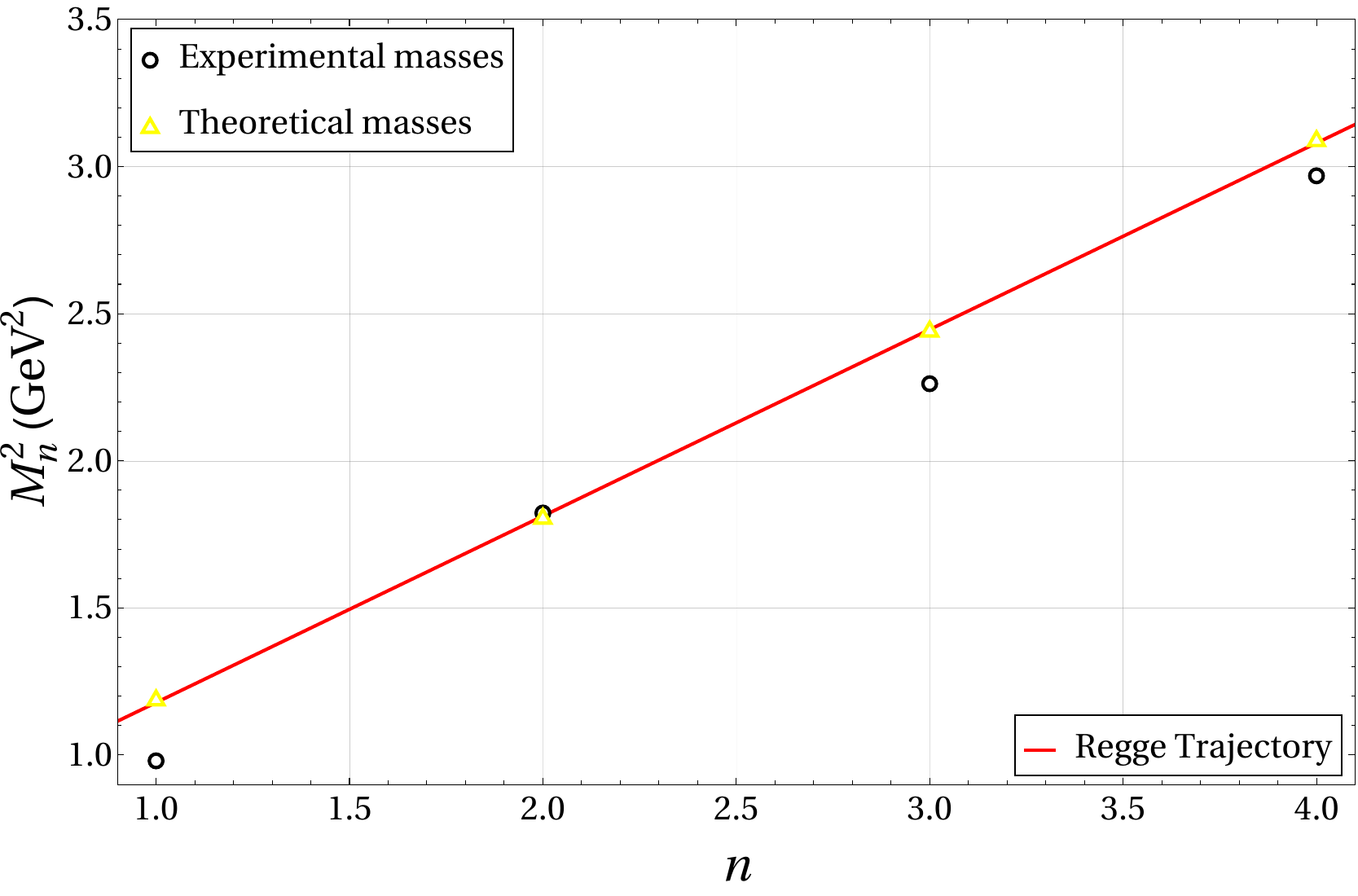}
        \caption{\small Regge trajectory \(M_n^2\) for the mesonic \( f_0 \) family obtained from the soft-wall model.}
        \label{numRegge}
    \end{minipage}%
    \hfill
    \begin{minipage}[t]{0.49\textwidth}
        \centering
        \includegraphics[width=\textwidth]{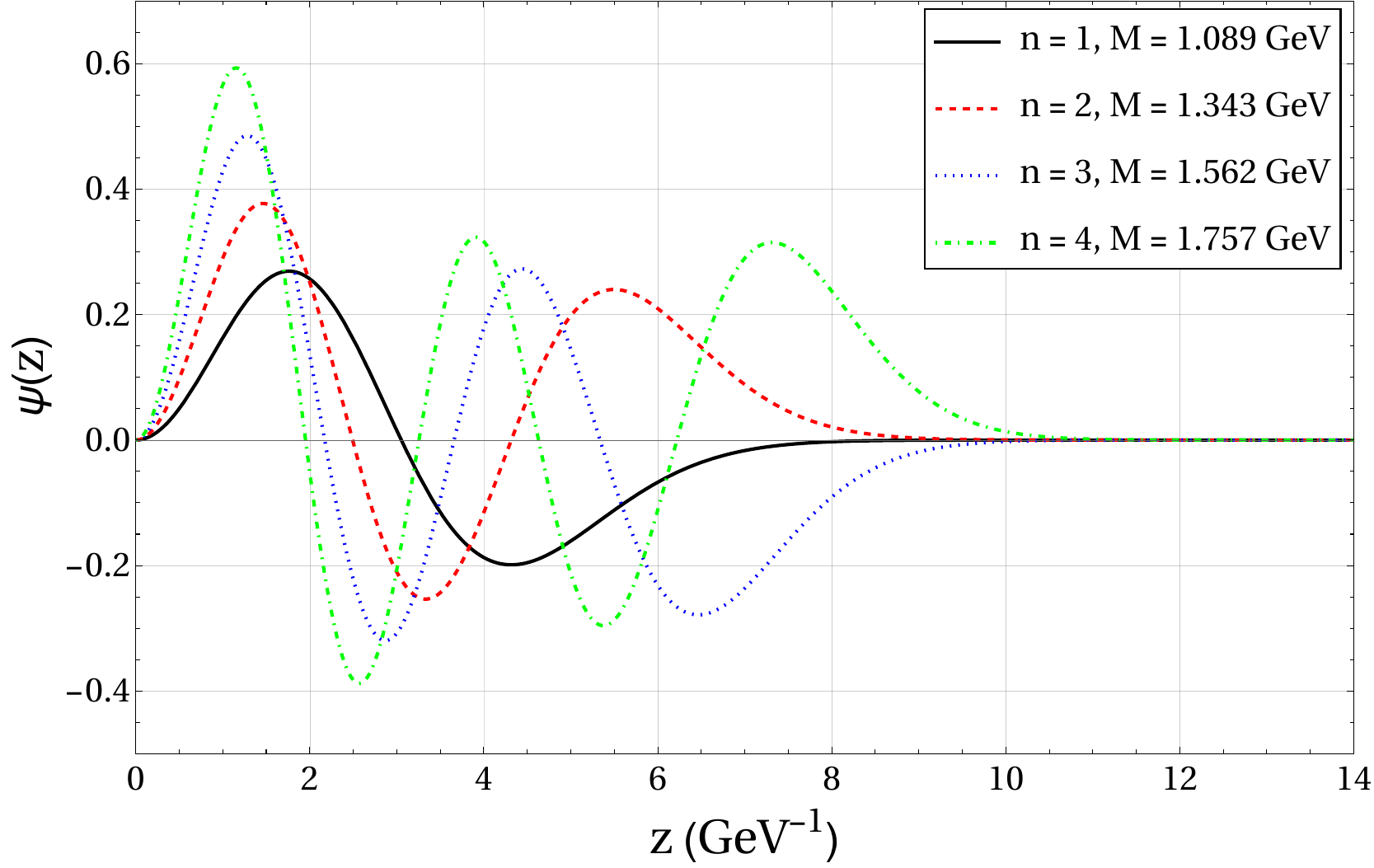}
        \caption{\small Numerical holographic wavefunctions for the mesonic \( f_0 \) family.}
        \label{numWF}
    \end{minipage}
\end{figure}

In this work, Eq.~(\ref{eq:softwall_sch}) was solved for the scalar \(f_0\) channel in the soft-wall background using
the dilaton parameter \(\kappa=0.322~\mathrm{GeV}^{2}\) (equivalently \(\sqrt{\kappa}=0.568~\mathrm{GeV}\)) on the
interval \(z\in[z_{\min},z_{\max}]=[0,10]~\mathrm{GeV}^{-1}\), with a uniform grid and the matrix-Numerov scheme
\eqref{finalnum}. Fig.~\ref{numRegge} shows the resulting Regge-like trajectory \(M_n^2\) versus the radial quantum
number \(n\), and Fig.~\ref{numWF} displays representative eigenfunctions.

As seen in Fig.~\ref{numRegge}, the spectrum exhibits an approximately linear behavior in \(M_n^2\) as a function of
\(n\), consistent with the analytic soft-wall expectation of linear trajectories. Moreover, the numerical
wavefunctions agree qualitatively with the analytic Laguerre-polynomial form in Eq.~(\ref{eq:softwall_wf}), as
illustrated in Fig.~\ref{numWF}. This provides a useful validation of the matrix-Numerov implementation within the
chosen discretization and cutoff, and it can be applied in the next chapters to backgrounds where no closed-form
solutions are available.

\newpage

\section{Heavy and exotic mesonic mass spectra in the EMD model}\label{EMDspec}

In this Section, we will comprehensively investigate the mass spectroscopy of exotic mesons featuring heavy quark flavors by employing the Einstein-Maxwell-dilaton (EMD) holographic model, explicitly under conditions of zero temperature and zero chemical potential. 

It is worth mentioning that Ref.~\cite{Dudal:2017max} previously utilized a similar EMD framework to explore a first-order phase transition between small and large black hole phases, which correspond respectively to ``almost confinement'' and deconfinement regimes. Such an approach offers a compelling method for studying near-confinement phenomena at finite temperatures within the small black hole phase. In this work, we present analytical and self-consistent solutions to the gravitational equations of motion formulated in terms of the warp factor. We then apply these solutions to characterize exotic mesons, carefully discussing the computed results and systematically comparing them with recent studies in the literature. The analysis presented here significantly advances the predictive capability and phenomenological accuracy of soft-wall holographic QCD models.

\subsection{The EMD action}
The procedure adopted here involves starting from the Einstein-Maxwell-dilaton (EMD) action to systematically develop a self-consistent holographic framework capable of accurately describing the properties of heavy and exotic vector mesons. From now on, in order to converse with most of the modern literature in the holographic QCD field, we will adopt the mostly-plus metric signature. In a five-dimensional spacetime, the corresponding EMD action can be expressed explicitly as follows
\begin{equation}\label{emd_action}
    S_{\text{EMD}} = -\frac{1}{16 \pi G_5} \int d^5x \, \sqrt{-g} 
    \left[ 
        R - \frac{f(\phi)}{4} F_{MN}F^{MN} 
        - \frac{1}{2} \partial_M \phi \partial^M \phi 
        - V(\phi) 
    \right]\,,
\end{equation}
where $F_{MN} = \partial_M A_N - \partial_N A_M$ is the field strength, $f(\phi)$ is a gauge kinetic function that couples the dilaton and the gauge field $A_M$, $V(\phi)$ is the dilaton field potential, and $G_5$ is Newton's gravitational constant in five dimensions. 

The variation of the action (\ref{emd_action}) should yield an equation for the Maxwell field $A_M$, one for the dilaton field $\phi$ and one for the metric $g_{MN}$ (taking into account the other field's back-reaction, in contrast to the original soft-wall model). 

First, let us take the variation of (\ref{emd_action}) with respect to $A_M$. The relevant piece of the Lagrangian is $\mathcal{L}_A=-\frac14f(\phi)F_{MN}F^{MN}$, so one finds
\begin{equation}
    \delta F_{MN} = \nabla_M\delta A_N - \nabla_N\delta A_M = 2 \nabla_{[M}\delta A_{N]}\,,
\end{equation}
the connection coefficients inside the covariant derivatives $\nabla_M$ are unchanged because we are not varying the metric, of course. This implies
\begin{equation}
\begin{split}
    \delta \mathcal{L}_A &= -\frac12 f(\phi) F^{MN}\Bigl(\nabla_M\delta A_N - \nabla_N\delta A_M \Bigl) \\
    &=   -\frac12 f(\phi) \Bigl(F^{MN}\nabla_M\delta A_N - F^{NM}\nabla_M\delta A_N \Bigl) \\
    &=   -f(\phi) F^{MN}\nabla_M\delta A_N\,,
\end{split} \label{varyA}
\end{equation}
it was used that $M$ and $N$ are dummy indices under integration, for they are to be summed over all coordinates and $F^{NM}=-F^{MN}$. We recognize that
\begin{equation}
\begin{split}
        f(\phi) F^{MN} \nabla_M \delta A_N &= \nabla_M\left[f(\phi)F^{MN}\delta_N\right] \\ &- \nabla_M\left[f(\phi)\right] F^{MN} \delta A_N - f(\phi) \nabla_M\left[F^{MN}\right]  \delta A_N \,,
\end{split}
\end{equation}

So, integrating by parts, the variation gives us 
\begin{equation}
\begin{split}
    \delta S_A &= -\frac{1}{16 \pi G_5} \int d^5x \, \sqrt{-g}  \left(-f(\phi) F^{MN}\nabla_M\delta A_N  \right) \\
        &= \frac{1}{16 \pi G_5}  \int d^5x \, \sqrt{-g}  \Bigl(\nabla_M\left[f(\phi)F^{MN}\delta_N\right]\\  &- \left[\nabla_Mf(\phi) F^{MN}+ f(\phi) \nabla_MF^{MN}  \right]  \delta A_N \Bigl) \\
        &=  \frac{1}{16 \pi G_5}  \int d^5x \, \sqrt{-g}\, \delta A_N \nabla_M\left[f(\phi)F^{MN}\right] = 0\,, \label{ASDs}
\end{split} 
\end{equation}
where we used the divergence theorem 
\begin{equation}
    \int d^5x \, \sqrt{-g} \, \nabla_M X^M=\oint_{\partial \mathcal{M}} d^4 \Sigma_M X^M\,,
\end{equation}
to conclude that the term $\sim \int d^5x \, \sqrt{-g} \,\nabla_M\left[f(\phi)F^{MN}\delta_N\right]$ must vanish as it is a surface term. Since $\delta A_N $ is arbitrary in (\ref{ASDs}), we get the following Maxwell's equations
\begin{equation}\label{maxwell_eq}
    \boxed{\nabla_M\left[f(\phi)F^{MN}\right] = 0\,.}
\end{equation}

Now, let us take the variation of (\ref{emd_action}) with respect to $\phi$. The relevant piece of the Lagrangian is $\mathcal{L}_\phi=-\frac12\partial^M\phi\partial_M\phi- V(\phi)-\frac14f(\phi)F_{MN}F^{MN}$, so one finds
\begin{equation}
    \delta \mathcal{L}_\phi = -g^{MN} \partial_M \phi \partial_N \phi \, \delta\phi - V'(\phi) \delta\phi - \frac14 f'(\phi) F_{MN}F^{MN} \delta \phi \label{opds}\,,
\end{equation}

If we apply the relations $\nabla_M X^M = \frac{1}{\sqrt{-g}}\partial_M\left(\sqrt{-g}X^M \right) $ and $g^{MN} \partial_M \phi \partial_N=\nabla_M\left(\partial^M\phi \delta \phi\right) + \left(\nabla_M \partial^M\phi \right) \delta\phi$ to (\ref{opds}), we find
\begin{equation}
    \delta \mathcal{L}_\phi = \nabla_M\left( \partial^M\phi\delta\phi \right)  + \delta \phi \left[ \nabla_M \partial^M\phi-V'(\phi) -\frac14f'(\phi)F_{MN}F^{MN}\right]\,,
\end{equation}

The integration by parts therefore results in
\begin{equation}
    \delta S_\phi = -\frac{1}{16 \pi G_5} \int d^5x \, \sqrt{-g} \, \delta \phi \left[ \nabla_M \partial^M\phi-V'(\phi) -\frac14f'(\phi)F_{MN}F^{MN}\right]\,,
\end{equation}
where we once again discarded the surface term. The bracket must vanish identically, so we find the dilaton equation of motion
\begin{equation}\label{KG_eq}
    \boxed{\partial_M \left[ \sqrt{-g} \partial^M \phi \right]
    - \sqrt{-g} 
    \left[ 
        \frac{\partial V}{\partial \phi} 
        + \frac{F^2}{4} \frac{\partial f}{\partial \phi} 
    \right]=0\,.}
\end{equation}
where $F^2=F_{MN}F^{MN}$. 
Finally, let us take the variation of (\ref{emd_action}) with respect to $g_{MN}$. The matter $\mathcal{L}_m$ contributes to the energy-momentum tensor as
\begin{equation}
    -\frac{2}{\sqrt{-g}} \frac{\delta \left( \sqrt{-g} \mathcal{L}_m \right)}{\delta g^{MN}}\,,
\end{equation} 
The useful relations $\delta\left(g^{MN}X_MX_N\right)=-(X^MX^N)\delta g_{MN}$, $\delta g_{MN} = -g_{MP}g_{NQ}\delta g^{PQ}$ and $\delta\sqrt{-g}=-\frac{1}{2}\sqrt{-g}g_{MN}\delta g^{MN}$ will be used for the following computations. 

For the dilaton field, one part reads
\begin{equation}
    -\delta \left[\sqrt{-g} \left(-\frac12 g^{PQ}\partial_P \phi \partial_Q \right) \right] = \frac12 \sqrt{-g} \left[-\frac12 g_{MN} g^{PQ} \partial_P \phi \partial_Q \phi - \partial_M \partial_N \phi \right]\delta g^{MN}\,,
\end{equation}
Since $\delta \left(\sqrt{-g}V \right) = -\frac12\sqrt{-g}g_{MN}V\delta^{MN}$, we read 
\begin{equation}
    2T_{MN}^\phi = \partial_M \phi \partial_M \phi - \frac12 g_{MN}\partial^M \phi \partial_M \phi - g_{MN}V(\phi) \,,
\end{equation}
as we multiply by $-2/\sqrt{-g}$.

As for the Maxwell field, 
\begin{equation}
    \delta \left[-\frac14 \sqrt{-g} f(\phi) F_{PQ}F^{PQ} \right] = -\frac14  \sqrt{-g} \, f(\phi)  \left[-\frac12 g_{MN} F_{PQ}F^{PQ} + 2 F_{MP} F^{\;P}_{N}  \right]\delta g^{MN}\,,
\end{equation}
where we applied $F^{QP}=g^{PR}g^{QS}F_{RS} \implies \delta F^{QP} = -\left(F^{P}_{\;R}\delta g^{RQ} + F^{Q}_{\;R} \delta g^{RP} \right)$. If we multiply by $-2/\sqrt{-g}$,
\begin{equation}
    2T_{MN}^A = f(\phi) \left[F_{MP} F^{\;P}_{N} - \frac14 g_{MN}F_{PQ}F^{PQ}\right]\,,
\end{equation}

The variation of the Hilbert term, $\delta (\sqrt{-g}R)$ gives, of course, the Einstein tensor $\sqrt{-g} \, G_{MN} \delta g^{MN}$ after removing the surface term. As for the matter term, $\sqrt{-g} \, \left(T_{MN}^\phi+T_{MN}^A  \right) \delta g^{MN}$. Setting $\delta S=0$ and demanding this holding for an arbitrary $\delta g^{MN}$, gives us the Einstein's field equations 
\begin{equation}\label{einstein_eq}
     \boxed{R_{MN} - \frac{1}{2} g_{MN} R = T_{MN}\,.}
\end{equation}
for the energy-momentum tensor 
\begin{equation}\label{EM_tensor}
    T_{MN} = \frac{1}{2} \left[ \partial_M \phi \partial_N \phi - \frac{1}{2} g_{MN} \left(\partial \phi \right)^2 - g_{MN} V(\phi) \right]+\frac{f(\phi)}{2} \left[ F_{MP} F^{\;P}_{N} - \frac{1}{4} g_{MN} F^2 \right]\,.
\end{equation}

To have a self-consistent holographic model, Eqs.~\eqref{maxwell_eq}, \eqref{KG_eq}, and  \eqref{einstein_eq}, which are coupled, must be solved simultaneously. This can be done using the potential reconstruction method \cite{Mahapatra:2020wym,Mahapatra:2018gig,Priyadarshinee:2023cmi,Priyadarshinee:2021rch,Daripa:2024ksg}. To do so, we take the following Ans\"atze for the metric, gauge, and dilaton fields,
\begin{align}
    \begin{split}
            ds^2 &= \frac{L^2 e^{2A(z)}}{z^2} 
            \left(-g(z) \, dt^2 + \frac{dz^2}{g(z)} + dx_i dx^i \right)\,, \\
            A_M &= (A_t = A_t(z), \, A_i = 0, \, A_z = 0)\,, \\
    \phi &= \phi(z)\,.
    \end{split}
    \label{ansatz_metric}
\end{align}
where $L$ is the AdS radius here, $A(z)$ is the warp factor and $g(z)$ is the blackening function. It is assumed that the nonzero component of the gauge field $A_t$ and the dilaton field $\phi$ depend only on the holographic coordinate $z \in [0,z_h]$. Here $z_h$ is the horizon radius defined by $g(z_h)=0$. The strongly coupled gauge theory lives at the asymptotic boundary of spacetime, which is defined by the $z\to0$ limit. If the Ans\"atze (\ref{ansatz_metric}) is plugged into  Eqs.~\eqref{maxwell_eq}, \eqref{KG_eq}, and  \eqref{einstein_eq}, one is expected to find five resulting differential equations. 

For reference, the geometric quantities $\sqrt{-g}=\frac{L^5e^{5A}}{z^5}$, $g^{tt}=-\frac{z^2}{L^2e^{2A}g}$, $g^{zz}=\frac{z^2g}{L^2e^{2A}}$, $F^{zt}=g^{zz}g^{tt}F_{zt}=-\frac{z^4 A'_t}{L^4e^{4A}}$ and $A^t=g^{tt}A_t=-\frac{z^2A_t}{L^2e^{2A}g}$ stem out of (\ref{ansatz_metric}).

From Maxwell's equation \eqref{maxwell_eq}, since the only non‑trivial equation comes from $N=t$ as $F_{MN}$ only has the $zt$ component, we can write then
\begin{equation}
    \nabla_M\left[f(\phi)F^{MN}\right] = \frac{1}{\sqrt{-g}} \partial_M \left(\sqrt{-g} f(\phi) F^{MN} \right) = 0\,,
\end{equation}
\begin{equation}
    \implies \frac{1}{\sqrt{-g}} \partial_z \left(\sqrt{-g} f(\phi) F^{zt} \right) =- \frac{z^5}{L^5 e^{5A}} \partial_z \left[\frac{e^A}{z}f(\phi) A_t' \right] = 0\,,
\end{equation}
\begin{equation}\label{Aequation}
    \boxed{\therefore A_t'' + A_t' \left(\frac{f_{,\phi}}{f}\phi' + A' -\frac{1}{z} \right) = 0\,.}
\end{equation}

From dilaton equation \eqref{KG_eq},
\begin{equation}
    \partial_z \left[ \sqrt{-g} \partial^z \phi \right]
    - \sqrt{-g} 
    \left[ 
        \frac{\partial V}{\partial \phi} 
        + \frac{F^2}{4} \frac{\partial f}{\partial \phi} 
    \right] = 0\,,
\end{equation} 
\begin{equation}
    \implies \partial_z\left[\frac{L^3e^{3A}}{z^3}\phi' \right] - \frac{L^5e^{5A}}{z^5} \left[\frac{\partial V}{\partial \phi} +\frac14\left( -2\frac{z^4}{L^4 e^{4A}}H_t^{\prime 2}\right)\frac{\partial f}{\partial \phi}  \right] =0\,,
\end{equation}
\begin{equation} \label{EMD_eqs}
    \boxed{
    \therefore \phi'' + \phi' \left(\frac{g'}{g}  + 3A' -\frac{3}{z}  \right) 
    - \frac{L^2 e^{2A}}{g z^2} \frac{\partial V}{\partial \phi} +A_t^{\prime 2} \frac{z^2 e^{-2A}}{2gL^2} \frac{\partial f}{\partial \phi} =0 \,.}
\end{equation}
in this case, $F_{MN}F^{MN} = 2F^{zt}F^{zt} = -2\frac{z^4}{L^4 e^{4A}}H_t^{\prime 2}$ and $\partial^z=g^{zz}\partial_z=\frac{z^2g}{L^2e^{2A}}\partial_z$ were used.

Now, we focus on the resulting differential equations that come out of Einstein's equation \eqref{einstein_eq}. We first define $W(z)=L\,e^{A(z)}/z$ and $\alpha(z)=A(z)-\ln z$, so that
\(
g_{tt}=-W^2\,g, g_{zz}=\frac{W^2}{g}, g_{ij}=W^2\delta_{ij}
\).

The nonzero Christoffel symbols $\Gamma^P_{\;MN}=\frac12g^{PQ}\left(\partial_Mg_{QN}+\partial_Ng_{QM}-\partial_Qg_{MN}\right)$ are
\begin{equation}
    \begin{aligned}
    \Gamma^t_{\;tz}&=\frac12 g^{tt}\partial_z g_{tt}=\alpha'+\frac{g'}{2g}\,,\quad
    \Gamma^i_{\;iz}=\frac12 g^{ii}\partial_z g_{ii}=\alpha'\,,\\
    \Gamma^z_{\;tt}&=-\frac12 g^{zz}\partial_z g_{tt}=\alpha'\,g^2 + \tfrac12g\,g'\,,\quad
    \Gamma^z_{\;zz}=\frac12 g^{zz}\partial_z g_{zz}= \alpha' - \tfrac{g'}{2g}\,,\\
    \Gamma^z_{\;ii}&=-\frac12 g^{zz}\partial_z g_{ii}=-\alpha'\,g\,.
\end{aligned}
\end{equation}

Now for the Ricci tensor,
\begin{equation}
    R_{MN}=\partial_P\Gamma^P_{\;MN}-\partial_N\Gamma^P_{\;PM}
      +\Gamma^P_{\;MN}\Gamma^Q_{\;PQ}-\Gamma^P_{\;MQ}\Gamma^Q_{\;PN}\,,
\end{equation}
one obtains $R_{tt}$ as 
\begin{equation}\label{der1}
    \partial_P \Gamma^P_{\;tt} = \partial_z \Gamma^z_{\;tt}\,, 
\quad
\partial_t \Gamma^P_{\;Pt} = 0 
\quad(\text{all }\Gamma\propto z)\,,
\end{equation}
\begin{equation}\label{pro1}
    \Gamma^P_{\;tt}\,\Gamma^Q_{\;PQ}
  = \Gamma^z_{\;tt}\bigl[\Gamma^t_{\;tz} + \Gamma^z_{\;zz} + 3\,\Gamma^i_{\;iz}\bigr]\,,
\quad
\Gamma^P_{\;tQ}\,\Gamma^Q_{\;Pt}
  = \Gamma^z_{\;tt}\,\Gamma^t_{\;tz}\,,
\end{equation}

Putting these derivative terms \eqref{der1} and product terms \eqref{pro1}, one writes
\begin{equation}
\begin{aligned}
    R_{tt} &= \partial_z\Gamma^z_{\;tt} + \Gamma^z_{\;tt}\bigl[\Gamma^t_{\;tz}+\Gamma^z_{\;zz}+3\Gamma^i_{\;iz}\bigr] -\Gamma^z_{\;tt}\,\Gamma^t_{\;tz}\\
          &= \frac{L^2e^{2A}}{z^2}\left[\tfrac12\,g'' +\left(3A'-\tfrac3z\right)\tfrac{g'}2\right]\,.
\end{aligned}
\end{equation}

Now for the $R_{zz}$, we have
\begin{equation}\label{der2}
    \partial_P \Gamma^P_{\;zz} = \partial_z \Gamma^z_{\;zz}\,,
\quad
\partial_z \Gamma^P_{\;Pz}
  = \partial_z\bigl[\Gamma^t_{\;tz}+\Gamma^z_{\;zz}+3\Gamma^i_{\;iz}\bigr]\,,
\end{equation}
\begin{equation}\label{pro2}
    \Gamma^P_{\;zz}\,\Gamma^Q_{\;PQ}
  = \Gamma^z_{\;zz}\bigl[\Gamma^t_{\;tz}+\Gamma^z_{\;zz}+3\Gamma^i_{\;iz}\bigr]\,,
\quad
\Gamma^P_{\;zQ}\,\Gamma^Q_{\;Pz}
  = (\Gamma^z_{\;zz})^2 +3(\Gamma^i_{\;iz})^2\,,
\end{equation}

Combining these derivative terms \eqref{der2} and product terms \eqref{pro2}, one writes
\begin{equation}
    \begin{aligned}
R_{zz}
&= \partial_z\Gamma^z_{\;zz}
  -\partial_z\bigl[\Gamma^t_{\;tz}+\Gamma^z_{\;zz}+3\Gamma^i_{\;iz}\bigr]
  +\Gamma^z_{\;zz}\bigl[\Gamma^t_{\;tz}+\Gamma^z_{\;zz}+3\Gamma^i_{\;iz}\bigr]\\
&\quad
  -\left[(\Gamma^z_{\;zz})^2+3(\Gamma^i_{\;iz})^2\right]
= \frac{L^2e^{2A}}{z^2g}\Bigl[-\tfrac12\,g''-3A'\,g'+\tfrac3z\,g'\Bigr]\,.
\end{aligned}
\end{equation}

Finally, for the $R_{ii}$, we have
\begin{equation}\label{der3}
    \partial_P \Gamma^P_{\;ii} = \partial_z \Gamma^z_{\;ii}\,, 
\quad
\partial_i \Gamma^P_{\;Pi} = 0\,,
\end{equation}
\begin{equation}\label{pro3}
    \Gamma^P_{\;ii}\,\Gamma^Q_{\;PQ}
  = \Gamma^z_{\;ii}\bigl[\Gamma^t_{\;tz}+\Gamma^z_{\;zz}+3\Gamma^j_{\;jz}\bigr]\,,
\quad
\Gamma^P_{\;iQ}\,\Gamma^Q_{\;Pi}
  = 2\,\Gamma^z_{\;ii}\,\Gamma^i_{\;iz}\,,
\end{equation}

Gathering these derivative terms \eqref{der3} and product terms \eqref{pro3}, one writes
\begin{equation}
    \begin{aligned}
R_{ii}
&= \partial_z\Gamma^z_{\;ii}
  +\Gamma^z_{\;ii}\bigl[\Gamma^t_{\;tz}+\Gamma^z_{\;zz}+3\Gamma^j_{\;jz}\bigr]
  -2\,\Gamma^z_{\;ii}\,\Gamma^i_{\;iz}\\
&= \frac{L^2e^{2A}}{z^2}\left[
  -A''(z)\,g(z)
  -\left(3A'(z)-\tfrac3z+\tfrac{g'}{2g}\right)A'g
\right].
\end{aligned}
\end{equation}

Now, the Ricci scalar is simply
\begin{equation}
    R = g^{tt}R_{tt}+g^{zz}R_{zz}+3\,g^{ii}R_{ii} \,.
\end{equation}

From the energy-momentum tensor, shown in \eqref{EM_tensor}, we find that the nonzero components are
\begin{equation}
\begin{aligned}
    T_{tt}
      &=-\frac14\,g_{tt}\,\phi^{\prime 2}
      -\frac12\,g_{tt}V
      +\frac{f}{4}\,\frac{z^2g}{L^2e^{2A}}A_t^{\prime 2}\,,\\
      T_{zz}
      &=\frac12\,\phi^{\prime 2}
      -\frac12\,g_{zz}V
      -\frac{f}{4}\,\frac{z^2}{L^2e^{2A}g}A_t^{\prime 2}\,,\\
      T_{ii}
      &=-\frac14\,g_{ii}\,\phi^{\prime 2}
      -\frac12\,g_{ii}V
      -\frac{f}{4}\,\frac{z^2}{L^2e^{2A}}A_t^{\prime 2}\,.
\end{aligned}
\end{equation}

This gives us all the ingredients to find the remaining equations after inserting the metric Ans\"atze \eqref{ansatz_metric} into the found EMD equations of motion. First, the $tt-ii$ combination gives
\begin{equation}
    \boxed{0 = g'' + g' \left(  3A' -\frac{3}{z} \right) 
    - A_t^{\prime 2} \frac{z^2 e^{-2A} f}{L^2}\,.}\label{2.9}
\end{equation}
where we subtracted the $ii$ equation from the $tt$ equation to eliminate the scalar potential term. Now, the $zz-tt$
\begin{equation}\label{equA}
    \boxed{0 = A'' - A' \left( A' -\frac{2}{z} \right) 
    + \frac{\phi^{\prime 2}}{6}\,.}
\end{equation}
where we subtracted the $tt$ equation from the $zz$ equation to isolate $A''$ and $\phi^{\prime 2}$. Lastly, the $ii$ gives us
\begin{equation}
    \boxed{0 = A'' + \frac{g''}{6g} 
    + A' \left(\frac{3g'}{2g}  -\frac{6}{z}  \right) 
    - \frac{1}{z} \left(\frac{3g'}{2g} -\frac{4}{z}  \right) 
    + 3A^{\prime 2} + V \frac{L^2 e^{2A}}{3gz^2}\,.\label{2.10}}
\end{equation}
where we used all three Ricci components together to give the Hamiltonian constraint. Given the five equations and four unknown functions, we choose Eq. (\ref{EMD_eqs}) to act as a constraint.

As demonstrated in \cite{Dudal:2017max}, treating \( A(z) \) and \( f(z) \) as arbitrary functions allows us to determine all unknown functions in closed form. In particular, by applying the following boundary conditions
\begin{equation}
    \lim_{z \to 0} g(z) = 1 \,, \qquad\qquad  \lim_{z \to z_h} g(z) = 0\,,  \qquad  \qquad \lim_{z \to z_h} A_t(z) = 0\,,
\end{equation}
we obtain, in terms of $A(z)$ and $f(z)$, the following solutions \cite{Dudal:2017max}
\begin{align}
    \label{EMD_sol1}
    g(z) &= 1 - 
    \frac{\int_0^{z} dx \, x^3 e^{-3A(x)} 
    \int_{C}^{x} dy \, \frac{y e^{-A(y)}}{f(y)}}
    {\int_0^{z_h} dx \, x^3 e^{-3A(x)}
    \int_{C}^{x} dy \, \frac{y e^{-A(y)}}{f(y)}}\,, \\[10pt]
    \phi'(z) &= \sqrt{6 \left( A^{'2} - A'' - \frac{2A'}{z} \right)}\,, \\[10pt]
    A_t(z) &= 
            -\frac{1}
        {\sqrt{\int_0^{z} dx \, x^3 e^{-3A(x)}
        \int_{C}^{x} dy \, \frac{y e^{-A(y)}}{f(y)}}} 
    \int_{z_h}^{z} dx \, \frac{x e^{-A(x)}}{f(x)}\,, \\[10pt]
    V(z) &= - \frac{3g z^2 e^{-2A}}{L^2} 
    \bigg(
        A'' + A' \left[ 3A' + \frac{3g'}{2g} - \frac{6}{z}  \right] 
        - \frac{1}{z} \left[\frac{3g'}{2g}  -\frac{4}{z} \right] 
        + \frac{g''}{6g} 
    \bigg)\,.
    \label{EMD_sol2}
\end{align}

The expressions (\ref{EMD_sol1})--(\ref{EMD_sol2}) can be put back into Eqs.~(\ref{einstein_eq})--(\ref{KG_eq}) to verify that the equations are satisfied and the system is mathematically self-consistent. In \cite{Dudal:2017max}, the integration constant $C$ can be fixed in terms of the chemical potential $\mu$. If $\mu$ is set to zero, we have the following analytical solutions to Eqs.~\eqref{EMD_sol1}--\eqref{EMD_sol2}:
\begin{align}
    \label{EMD_muzero}
    g(z) &= 1 - 
    \frac{\int_0^{z} dx \, x^3 e^{-3A(x)}}
         {\int_0^{z_h} dx \, x^3 e^{-3A(x)}}\,, \\[10pt]
    \phi'(z) &= \sqrt{6 \left( A^{'2} - A'' - \frac{2A'}{z} \right)}\,, \\[10pt]
    A_t(z) &= 0\,, \\[10pt]
    V(z) &= - \frac{3g z^2 e^{-2A}}{L^2} 
    \bigg(
        A'' + A' \left[ 3A' + \frac{3g'}{g} - \frac{6}{z}  \right] 
        - \frac{1}{z} \left[\frac{3g'}{2g}  -\frac{4}{z} \right] 
        + \frac{g''}{6g} 
    \bigg)\,.
\end{align}

Without loss of generality, we take the AdS radius to be $L=1$ when performing numerical computations for the rest of this work.

We see that the solution presented above describes a black hole with a horizon located at $z = z_h$. However, there exists another distinct solution to the EMD field equations, corresponding to a thermal AdS geometry without a horizon. This horizonless solution can be explicitly recovered by taking the limit $z_h \rightarrow \infty$, leading to the simple form $g(z) = 1$. As it will be elaborated upon later, these two gravitational solutions, the black hole and thermal AdS, compete in the free energy landscape and undergo a Hawking-Page-type first-order phase transition as temperature changes. Specifically, at low temperatures, the thermal AdS solution is energetically preferred, while the black hole configuration dominates at higher temperatures. Translating this into the language of the dual field theory, the thermal AdS phase corresponds to a confined state, whereas the black hole solution corresponds to the deconfined state.  In the subsequent analysis, we will fix the parameters of the holographic model by employing the thermal AdS (black hole) background to study meson properties at zero (finite) temperature.

In order to completely solve the EMD equations, it is necessary to specify the forms of the initially arbitrary functions \( f(z) \) and \( A(z) \). In the next Section, the selection of \( f(z) \) is guided by established holographic QCD models previously studied in the literature \cite{MartinContreras:2020cyg, MartinContreras:2021bis}, which incorporate a carefully constructed deformation of the dilaton field to effectively describe heavy and exotic mesons. Meanwhile, following \cite{Dudal:2017max}, the function \( A(z) \) is chosen to have a simple analytical form that clearly exhibits a confinement-deconfinement phase transition within the boundary theory when analyzed in this EMD framework. With these choices, the goal is to calculate critical physical quantities—such as mass spectra and spectral functions—for various heavy and exotic QCD states. In particular, we aim to investigate the behavior of spectral functions at finite temperature and density, providing insights into the melting dynamics of these states and consequently shedding light on the nature of the QCD deconfinement phase transition occurring in the QGP.

\subsection{Heavy and exotic mesons}

The quadratic dilaton-field profiles are frequently employed in AdS/QCD models to reproduce the meson mass spectra for light-flavor states, resulting in linear Regge trajectories as functions of the radial quantum number \cite{MartinContreras:2020cyg,Song,Rinaldi:2020ybv,Bernardini:2018uuy}. This was already shown in this work, as seen in Sec. \ref{spectra}, where the smooth background dilaton introduced a mass spectrum to the previous continuous conformal system. However, when heavy quarks are introduced as constituents, this linearity in the Regge trajectories is disrupted, although the spectra of heavy mesons can still be accurately described \cite{Afonin:2014nya,jk,Karapetyan:2023sfo}. 

Heavy quarks also form the constituents of various exotic hadronic states, collectively termed heavy-quark exotica, which include tetraquarks, pentaquarks, hadrocharmonia, hybrid mesons, and hadronic molecules \cite{Guo,Lebed1,Oncala:2017hop,Karapetyan:2021ufz}. Among the wide array of detected hadronic states, exotic mesons remain particularly challenging to understand within QCD, and significant theoretical efforts continue in modeling and interpreting their fundamental properties. These exotic mesonic states deviate from the conventional \( q\bar{q} \) quark-antiquark structure, where \( q \) represents a quark \cite{Lebed1,Jaffe1}. In general, mesons possessing quantum numbers forbidden by the standard \( q\bar{q} \) framework are classified as exotic mesons \cite{Jaffe1}.

The standard \( q\bar{q} \) meson composition relies on the empirical observation that no mesons exist with either isospin or strangeness greater than one. The known mesons, cataloged by the Particle Data Group (PDG) \cite{ParticleDataGroup:2024cfk}, are typically interpreted as bound \( q\bar{q} \) states organized into \( SU(N_f) \) flavor multiplets, characterized by parity \( P = (-1)^{L+1} \) and charge conjugation \( C = (-1)^{L+S} \). Here, \( S \) denotes the meson spin, \( L \) represents the orbital angular momentum, and \( J \) is the total angular momentum.

In contrast, exotic mesons do not fit into the conventional \( q\bar{q} \) quark model, as certain resonances—particularly those with quantum numbers \( J^{PC} = 0^{--} \) and states like \( J^{PC} = 0^{+-}, 1^{-+}, 2^{+-} \), characterized by \( P = (-1)^J \) and \( C = (-1)^{J+1} \)—cannot be accommodated within this standard framework \cite{Lebed1}. However, some mesonic states that deviate from the pure \( q\bar{q} \) structure can still incorporate valence gluons, forming hybrid exotic mesons \cite{Meyer}. These hybrid states support non-perturbative descriptions of meson confinement in QCD, as gluonic degrees of freedom are explicitly permitted.

Hybrid mesons consist of \( \bar{q}q \)-pairs bound by gluonic color flux tubes, enabling them to possess quantum numbers forbidden in pure \( q\bar{q} \) models. Notable examples include the exotic meson state \( \pi_1 \), having quantum numbers \( I^G(J^{PC})=1^-(1^{-+}) \), and the bottomonium-like state \( Z_b \), with \( I^G(J^{PC})=1^+(1^{+-}) \). Ref.~\cite{Bellantuono:2014lra} examined the hybrid exotic meson \( \pi_1 \) via gauge/gravity duality, demonstrating that these states deconfine at lower temperatures compared to ordinary light-flavor mesons.

Exotic mesons also include charged resonances such as the charmonium-type \( Z_c \)-family states with quantum numbers \( I^G(J^{PC})=1^+(1^{+-}) \). These states feature additional valence quarks, as exemplified by the charged state \( Z_c(4430) \), contrasting with neutral \( c\bar{c} \) charmonium states. The charged exotic meson \( Z_c(3900)^+ \) was among the first such states discovered, consisting of a \( c\bar{c}u\bar{d} \) tetraquark structure \cite{BESIII:2015cld}.

Tetraquark states can be viewed as hadronic molecules composed of four valence quarks organized either into two heavy quarkonia or one heavy quarkonium bound to a lighter meson. For instance, the exotic state \( X(3872) \) can be classified as a hadronic molecule \cite{ParticleDataGroup:2024cfk}. Moreover, tetraquarks can form as color-singlet combinations such as \( qq\bar{q}\bar{q} \) or \( Qq\bar{Q}\bar{q} \), where \( Q \) denotes either a charm or bottom heavy-flavor quark \cite{Lebed1}. Color-singlet multiquark states can similarly encompass both tetraquark structures and hadronic molecules \cite{Liu}. Notably, tetraquark states involving \( c\bar{c} \) pairs differ from standard charmonia due to the possibility of charged resonances, like \( c\bar{c}u\bar{d} \), as well as states carrying nonzero strangeness, such as \( c\bar{c}d\bar{s} \), or both features simultaneously, like \( c\bar{c}u\bar{s} \) \cite{Godfrey:2008nc}.

Recent experimental discoveries further expanded the known spectrum of exotic states. The LHCb collaboration recently observed the \( qq\bar{q}\bar{q} \) and \( Qq\bar{Q}\bar{q} \) tetraquark states \cite{Fang:2022mks,LHCb:2021uow}. The QGP, produced in heavy-ion collisions, serves as a rich environment for deconfined quarks, subsequently recombining into hadrons upon cooling. Measuring exotic mesons in heavy-ion collisions provides valuable insights into their formation mechanisms and contributes to understanding the dynamics of enhancement or suppression of such states within the QCD confinement-deconfinement transition.

The terminology of exotic mesons also encompasses states known as hadrocharmonia, characterized by a heavy-quark core \( c\bar{c} \) surrounded by a light-flavor \( q\bar{q} \) meson cluster, bound together through weak color van der Waals interactions. From a phenomenological perspective, typical charmonium decay channels, such as those involving resonances like \( J/\psi \), \( \psi(2S) \), and \( \chi_c \), reflect the dissociation of the charmonium core from the light-flavor meson, leading to prominent measurable effects \cite{Campanella:2018xev}. A notable example is the exotic meson \( Z_c(4430) \), where the distinct decay patterns into states like \( J/\psi \) and \( \psi(2S) \) strongly suggest the existence of a \( c\bar{c} \) heavy-quark core. Similarly, the meson family \( Y(4008) \), \( Y(4230) \), \( Y(4260) \), \( Y(4360) \), and \( Y(4660) \) exemplifies heavy-quark exotica formed by standard charmonium states bound to lighter mesonic configurations.

In the next Section then we will look into incorporating some phenomenological description of such heavy and exotic mesons into the self-consistent EMD model presented. 

\subsection{$A(z)$ and $f(z)$ as phenomenological tools}

The function \(f(z)\) will first be determined by analyzing the mass spectra of heavy and exotic vector mesons at zero temperature and chemical potential. As was already shown in other works \cite{Mamani:2022qnf, Dudal:2017max}, a particularly good choice for this function is 
\begin{equation}
    f(z) = e^{-\left[A(z)+B(z) \right]}\,,
\end{equation}
where $B(z)$ is chosen to match the mesonic mass spectra. In the context of the original soft-wall model, Sec. \ref{spectra}, $B(z)$ is defined to be the dilaton field $\phi(z)$: $B(z) \equiv \phi(z) = (az)^2$, where $a$ is a constant that defines the slope of the Regge trajectory.\footnote{Note that the dilaton field $\phi(z)$ used in the soft-wall model should not be confused with the dilaton field $\phi(z)$ used in the holographic EMD model in Eq.~(\ref{emd_action}).} 

As already stated, instead of employing the conventional quadratic dilaton, we adopt the non-quadratic dilaton profile introduced in Refs.~\cite{MartinContreras:2020cyg, MartinContreras:2021bis}. While linear Regge trajectories successfully reproduce the hadronic mass spectra for mesons composed of light-flavor quarks with small constituent masses, this linearity is no longer valid when considering hadrons containing heavier quarks, such as the strange quark \( s \). Indeed, Bethe trajectories lose linearity as heavier constituent quarks are introduced, with the excitation spectra becoming increasingly nonlinear. As argued in \cite{MartinContreras:2020cyg, MartinContreras:2021bis}, Bethe--Salpeter type analyses indicate that heavy hadrons follow the scaling of trajectories as \(n^{3/2}\), where \(n\) is the excitation number.

This behavior implies a direct relationship between the Regge trajectory shape and constituent quark masses, exhibiting linear trajectories for light-flavor mesons but significant deviations for mesons containing heavier quarks. Consequently, the corresponding dilaton field profile should also reflect this mass dependence.

In line with Ref.~\cite{MartinContreras:2020cyg}, we incorporate constituent quark masses explicitly, utilizing the averaged quark mass values: \( m_u = 0.336\,\text{GeV} \), \( m_d = 0.340 \, \text{GeV} \), \( m_s = 0.486 \, \text{GeV} \), \( m_c = 1.550 \, \text{GeV} \), and $m_b = 4.730 \, \text{GeV}$, which correspond to phenomenologically motivated constituent quark masses commonly adopted in holographic QCD frameworks. These quark masses are taken to be effectively dressed by the QCD non-perturbative binding energy \cite{Scadron:2006dy}.

Then, we introduce a modified dilaton profile of the form
\begin{equation}
    \Phi(z) = (\kappa z)^{2-\alpha}\,,
\end{equation}
where the exponent $\alpha$ encodes the effects of the constituent quark mass. For $\alpha = 0$, one recovers the original quadratic dilaton and the corresponding linear Regge trajectory. In \cite{MartinContreras:2020cyg} both $\alpha$ and $\kappa$ vary with the average quark masses
\begin{align}
    \alpha(\bar{m}) & \equiv \alpha = a_\alpha - b_\alpha e^{-c_\alpha \bar{m}^2}\,, \\
    \kappa(\bar{m}) & \equiv \kappa = a_\kappa - b_\kappa e^{-c_\kappa \bar{m}^2}\,,
\label{kappam}\end{align}
with parameters $a_\alpha = 0.8454$, $b_\alpha = 0.8484$, $c_\alpha = 0.4233$, $a_\kappa = 15.2085$, $b_\kappa = 14.8082$ and $c_\kappa = 0.0524$ \cite{MartinContreras:2020cyg}. Larger values of $\alpha$ and $\kappa$ correspond to heavier states. Thus, if we know (or model) a meson with some combination of constituent quarks, we can find the $\alpha$ and $\kappa$ parameters that suit the problem at hand best. 

In addition to the non-quadratic behavior introduced in \cite{MartinContreras:2020cyg}, \cite{MartinContreras:2021bis} presents small-$z$ corrections to $\Phi(z)$. Consequently, we adopt the dilaton profile proposed in \cite{MartinContreras:2021bis} and identify their $\Phi(z)$ dilaton field with the $B(z)$ function, which is incorporated into the arbitrary function $f(z)$ as
\begin{equation}\label{f_definition}
f(z) = e^{-\left[A(z)+B(z) \right]} = e^{-\left[A(z) + (\kappa z)^{2-\alpha} + Mz + \tanh{\left(\frac{1}{Mz} - \frac{\kappa}{\sqrt{\Gamma}}\right)}\right]} \,.
\end{equation}
in which $Mz$ and $\tanh{\left(\frac{1}{Mz} - \frac{\kappa}{\sqrt{\Gamma}}\right)}$ are the small-$z$ corrections to $\Phi(z)$, see also Refs. \cite{Braga:2017bml,Braga:2018fyc}. 

The other term, $A(z)$, can be chosen as a simple quadratic function, whose consequences were studied in \cite{Dudal:2017max}. In this model, such $A(z)$ can show the confined-deconfined phases in the boundary theory, mimicking QCD. Therefore here we adopt the following $A(z)$:
\begin{equation}
    A(z) = -\frac{\mathcal{C}}{8} z^2 \,.
\end{equation}
where $\mathcal{C} = 1.16 \, \text{GeV}^2$ is fixed to yields the Hawking-Page geometric transition at around $T \sim 270 \, \text{MeV}$ at zero chemical potential \cite{Dudal:2017max}, based on the lattice estimate of Ref. \cite{Lucini:2003zr}. 

\subsection{Description of vector mesons}

Given that all parameters and functions are fixed, we can first compute the mesonic mass spectra at zero temperature. A five-dimensional gauge field describes the vector mesons in the dual field theory. From the EMD action (\ref{emd_action}), we can write
\begin{equation} \label{Meson_Action}
S = -\frac{1}{16 \pi G_5} \int d^5x \sqrt{-g} \frac{f(\phi)}{4} F_{MN}F^{MN}\,,
\end{equation}

Varying the action (\ref{Meson_Action}), we find the following equations of motion (the Maxwell equation, already computed before)
\begin{equation} \label{EOM_1}
\partial_M \left( \sqrt{-g} f(\phi) F^{MN} \right) = 0\,.
\end{equation}

In the zero temperature limit, $g(z) \rightarrow 1$ and the spacetime interval becomes
\begin{align}
    \label{metric_T0}
    ds^2 &= \frac{L^2 e^{2A(z)}}{z^2} 
    \left( 
        -\, dt^2 + dz^2 + dx_i dx^i 
    \right)\,,
\end{align}

By plugging (\ref{metric_T0}) into (\ref{EOM_1}) and choosing the radial gauge $A_z = 0$, supplemented with the transversal Landau (Lorenz) gauge $\partial_\nu A^\nu = 0$, Eq.~(\ref{EOM_1}) turns into
\begin{equation} \label{EOM_2}
\frac{z e^{-A(z)}}{L f(z)} \partial_z \left(\frac{L e^{A(z)} f(z)}{z} \partial_z A^\mu \right) + \Box A^\mu = 0\,,
\end{equation}
which can be led into a Schr\"odinger-like form by first implementing a Fourier transform 
\begin{equation} 
\label{fourier}
A^\mu(z, x^\nu) = \int \frac{d^4k}{(2\pi)^4} e^{i k_\alpha x^\alpha} A^\mu(z, k)\,,
\end{equation}
and performing the following change of variables   
\begin{align} \label{B_def}
    A^\mu &= e^{-H} \psi^\mu\,, \\
    H &= \frac{1}{2} \log\left[\frac{L e^{A(z)} f(z)}{z}\right]\,.
\end{align}

This turns Eq.~(\ref{EOM_2}) into
\begin{equation} \label{Mass_Spectra_T0}
-\partial_z^2 \psi(z) + U(z) \psi(z) = m_n^2 \psi(z)\,,
\end{equation}
for 
\begin{equation} \label{Potential_T0}
\boxed{U(z) = \left(\partial_z \left[\frac{1}{2} \log\left(\frac{L e^{A(z)} f(z)}{z}\right)\right]\right)^2 
+ \partial_z^2 \left[\frac{1}{2} \log\left(\frac{L e^{A(z)} f(z)}{z}\right)\right]\,.}
\end{equation}
where we identify the eigenmasses with the relation $k^2=m_{n}^{2}=-\omega^2+p^2$. 

As already seen on Sec. \ref{adsover}, for a massive perturbation in the bulk theory, the AdS/CFT dictionary dictates that a mass term of the form (\ref{bulkmass}) must be added to the potential $U(z)$. We can see right away that for isovector mesons, such as the charmonium (\( c\bar{c} \))  and bottomonium (\( b\bar{b} \)), $M^2 L^2 = 0$ as $\Delta=3$ and $p=1$.

\subsection{Mass spectra: Numerical results}

An important application of this nonquadratic dilaton model is in studying hadronic states that fall outside the standard \( q\bar{q} \) meson classification. As previously mentioned in Sec.~\ref{adsover}, holographic QCD provides a natural framework to describe various exotic states through the scaling dimension \( \Delta \) (and spin $S$). By appropriately identifying \( \Delta \) and estimating the associated parameter \( \bar{m} \), this approach enables a systematic investigation of these unconventional meson states. However, given that hadronic bound states may exhibit degeneracies in terms of \( \Delta \), depending on their specific quark and meson content, a weighted averaging procedure—as described earlier—can be utilized to accurately determine the corresponding parameter \( \bar{m} \), as seen in \cite{MartinContreras:2020cyg}
\begin{equation}\label{exp13}
\bar{m}_{\mathrm{multi}} = \sum_{i=1}^{N} \left( P_i^{\mathrm{quark}} \bar{m}_i^{\mathrm{quark}} + P_i^{\mathrm{meson}} m_i^{\mathrm{meson}} \right)\,,\quad \text{with} \,\ \sum_{i=1}^{N} \left( P_i^{\mathrm{quark}}  + P_i^{\mathrm{meson}}  \right)=1\,.
\end{equation}

The considered exotic state is then modeled as a collection of \( N \) constituent quarks and/or mesons. Each constituent is assigned a weight \( P_i \) associated with its mass \( m_i \). These masses are then utilized to determine the constants \( \alpha \) and \( \kappa \). We have per case also included the corresponding weight factors and scale dimension $\Delta$ that we borrowed from \cite{MartinContreras:2020cyg}. 

In this work, we consider the exotic states $Z_c$ and $\pi_1$. $Z_c$ is modeled as a tetraquark which in turn is constructed as a ``hadronic molecule'', two mesons bounded by a residual colorless QCD interaction, formed by a $J/\psi$ and a $\rho$: $\bar{m}_{tetra} = 0.283 m_{J/\psi} + 0.717 m_{\rho} $. Whereas $\pi_1$ is modeled as a ``hybrid meson'', a state consisting of valence quarks and a constituent gluon: $\bar{m}_{hybrid} = 0.497 m_{u} + 0.497 m_{d} + 0.006 m_{G} $, where $m_G$ is the constituent gluon mass (taken to be $m_G=0.7\, \text{GeV}$ \cite{Hou:2001ig}). Both of these exotic states are massive perturbations with $\Delta=5$ for the hybrid meson and $\Delta=6$ for the tetraquark.

We provide therefore some Tables~\ref{tcharm}--\ref{thybrid} with the heavy mesons' masses, including charmonium and bottomonium, as well the considered exotic states. The masses are numerically obtained by solving the eigenvalue problem given by Eq.~(\ref{Mass_Spectra_T0}) for the shown parameters. The numerical solutions were obtained by applying the developed Matrix-Numerov scheme (\ref{finalnum}) to (\ref{Mass_Spectra_T0}).

\begin{table}[h!]
\centering
\begin{tabular}{||c||c|c|c|c||}
    \hline\hline
    \multicolumn{5}{|c|}{Charmonium $J/\psi$, $I^G (J^{PC}) = 0^+(1^{--})$} \\
    \hline\hline
    $n$ & State & $m_{\text{exp}}$ (\text{MeV}) & $m_{\text{th}}$ (\text{MeV}) & Relative error $\delta M$ \\
    \hline
    1 & $J/\psi$ & $3096.900 \pm 0.006$ & $3140.1$ & $1.39 \%$ \\\hline
    2 & $\psi(2S)$ & $3686.097 \pm 0.011$ & $3656.9$ & $0.79 \%$ \\\hline
    3 & $\psi(4040)$ & $4039.6 \pm 4.3$ & $4055.8$ & $0.40 \%$ \\\hline
    \hline
    \multicolumn{5}{||c||}{Experimental linear fit: $M^2 = 3.364(1.914 + n)$, $R^2 = 0.999508$} \\\hline
    \multicolumn{5}{||c||}{Experimental nonlinear fit: $M^2 = 10.947(-0.282 + n)^{0.399}$, $R^2 = 1.$} \\\hline
    \multicolumn{5}{||c||}{Theoretical linear fit: $M^2 = 3.295(2.015 + n)$, $R^2 = 0.999942$} \\\hline
    \multicolumn{5}{||c||}{Theoretical nonlinear fit: $M^2 = 7.001(0.685 + n)^{0.654}$, $R^2 = 1.$} \\\hline
\end{tabular}
\caption{\footnotesize Mass spectrum and Regge trajectories of the $J/\psi$ charmonium for $\kappa = 1.8 \, \text{GeV}$, $M = 1.7 \, \text{GeV}$, $\sqrt{\Gamma} = 0.53 \, \text{GeV}$, and $\alpha = 0.54$.\label{tcharm}}
\end{table}

\begin{table}[h!]
\centering
\begin{tabular}{||c||c|c|c|c||}
    \hline\hline
    \multicolumn{5}{|c|}{Bottomonium $\Upsilon$, $I^G (J^{PC}) = 0^+(1^{--})$} \\
    \hline\hline
    $n$ & State & $m_{\text{exp}}$ (\text{MeV}) & $m_{\text{th}}$ (\text{MeV}) & Relative error $\delta M$ \\
    \hline
    1 & $\Upsilon (1S)$ & $9460.4 \pm 0.09\pm0.04$ & $9506.5$ & $0.49 \%$ \\\hline
    2 & $\Upsilon (2S)$ & $10023.4 \pm 0.5$ & $9892.9$ & $1.30 \%$ \\\hline
    3 & $\Upsilon (3S)$ & $10355.1 \pm 0.5$ & $10227.2$ & $1.24 \%$ \\
    \hline\hline
    \multicolumn{5}{||c||}{Experimental linear fit: $M^2 = 8.864(9.175 + n)$, $R^2 = 0.9999$} \\\hline
    \multicolumn{5}{||c||}{Experimental nonlinear fit: $M^2 = 91.165(-0.113 + n)^{0.153}$, $R^2 = 1.$} \\\hline
    \multicolumn{5}{||c||}{Theoretical linear fit: $M^2 = 7.111(11.727 + n)$, $R^2 = 0.999997$} \\\hline
    \multicolumn{5}{||c||}{Theoretical nonlinear fit: $M^2 = 49.120(3.559 + n)^{0.402}$, $R^2 = 1.$} \\\hline
\end{tabular}
\caption{\footnotesize Mass spectrum and Regge trajectories of the $\Upsilon$ bottomonium for $\kappa = 9.9 \, \text{GeV}$, $M = 2.74 \, \text{GeV}$, $\sqrt{\Gamma} = 1.92 \, \text{GeV}$, and $\alpha = 0.863$.\label{tbottom}}
\end{table}

\begin{table}[h!]
\centering
\begin{tabular}{||c||c|c|c|c||}
    \hline\hline
    \multicolumn{5}{|c|}{Tetraquark $Z_c$, $I^G (J^{PC}) = 1^+(1^{+-})$} \\
    \hline\hline
    $n$ & State & $m_{\text{exp}}$ (\text{MeV}) & $m_{\text{th}}$ (\text{MeV}) & Relative error $\delta M$ \\
    \hline
    1 & $Z_c (3900)$ & $3887.2 \pm 2.3$ & $3890.1$ & $0.07 \%$ \\\hline
    2 & $Z_c (4200)$ & $4196 \substack{+35 \\ -32}$ & $4189.5$ & $0.15 \%$ \\\hline
    3 & $Z_c (4430)$ & $4478 \substack{+15 \\ -18}$ & $4444.6$ & $0.74 \%$ \\
    \hline\hline
    \multicolumn{5}{||c||}{Experimental linear fit: $M^2 = 2.471(5.112 + n)$, $R^2 = 1.$} \\\hline
    \multicolumn{5}{||c||}{Experimental nonlinear fit: $M^2 = 3.554(4.234 + n)^{0.874}$, $R^2 = 1.$} \\\hline
    \multicolumn{5}{||c||}{Theoretical linear fit: $M^2 = 2.311(5.564 + n)$, $R^2 = 0.999992$} \\\hline
    \multicolumn{5}{||c||}{Theoretical nonlinear fit: $M^2 = 6.836(2.665 + n)^{0.612}$, $R^2 = 1.$} \\\hline
\end{tabular}
\caption{\footnotesize Mass spectrum and Regge trajectories of the $Z_c$ tetraquark, for $\kappa = 1.75 \, \text{GeV}$, $M = 1.44 \, \text{GeV}$, $\sqrt{\Gamma} = 0.30 \, \text{GeV}$, and $\alpha = 0.539 $. \label{ttetra}}
\end{table}

\begin{table}[h!]
\centering
\begin{tabular}{||c||c|c|c|c||}
    \hline\hline
    \multicolumn{5}{|c|}{Hybrid meson $\pi_1$, $I^G (J^{PC}) = 0^-(1^{+-})$} \\
    \hline
    $n$ & State & $m_{\text{exp}}$ (\text{MeV}) & $m_{\text{th}}$ (\text{MeV}) & Relative error $\delta M$ \\
    \hline
    1 & $\pi_1 (1400)$ & $1354 \pm 25$ & $1354.7$ & $0.05 \%$ \\\hline
    2 & $\pi_1 (1600)$ & $1660 \substack{+15 \\ -11}$ & $1618.1$ & $2.52 \%$ \\\hline
    3 & $\pi_1 (2015)$ & $2014 \pm 20 \pm 16$ & $1849.9$ & $8.15 \%$ \\
    \hline\hline
    \multicolumn{5}{||c||}{Experimental linear fit: $M^2 = 1.111(0.593 + n)$, $R^2 = 0.99913$} \\\hline
    \multicolumn{5}{||c||}{Experimental nonlinear fit: $M^2 = 1.214(0.484 + n)^{0.953}$, $R^2 = 1.$} \\\hline
    \multicolumn{5}{||c||}{Theoretical linear fit: $M^2 = 0.793(1.308 + n)$, $R^2 = 0.999997$} \\\hline
    \multicolumn{5}{||c||}{Theoretical nonlinear fit: $M^2 = 0.816(1.268 + n)^{0.987}$, $R^2 = 1.$} \\\hline
\end{tabular}
\caption{\footnotesize Mass spectrum and Regge trajectories of the $\pi_1$ hybrid meson, for $\kappa = 0.468 \, \text{GeV}$, $M = 0.20 \, \text{GeV}$, $\sqrt{\Gamma} = 0.12 \, \text{GeV}$, and $\alpha = 0.034$. \label{thybrid}}
\end{table}

From the shown Tables \ref{tcharm}--\ref{thybrid}, we can conclude that a nonlinear Regge trajectory as parameterized in \cite{MartinContreras:2020cyg}, 
\begin{align}\label{nu_index}
    M_n^2 &= a(n+b)^\nu\,,\\
    \nu (\Bar{m}) &\equiv \nu = a_\nu + b_\nu e^{-c_\nu \bar{m}}\,.
\end{align}
tends to better fit the experimental masses. The computed masses, in general, agree well with the experimental data. We stress here again that the predictions for these masses, and associated nonlinear Regge behavior, are a consequence of the developed self-consistent EMD model.

\begin{figure}[htb!]
    \centering
    \begin{minipage}[t]{0.49\textwidth}
        \centering
        \includegraphics[width=\textwidth]{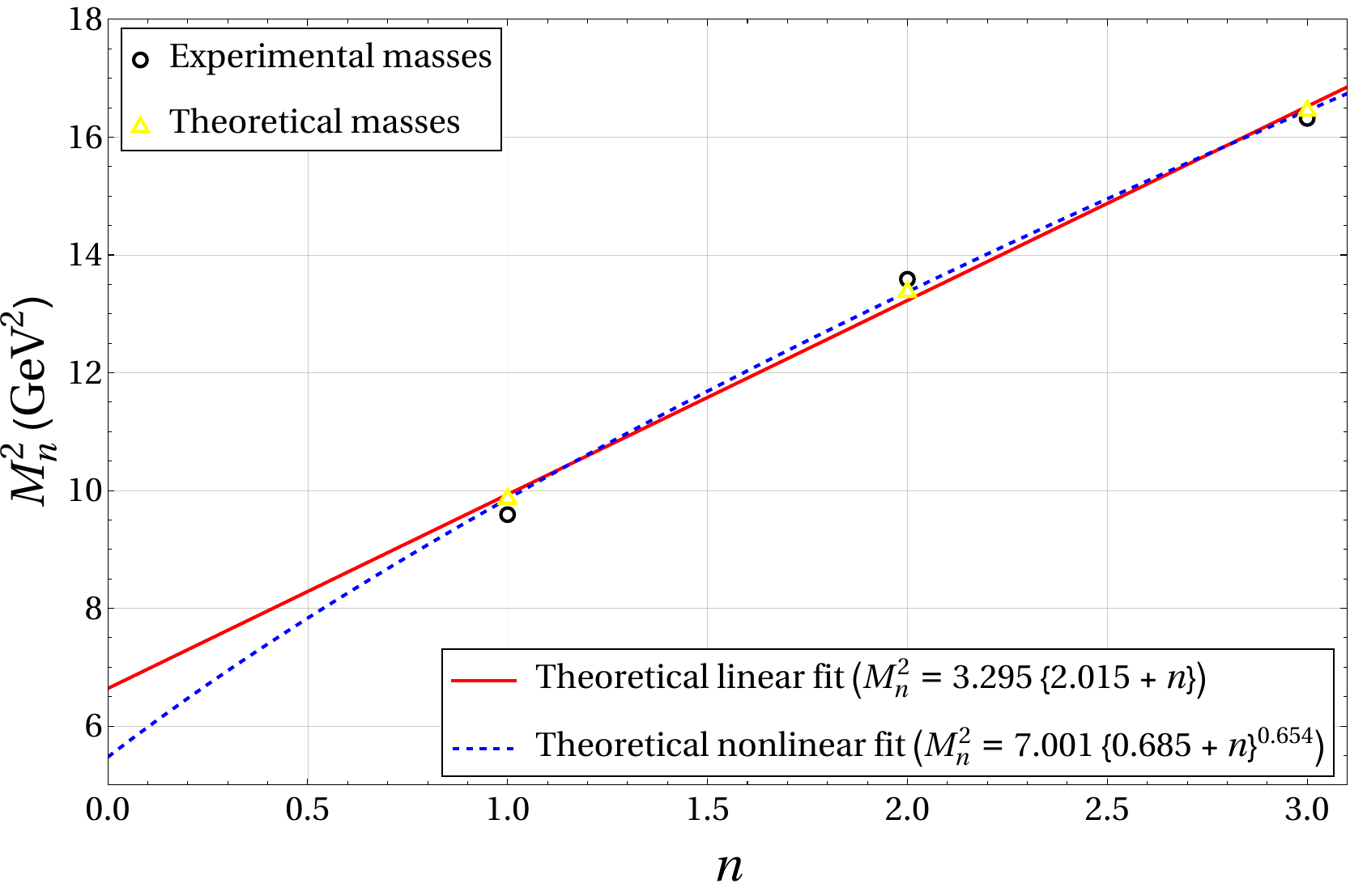}
        \caption{\small Linear and nonlinear fits for the charmonium.}
        \label{fig:RTcharm}  
    \end{minipage}%
    \hfill
    \begin{minipage}[t]{0.49\textwidth}
        \centering
        \includegraphics[width=\textwidth]{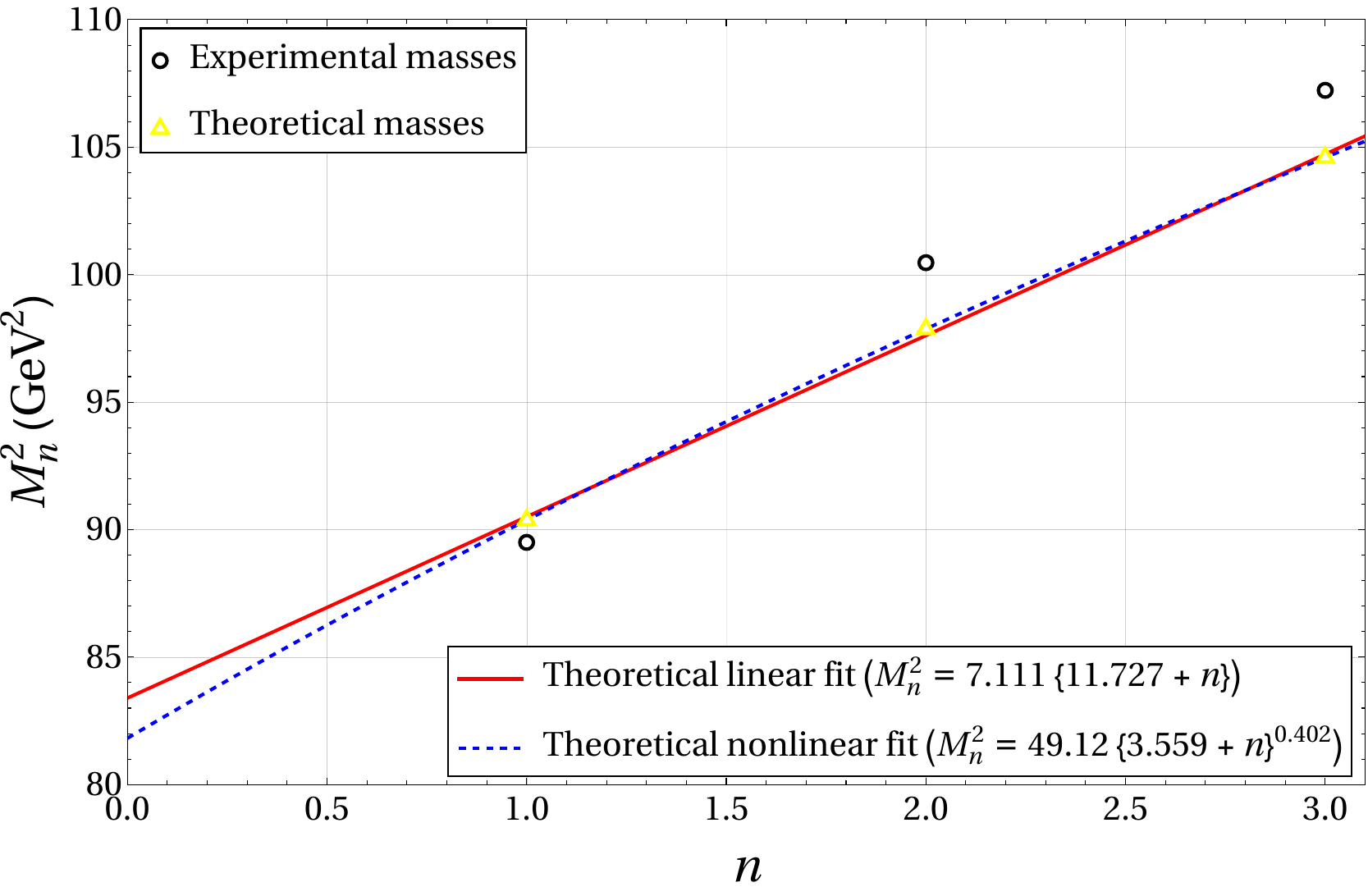}
        \caption{\small Linear and nonlinear fits for the bottomonium.}
        \label{fig:RTbottom} 
    \end{minipage}
\end{figure}

\begin{figure}[htb!]
    \centering
    \begin{minipage}[t]{0.49\textwidth}
        \centering
        \includegraphics[width=\textwidth]{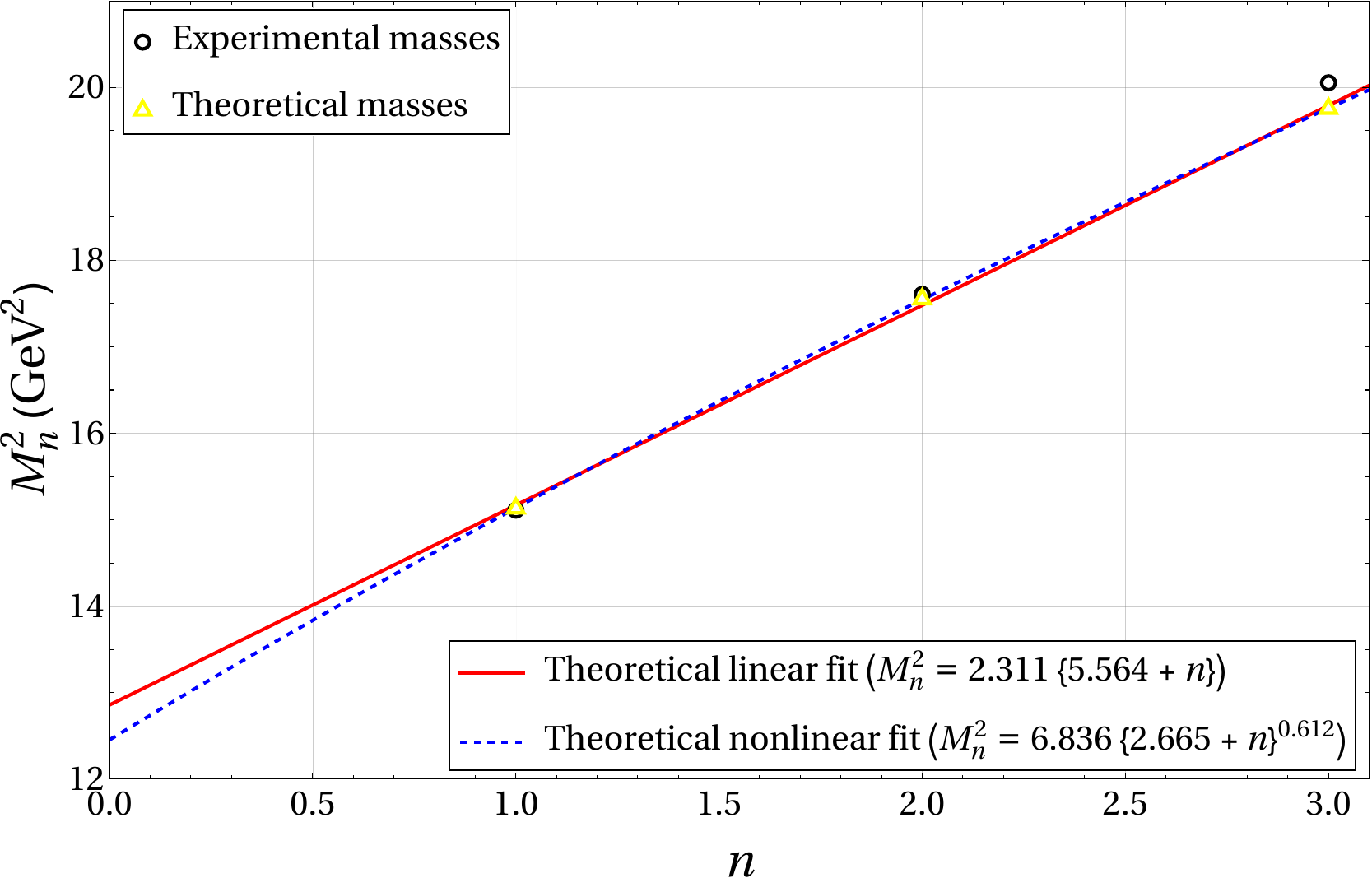}
        \caption{\small Linear and nonlinear fits for the tetraquark.}
        \label{fig:RTtetra}  
    \end{minipage}%
    \hfill
    \begin{minipage}[t]{0.49\textwidth}
        \centering
        \includegraphics[width=\textwidth]{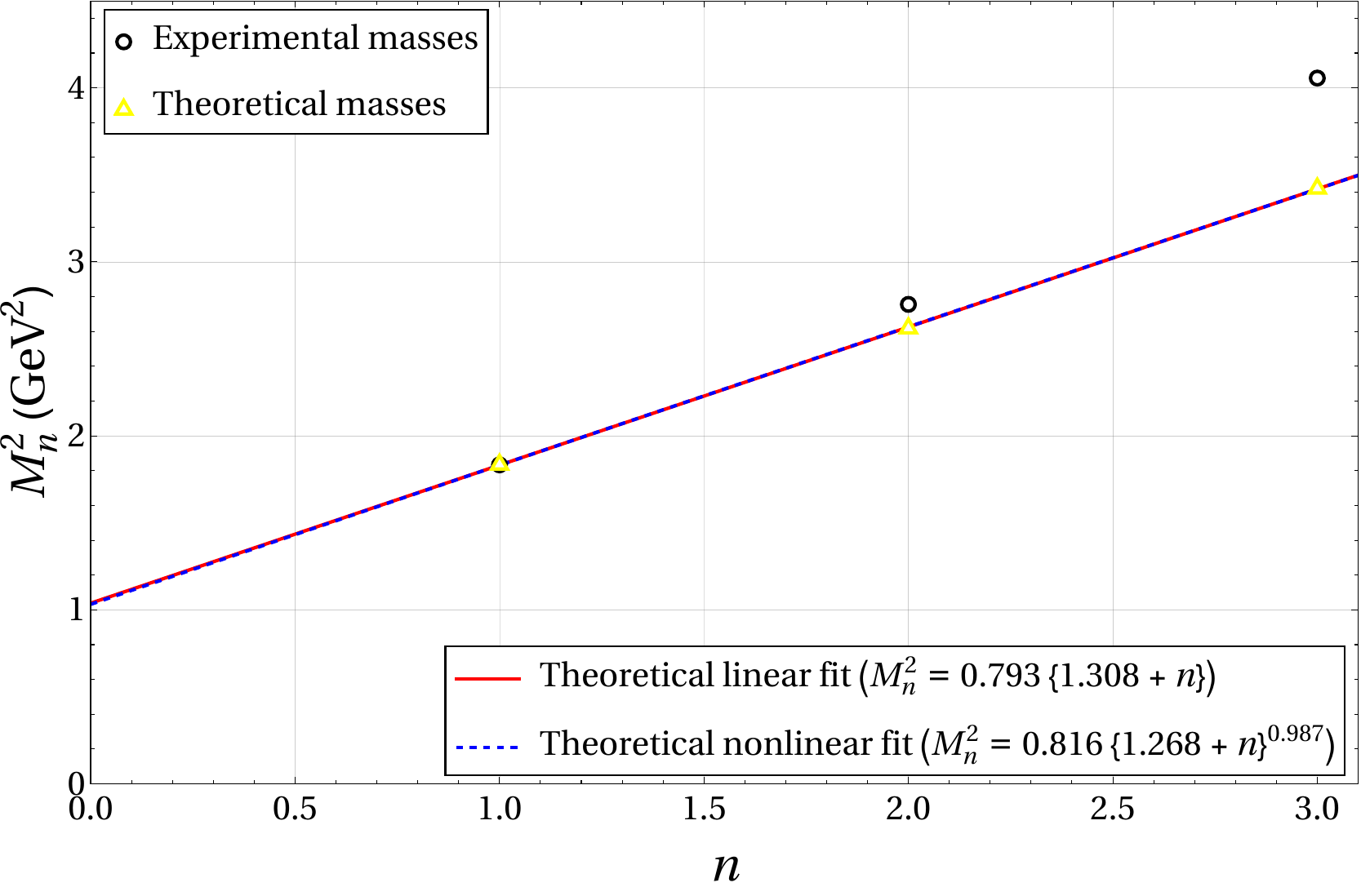}
        \caption{\small Linear and nonlinear fits for the hybrid meson.}
        \label{fig:RT hybrid}  
    \end{minipage}
\end{figure}

In the Figs.~\ref{fig:RTcharm}--\ref{fig:RT hybrid} we can better visualize the difference between a linear and a nonlinear Regge trajectory. We note that as the particle mass decreases, like the considered hybrid meson, the linear and nonlinear trajectories become nearly identical. 

We thus conclude that the derived equation of motion (\ref{Mass_Spectra_T0}) provides a consistent framework for approximating the mass spectra of heavy and exotic mesonic states. The presented results closely match the experimental data and show excellent agreement with previous findings reported in Ref.~\cite{MartinContreras:2020cyg}.

\begin{figure}[htb!]
    \centering
    \begin{minipage}[t]{0.49\textwidth}
        \centering
        \includegraphics[width=\textwidth]{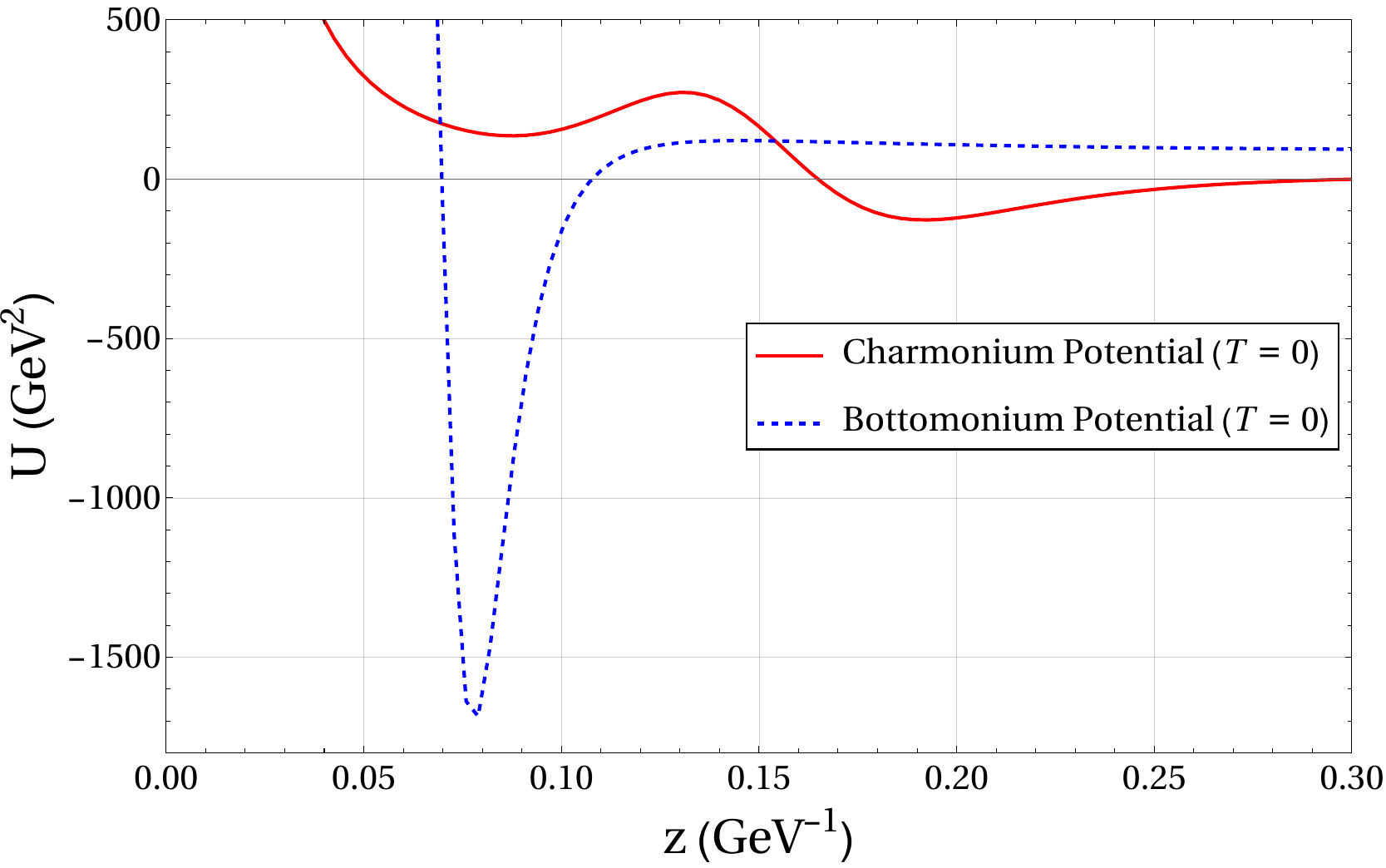}
        \caption{\small Charmonium and bottomonium potentials at $T=0$.}
        \label{fig:T0potcharm}
    \end{minipage}%
    \hfill
    \begin{minipage}[t]{0.49\textwidth}
        \centering
        \includegraphics[width=\textwidth]{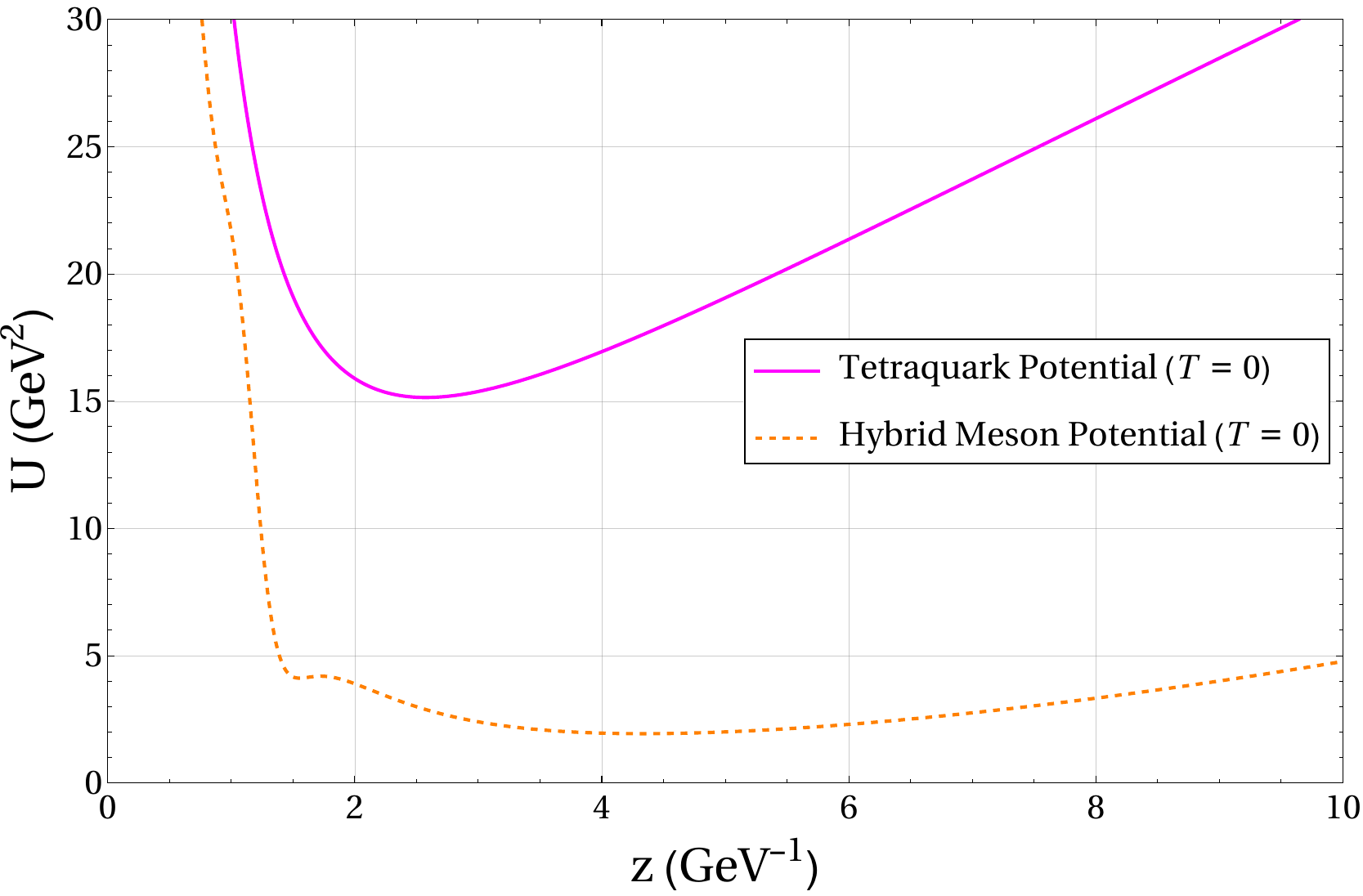}
        \caption{\small Tetraquark and hybrid meson potentials at $T=0$.}
        \label{fig:T0pottetra}
    \end{minipage}
\end{figure}

The resulting effective potential from the Schrödinger-like equation (\ref{Mass_Spectra_T0}) is illustrated in Figs.~\ref{fig:T0potcharm} and \ref{fig:T0pottetra} for the charm and tetraquark states, respectively. These plots highlight the dynamics of the hadronic structures within the adopted holographic framework. Analyzing this potential allows us to gain valuable insights into the stability, behavior, and mass spectra of tetraquark states based on the selected model parameters. In particular, we emphasize how the model parameters modify the hadronic dynamics within this holographic approach, especially under varying conditions. Later, we will look into how this potential is deformed by increasing the temperature $T$ and chemical potential $\mu$.

\begin{figure}[htb!]
    \centering
    \begin{minipage}[t]{0.49\textwidth}
        \centering
        \includegraphics[width=\textwidth]{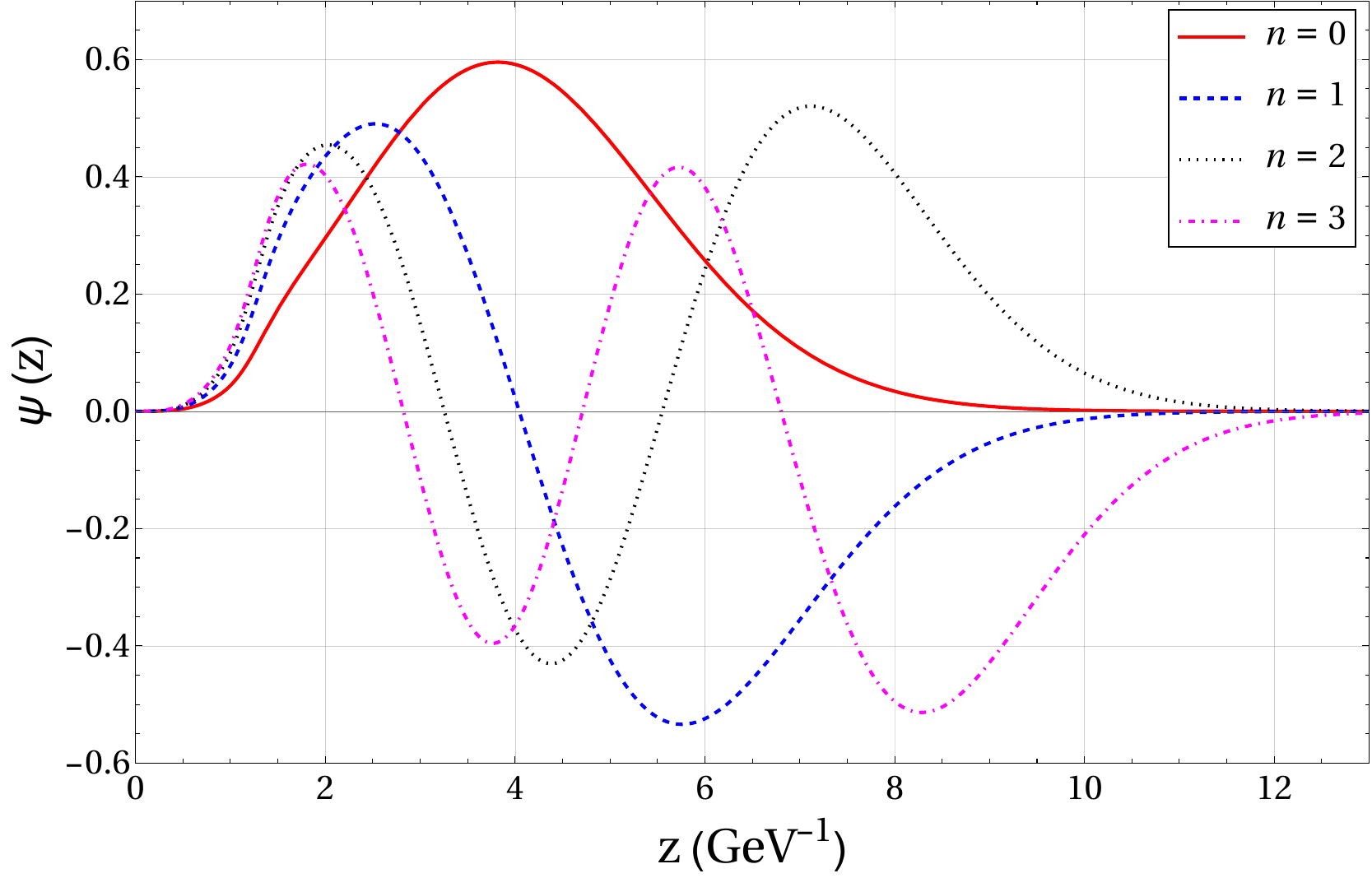}
        \caption{\small Hybrid meson wavefunctions.}
        \label{fig:T0wavehybrid}
    \end{minipage}%
    \hfill
    \begin{minipage}[t]{0.49\textwidth}
        \centering
        \includegraphics[width=\textwidth]{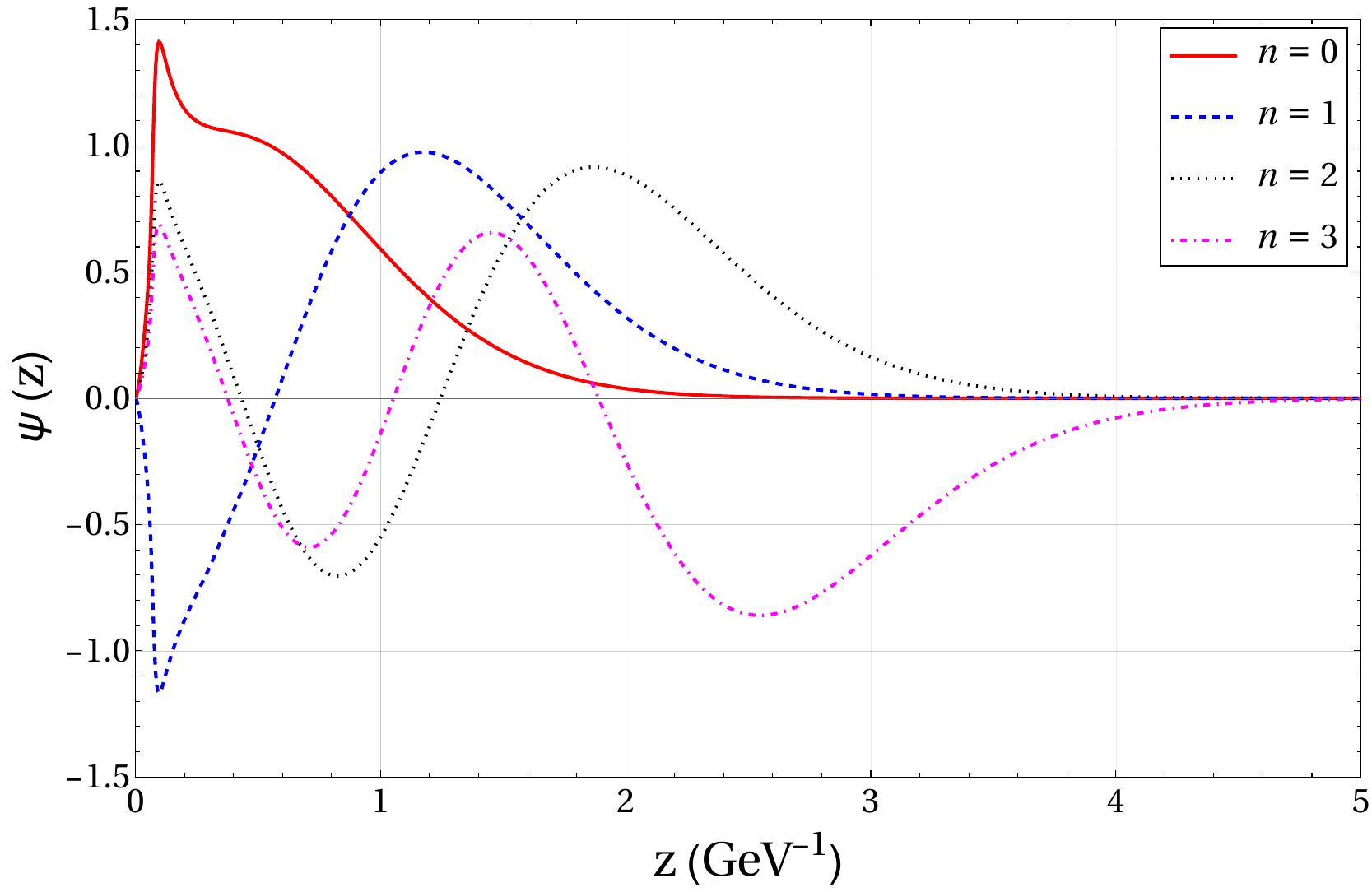}
        \caption{\small Bottomonium wavefunctions.}
        \label{fig:T0wavebottom}
    \end{minipage}
\end{figure}

In addition, the corresponding wavefunctions for hybrid meson and bottomonium states, obtained by numerically solving Eq.~(\ref{Mass_Spectra_T0}), are shown in Figs.~\ref{fig:T0wavehybrid} and \ref{fig:T0wavebottom}. It can be readily observed that, as the model parameters vary—specifically, lower values for the hybrid meson and higher values for bottomonium—the wavefunctions decrease more rapidly and become more distorted near the boundary region. 

\newpage

\section{Thermodynamical properties of the EMD model} \label{thermo}

This Section delves into the finite temperature ($\mu=0$) EMD background, relating the black hole horizon to the Hawking temperature and thermodynamic quantities. It shows the appearance of stable and unstable black hole branches and identifies a first-order Hawking--Page transition between thermal AdS (confined) and the black hole phase (deconfined). Finally, it sets up the vector-fluctuation equation in the black hole geometry in a form suitable for later spectral and melting analyses.

\subsection{Black hole temperature}

For the analysis of a system at finite temperature (where we set \( \mu = 0 \) in this calculation), the blackening function \( g(z) \) must first be determined for the chosen profile \( A(z) = -\frac{\mathcal{C}}{8} z^2 \). From Eq.~(\ref{EMD_muzero}), it is evident that \( f(z) \) does not contribute to \( g(z) \). By evaluating the integral, we obtain
\begin{equation} \label{g_T_mu_0}
g(z) = 1 + \frac{e^{\frac{3 \mathcal{C} z^2}{8}} \left(8 - 3 \mathcal{C} z^2\right) - 8}{e^{\frac{3 \mathcal{C} z_h^2}{8}} \left(3 \mathcal{C} z_h^2-8 \right) + 8}\,,
\end{equation}

The black hole temperature is calculated as
\begin{equation} \label{BH_T_mu0}
\boxed{T = -\frac{g'(z_h)}{4 \pi} = \frac{9  z_h^3 \mathcal{C}^2 e^{\frac{3 \mathcal{C} z_h^2}{8}}}{16 \pi \left[8 + e^{\frac{3 \mathcal{C} z_h^2}{8}} \left(3 \mathcal{C} z_h^2-8\right)\right]}\,.}
\end{equation}

\begin{figure}[h!]
    \centering
    \includegraphics[width=0.75\textwidth]{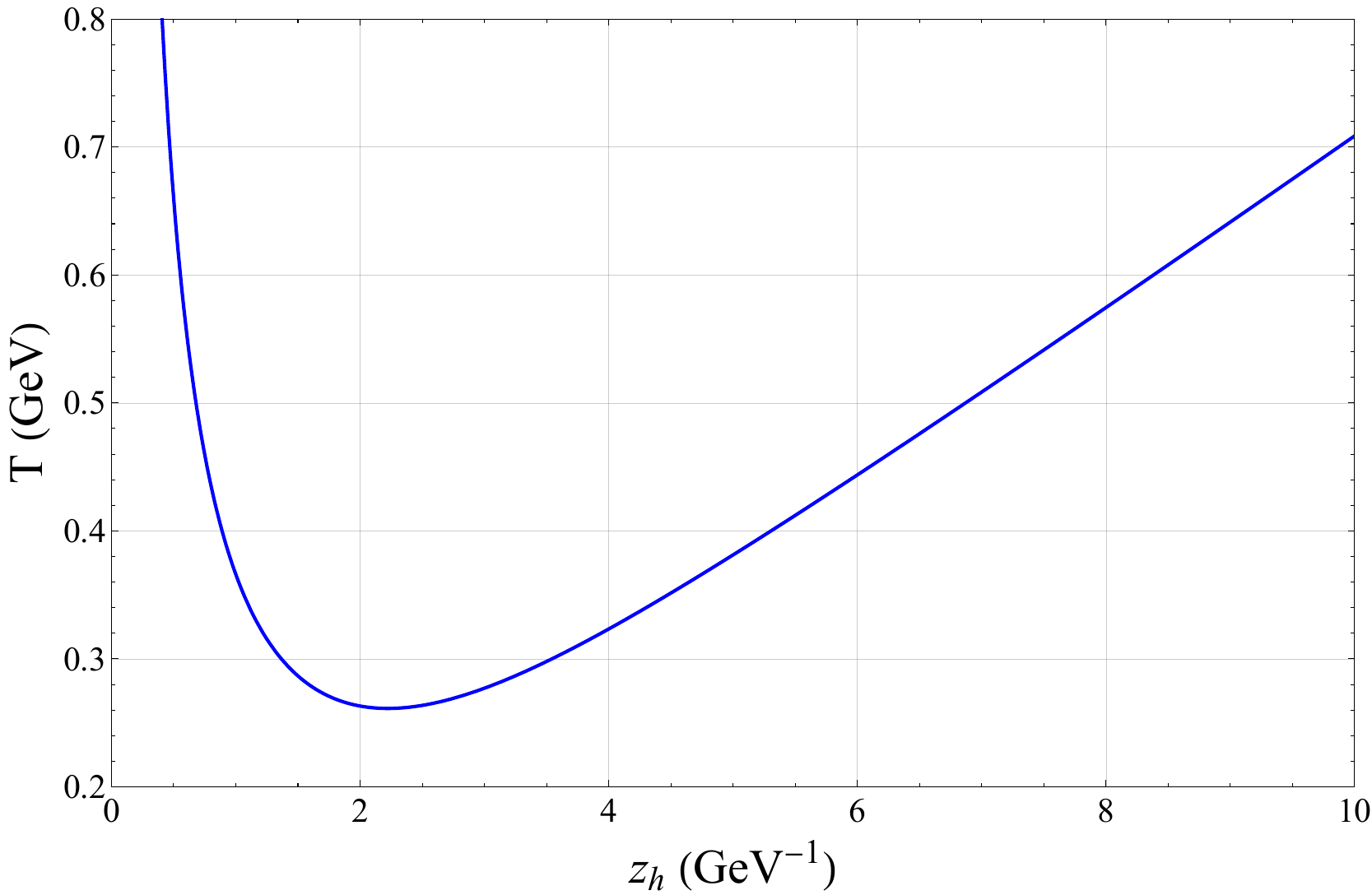}
    \caption{\small The Hawking temperature \( T \) as a function of horizon radius \( z_h \).}
    \phantomsection
    \label{fig:Tempmu0}
\end{figure}

Fig.~\ref{fig:Tempmu0} is presented and shows the Hawking temperature $T$ as a function of the black hole horizon $z_h$ for the model parameter $\mathcal{C} = 1.16 \, \text{GeV}^2$. In this context, the thermal AdS phase exists at all temperatures. However, there is a minimum temperature below which a black hole solution does not exist. Above this minimum temperature, two black hole solutions exist: a large black hole and a small one. The large black hole (small $z_h$) has a temperature that decreases as the horizon radius $z_h$ increases. It is thermodynamically stable due to its positive specific heat. In contrast, the small black hole (large $z_h$) has a temperature that rises with the horizon radius. It is thermodynamically unstable because of its negative specific heat. 

\subsection{Free energy difference $\Delta F$}

The black hole entropy is given by the usual Bekenstein–Hawking formula $S=\mathcal{A}_h/4G_5$, where $\mathcal{A}_h$ is the horizon area. The induced spatial metric, $t=\text{constant}$, for our metric Ans\"atze~(\ref{ansatz_metric}), on the horizon slice ($z=z_h$) is given by
\begin{equation}
    \gamma_{ij} = \frac{L^2 e^{2A(z_h)}}{z_h^2} \delta_{ij} \implies \sqrt{\gamma}= \frac{L^3 e^{3A(z_h)}}{z_h^3} \,,
\end{equation}
where $\gamma=\det{\gamma_{ij}}$. So we conclude that
\begin{equation}
    \mathcal{A}_h = \int d^3x \sqrt{\gamma} = V_3 \frac{L^3 e^{3A(z_h)}}{z_h^3} \implies S = \frac{V_3}{4G_5} \frac{L^3 e^{3A(z_h)}}{z_h^3} \,.
    \label{BHentropy}
\end{equation}

In order to compute the Helmholtz free energy of the black hole system, we start from the fundamental relation $dF=-S\,dT$ and, as both $S$ and $T$ are functions of $z_h$, write the following relation
\begin{equation}
\boxed{F(z_h) = \int_{z_h}^{\infty}dx\,S(x)\,dT'(x)= -S(z_h)\,T(z_h) - \int_{z_h}^{\infty}dx\,S'(x)T(x) \,.
}
    \label{freeE}
\end{equation}
where we integrated by parts and chose $F$ such that $F(\infty)\to0$. 

Here, the two phases considered are the thermal AdS and black hole. Therefore, we can define the free energy difference as $\Delta F=F_{\text{BH}}-F_{\text{AdS}}$, for the computed free energy~(\ref{freeE}). So, $\Delta F$ is the free energy difference  between the  black hole and thermal AdS geometries. Usually, such a quantity is computed for a range of temperature values $T$ with some chemical potential $\mu$ fixed. For the results that follows, we adopt a normalization in which the thermal AdS free energy is set to zero.

\begin{figure}[h!]
    \centering
    \includegraphics[width=0.75\textwidth]{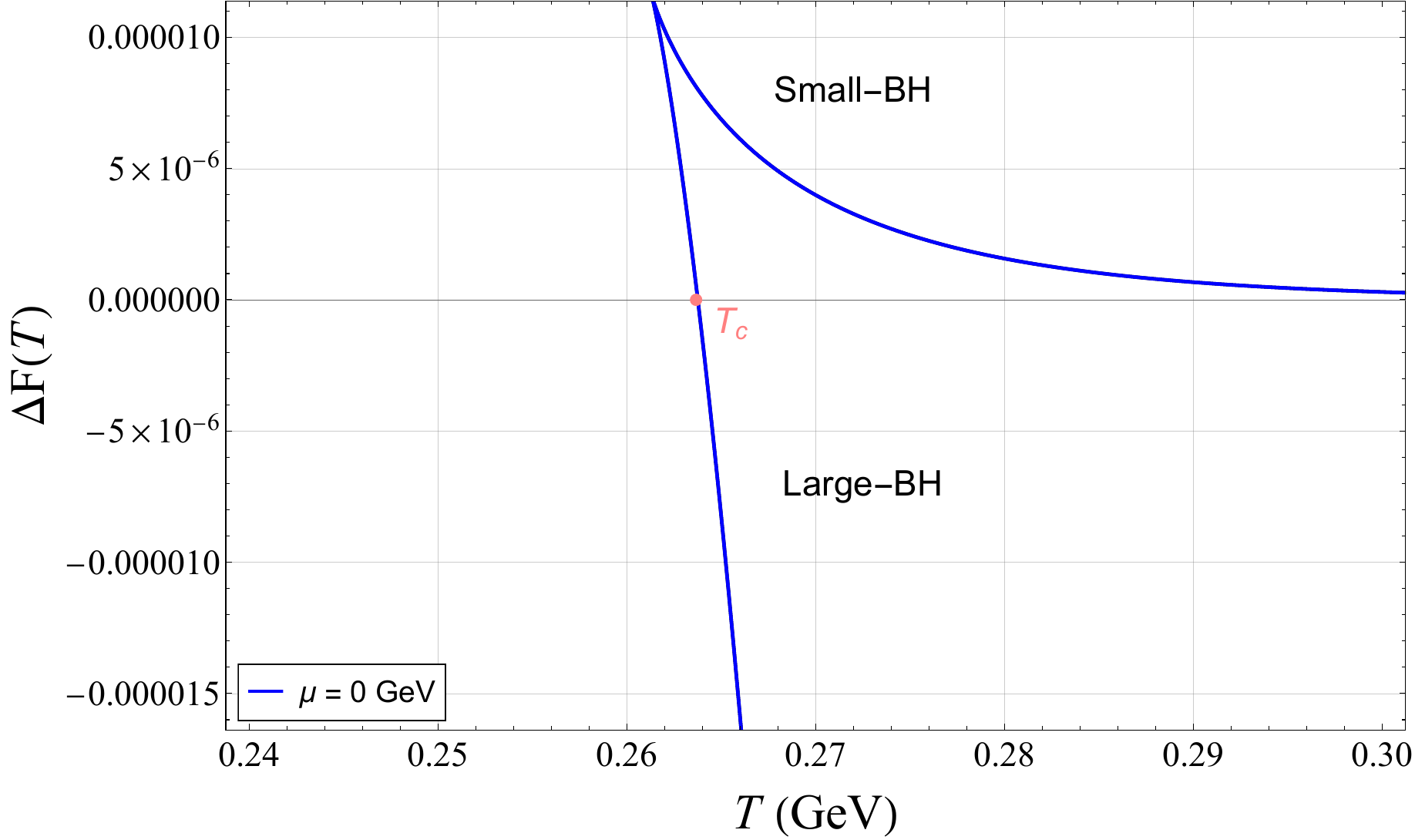}
    \caption{\small Free energy difference $\Delta F(T)$ for $\mu=0$.}
    \phantomsection
    \label{fig:freemu0}
\end{figure}

The corresponding free-energy diagram is shown in Fig.~\ref{fig:freemu0} for the system, at finite temperature, with $\mu=0$. As the temperature varies, the dominance between the large black hole phase and the thermal AdS phase shifts. Specifically, at high temperatures, the large black hole phase possesses the lowest free energy, whereas at low temperatures, the thermal AdS phase is energetically favored. Consequently, a first-order Hawking-Page phase transition occurs between these two phases. For $\mu=0$, this phase transition appears at $T_{c}=0.264~\text{GeV}$. The thermal AdS phase corresponds to the confined phase in the boundary theory, while the black hole phase represents the deconfined phase. In later Sections, we will examine how introducing a nonzero chemical potential influences this transition temperature. It is important to note that the small black hole phase consistently exhibits a higher free energy than both the large black hole and thermal AdS phases. As a result, it corresponds to the global maximum among the solutions and remains thermodynamically disfavored at all temperatures. Therefore, in the following Sections, we will utilize the large black hole phase to analyze the finite temperature behavior of various observables.

\subsection{Equation of motion at finite temperature}

In order to determine the perturbation's equation of motion in the presence of a black hole background geometry, described by a generic \( g(z) \), we put back the factors of $g(z)$ in the equation of motion (\ref{EOM_2}) to find

\begin{equation} \label{FT1}
\frac{z e^{-A(z)}g(z)}{L f(z)} \partial_z \left(\frac{L e^{A(z)} f(z) g(z)}{z} \partial_z A^\mu \right) + \Box A^\mu = 0\,,
\end{equation}

From now on, we proceed in the particle's rest frame, where the four-momentum is given by \( q^\mu = (\omega, 0, 0, 0) \). Employing the radial gauge \( A_z = 0 \), we consider plane wave solutions of the form
\begin{equation}
A^\mu (z, x^\nu) = e^{i q_\alpha x^\alpha} A^\mu(z, \omega) = e^{-i \omega t} A^\mu(z, \omega)\,,
\end{equation}

By applying a Fourier transform, as shown in (\ref{fourier}), and doing the change of variable $A^\mu=e^{-H}\psi^\mu$ to (\ref{FT1}) we transform the finite temperature equation of motion into 
\begin{equation} \label{EOM_T}
-\partial_{r_*}^2 \psi^\mu(z) + V(z) \psi^\mu(z) = \omega^2 \psi^\mu(z)\,,
\end{equation}
where the tortoise coordinate $\partial_{r_*} = -g(z) \partial_z$ was applied to (\ref{FT1}) in order to redefine it into a Schr\"odinger-like form. The associated effective potential is given by 
\begin{equation} \label{Potential_T}
\boxed{V(z) = \left(\partial_{r_*} \left[\frac{1}{2} \log\left(\frac{L e^{A(z)} f(z)}{z}\right)\right]\right)^2 
+ \partial_{r_*}^2 \left[\frac{1}{2} \log\left(\frac{L e^{A(z)} f(z)}{z}\right)\right]\,.}
\end{equation}
in which \( H \) is the same function as defined in Eq.~(\ref{B_def}). Equipped with the equation of motion formulated at finite temperature, we are able to calculate the spectral functions of heavy and exotic mesons, enabling the analysis of their melting behavior. Considering that (\ref{EOM_T}) is generic in $g(z)$, this is also the equation employed to study the states in a finite density plasma ($\mu>0$) as the chemical potential only alters the black hole metric. In the next Section, we lay out the numerical procedure to do such computation.

An important remark concerns the supplementary Landau gauge choice used to derive the equation of motion for the perturbation (\ref{EOM_T}). As discussed in \cite{Dudal:2015kza}, introducing a nonzero spatial momentum breaks the system’s isotropy, potentially rendering the chosen gauge unsuitable for a given metric. In such a case, considering a finite $q$ would result in two distinct equations of motion—one for transverse and one for longitudinal field propagation relative to the momentum—leading to different predictions for the spectral functions. However, since we have set $q$ beforehand, the effective potentials for these equations coincide, leaving us with Eq.~(\ref{EOM_T}).

\newpage

\section{Computation of the spectral functions} \label{procespec}

In this Section, the numerical procedure for computing the spectral functions is outlined. This involves certain approximations, such as the use of a tortoise coordinate and a Padé-like approximant for $B(z)$, to facilitate the calculation of the retarded Green's function.

\subsection{Approximation to the tortoise coordinate $r_*$}
The procedure outlined in \cite{Jena:2024cqs,Toniato:2025gts} is followed here to compute the spectral functions. Usually, in a non-dynamical soft-wall model, where Einstein's equations are not necessarily satisfied, \( 1/g(z) \) can be explicitly integrated, allowing for a closed-form expression for \( r_* \). For instance, in the case of a simple blackening function \( g(z) = 1 - \frac{z^4}{z_h^4} \), the expression for \( r_* \) is given by
\begin{eqnarray}
	r_*(z) = \frac{z_h}{2} \left[\log\left({\frac{z_h-z}{z_h+z}}\right) -\arctan\left({\frac{z}{z_h}}\right) \right]\,,
\end{eqnarray}

Moreover, as the complexity of the expression for \( g(z) \) prevents one from obtaining a closed-form expression for \( r_* \), it is necessary resort to some sort of approximation. To proceed with the computations, we perform a series expansion of \( 1/g(z) \), around $z=z_h$, up to a specified order and then integrate the resulting expression. Expanding up to the second order, as an example, this approach yields an approximate expression for \( r_* \) as
\begin{equation}\label{tortEMD}
\boxed{\begin{split}
          r_*(z) &\approx  \frac{3 \mathcal{C} z_h^2+ 8 e^{-\frac{3 \mathcal{C} z_h^2}{8}}-8}{864 \mathcal{C}^2 z_h^6} z \Bigl[z^2 \left(80-3 \mathcal{C}^2 z_h^4\right)-48 z z_h \left(\mathcal{C} z_h^2+10\right)\\&+3 z_h^2 \left(\mathcal{C} z_h^2 \left[3 \mathcal{C} z_h^2+80\right]+432\right)\Bigl]-384
   z_h^3 \log \left(1-\frac{z}{z_h}\right)\,.
    \end{split}}
\end{equation}

\begin{figure}[h!]
    \centering
    \includegraphics[width=0.75\textwidth]{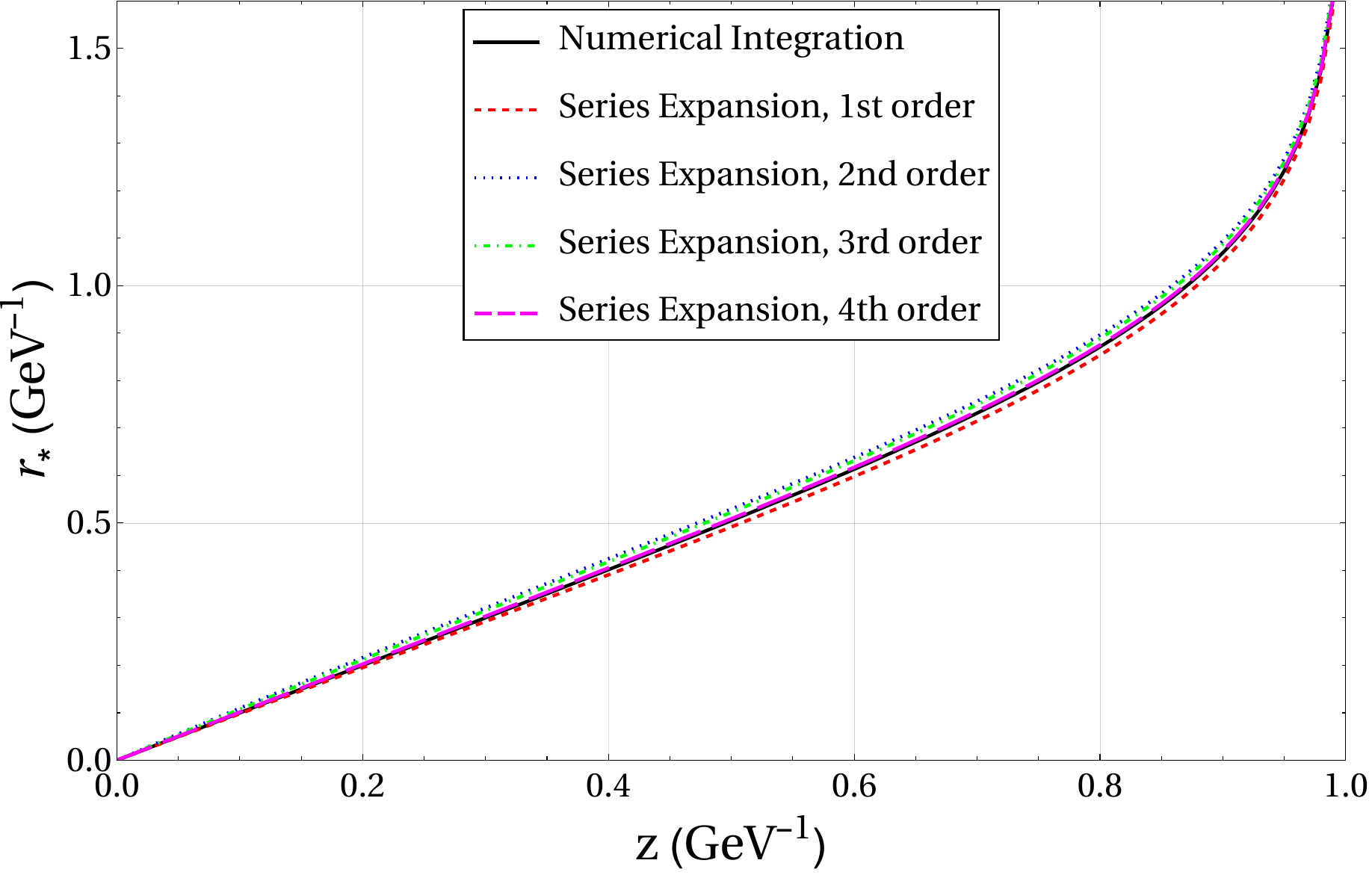}
    \caption{\small Tortoise coordinate \( r_* \) approximations.}
    \phantomsection
    \label{fig:Tortapprox}
\end{figure}

A comparison between the numerical approximation (integrating  \( 1/g(z) \) numerically) and the series expansion is plotted in Fig.~\ref{fig:Tortapprox} for the tortoise coordinate \( r_* \). We can see that even at a low order, the series expansion of the tortoise coordinate agrees quite well with the numerical result.

\subsection{Construction of a Pad\'{e}-like approximant}

Another key approximation employed throughout this Section to enhance the stability of the numerical scheme and improve the Frobenius analysis is the Padé-like approximation applied to the hyperbolic tangent term in the definition of $f(z)$, which is embedded in the $B(z)$ function in Eq.~(\ref{f_definition}). 

The Pad\'{e} approximation is a method for approximating a function using a rational function, expressed as the ratio of two polynomials \cite{Baker_Graves-Morris_1996}. This technique often provides superior convergence and accuracy compared to standard polynomial approximations, particularly for functions with singularities or complex behavior. A Pad\'{e} approximant is characterized by the degrees of its numerator and denominator polynomials, denoted as \( [n, m] = P_n(x)/Q_m(x) \), where \( n \) and \( m \) specify the respective polynomial degrees. The Pad\'{e} approximant of order $[n,m]$ is unique and shares its Taylor expansion with the function it approximates up to order $m+n$. 

As we have that \( \tanh{x} = \frac{e^{2x} - 1}{e^{2x} + 1} \), the Pad\'{e} approximant for \( e^x \) can be computed and substituted into the formula for \( \tanh{x} \) to find a good rational approximation to $\tanh{x}$ (what we call Pad\'{e}-like approximant). To achieve a more accurate approximation, we focus on deriving a Pad\'{e}-like approximant for \( \tanh{x} \) centered at a generic point \( x = a \). This process begins with the observation that the Taylor series expansion of \( e^x \) around \( x = a \) is given by
\begin{equation}\label{expexp}
    e^x = e^a \sum_{k=0}^\infty \frac{1}{k!} (x - a)^k\,,
\end{equation}

The constructed expansion (\ref{expexp}) allows one to simply extend the Pad\'{e} approximant for $e^x$, computed at $x=0$  \cite{05fc0d9b-f28e-3545-ab7b-94a0385f9451}. We find then that the Pad\'{e} approximant $[n, m]$ for $e^x$ centered at $x=a$ is given by
\begin{equation} \label{padeexp}
    e^x_{[n,m]} = \frac{P_n(x)}{Q_m(x)} = \frac{
e^a \sum_{j=0}^n \frac{(n+m-j)!}{n!} \binom{n}{j} (x-a)^j
}{
\sum_{j=0}^m \frac{(n+m-j)!}{n!} \binom{m}{j} (a-x)^j
}\,,
\end{equation}

As an example, we set $[n,m] = [2,2]$ to get
\begin{equation} \label{exp22}
    e^x_{[2,2]} = \frac{
e^a \left[
1 + \frac{1}{2}(x-a) + \frac{1}{12}(x-a)^2
\right]
}{
1 + \frac{1}{2}(a - x) + \frac{1}{12}(a - x)^2
}\,.
\end{equation}

We then apply Eq. (\ref{exp22}) to \( \tanh{x} = \frac{e^{2x} - 1}{e^{2x} + 1} \), yielding the $[2,2]$ Pad\'{e}-like\footnote{Note that the Pad\'{e}-approximant of a ratio is not the ratio of the Pad\'{e} approximants. However, for the eventual needs, this Pad\'{e}-like approximant serves perfectly well, see later in the text.} approximant for $\tanh{x}$, centered at $x=a$, as:
\begin{equation} \label{tanh22}
\tanh{x}_{[2,2]} = 1 + \frac{
8x(a+3) - 2a(a+6) - 8x^2 - 24
}{
e^a (a - 2x)\left[a - 2(3 + x)\right] + 12(1 + e^a - x) + 4x^2 - 4ax + a^2 + 6a
}\,,
\end{equation}

For the chosen parameters, $a=3$ gives a good interpolation between the behavior of $B(z)$ near zero and for higher values of $z$. We also set $x \rightarrow \left(\frac{1}{Mz} - \frac{\kappa}{\sqrt{\Gamma}}\right)$ in (\ref{tanh22}) by looking at (\ref{f_definition}). Given so, we can write the $[2,2]$ Pad\'{e}-like approximant for $B(z)$ as
\begin{multline}\label{Pade22}
B(z) \approx B(z)_{[2,2]} = (\kappa z)^{2-\alpha} + Mz
\\
\!\!\!\!\!\!\!\scalemath{.95}{+ \frac{
 M^2 z^2 \left[3\Gamma ( e^3 -13) + 4 \kappa^2 ( e^3 -1 ) - 24 \sqrt{\Gamma} \kappa \right] + 8 \sqrt{\Gamma} Mz (3 \sqrt{\Gamma} + \kappa- \kappa e^3 )  + 4 \Gamma (e^3 -1)
}{
M^2 z^2 \left[3\Gamma (e^3 + 13)+ 4 \kappa^2 (e^3 + 1) + 24 \sqrt{\Gamma} \kappa \right] - 
8 \sqrt{\Gamma}  M z ( 3 \sqrt{\Gamma} + \kappa + \kappa e ^ 3) + 4 \Gamma (e^ 3 + 1)
}}.
\end{multline}

\begin{figure}[h!]
    \centering
    \includegraphics[width=0.75\textwidth]{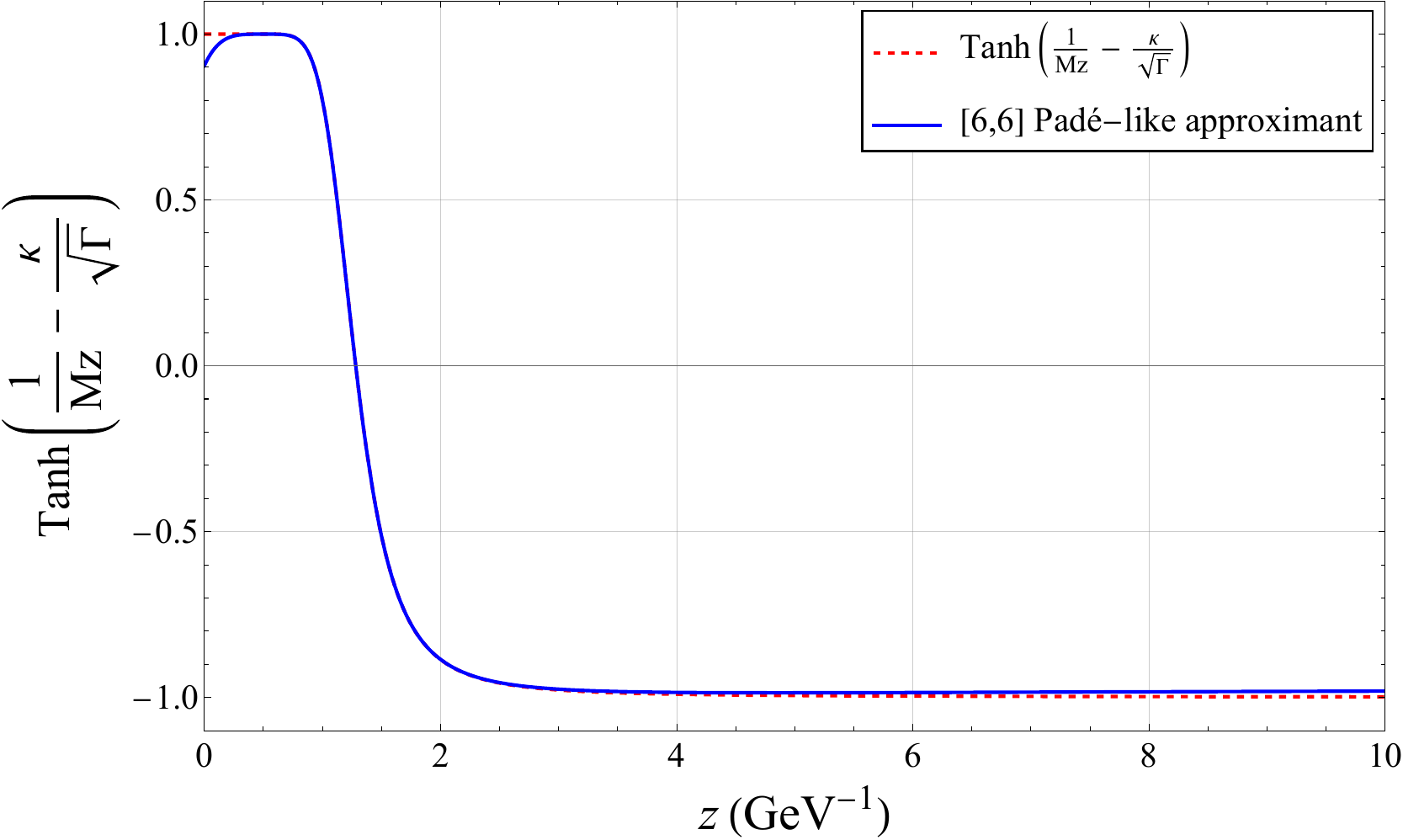}
    \caption{\small Pad\'{e}-like approximant to $\tanh{\left(\frac{1}{Mz}-\frac{\kappa}{\sqrt{\Gamma}}\right)}$ for hybrid meson parameters.}
    \phantomsection
    \label{fig:Padetan}
\end{figure}

Fig.~\ref{fig:Padetan} shows this Pad\'{e}-like approximant, using the $[6,6]$ Pad\'{e} approximant to the exponential (centered at $a=3$), to the desired function $\tanh{\left(\frac{1}{Mz}-\frac{\kappa}{\sqrt{\Gamma}}\right)}$. We can clearly see that the constructed rational function does an excellent job at approximating $\tanh{\left(\frac{1}{Mz}-\frac{\kappa}{\sqrt{\Gamma}}\right)}$.

\begin{figure}[h!]
    \centering
    \includegraphics[width=0.75\textwidth]{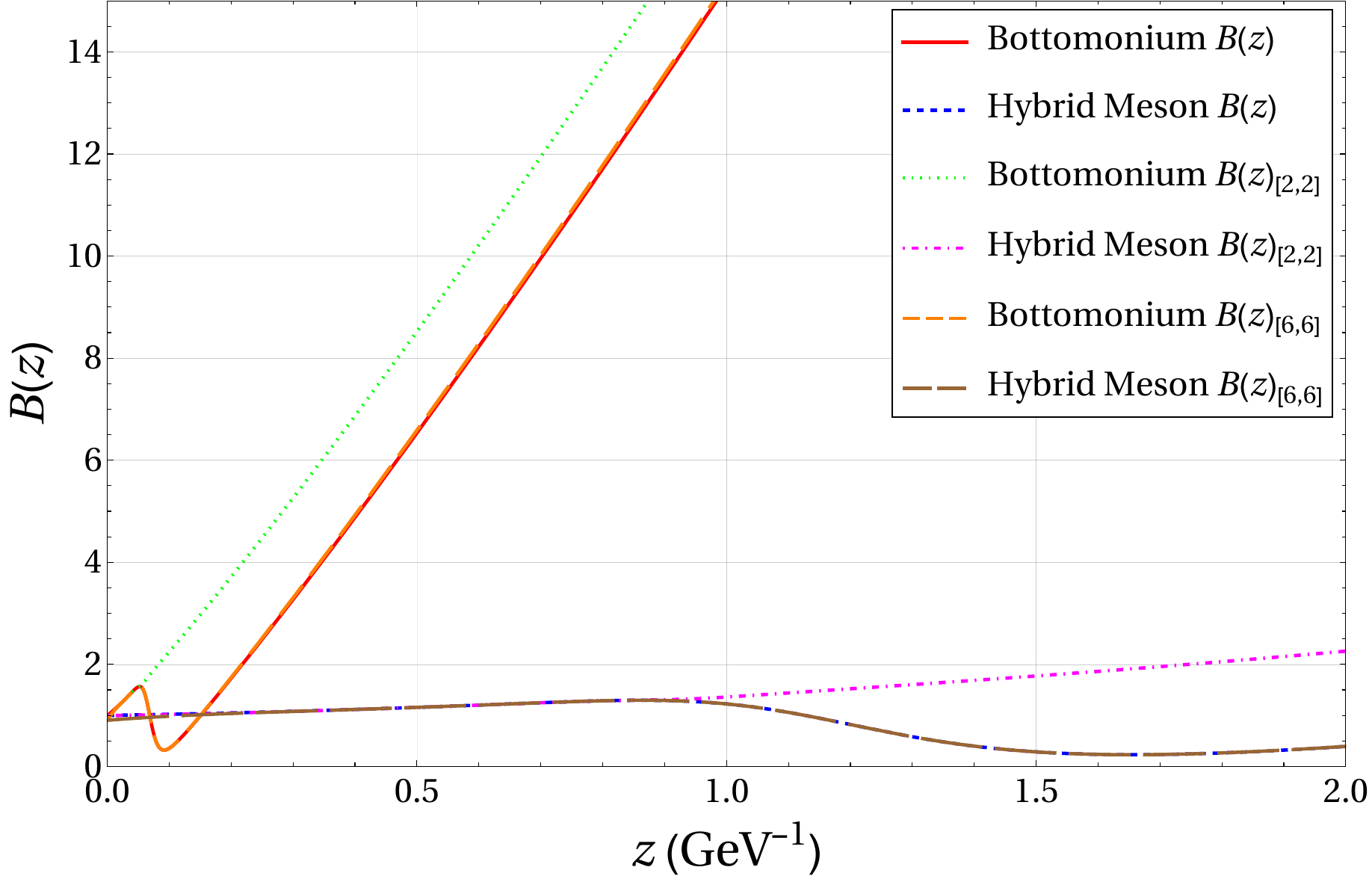}
    \caption{\small Padé-like approximants $B(z)_{[2,2]}$ and $B(z)_{[6,6]}$.}
    \phantomsection
    \label{fig:PadeB}
\end{figure}

Fig.~\ref{fig:PadeB} illustrates the Pad\'{e}-like approximants for $B(z)$ of the bottomonium and hybrid meson states parameters. 

In addition to \( B(z)_{[2,2]} \), the Pad\'{e}-like approximant \( B(z)_{[6,6]} \) was also computed following the exact same procedure as previously described in this work and subsequently plotted. It is evident that increasing the degrees of the polynomials \( P_n(x) \) and \( Q_m(x) \) significantly improves the quality of the approximations. The approximant \( B(z)_{[6,6]} \) accurately captures the behavior of \( B(z) \) across both lower and higher values of \( z \). Consequently, \( B(z)_{[6,6]} \) will be employed in the computations presented in the subsequent calculations. For reference, \( B(z)_{[6,6]} \) is given by
\begin{equation}\label{Pade66}
\boxed{B(z) \approx B(z)_{[6,6]} = (\kappa z)^{2-\alpha} + Mz + \frac{\sinh{\left(\frac{3}{2}\right)} \, \mathcal{K}(z) - \cosh{\left(\frac{3}{2}\right)} \, \mathcal{J}(z)}{\sinh{\left(\frac{3}{2}\right)} \, \mathcal{J}(z) - \cosh{\left(\frac{3}{2}\right)} \, \mathcal{K}(z)}\,.}
\end{equation}
where
\begin{eqnarray}
\!\!\!\!\!\!\!\!\!\!\!\!\!\!\!\!\!\!\!\!\mathcal{J}(z) &=& 42 \left(\frac{2 \kappa}{\sqrt{\Gamma}}-\frac{2}{M z}+3\right) \left[\left(\frac{2 \kappa}{\sqrt{\Gamma}}-\frac{2}{M z}+3\right)^4+240 \left(\frac{2 \kappa}{\sqrt{\Gamma}}-\frac{2}{M z}+3\right)^2+7920\right]\,, \\
\!\!\!\!\!\!\!\!\!\!\!\!\!\!\!\!\!\!\!\!\!\!\mathcal{K}(z)\!&\!=\!&\left(\frac{2 \kappa}{\sqrt{\Gamma}}\!-\!\frac{2}{M z}\!+\!3\right)^6\!+\!840 \left(\frac{2 \kappa}{\sqrt{\Gamma}}\!-\!\frac{2}{M z}\!+\!3\right)^4\!+\!75600 \left(\frac{2
   \kappa}{\sqrt{\Gamma}}\!-\!\frac{2}{M z}\!+\!3\right)^2\!+\! 665280\,.
\end{eqnarray}

\subsection{Near-horizon power expansion}

From (\ref{g_T_mu_0}), one can readily notice that \( g(z_h) = 0 \), which implies that the equations of motion simplify to a free-particle form near the horizon. As a consequence of this, the solutions close to the horizon can be expressed as ingoing and outgoing plane waves traveling into and out of the black hole interior as
\begin{eqnarray}
    \psi \sim \mathbb{C} e^{-i\omega r_*} + \mathbb{D} e^{i\omega r_*}\,,
\end{eqnarray}
where \( \mathbb{C} \) and \( \mathbb{D} \) are constants. Building on this, the solutions to the equations of motion near the horizon can be expanded in a power series as
\begin{eqnarray}\label{exppower}
\boxed{    \psi_{\pm} = e^{\pm i\omega r_*} \left[ a_0^{\pm} + a_1^{\pm} (z_h - z) + a_2^{\pm} (z_h - z)^2 + \dots \right]\,.
}
\end{eqnarray}
where we set \( a_0^{\pm} = 1 \). By substituting this expansion into the equations of motion, we can determine the coefficients \(a_1^{\pm}\) and \(a_2^{\pm}\). These coefficients depend on \(\omega\) and model parameters such as \(\mathcal{C}\), \(\sqrt{\Gamma}\), among others. For the purposes of subsequent computations, the solutions \(\psi_{\pm}\) are truncated at the second-order term, \((z_h - z)^2\). Using the tortoise coordinate approximation shown in (\ref{tortEMD}) applied to (\ref{EOM_T}), the computed coefficients $a_1^-$ and $a_2^-$ for the expansion (\ref{exppower}) are given by
\begingroup
\allowdisplaybreaks
\begin{align}
a_1^- &= 9 z_h \mathcal{C}^2 e^{\frac{3 \mathcal{C} z_h^2}{8}} \left(k z_h \right)^{- \alpha} \nonumber \\ &\times \frac{(\kappa z_h)^\alpha \left[M z_h (M z_h +1) - \sech{\left(\frac{\kappa}{\sqrt{\Gamma}} - \frac{1}{M z_h} \right)^2} \right] -(\alpha-2)z_h^3 M \kappa^2 }{2 M \left[ 64 i \omega + e^{\frac{3 \mathcal{C} z_h^2}{8}} \left(9 \omega \mathcal{C}^2 z_h^3 + 8 i \omega \left[3 \mathcal{C} z_h^2 -8 \right] \right) \right]} \,,  \\
a_2^- &= \frac{9 \mathcal{C}^2 e^{\frac{3 \mathcal{C} z_h^2}{8}}}{8z_h M^2} \nonumber\ \\ &\times \Biggl( M^2 z_h^2 (\kappa z_h)^{-2 \alpha} \Biggl[ -32 i\omega \Bigl( e^{\frac{3 \mathcal{C} z_h^2}{8}}- 1\Bigl) \Bigl[(\kappa z_h)^{2 \alpha} \{ M^2 z_h^2 - 4 Mz_h -1 \} \nonumber\ \\
     &- 2 (\alpha-2) \kappa^{\alpha +2}z_h^{\alpha+3}M + (\alpha-2)^2 \kappa^4 z_h^4 - 4 (\alpha-2)^2 \{\kappa z_h \}^{\alpha+2} \Bigl] \nonumber\ \\
     &+ 12 i \omega \mathcal{C}z_h^2 \Bigl[- 4(\kappa z_h)^{\alpha}\Bigl[(\alpha-2)\kappa^2 z_h^2 + (\kappa z_h)^{\alpha}\{Mz_h +1 \} \Bigl]+ e^{\frac{3 \mathcal{C} z_h^2}{8}}\Bigl(\{\kappa z_h \}^\alpha [M^2z_h^2 + 3] \nonumber\  \\
     &- 2M (\alpha-2) \kappa^{\alpha+2}z_h^{\alpha+3} + (\alpha-2)^2 \kappa^4 z_h^4 - 4 (\kappa z_h)^{\alpha+2} \{\alpha^2 -3\alpha +2 \} \Bigl) \Bigl] \nonumber\ \\
     &+ 9 z_h^3 \mathcal{C}^2 e^{\frac{3 \mathcal{C} z_h^2}{8}} \Bigl( [\alpha-2]^2 \kappa^4 z_h^4 - 2 \alpha \{ \alpha-2 \} (\kappa z_h)^{\alpha+2} - 2 [M-i\omega] (\alpha-2) \kappa^{\alpha+2}z_h^{\alpha+3} \nonumber\ \\
     &+ (\kappa z_h )^{2\alpha} \{2 M z_h (1 - i \omega z_h) - 2 i \omega z_h + M^2 z_h^2 + 3 \} \Bigl) \Biggl] + \sech{\left(\frac{\kappa}{\sqrt{\Gamma}}-\frac{1}{M z_h}\right)}^4 \\
     &\times \Bigl[9 z_h^3 \mathcal{C}^2 e^{ \frac{3 \mathcal{C} z_h^2}{8}} + 12 i \omega \mathcal{C} e^{ \frac{3 \mathcal{C} z_h^2}{8}} - 32 i\omega \Bigl(e^{ \frac{3 \mathcal{C} z_h^2}{8}} - 1 \Bigl) \Bigl] - 2 (\kappa z_h)^{-\alpha} \sech{\left(\frac{\kappa}{\sqrt{\Gamma}}-\frac{1}{M z_h}\right)}^2 \nonumber \\
     &\times \Bigl[ M z_h \Bigl( 12i \omega \mathcal{C} z_h^2 \Bigl\{ e^{ \frac{3 \mathcal{C} z_h^2}{8}}[(\kappa z_h)^\alpha\{M z_h +4\} - (\alpha-2) \kappa^2 z_h^2 ] -2 (\kappa z_h)^\alpha \Bigl\} - 32 i \omega \Bigl(e^{ \frac{3 \mathcal{C} z_h^2}{8}} -1 \Bigl) \nonumber \\
     &\times \{ (\alpha-2) \kappa^2 z_h^2 (\kappa z_h)^\alpha [Mz_h + 2] \} + 9 z_h^3 \mathcal{C}^2 e^{ \frac{3 \mathcal{C} z_h^2}{8}} [(\kappa z_h)^\alpha \{ Mz_h + 3 - i \omega z_h \} \nonumber \\
     &-(\alpha-2)\kappa^2z_h^2] \Bigl)  + 2  \{\kappa z_h \}^\alpha \tanh{\left(\frac{\kappa}{\sqrt{\Gamma}}-\frac{1}{M z_h}\right)} \Bigl( 9 z_h^3 \mathcal{C}^2 e^{ \frac{3 \mathcal{C} z_h^2}{8}} + 21i\omega \mathcal{C} z_h^2e^{ \frac{3 \mathcal{C} z_h^2}{8}} \nonumber  \\
     &- 64i\omega \Bigl\{e^{ \frac{3 \mathcal{C} z_h^2}{8}}-1 \Bigl\} \Bigl) \Bigl] \Bigl[ \Bigl( 9 z_h^3 \mathcal{C}^2 e^{ \frac{3 \mathcal{C} z_h^2}{8}} - 32i\omega \Bigl\{e^{ \frac{3 \mathcal{C} z_h^2}{8}} -1 \Bigl\} +12 i \omega \mathcal{C} z_h^2 e^{ \frac{3 \mathcal{C} z_h^2}{8}} \Bigl) \nonumber \\
     &\times \Bigl\{9 z_h^3 \mathcal{C}^2 e^{ \frac{3 \mathcal{C} z_h^2}{8}} - 64 i \omega \Bigl(e^{ \frac{3 \mathcal{C} z_h^2}{8}} -1 \Bigl) + 24 i \omega \mathcal{C} z_h^2 e^{ \frac{3 \mathcal{C} z_h^2}{8}} \Bigl\} \Bigl]^{-1} \nonumber \,.
\end{align}
\endgroup
whereas the coefficients $a_{1,2}^+$ are quite similar to their counterparts shown above and are not actually employed in the found formula (\ref{green}) to compute the numerical spectral functions.

Evidently, ingoing and outgoing solutions form a basis for any wave function that may be considered. Consequently, renormalizable and non-renormalizable solutions can be expressed as linear combinations of the ingoing and outgoing solutions
\begin{align}    
    \psi_{2} &= \mathbb{C}_2 \psi_- + \mathbb{D}_2 \psi_+ \,, \\
    \psi_{1} &= \mathbb{C}_1 \psi_- + \mathbb{D}_1 \psi_+ \,.
\end{align}

\subsection{Near-boundary Frobenius analysis}

If we take the limit \( z \to 0 \), we can see that the leading term in the power series solution is assumed to take the form \( z^\beta \), following the prescription for a Frobenius expansion near \( z = 0 \). The asymptotic analysis of Eq.~(\ref{EOM_T}) yields the following indicial equation
\begin{equation} \label{frobEMD}
    \beta^2 - \beta - \frac{3}{4} = 0\,,
\end{equation}
which has solutions \( \beta = -\frac{1}{2} \) and \( \beta = \frac{3}{2} \). To solve the differential equation for the perturbation by evolving from the boundary to the horizon, it is also necessary to determine the derivative of \( \psi_i \) at the boundary, ensuring a consistent numerical setup. Obtaining the first derivative at the holographic boundary requires a detailed examination of the asymptotic structure of \( \psi_i \) via Frobenius analysis. 

Since the indicial equation produces two solutions differing by an integer, the larger root \( \beta = \frac{3}{2} \) corresponds to a regular series expansion, whereas the smaller root \( \beta = -\frac{1}{2} \) leads to a logarithmic series. These two solutions can be expressed as
\begin{align}\label{expEMD}
    \psi_1(z) &= z^{3/2} \sum_{k=0}^{+\infty} a_k z^k \,, \\
    \psi_2(z) &= K \ln(z) \psi_1(z) + z^{-1/2} \sum_{k=0}^{+\infty} b_k z^k \,.
 \label{power}
\end{align}

The explicit forms of \( a_k \) and \( b_k \) can be determined by substituting the series solutions into Eq.~(\ref{EOM_T}) and matching terms order by order. For normalization, we choose \( a_0 = b_0 = 1 \). The coefficients of these expansions depend on \( \omega \) and the parameters of the model, such as $\mathcal{C}$ and $M$. In this work, the presented expansion (\ref{power}) for \( \psi_1(z) \) and \( \psi_2(z) \) was truncated at $k=2$. Given so, for the considered equation of motion (\ref{EOM_T}), taking the full non-approximate $B(z)$, the computed coefficients are given by
\begin{align}
    a_1 &= \frac{1}{32} \left(M^2 - 4 \omega^2 \right) \,, \\
    a_2 &= \frac{1}{776} \left(32 \kappa^4 + \frac{153 \mathcal{C}^2}{8 + e^{\frac{3 \mathcal{C}z_h^2}{8}}\left[3 \mathcal{C} z_h^2 - 8\right]} - M^2 \omega^2 + 4\omega^2 \right) \,,  \\
    b_1 &= 0 \,,  \\
    b_2 &= \frac{1}{1056} \left(128 \kappa^4 + \frac{36 \mathcal{C}^2}{8 + e^{\frac{3 \mathcal{C}z_h^2}{8}}\left[3 \mathcal{C} z_h^2 - 8\right]} - 3\left[M^2 - 4\omega^2 \right]^2 \right) \,,  \\
    K &= \frac{1}{8} \left(M^2 - 4 \omega^2 \right) \,.
 \label{power1}
\end{align}
in which the coefficient $b_1$ can be chosen freely, as showed in \cite{Mamani:2022qnf}, and here is taken to be null.

In general, the coefficients depend on the momentum $q^2$, but we did set $q=0$ beforehand. These solutions, starting near the boundary, also form a basis for constructing wave functions, which can be written as
\begin{align}
    \psi_{+} &= \mathcal{A}_+ \psi_2 + \mathcal{B}_+ \psi_1 \,, \\
    \psi_{-} &= \mathcal{A}_- \psi_2 + \mathcal{B}_- \psi_1 \,.
\end{align}

\subsection{Formula for the retarded Green's function}

As seen, both the near-boundary and near-horizon expansions provide a basis in which wave functions can be expressed as linear combinations. From the relations between \( \psi_{\pm} \), \( \psi_1 \), and \( \psi_2 \), it follows that the coefficients satisfy
\begin{gather}
 \begin{pmatrix} \mathcal{A}_{-} & \mathcal{B}_{-} \\ \mathcal{A}_{+} & \mathcal{B}_{+} \end{pmatrix}
 =
  \begin{pmatrix} \mathbb{C}_{2} & \mathbb{D}_{2} \\ \mathbb{C}_{1} & \mathbb{D}_{1} \end{pmatrix}^{-1}\,.
\end{gather}

Within this approach, the calculation of the spectral functions requires determining the retarded Green's function and subsequently extracting its imaginary part  \cite{Mamani:2022qnf,Jena:2024cqs, Toniato:2025gts}.

The procedure goes as: The retarded Green's function is computed by numerically integrating the equations of motion over a regular lattice extending from a point near the boundary, \( z = \epsilon \), to a point close to the horizon, \( z = z_h - \epsilon \). The initial conditions are derived from the solution near the boundary at \( z = 0 \), specifically \( \psi_1(\epsilon) \), obtained through Frobenius analysis, along with its derivative \( \partial_z \psi_1(\epsilon) \). Similarly, the second solution near the boundary, \( \psi_2(\epsilon) \), along with its derivative \( \partial_z \psi_2(\epsilon) \), is used to compute the other independent solution for the equations of motion using the same numerical scheme.

As a result, two numerical solutions for \( \psi \) are obtained. These functions are then evaluated near the horizon, \( z_h - \epsilon \), where the following relationship is constructed
\begin{gather}
\begin{pmatrix} \psi_a(z_h - \epsilon) \\ \partial_z \psi_a(z_h - \epsilon) \end{pmatrix}
=
\begin{pmatrix} \psi_-(z_h - \epsilon) & \psi_+(z_h - \epsilon) \\ \partial_z \psi_-(z_h - \epsilon) & \partial_z \psi_+(z_h - \epsilon) \end{pmatrix}
\begin{pmatrix} \mathbb{C}_a \\ \mathbb{D}_a \end{pmatrix}\,,
\label{coef}
\end{gather}
where \( a = 1, 2 \) correspond to the numerical results obtained from solving the equations of motion, and \( \psi_{\pm} \) represent the asymptotic solutions near the horizon as given by the power expansions in Eqs.~(\ref{power}) and (\ref{power1}). All quantities are evaluated at \( z_h - \epsilon \). The coefficients \( \mathbb{C}_a \) and \( \mathbb{D}_a \) are computed by inverting the above square matrix. 

Combining this with the coefficient relations in Eq.~(\ref{coef}), we arrive at
\begin{eqnarray} \label{green}
    \frac{\mathcal{B}_-}{\mathcal{A}_-} = - \frac{\mathbb{D}_2}{\mathbb{D}_1} = \frac{\partial_z \psi_-(z_h - \epsilon) \psi_2(z_h - \epsilon) - \psi_-(z_h - \epsilon) \partial_z \psi_2(z_h - \epsilon)}{\partial_z \psi_-(z_h - \epsilon) \psi_1(z_h - \epsilon) - \psi_-(z_h - \epsilon) \partial_z \psi_1(z_h - \epsilon)}\,.
\end{eqnarray}
This provides a stable numerical method for calculating the ratio \( \frac{\mathcal{B}_-}{\mathcal{A}_-} \), which is sufficient to compute the retarded Green's function and, consequently, the spectral function. As demonstrated in \cite{Miranda:2009uw, Mamani:2018uxf, Mamani:2022qnf}, the spectral function is directly related to \( \frac{\mathcal{B}_-}{\mathcal{A}_-} \) via the expression
\begin{eqnarray}
\boxed{\rho(\omega, q = 0) = - 2 \Im G^R(\omega, q = 0) \propto \Im \frac{\mathcal{B}_-}{\mathcal{A}_-}\,.} 
\end{eqnarray}
where \( \rho(\omega, q = 0) \) represents the spectral function calculated numerically using the described prescription (with \( q = 0 \) set initially), and \( \Im \) denotes the imaginary part. The factor of $2$ is introduced for it is the standard in the literature.

The numerical integration of the equation of motion over the interval \( z \in [\epsilon, z_h - \epsilon] \), necessary to compute \( \psi_i(z_h - \epsilon) \) and \( \partial_z \psi_i(z_h - \epsilon) \) as they appear in the Green's function formula (\ref{green}), is carried out using Mathematica's \texttt{NDSolve} with the \texttt{StiffnessSwitching} numerical scheme. This method is specifically chosen to address the stiff nature of the problem, as explicit methods like the standard fourth-order Runge-Kutta or Euler methods proved inadequate for producing reliable results. 

The constructed Matrix-Numerov method (\ref{finalnum}) applied in previous numerical evaluations to the differential numerical eigenproblems is not easily implemented for the numerical procedure described in this Section. As we have to solve the equation of motion (\ref{EOM_T}) in a fine $\epsilon$-lattice for thousands of values of $\omega$, for time-related reasons, it is simpler to use the \texttt{StiffnessSwitching} method. In any case, the described Matrix-Numerov method could be refined to deal with situations like this (including higher-order terms to approximate derivatives, for example). This is something to be explored in the future.

The remaining components in Eq.~(\ref{green}), specifically \( \psi_{\pm}(z_h - \epsilon) \) and \( \partial_z \psi_{\pm}(z_h - \epsilon) \), are obtained analytically based on the results of the Frobenius analysis and the associated power series expansions.

The small positive parameter \( \epsilon \) functions as both the lattice spacing and the boundary cutoff. For the computations performed, a value of \( \epsilon = 0.001 \) was used. The value of \( z_h \), corresponding to a specific temperature \( T \), is determined by solving the transcendental equation (\ref{BH_T_mu0}) for \( z_h \) using Newton's method.

\newpage

\section{Analysis of the mesonic melting of heavy and exotic states} \label{meltingmu0}

In this Section, the numerical scheme developed in Sec. \ref{procespec} is applied to the model parameters for the chosen mesons. The resulting spectral functions are analyzed to determine the mesonic melting.

\subsection{Numerical spectral functions ($\mu=0$)}

One quantity of most importance in the melting analysis is the spectral function as it describes the distribution of states that can be excited as a quarkonium state dissolves in a hot and dense QCD medium. It provides crucial insights into the dissociation of quarkonium states and the formation of the QGP. In particular, the spectral peaks signify the presence of bound states of heavy quarks and antiquarks, such as charm and bottom quarks, within the QGP. The emergence, disappearance, or modification of these peaks reveals essential properties of the QGP. As shown earlier, taking the small $\epsilon$ limit allows for the extraction of holographically significant information about the spectral function.

\begin{figure}[htb!]
    \centering
    \begin{minipage}[t]{0.49\textwidth}
        \centering
        \includegraphics[width=\textwidth]{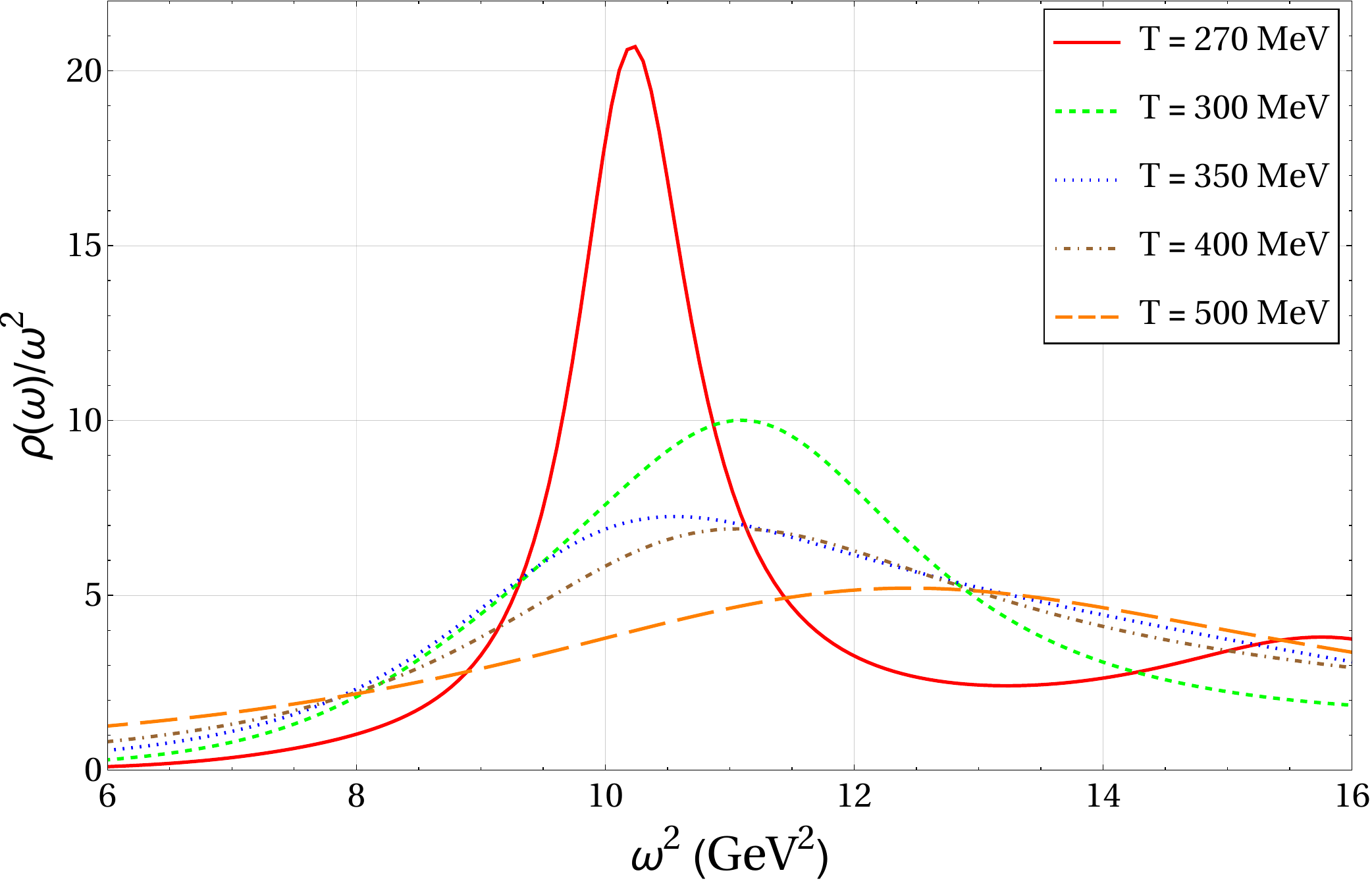}
        \caption{\small Charmonium spectral function.}
        \phantomsection
        \label{fig:SpecTcharm}
    \end{minipage}%
    \hfill
    \begin{minipage}[t]{0.49\textwidth}
        \centering
        \includegraphics[width=\textwidth]{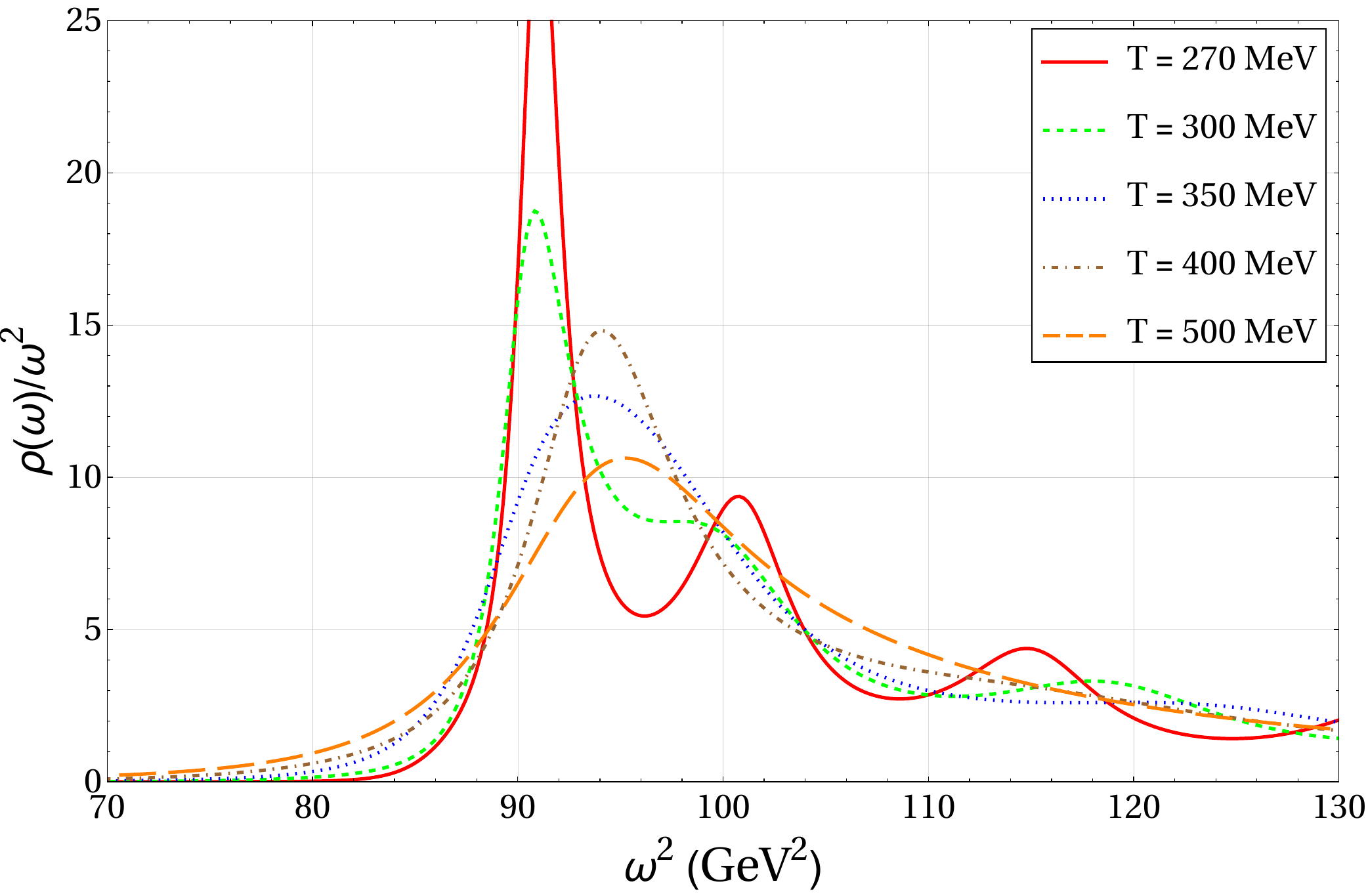}
        \caption{\small Bottomonium spectral function.}
        \phantomsection
        \label{fig:SpecTbottom}
    \end{minipage}
\end{figure}

\begin{figure}[htb!]
    \centering
    \begin{minipage}[t]{0.49\textwidth}
        \centering
        \includegraphics[width=\textwidth]{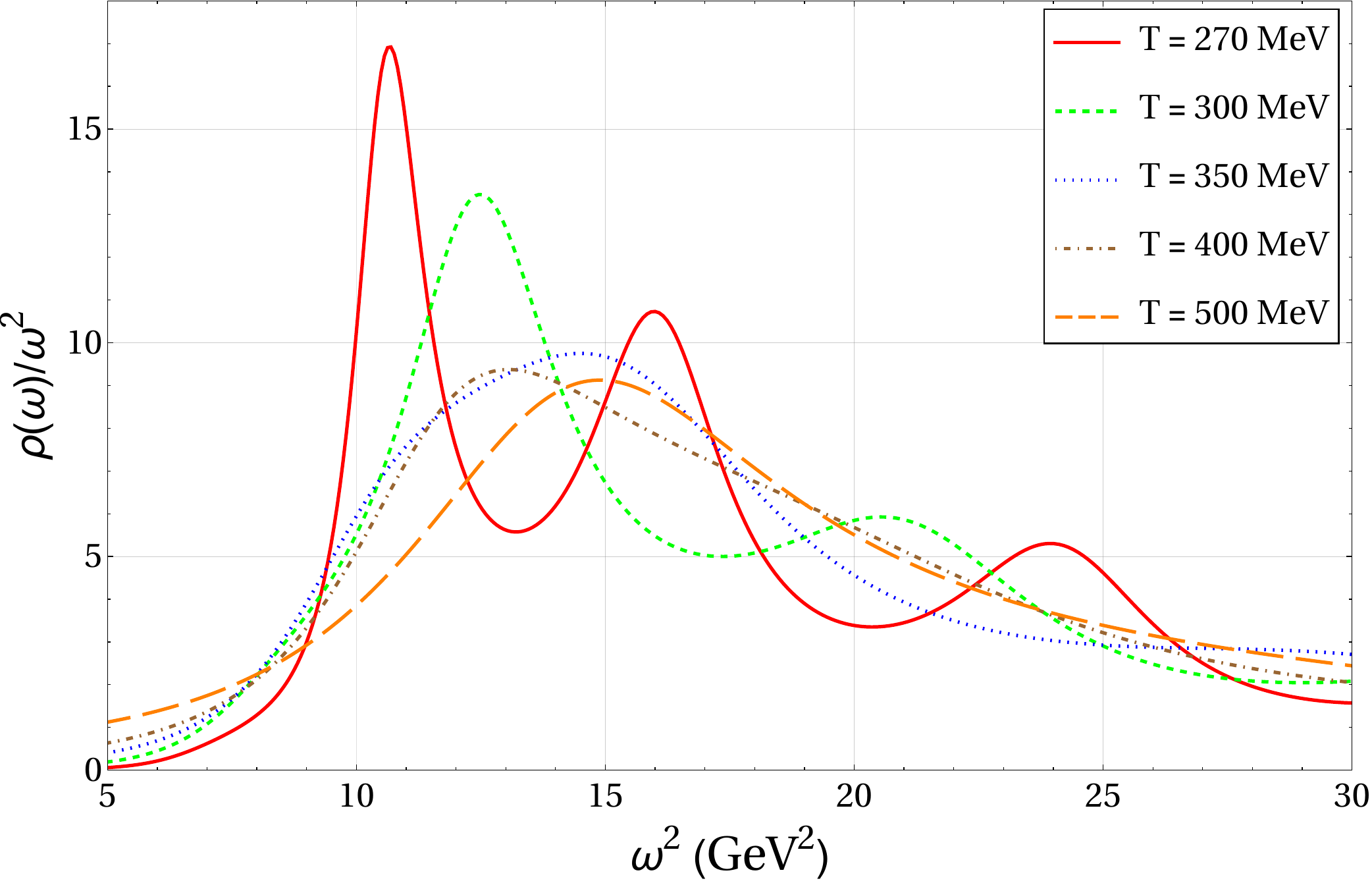}
        \caption{\small Tetraquark spectral function.}
        \phantomsection
        \label{fig:SpecTtetra}
    \end{minipage}%
    \hfill
    \begin{minipage}[t]{0.49\textwidth}
        \centering
        \includegraphics[width=\textwidth]{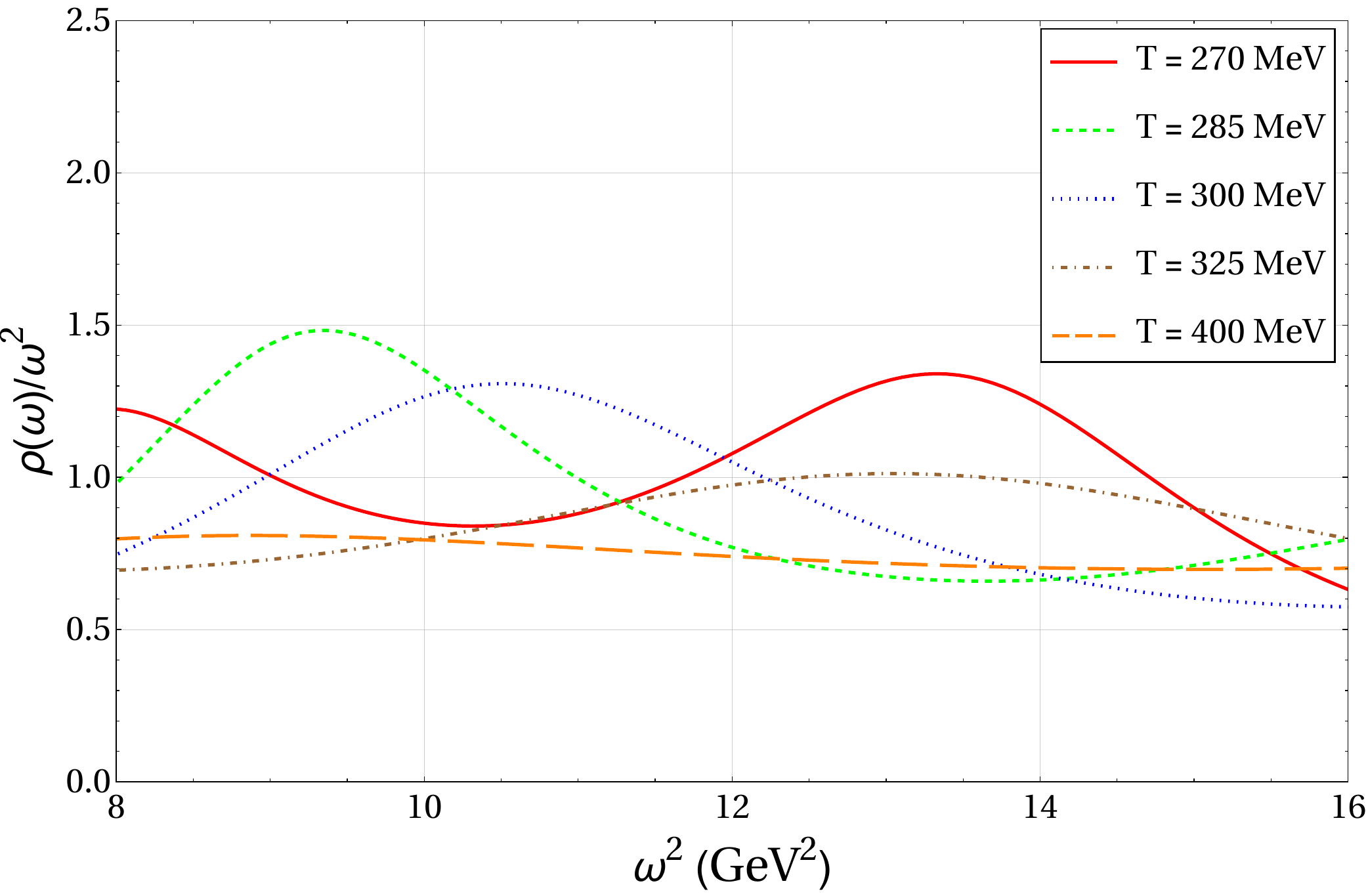}
        \caption{\small Hybrid meson spectral function.}
        \phantomsection
        \label{fig:SpecThybrid}
    \end{minipage}
\end{figure}

It is seen that experimental measurements and theoretical studies of quarkonium production and suppression in heavy-ion collisions serve as powerful tools for probing the properties of the QGP and investigating the behavior of strong interactions under extreme conditions. In this work, we focus on exploring the melting of quarkonium states by examining the evolution of spectral peaks as a function of temperature. Utilizing the numerical procedure outlined in the previous Section, we compute the spectral functions for charmonium, bottomonium, tetraquark, and hybrid states. The numerical results are presented in Figs.~\ref{fig:SpecTcharm}--\ref{fig:SpecThybrid} across a range of temperatures.

As promised, one can notice the melting process through the spectral functions. As the temperature increases, the peaks tend to disappear and the function broadens out. This indicates the melting of the mesonic states into a hot medium. We can do a very rough estimate of the melting temperature for the charmonium, bottomonium, and tetraquark states to be around $T\sim 400-450~ \text{MeV}$ as their spectral function peaks disappear at these temperatures. However, we can clearly see that the hybrid meson melts at a lower temperature than the rest, as the peaks disappear more quickly as we increase the temperature. For this state, we estimate this temperature to be around $T\sim 325-400~ \text{MeV}$.

\subsection{The effective potential $V(z)$ at finite temperature  ($\mu=0$)}

\begin{figure}[htb!]
    \centering
    \begin{minipage}[t]{0.49\textwidth}
        \centering
        \includegraphics[width=\textwidth]{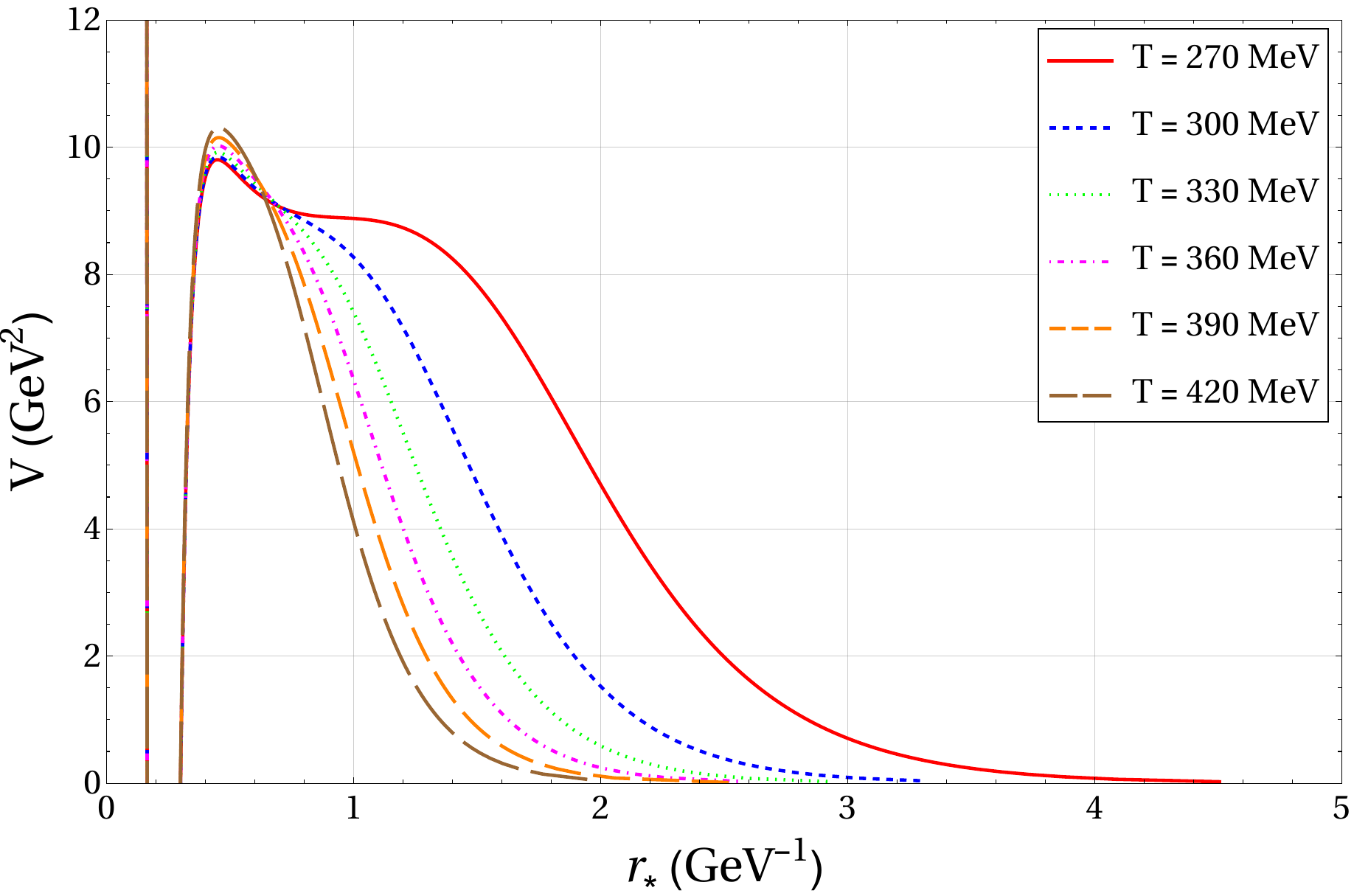}
        \caption{\small Charmonium effective potential $V(z)$.}
        \phantomsection
        \label{fig:Tpotcharm}
    \end{minipage}%
    \hfill
    \begin{minipage}[t]{0.49\textwidth}
        \centering
        \includegraphics[width=\textwidth]{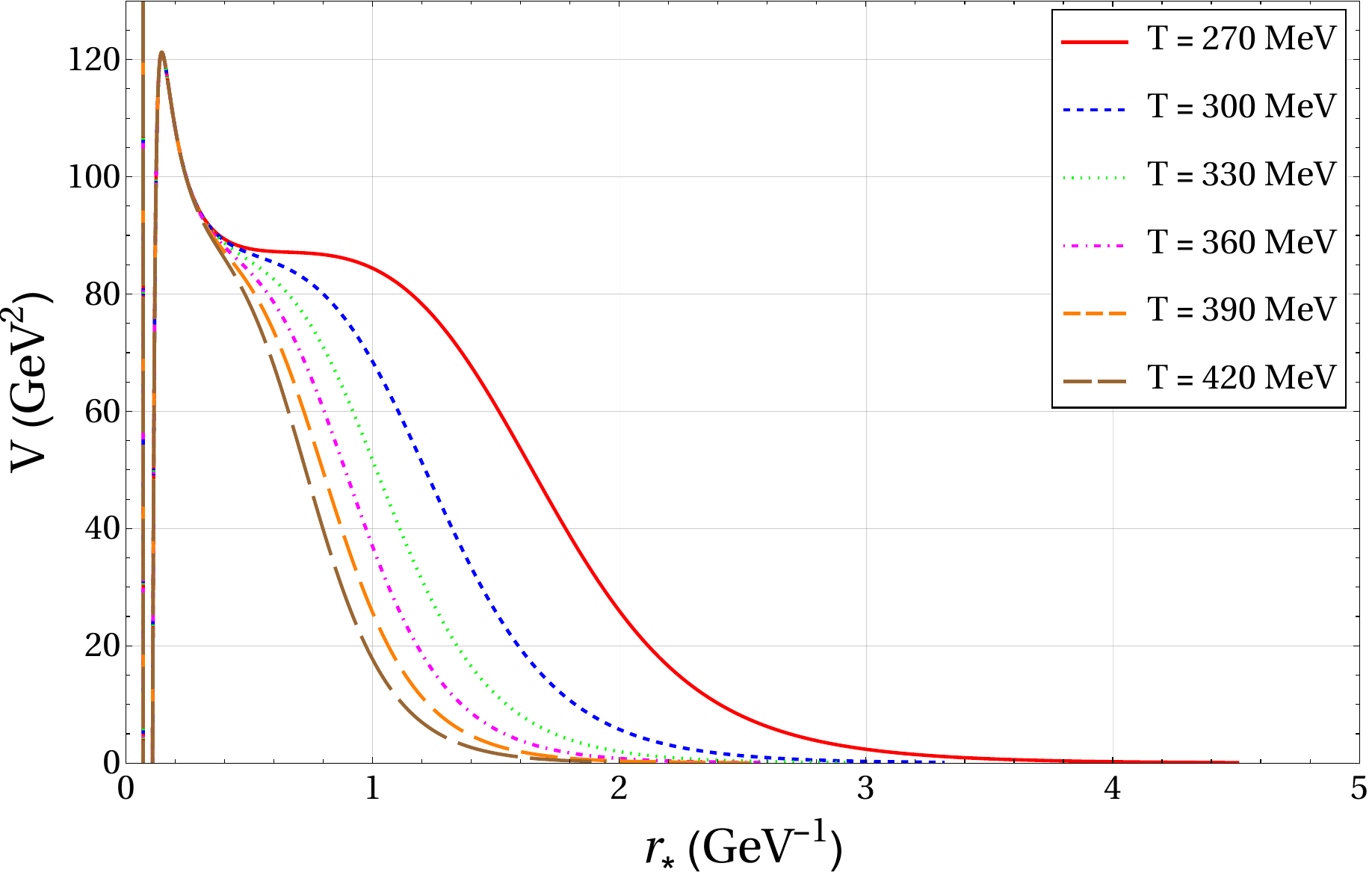}
        \caption{\small Bottomonium effective potential $V(z)$.}
        \phantomsection
        \label{fig:Tpotbottom}
    \end{minipage}
\end{figure}

\begin{figure}[htb!]
    \centering
    \begin{minipage}[t]{0.49\textwidth}
        \centering
        \includegraphics[width=\textwidth]{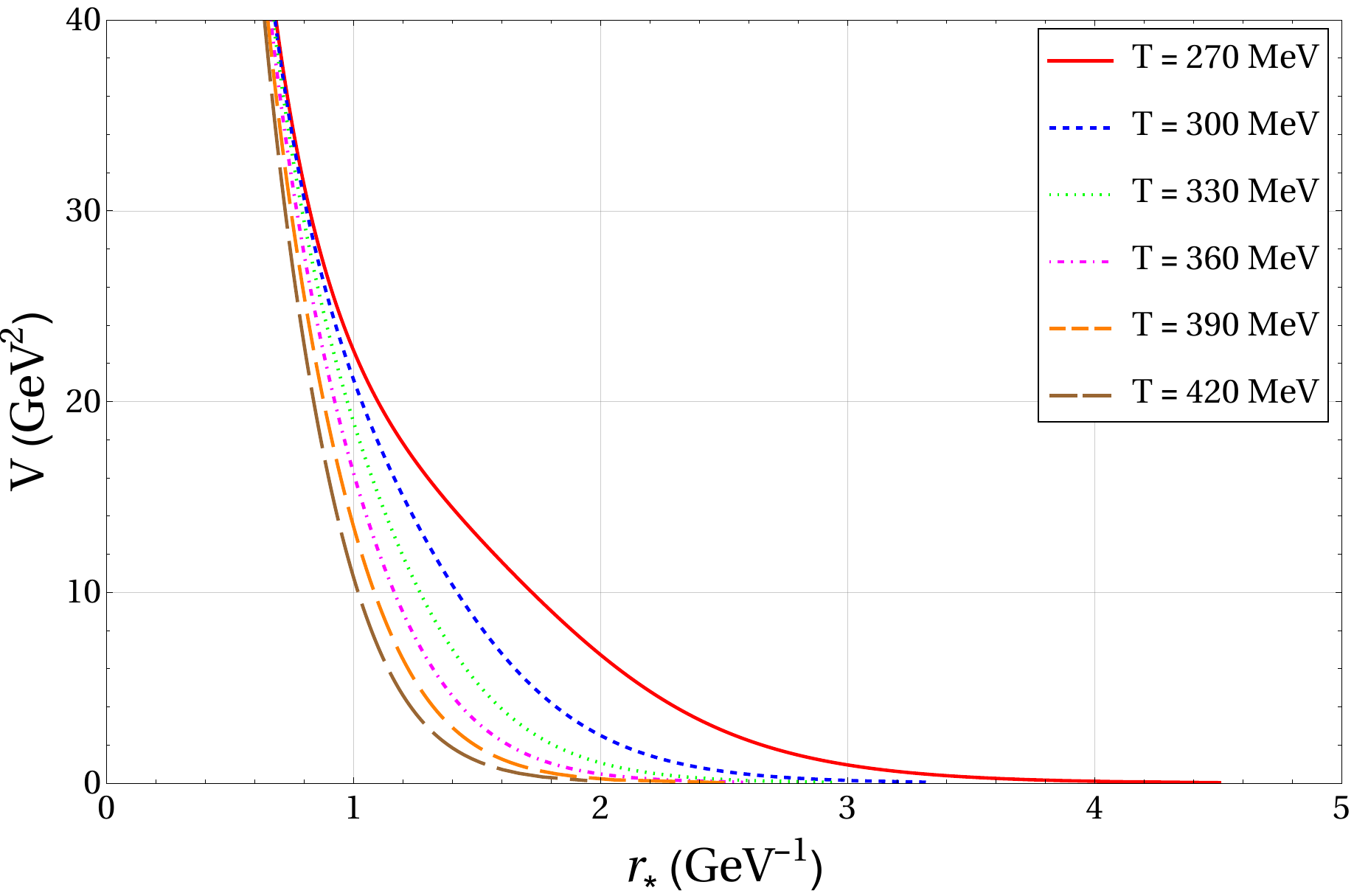}
        \caption{\small Tetraquark effective potential $V(z)$.}
        \phantomsection
        \label{fig:Tpottetra}
    \end{minipage}%
    \hfill
    \begin{minipage}[t]{0.49\textwidth}
        \centering
        \includegraphics[width=\textwidth]{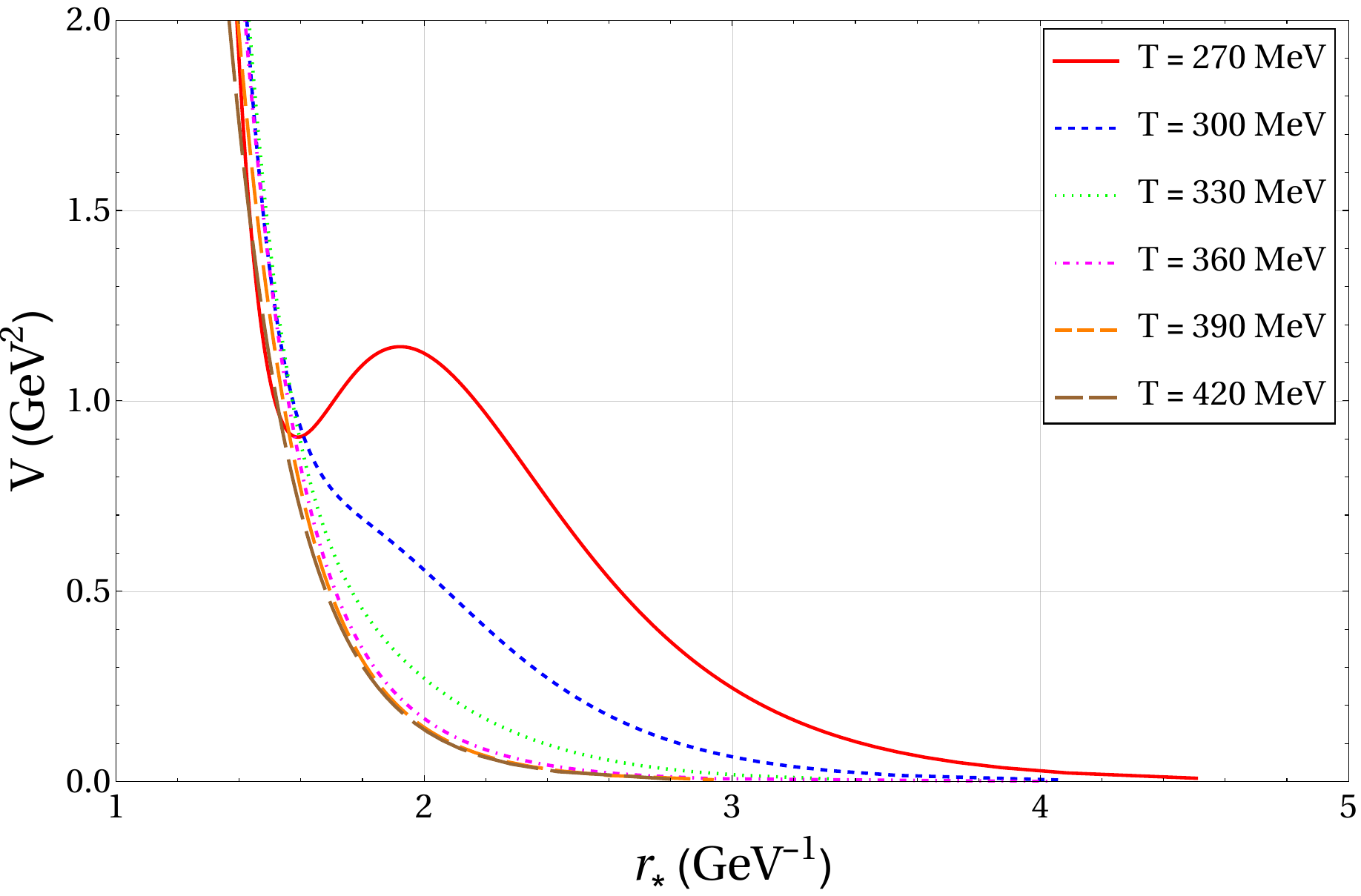}
        \caption{\small Hybrid meson effective potential $V(z)$.}
        \phantomsection
        \label{fig:Tpothybrid}
    \end{minipage}
\end{figure}

The other approach to verifying the melting process is to analyze the evolution of the effective potential in the Schrödinger-like equation, Eq.~(\ref{EOM_T}), at different temperatures and observe how it changes with increasing temperature. Accordingly, in Figs.~\ref{fig:Tpotcharm}--\ref{fig:Tpothybrid}, we present the holographic potential, as defined in Eq.~(\ref{Potential_T}), for various temperatures across the considered states.

As illustrated in Figs.~\ref{fig:Tpotcharm}--\ref{fig:Tpothybrid}, we can conclude that for temperatures approaching the minimum accessible value in this model ($T \sim T_c$), the potential generally exhibits a single global minimum (except in the case of the tetraquark), signifying that the potential exhibits a pronounced well, supporting long-lived quasi-bound resonances and the stability of mesonic states. As the temperature rises, the potential undergoes significant transformations, with the global minimum vanishing and the well becoming shallower, indicating the weakening of confinement and the thermal dissociation of mesons. This evolution of the potential underscores the holographic representation of thermal melting, where the interplay between the dilaton structure and increasing temperature dictates the confinement and dissociation processes of mesonic states.

To avoid confusion about the temperature range shown in Figs.~\ref{fig:Tpotcharm}--\ref{fig:Tpothybrid}, it is worth recalling that in the present
setup the black hole geometry exists only above a minimum temperature, and the thermodynamically preferred saddle
switches from thermal AdS to the large black hole at the Hawking--Page temperature $T_c$ discussed in
Sec.~\ref{thermo}.  Consequently, the lowest temperatures displayed in the effective-potential plots correspond to
the smallest stable black holes (i.e.\ temperatures just above $T_c$), while higher curves probe progressively hotter
plasmas.  In this sense, the disappearance of the potential well as $T$ increases should be interpreted as the
gradual loss of quasi-bound resonant structures in the deconfined phase, rather than as a statement about the
existence of a confining background below $T_c$.

\newpage

\section{Holographic plasma at finite density} \label{metricmu}
As we have already examined the system at zero chemical potential and analyzed its thermal behavior, we now turn to the investigation of the dynamics of a holographic finite density plasma. To incorporate finite density effects, we introduce the chemical potential $\mu$ by solving Eq.~(\ref{Aequation}) under a specific boundary condition. The chemical potential serves as a critical parameter that modifies the holographic setup and allows the study of systems where baryonic or charge density plays a significant role.

\subsection{Equation for $A_t(z)$}

We begin by noticing that Eq.~(\ref{Aequation}) governs the behavior of the temporal component of the gauge field, $A_t(z)$, which is directly related to the chemical potential in holography. By solving this equation with appropriate boundary conditions, we establish the profile of $A_t(z)$ across the bulk spacetime. The boundary conditions are as follows:
\begin{enumerate}
    \item[(I)] $A_t(z)$ must vanish at the black hole horizon ($z = z_h$), ensuring a regular solution at this point.
    \item[(II)] At the asymptotic boundary ($z \to 0$), the value of $A_t(z)$ is proportional to the chemical potential $\mu$, linking the bulk dynamics to the boundary field theory.
\end{enumerate}

The gauge field $A_t(z)$ allows one to introduce finite density effects into the EMD holographic system, modifying the equations of motion and the resulting physical quantities. Specifically, the presence of a nonzero chemical potential impacts the blackening function $g(z)$. The change in the metric, in turn, influences the thermodynamic properties and phase structure of the system, giving us also different spectral functions found in the last Section. After computing the gauge field $A_t(z)$, we can calculate the metric $g(z)$ by solving Eq.~(\ref{2.9}). 

It was not possible, however, to compute a general analytic solution of $g(z)$ unless we excluded the low-$z$ corrections to the function $f(z)$ presented in Eq.~(\ref{kappam}). This limitation arises due to the intricate coupling between the equations of motion, which becomes analytically intractable when the full complexity of $f(z)$, including its low-$z$ corrections, is incorporated. Importantly, this means that we restrict the analysis to mesons with smaller values for the constants $\kappa$ and $\alpha$ (like the hybrid meson).

Using this new definition, we find the equation of motion for $A_t(z)$, from Eq.~(\ref{2.9}):
\begin{align} \label{gaugemotion}
    f(z) &= e^{-\left[-\frac{c}{8} z^2 + (\kappa z)^{2-\alpha} \right]} \,, \\
    0 &= A_t'' + A_t' \left( -\frac{1}{z} + \frac{f'}{f} + A' \right) \Rightarrow 0 = \left[{(\alpha -2) (\kappa z)^{2 - \alpha}  -1}\right] \, A_{t}' + z\,A_{t}'' \label{Atequation} \,.
\end{align}

In the next Section, we look for a general solution for (\ref{Atequation}).

\subsection{General solution for $A_t(z)$}

As a first step, we define 
\begin{equation}
    \chi(z) = A_t'(z) \,, \quad \chi'(z) = A_t''(z)\,, 
\end{equation}

Applying this to (\ref{Atequation}) let us rewrite it as a first-order ODE in $\chi(z)$:
\begin{equation}\label{firstchi}
    \frac{\chi'(z)}{\chi(z)} = - \frac{(\alpha-2)(\kappa z)^{2-\alpha}-1}{z} \,, 
\end{equation}

Integrating both sides of (\ref{firstchi}) gets us
\begin{equation}
    \begin{split}
    \ln{\chi(z)} &= - \int dz \frac{(\alpha-2)(\kappa z)^{2-\alpha}-1}{z} + C_1\\
                 &= (\kappa z)^{2-\alpha} + \ln{z}+ C_1 \quad (\alpha \neq 2) \,, 
\end{split}
\end{equation}

After exponentiation, $A_t'(z) = z \exp{\left[(kz)^{2-\alpha}\right]}  C_2$, with $C_2=e^{C_1}$. Thus, after a second integration, we find
\begin{equation} \label{secondint}
    A_t(z) = C_3+C_2\int z \exp{\left[(kz)^{2-\alpha}\right]}  dz \,,
\end{equation}

By a change of variable, $u=(kz)^{2-\alpha} \implies z \,dz=\frac{u^{\frac{2}{2-\alpha}}}{(2-\alpha)\kappa^2}$, we can rewrite the integral (\ref{secondint}) as
\begin{equation}
     \int dz \, z \, e^{(kz)^{2-a}} = \frac{1}{(2-\alpha)\kappa^2} \int  u^{\frac{\alpha}{2-\alpha}} e^u du \,,
\end{equation}

By a second change of variable, $v=u^{\gamma+1}$, where $\gamma=\frac{\alpha}{2-\alpha}$, we find
\begin{equation} \label{incgamma}
\begin{split}
    \int du \, u^\gamma \, e^u &= \frac{1}{1+\gamma} \int dv \, \exp{\left[v^{(1+\gamma)^{-1}}\right]} \\
    &= - (-1)^{-(1+\gamma)} (1+\gamma) \, \Gamma{\left(1+\gamma,-v^{(1+\gamma)^{-1}}\right)}\\
    &= (-1)^{-\gamma} \, \Gamma{\left(1+\gamma,-u\right)}\,.
\end{split}
\end{equation}

Whereas the integral (\ref{incgamma}) was recognized as the incomplete gamma function $\Gamma(u, v) = \int_v^{\infty} t^{u-1} e^{-t} \, dt$. We therefore see that the solution to the gauge field $A_t(z)$ is given by the incomplete gamma function. 

\subsection{$A_t(z)$: Boundary conditions}

The general solution to Eq.~(\ref{Atequation}) has been determined in the last Section. If we redo the change of variables and include back the constants, after some algebra, we get
\begin{equation} \label{generalAt}
    A_t(z) = \mathcal{Y}_2 + \mathcal{Y}_1 \frac{1}{\alpha - 2}\left[(-1)^{\frac{2}{\alpha - 2}} z^2 (\kappa z)^{\frac{2 (2 - \alpha)}{\alpha - 2}} \, \Gamma\left(-\frac{2}{\alpha - 2}, -(\kappa z)^{2 - \alpha}\right)\right]\,,
\end{equation}g
where $\mathcal{Y}_{1,2}$ are the overall integration constants. As a consistency check, we consider the general solution (\ref{generalAt}) in the limit $\alpha \rightarrow 0$ and apply the boundary conditions outlined earlier in this Section, yielding:
\begin{equation}
    A_t(z) = \mathcal{Y}_2 + \mathcal{Y}_1 \frac{e^{\kappa^2 z^2}}{2 \kappa^2} \Rightarrow A_t(z) = \mu\frac{e^{\kappa^2 z_h^2} -e^{\kappa^2 z^2} }{e^{\kappa^2 z_h^2} -1}\,.
\end{equation}

We readily see that, in this limit, the general solution to $A_t(z)$ reproduces what was found in \cite{Mamani:2022qnf}, where a quadratic dilaton profile was implied. 

In order to apply the given boundary conditions to Eq.~(\ref{generalAt}), with $\alpha \neq 0$, we found it was better to introduce a small positive number $\delta$ (to impose the asymptotic boundary condition, where $z \rightarrow \delta$) and after the computations take the limit $\delta \rightarrow 0$. 
This procedure gives us the following constants of integration: 
\begin{equation}
\mathcal{Y}_2 = \frac{\mu}{1 - \frac{\Gamma\left(-\frac{2}{\alpha-2}, -(\kappa \delta)^{2 - \alpha}\right)}{\Gamma\left(-\frac{2}{\alpha-2}, -(\kappa z_h)^{2 - \alpha}\right)}} \,,   
\end{equation}
\begin{equation}
\scalemath{.95}{\mathcal{Y}_1 = \frac{- \mu (-1)^{-\frac{2}{\alpha-2}} (\alpha-2) (\kappa^2 z_h \delta)^{\frac{2\alpha}{ \alpha-2}}}{z_h^2 (\kappa z_h)^{\frac{4}{ \alpha -2}} (\kappa \delta)^{\frac{2\alpha}{\alpha-2}} \Gamma\left(-\frac{2}{\alpha-2}, -(\kappa z_h)^{2 - \alpha}\right) - \kappa^{\frac{4}{\alpha-2}} (\kappa z_h \delta)^{\frac{2\alpha}{\alpha-2 }} \Gamma\left(-\frac{2}{\alpha-2 }, -(\kappa \delta)^{2 - \alpha}\right)}} \,. 
\end{equation}

Substituting these constants into (\ref{generalAt}) and taking the limit $\delta \rightarrow 0$ gives the full solution to $A_t(z)$: 
\begin{equation}
    \begin{split}
        A_t(z) &= \mu \lim_{\delta \rightarrow 0} \frac{\Gamma\left(-\frac{2}{ \alpha -2}, -(\kappa z)^{2 - \alpha}\right) - \Gamma\left(-\frac{2}{\alpha -2 }, -(\kappa z_h)^{2 - \alpha}\right)}{\Gamma\left(-\frac{2}{\alpha-2 }, -(\kappa \delta)^{2 - \alpha}\right) -\Gamma\left(-\frac{2}{\alpha-2}, -(\kappa z_h)^{2 - \alpha}\right)} \\ 
    &= \mu + \mu \frac{ \Gamma\left(-\frac{2}{\alpha -2 }\right) - \Gamma\left(-\frac{2}{\alpha -2 }, -(\kappa z)^{2 - \alpha}\right)}{\Gamma\left(-\frac{2}{ \alpha-2 }, -(\kappa z_h)^{2 - \alpha}\right) -\Gamma\left(-\frac{2}{ \alpha-2 }\right)} \,. \label{Atsolution}
    \end{split}
\end{equation}

$A_t(z)$, as seen in (\ref{Atsolution}), is quite intricate and we need to approximate the incomplete gamma function in some way to introduce $A_t(z)$ in Eq.~\eqref{2.9} to solve for $g(z)$. To do so, we use the following series expansion (about $v=0$): 
\begin{equation}
    \Gamma(u, v) = \Gamma(u) - \sum_{n=0}^{\infty} \frac{(-1)^n v^{n + u}}{n! (u + n)}\,,
\end{equation}

\begin{figure}[h!]
    \centering
    \includegraphics[width=0.75\textwidth]{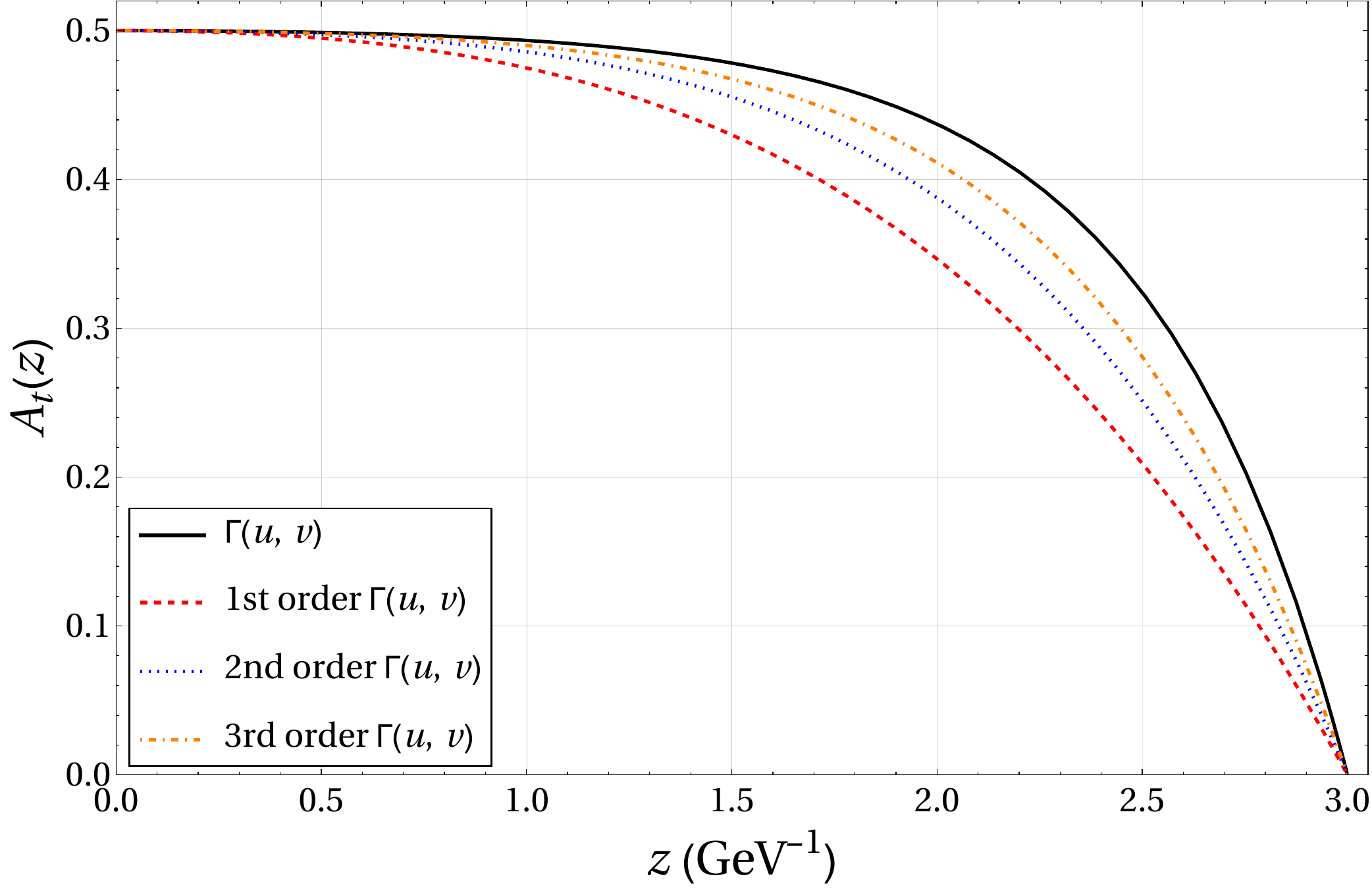}
    \caption{\small $A_t(z)$ approximation to 3rd order in $\Gamma(u, v)$ with parameters $z_h = 3 \, \text{GeV}^{-1}$, $\mu = 0.5 \, \text{GeV}$, $k = 0.8 \,  \text{GeV}$, and $\alpha = 0.5$.}
    \phantomsection
    \label{fig:GaugeA}
\end{figure}

In Fig.~\ref{fig:GaugeA}  we plot $A_t(z)$ for different orders of approximation to $\Gamma(u, v)$. We can see then that higher orders in $\Gamma(u, v)$ produce better approximations to $A_t(z)$. If we choose lower values for the model parameters, like the hybrid meson one, the first order is considered sufficient to continue the calculations. From now on, we use the following $A_t(z)$:
\begin{equation} \label{At1st}
\boxed{A_t(z) = \mu -  \mu\frac{\left(\frac{z}{z_h}\right)^{2 - \alpha} \left[(\alpha -4)(\kappa z)^\alpha  - 2 \kappa^2 z^2 \right]}{ (\alpha-4 )(\kappa z_h)^\alpha -2 \kappa^2 z_h^2}\,.}
\end{equation}

\subsection{Analytic formula for $g(z)$}

By substituting what was found for $A_t(z)$, in Eq.~(\ref{At1st}), the equation of motion for the metric, Eq.~\eqref{2.9}, becomes
\begin{equation} \label{gequation}
    0 = g'' - \frac{3 (4 + c z^2)}{4 z} g' - 4\mu^2 z^2 e^{\frac{3 c z^2}{8} - (\kappa z)^{2 - \alpha}} \frac{ z^{2 - 2\alpha} \left[\kappa^2 z^2 + (\kappa z)^\alpha\right]^2 z_h^{ 2\alpha-4} (\alpha-4 )^2 }{\left[ (\kappa z_h)^\alpha ( \alpha-4)-2 \kappa^2 z_h^2 \right]^2}\,.
\end{equation}

The boundary conditions of Eq.~(\ref{gequation}) are so that $g(z)$ must vanish at the horizon, $g(z_h) = 0$, and become unity at the holographic boundary, $g(0) = 1$. 
To compute Eq.~(\ref{gequation}) with such boundary conditions,  the Taylor series expansion of $\exp{\left[\frac{3 c z^2}{8} - (\kappa z)^{2 - \alpha}\right]}$ can be implemented to first order, to eliminate such exponential from the differential equation. After doing all these approximations, the problem remains barely analytically tractable. 

It was possible then to find a solution, with the help of Mathematica, to $g(z)$ that depends on the generic set of parameters $\mu$, $\alpha$, $\kappa$, $c$, and $z_h$. Even so, we should say that for larger values of these parameters, especially $\kappa$ (which correlates with bigger $\alpha$), the derived quantities may be inflicted with larger errors. We then focus, for the rest of this Section, on computing the finite density corrections to the hybrid meson ($\kappa = 0.468 \, \text{GeV} $ and $\alpha=0.034$) spectral functions. The calculated $g(z)$ contains hundreds of terms. Just for reference, we show a few of those terms: 
\begin{align} \label{g_fullmu}
    g(z) &= \alpha \mu^2 e^{\frac{3 c z^2}{8}} \frac{   2^{21 - \frac{9\alpha}{2}} 3^{-6 + \frac{3\alpha}{2}} c^{-6 + \frac{\alpha}{2}}  \left(\frac{c}{k}\right)^\alpha k^6 z_h^{2\alpha - 4}  \Gamma\left(4 - \frac{3\alpha}{2}, \frac{3 c z^2}{8}\right)}{\left[2 k^2 z_h^2 - (k z_h)^\alpha (\alpha - 4)\right]^2} \nonumber \\
    &- \alpha^2 \mu^2  z^2  e^{\frac{3 c z^2}{8}} \frac{2^{10 - \frac{3\alpha}{2}} 3^{\frac{\alpha}{2} - 3} c^{\frac{\alpha}{2} - 3}  k^{2 - \alpha}\left(\frac{1}{k z_h}\right)^{-2\alpha}  \Gamma\left(3 - \frac{\alpha}{2}, \frac{3 c z^2}{8}\right)}{z_h^4 \left(2 k^2 z_h^2 - (k z_h)^\alpha (\alpha - 4)\right)^2} \\
    &+ \mu^2 z^{6 - 3\alpha} \frac{4096 k^{2 - \alpha}  (k z z_h)^{2\alpha} }{3 c z_h^4 \left[2 k^2 z_h^2 - (k z_h)^\alpha (\alpha - 4)\right]^2 (\alpha - 8) (\alpha - 6)} + \cdots \nonumber
\end{align}

\begin{figure}[h!]
    \centering
    \includegraphics[width=0.75\textwidth]{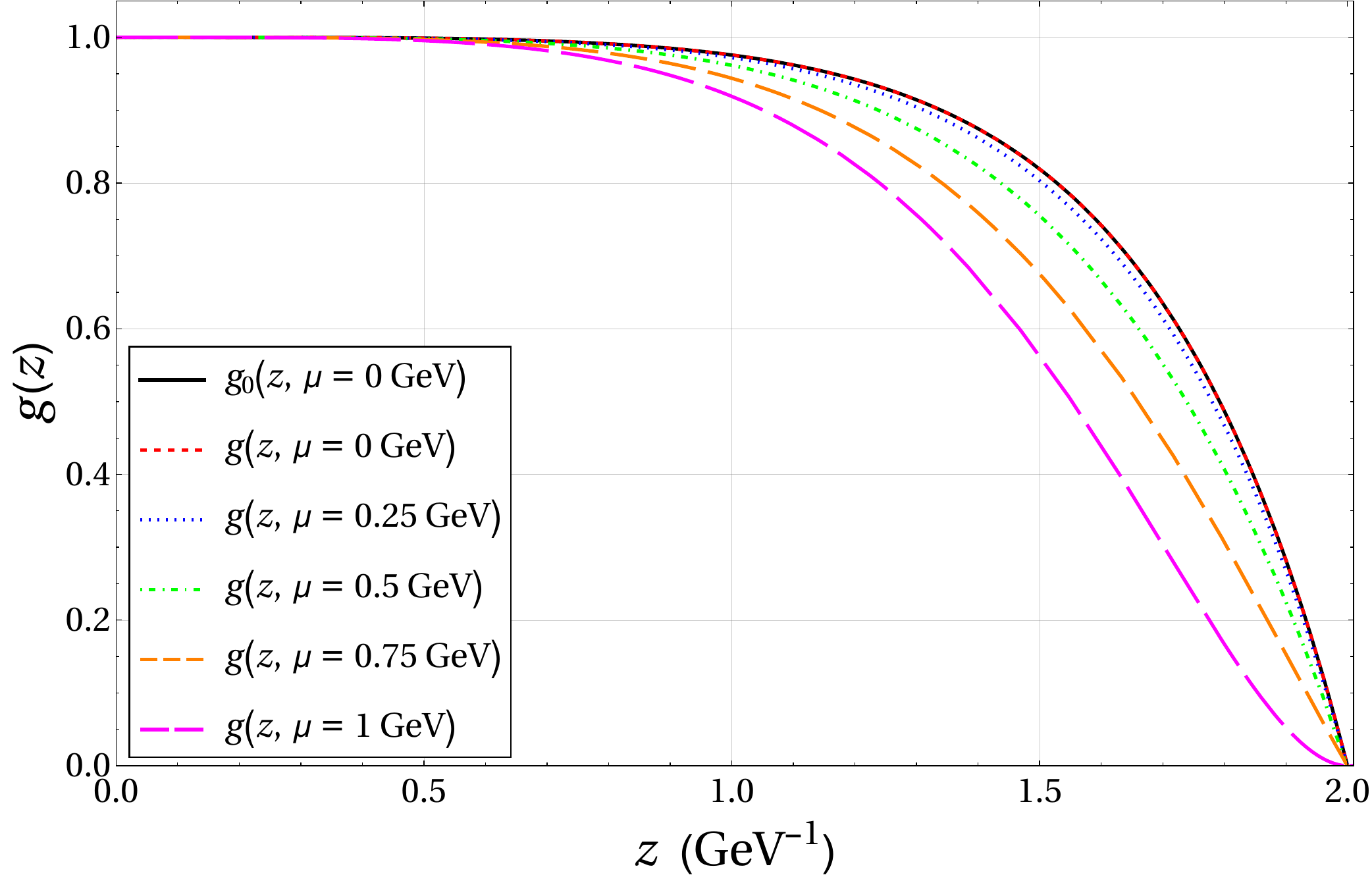}
    \caption{\small The blackening function $g(z)$ for various $\mu$ using the hybrid meson parameters. Here $z_h=2$ is used.}
    \phantomsection
    \label{fig:gmu}
\end{figure}

Fig.~\ref{fig:gmu} shows the blackening function $g(z)$ for some values of chemical potential $\mu$. We define $g_0(z)$ as the metric calculated earlier at zero chemical potential, shown in Eq.~(\ref{g_T_mu_0}). We refer to $g(z)$ as the now calculated metric shown in Eq.~(\ref{g_fullmu}). 

From Fig.~\ref{fig:gmu}, we can clearly see that the calculated metric at nonzero chemical potential reproduces the metric calculated at $\mu = 0$ as their curves overlap. We notice the deformation of the curves as we increase $\mu$. We use this computed metric (\ref{g_fullmu}) in the next Sections to analyze its thermodynamical properties and calculate how the $\mu$ affects the mesonic melting process.

\newpage

\section{Thermodynamical analysis of the EMD model at finite $\mu$}\label{fimuspec}

This Section extends the thermodynamical analysis to finite chemical potential by using the constructed blackening function $g(z)$ to obtain the Hawking temperature $T(z_h,\mu)$ and its branch structure. 

\subsection{Black hole temperature}
\begin{figure}[h!]
    \centering
    \includegraphics[width=0.75\textwidth]{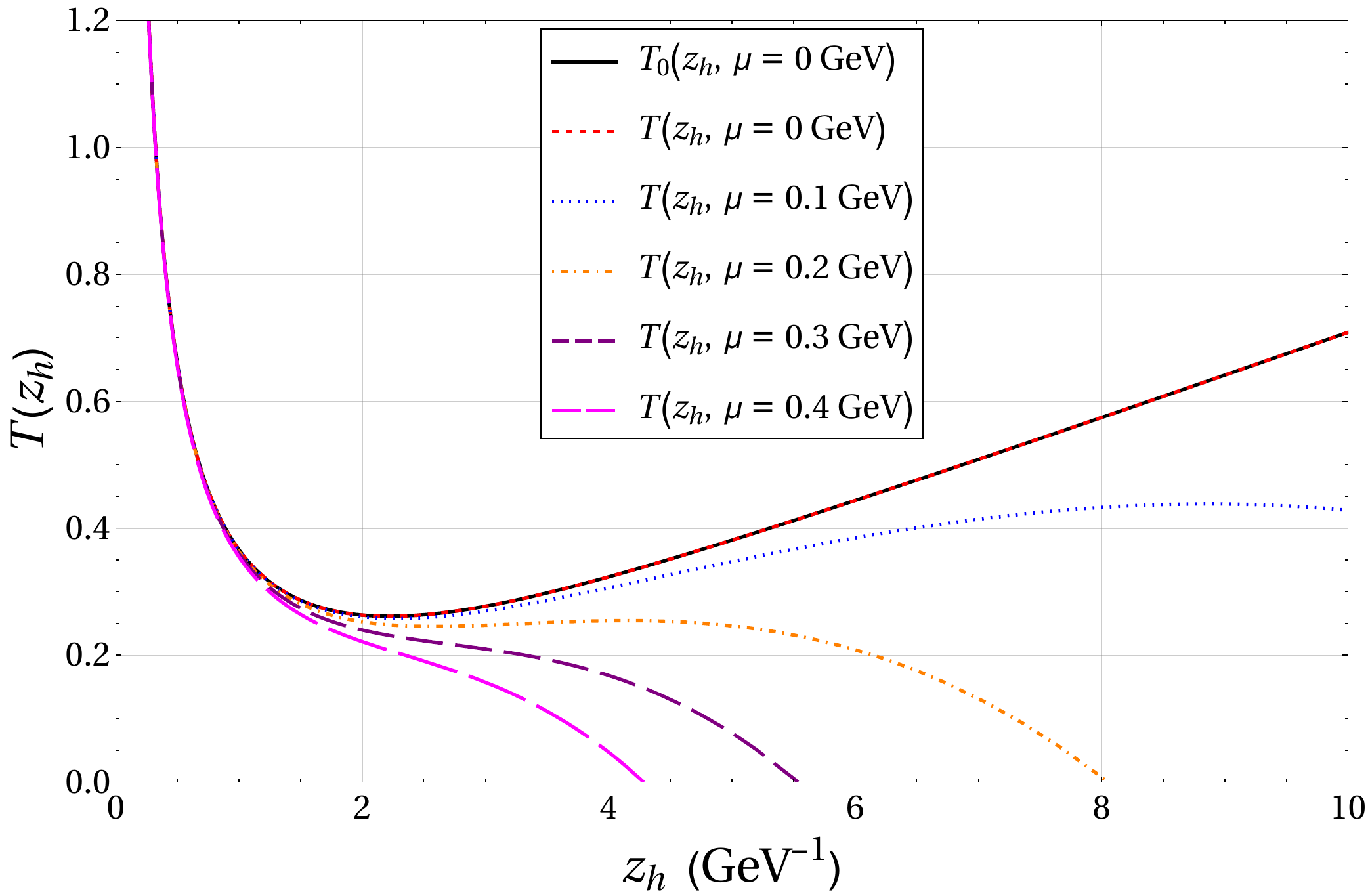}
    \caption{\small Temperature as a function of $z_h$ for various $\mu$ using the hybrid meson parameters.}
    \phantomsection
    \label{fig:tvsmu}
\end{figure}

\begin{figure}[h!]
    \centering
    \includegraphics[width=0.75\textwidth]{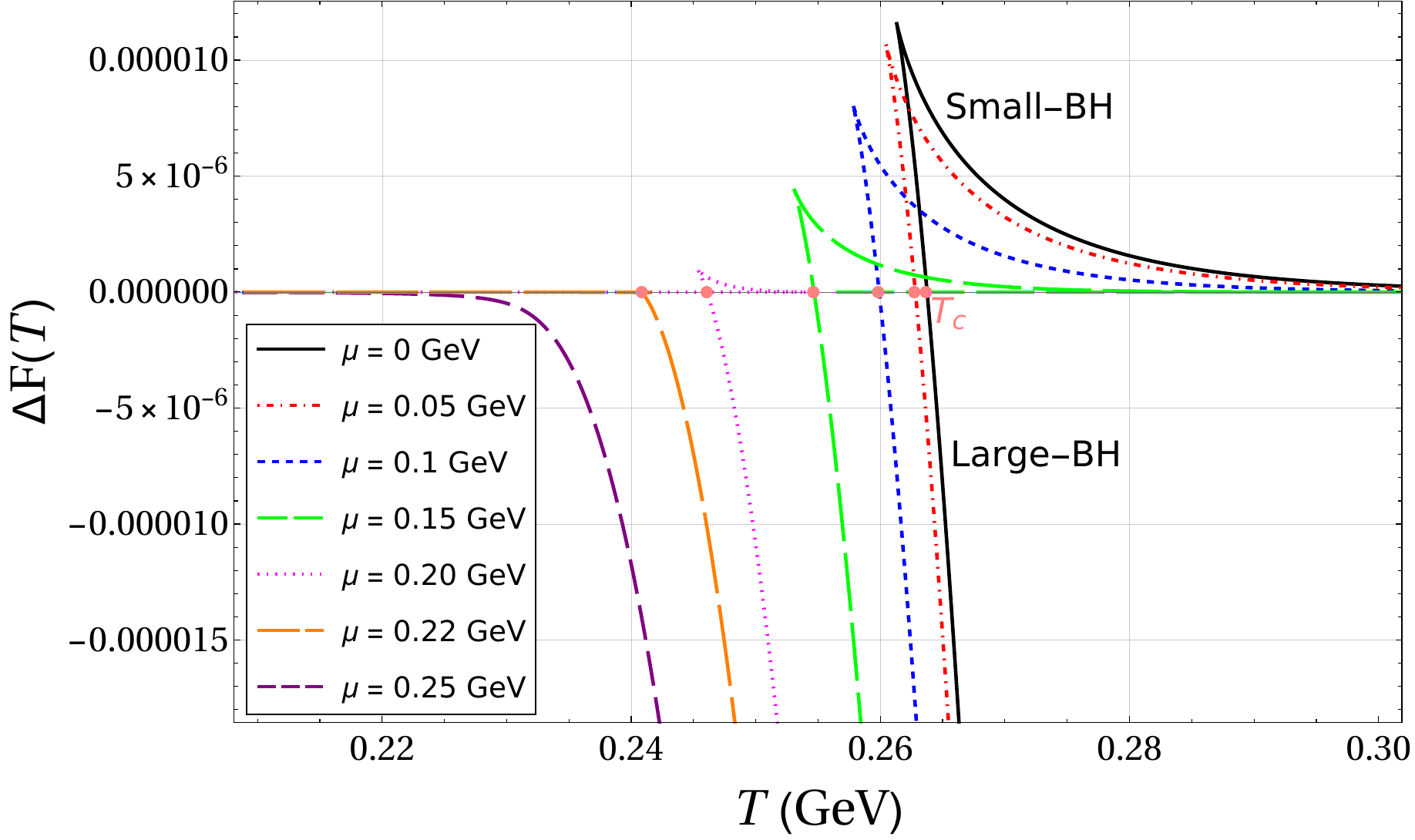}
    \caption{\small The thermal profile of the free energy difference $\Delta F (T)$ for various values of $\mu$. Here, hybrid meson parameters are used.}
    \phantomsection
    \label{fig:freemu}
\end{figure}

In the last Section, we analytically computed the metric for a holographic plasma at finite density. With an expression of $g(z)$ at hand, we can analyze the thermodynamical structure of the gravity solution at finite density. The computed quantities are quite similar to those of \cite{Dudal:2017max}. The results are shown in Figs.~\ref{fig:tvsmu} and \ref{fig:freemu}. Here, $T_0$ is defined as the temperature corresponding to the choice $g_0(z)$. Fig.~\ref{fig:tvsmu} shows that, again, the temperature agrees with the case of $g(z)$ if $\mu = 0$ is taken into account. 

By looking at Fig.~\ref{fig:tvsmu}, the temperature $T = -g'(z_h)/(4 \pi)$ for the newfound metric dependent on $\mu$, we conclude that the thermodynamic structure at finite chemical potential differs significantly from its uncharged counterpart. For small but nonzero $\mu$, the system now features three black hole branches—large, small, and intermediate—rather than two. In both the large and small black hole branches, the temperature decreases with $z_h$, and these branches exhibit positive specific heat, making them thermodynamically stable. In contrast, for the intermediate branch, the temperature increases with $z_h$, resulting in negative specific heat and, consequently, thermodynamic instability. As $\mu$ increases, the intermediate branch gradually shrinks, while the large and small black hole branches move closer to each other. At a critical value of the chemical potential, $\mu=\mu_c$, the intermediate branch vanishes entirely, and the small and large black hole branches merge into a single black hole branch. For $\mu>\mu_c$, only one black hole branch remains, which becomes extremal at a certain horizon radius.

\subsection{Free energy difference and the overall phase structure}

The following computations of the free energy are done as outlined in Sec. \ref{thermo}. Importantly, now, at finite density, we are actually computing the grand potential (not the Helmholtz). Again, the reference point is thermal AdS with null energy. The corresponding free energy behavior for the newfound metric solution is shown in Fig.~\ref{fig:freemu}. Here, the black hole free energy is normalized again with respect to the thermal AdS as said. For $\mu<\mu_c$, the free energy exhibits a swallowtail-like structure, indicating a first-order phase transition. In particular, the free energy of large/small black hole phases is minimal for high/low temperatures. This suggests a first-order phase transition between the large and small black hole phases as the temperature is varied. The corresponding transition temperature turns out to be decreasing with $\mu$.  

The intermediate phase, forming the base of the swallowtail, always has higher free energy than the large and small black hole phases and is thermodynamically disfavored at all temperatures. At $\mu=\mu_c$, the large and small black hole phases merge together and the first-order phase transition between them ceases to exist. The $\mu_c$, therefore, defines a second-order critical point at which the first-order phase transition line between the small/large black hole phases stops. This, at least qualitatively, reflects what one expects for genuine (lattice) QCD at finite density/temperature \cite{deForcrand:2002hgr,Stephanov:2004wx,Borsanyi:2025dyp}.
\begin{figure}[h!]
    \centering
    \includegraphics[width=0.75\textwidth]{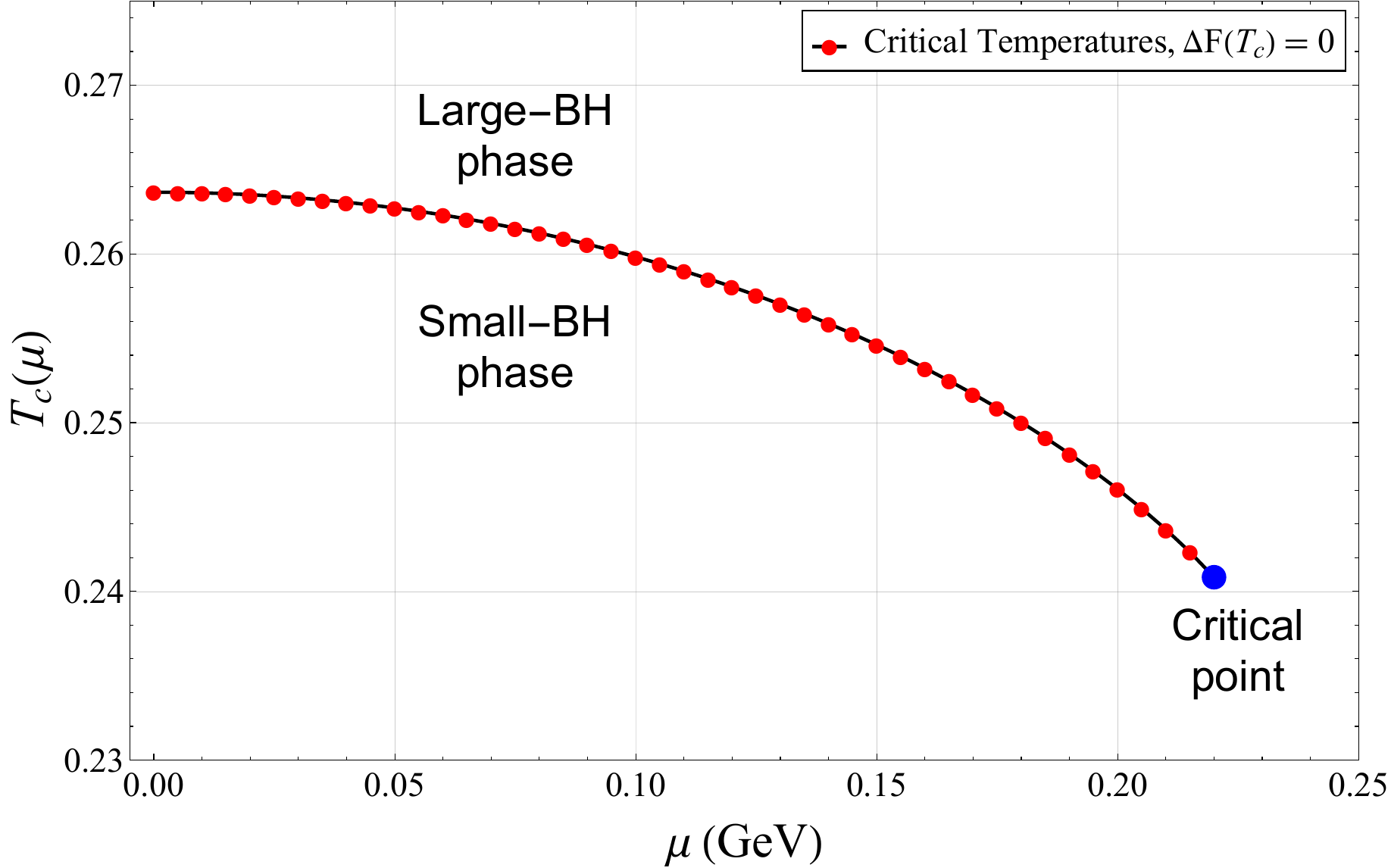}
    \caption{\small $T_{c}$ as a function of $\mu$: For small values of $\mu$, a first-order phase transition occurs between the large and small black hole phases. This first-order phase transition line terminates at a second-order critical point.}
    \phantomsection
    \label{fig:criticalphase}
\end{figure}

Therefore, in the free energy thermodynamic sense, the system is analogous to the famous van der Waals type phase transition observed in ordinary liquid-gas thermodynamic systems. For $\mu>\mu_c$, only one black hole phase exists, which remains thermodynamically favored at all temperatures, i.e., its free energy is always negative. The overall thermodynamic phase structure at finite $\mu$ is shown in Fig.~\ref{fig:criticalphase}. We readily see that the critical temperature for this small/large geometrical transition decreases monotonically with the increase of $\mu$.

In the dual boundary theory, the small and large black hole phases correspond to the specious-confined and deconfined phases \cite{Dudal:2017max}, respectively. In particular, the small black hole phase does not represent true confinement (which would typically be dual to a horizonless geometry at zero temperature) because its Polyakov loop expectation value, while extremely small, is still nonzero. It does, however, exhibit linear confinement at large distances and low temperatures. Due to this distinction, this phase is referred to as the specious-confined phase rather than simply the confined phase. On the gravity side, the presence of a small black hole in the specious-confined phase introduces a well-defined notion of temperature. Given this, it will be interesting to examine the thermal behavior of the melting process in both specious-confined and deconfined phases.

\newpage

\section{Melting process in a finite density plasma}\label{metingmu}

In this Section, we study the melting of mesonic excitations in a finite density holographic plasma by using the $\mu$-dependent black-hole background to compute the effective Schr\"odinger potential and the corresponding spectral functions.

\subsection{Effective potential}

\begin{figure}[h!]
    \centering
    \includegraphics[width=0.75\textwidth]{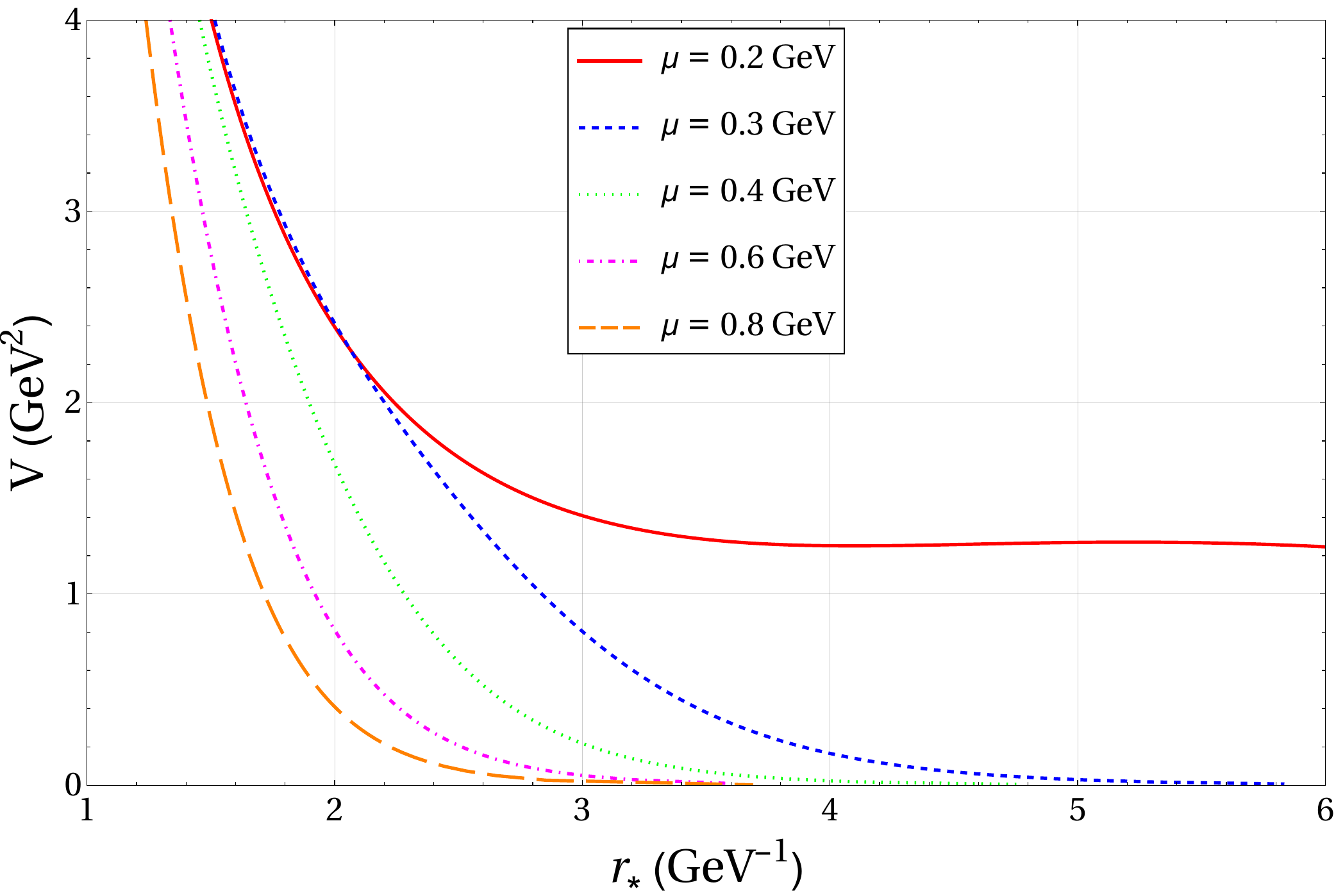}
    \caption{\small Effective potential $V$ at $T=270 \, \text{MeV}$ as a function of $r_*$ for various values of $\mu$. Here hybrid meson parameters are used.}
    \phantomsection
    \label{fig:potmu}
\end{figure}

Having obtained an explicit formula for $g(z)$ that depends on $\mu$, allows one to consider how a finite density configuration affects the melting process. For the hybrid meson parameters, we plug the newfound metric into the fluctuation's equation of motion (\ref{EOM_T}) to find how $\mu$ changes the effective potential (Fig.~\ref{fig:potmu}) for a given fixed temperature, $T=270 \, \text{MeV}$ in this case.

We observe that the increase in $\mu$ has a similar effect as the increase in temperature, evidenced by comparing Fig.~\ref{fig:potmu} with Figs.~\ref{fig:Tpotcharm}--\ref{fig:Tpothybrid}. Given so, the effective potential indicates that the introduction of a chemical potential speeds up the melting process. 

\begin{figure}[h!]
    \centering
    \includegraphics[width=0.75\textwidth]{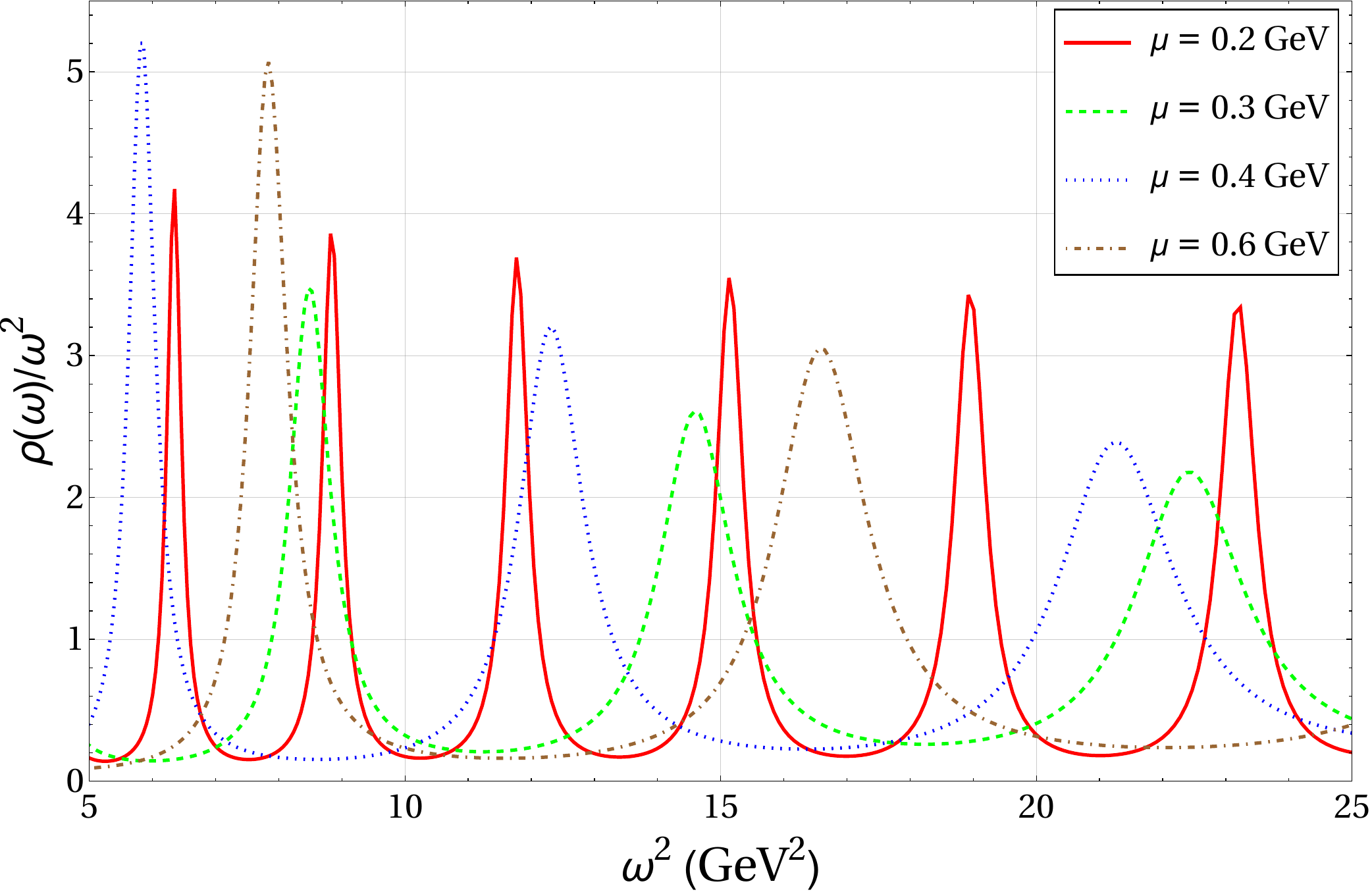}
    \caption{\small Spectral functions at $T=270~\text{MeV}$ for various values of $\mu$.}
    \phantomsection
    \label{fig:figmuspec}
\end{figure}

\subsection{Spectral functions}

\begin{figure}[h!]
    \centering
    \includegraphics[width=0.75\textwidth]{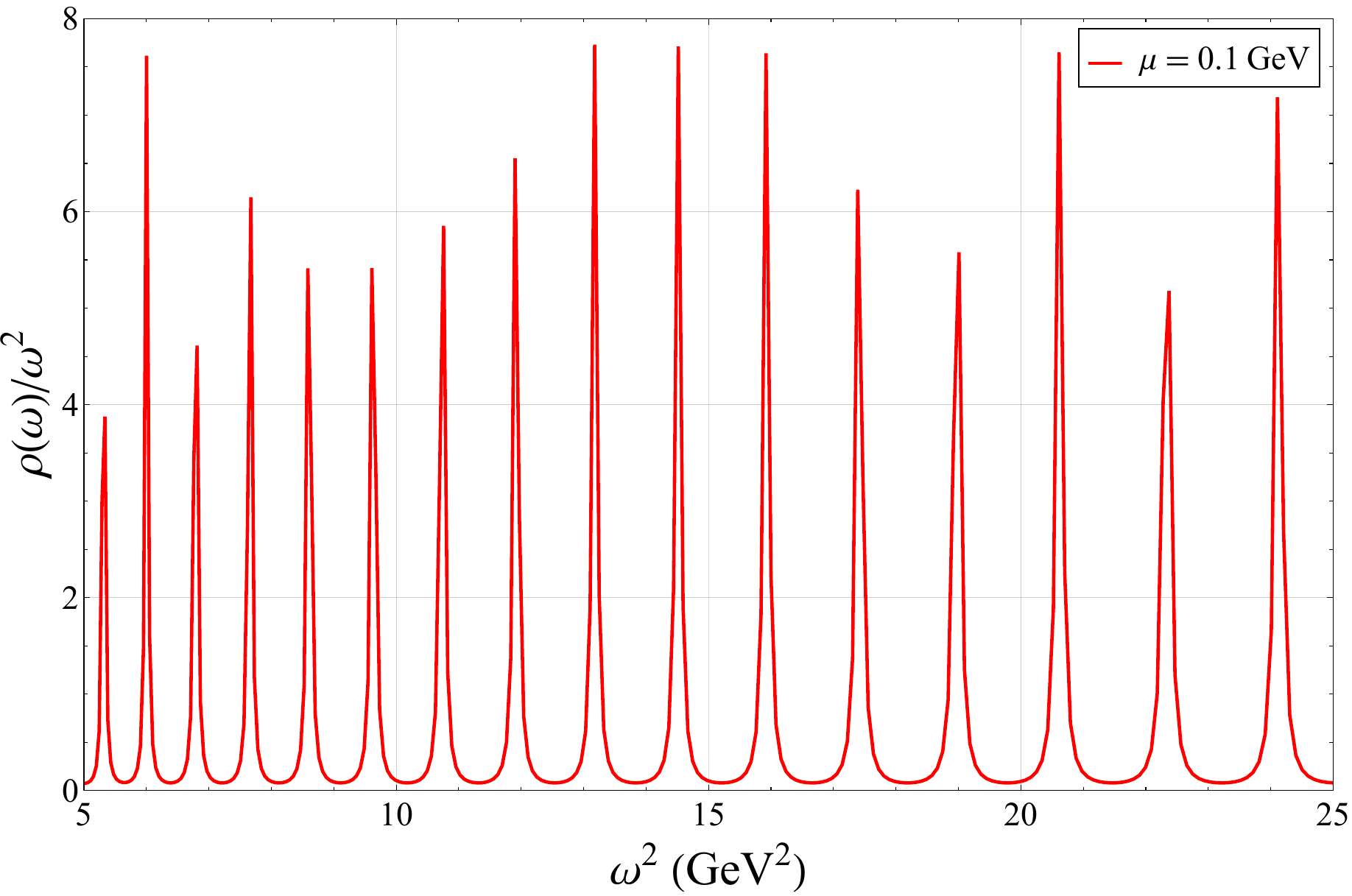}
    \caption{\small Spectral function in the small black hole geometry. Here $\mu=0.1~\text{GeV}$ and $T=0.1~\text{GeV}$ are used.}
    \phantomsection
    \label{fig:small}
\end{figure}

By applying the numerical procedure outlined in Section \ref{procespec} to compute the spectral functions\footnote{In this case, the Pad\'{e}-like approximant becomes unnecessary as the function $f(z)$ does not contain the hyperbolic tangent term.} for this finite density system, we are able to compute and plot the spectral functions, for a fixed $T=270 \, \text{MeV}$ temperature, and see how the peaks change as $\mu$ becomes larger (Fig.~\ref{fig:figmuspec}). 

Fig.~\ref{fig:figmuspec} shows that, as $\mu$ increases, the peaks not only broaden out but also the distance between the consecutive peaks increases. This indicates, as the effective potential demonstrates, the melting process is occurring more rapidly. This implies that the medium at higher baryon density also melts the resonances—much the same way a hot medium does at zero density. 

By fixing the temperature $T$ and changing $\mu$, one can see that an increase in chemical potential can cause these once‐sharp excitations to lose their strength and ultimately dissolve into the continuum. The fact that one still sees fairly well‐defined peaks for small to moderate $\mu$ means that the mesonic states are not immediately destroyed at low density, but eventually, at sufficiently large $\mu$, they do disappear, signaling the loss of a well‐defined bound state in a dense medium.

It is also interesting to investigate the spectral function in the specious-confined phase, corresponding to a small black hole phase on the dual gravity side. For this purpose, we choose the parameters $\mu=0.1~\text{GeV}$ and $T=0.1~\text{GeV}$, well within the small black hole geometry (large $z_h$), and apply the already developed numerical procedure. The result is shown in Fig.~\ref{fig:small}.

It is evident that the spectral function displays a large number of sharp and well-localized peaks in the specious-confined phase. This behavior is expected, as the confinement-mimicking nature of this model should produce well-defined and narrow states, consistent with what one anticipates from genuine QCD at low energy scales. A similar pattern is observed at low temperatures in the specious-confined phase for other values of $\mu$ as well. In the limit of zero temperature and chemical potential, the spectral function reduces to a collection of delta functions centered at the masses of the corresponding particles.

\begin{figure}[h!]
    \centering
    \includegraphics[width=0.75\textwidth]{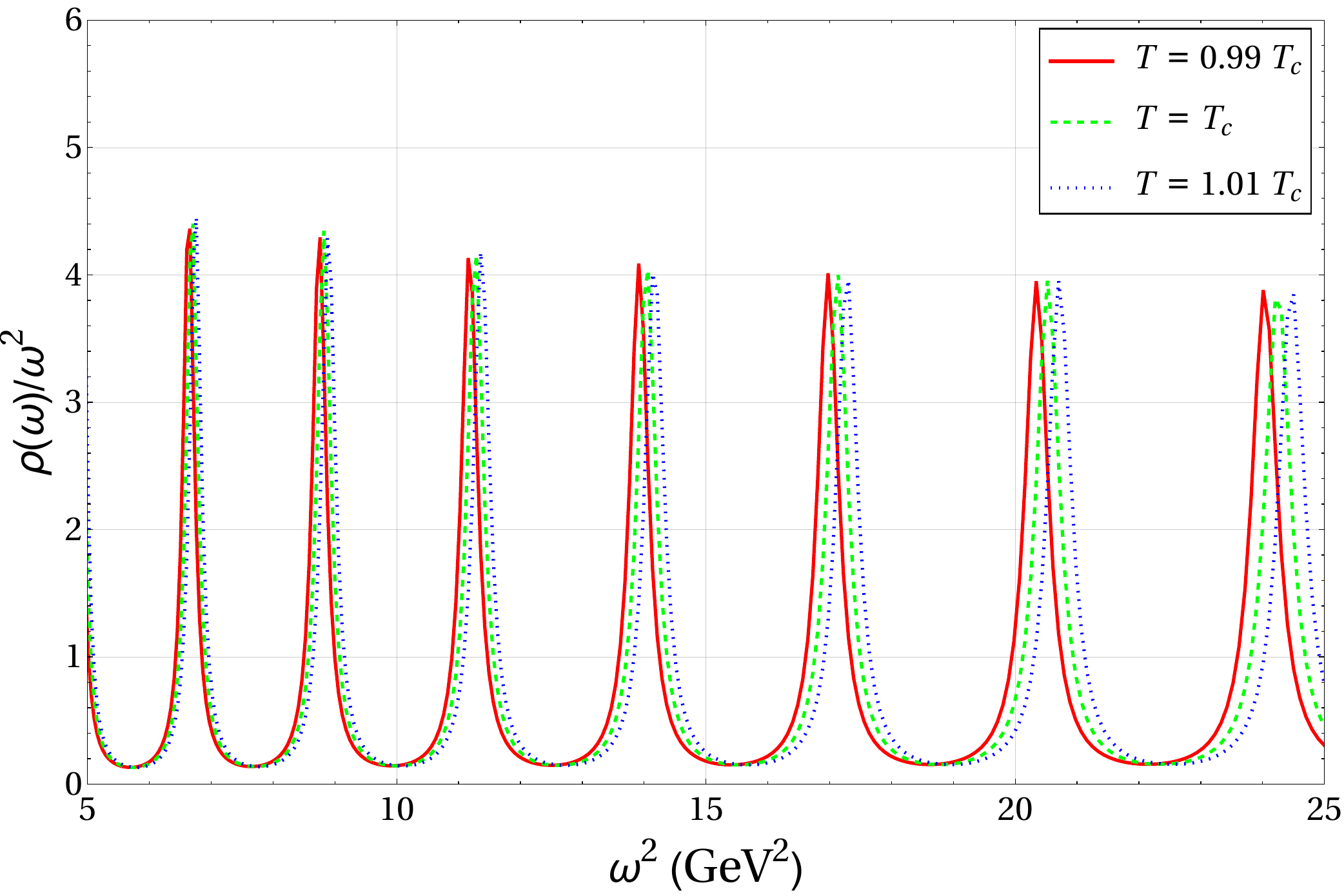}
    \caption{\small Spectral functions around the critical temperature for $\mu=0.1 ~\text{GeV}$, large and small black hole phase.}
    \phantomsection
    \label{fig:trans}
\end{figure}

The above specious-confined phase result is in contrast with what is seen in the deconfined phase. In Fig.~\ref{fig:figmuspec}, where we have a considerably smaller $z_h$, the peaks are more spurious and broader, compared to what is seen on Fig.~\ref{fig:small}. This signals the melting of the states in the finite density plasma. 

We conclude then, from the spectral functions computations, that the small black hole phase represents well defined, non-melted, states and the large black hole phase represents, at least for high enough temperatures and chemical potentials, not well-defined, melted, states. 

One important thing to be noticed is that despite the large/small black hole geometry transition being regarded as a first order phase transition, we see that this does not correspond to substantial variations to the spectral functions near the transition point. We observe a gradual and smooth transition from the confined states to the melted states. In other words, to be able to see the clear differences in geometries, at least from the perspective of the spectral functions, that reproduce the different QCD matter phases, we have to go ``deep'' into the large/small geometries so that we examine a ``very small'' (low temperature) and ``very large'' (high temperature) black hole. 

To illustrate this, in Fig.~\ref{fig:trans}, we plot the spectral functions near the transition temperature for $\mu=0.1~\text{GeV}$, that corresponds to $T_{c} = 0.26~\text{GeV}$, in both large ($T=1.01 T_{c}$) and small ($T=0.99 T_{c}$) black hole configurations. 

One other point of interest is the critical end point (the blue point at the end of the phase structure curve, shown in Fig.~\ref{fig:criticalphase}) that defines where, in parameter space, the two black hole phases merge to form a single stable black hole that exists for all temperatures. 

\begin{figure}[htb!]
    \centering
    \begin{minipage}[t]{0.49\textwidth}
        \centering
        \includegraphics[width=\textwidth]{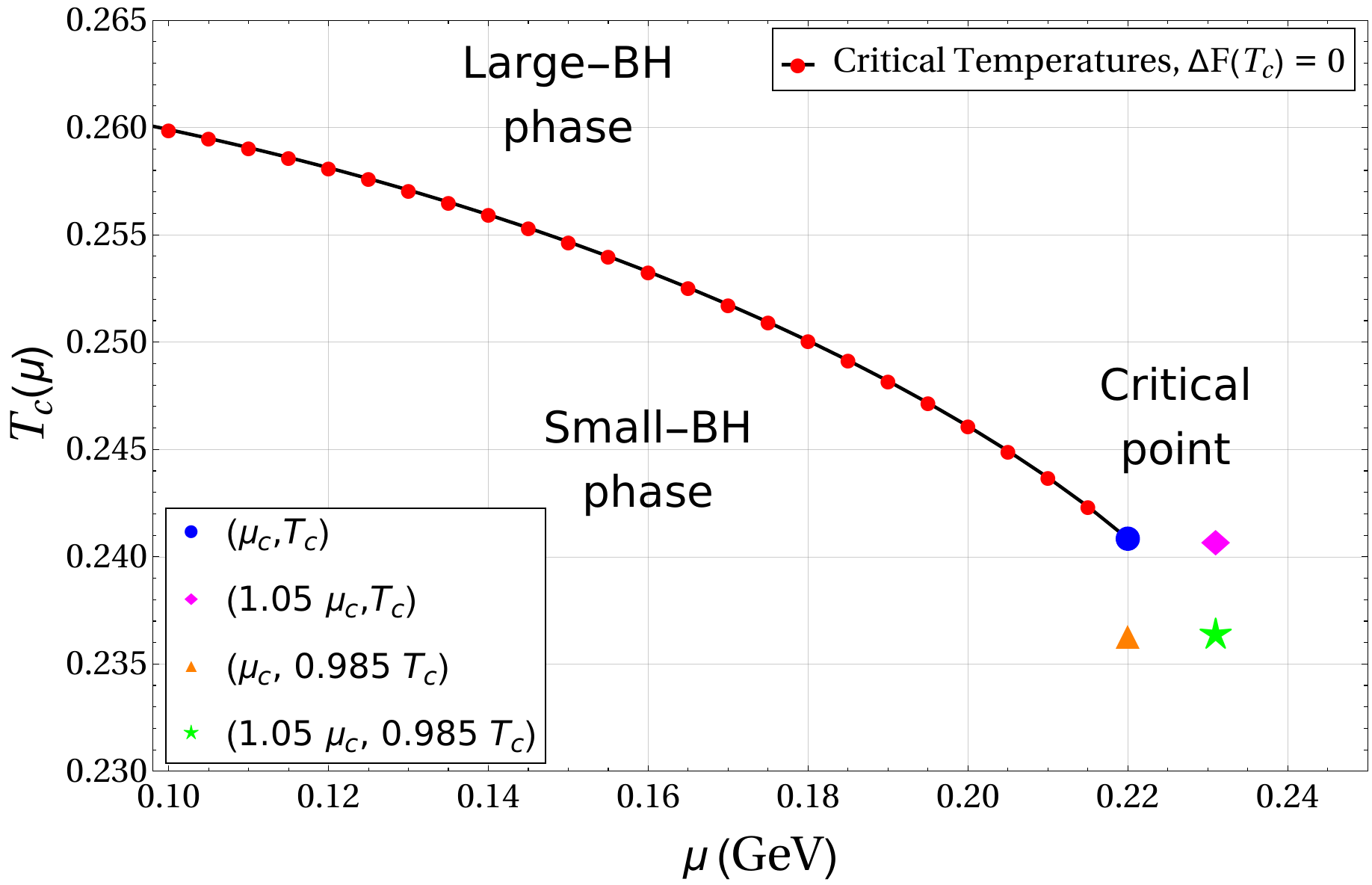}
        \caption{\small Phase structure near the critical end point.}
        \label{fig:squarephase}
    \end{minipage}%
    \hfill
    \begin{minipage}[t]{0.49\textwidth}
        \centering
        \includegraphics[width=\textwidth]{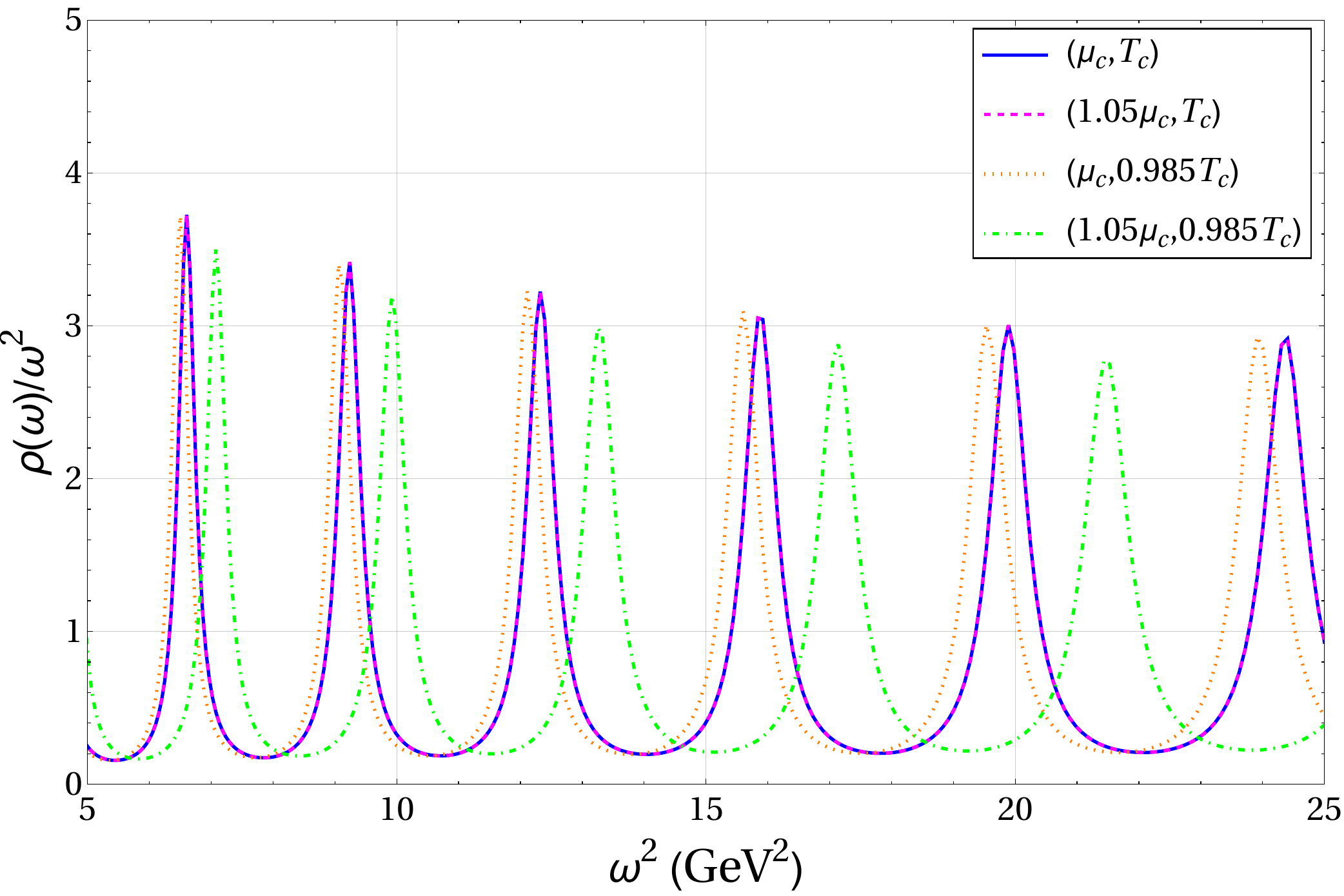}
        \caption{\small Spectral functions around the critical end point.}
        \label{fig:squarespec}
    \end{minipage}
\end{figure}

We then also investigate how the spectral functions change near this critical point $(\mu_c,T_c) \sim (0.22~\text{GeV},0.24~\text{GeV})$. In Fig.~\ref{fig:squarephase} we show, in the $\mu \times T_c$ plane, the chosen points of interest, near the critical end point, and, in Fig.~\ref{fig:squarespec}, the corresponding spectral functions at these points.  

Again, we observe a smooth change in the spectral functions near the point where the distinct black hole geometries merge together so the comments made earlier for Fig.~\ref{fig:trans} still persists.

\newpage

\section{Mesons in a magnetized holographic plasma} \label{modEBID}

Having established in the previous chapters a self-consistent Einstein--Maxwell--dilaton (EMD) framework for QCD-like
plasmas at finite temperature and density, we now extend the same philosophy to include \emph{external fields}.
Among them, a background magnetic field is of particular interest: in non-central heavy-ion collisions, very intense
fields of order $10^{14}\,\mathrm{T}$ can be generated at early times \cite{Skokov:2009qp,Bzdak2012Mar,Voronyuk2011May},
and although the field decays rapidly, it may remain sizable during the formation of the quark--gluon plasma or near
the deconfinement region \cite{Tuchin2013Aug,McLerran2014Sep,D'Elia2010Sep}. Such fields are expected to modify both
thermodynamic and real-time observables, with connections to phenomena such as the chiral magnetic effect
\cite{Fukushima:2008xe,Kharzeev:2007jp} and (inverse) magnetic catalysis \cite{Miransky:2002rp,Gatto:2010pt,Bali:2011qj},
as well as applications to primordial magnetism \cite{Grasso:2000wj,Vachaspati:1991nm}.

From the QCD side, early effective-model studies led to conflicting trends for the transition temperature $T_c(B)$:
some approaches predicted $T_c$ increasing with $B$ (magnetic catalysis) \cite{Gusynin1995Apr,Mizher:2010zb}, whereas
others predicted a decrease (inverse magnetic catalysis) \cite{Agasian2008Jun}. From here on, $B$ denotes the external magnetic field and should not be confused with the auxiliary function $B(z)$ introduced in earlier Sections. Modern unquenched lattice simulations
favor inverse magnetic catalysis around the crossover region \cite{Bali:2011qj}, emphasizing that the relevant
dynamics are inherently non-perturbative. This makes holography a natural arena to explore magnetized QCD-like matter
in a controlled strong-coupling setting, in the same spirit as our EMD analysis of thermodynamics and spectral
functions at $B=0$.

In this vein, we are set to investigate how a magnetic field influences the geometric transition (confinement/deconfinement), through
a Hawking--Page-type competition between thermal AdS and black hole saddles, and the in-medium dissociation
(``melting'') of heavy bound states, through the behavior of spectral functions. While top-down constructions provide
a firm string-theoretic underpinning, they often miss essential QCD features; in particular, the magnetized
Einstein--Maxwell AdS backgrounds of \cite{DHoker:2009mmn,DHoker:2009ixq} do not exhibit a Hawking--Page transition,
making a holographic definition of $T_c$ subtle. Bottom-up approaches, by contrast, can be engineered to reproduce
QCD-inspired thermodynamics and confinement physics more flexibly \cite{Bohra:2019ebj,Bohra:2020qom,Dudal:2021jav}.

Motivated by the EMD construction developed earlier, we therefore adopt a bottom-up, self-consistent model in which
the magnetic field is introduced through a nonlinear Born--Infeld gauge sector. This choice is also physically
well-motivated: in the soft-wall context, Born--Infeld dynamics provides an effective way to couple external fields
to the charged constituents of vector mesons and to account for nonlinear electromagnetic effects beyond the Maxwell
approximation \cite{Dudal:2014jfa,Dudal:2018rki}. Concretely, we work with an
Einstein-Born-Infeld-dilaton (EBID) theory and solve the coupled bulk equations analytically using the same
potential-reconstruction strategy employed in the EMD model \cite{Dudal:2017max,Mahapatra:2018gig,He:2013qq,Toniato:2025gts}.
This yields explicit magnetized geometries with a nontrivial dilaton profile and a magnetic-field-dependent
Hawking--Page transition. Finally, in the deconfined phase we compute heavy-quarkonium spectral functions by studying
Born--Infeld gauge fluctuations polarized parallel and transverse to the magnetic field, using both the real-time
holographic prescription and the membrane paradigm.

\subsection{EBID action}

In order to write a consistent dynamical version of the soft-wall model of \cite{Dudal:2014jfa}, we need to formulate a gravity model incorporating a background magnetic and a dilaton field. For this purpose, we consider the five-dimensional Einstein-Born-Infeld-dilaton (EBID) action,
\begin{equation}
    \begin{split}
        S_{EBI} = & -\frac{1}{16 \pi G_5} \int \mathrm{d^5}x  \biggl[\sqrt{-g} \ \biggl(R + \frac{f(\phi)}{(2 \pi \alpha')^2} \  -\frac{1}{2}\partial_{M}\phi \partial^{M}\phi -V(\phi)\biggr)  \\
        & - \frac{1}{(2 \pi \alpha')^2}\sqrt{-\det\biggl(g_{M N}+ 2 \pi \alpha' F_{M N}\biggr)}\biggr]\,,\label{actionEBI}
    \end{split}
\end{equation}
where $G_5$ is the five-dimensional Newton's constant, $R$ is the Ricci scalar, $F_{M N}$ is the field strength tensor of the $U(1)$ gauge field, and $\phi$ is the dilaton field. The $U(1)$ gauge field will be used to introduce a constant background magnetic field $B$. The coupling between the $U(1)$ gauge field and dilaton field is represented by the gauge kinetic function $f(\phi)$, and $V(\phi)$ is the dilaton potential, to be determined later on.

The Regge slope, string parameter, $\alpha'$ is a new dimensional parameter and is related to the Born-Infeld parameter $b$ by the relation $b=1/(2 \pi \alpha')$. By substituting the value of $b$ and rearranging the preceding Eq.~(\ref{actionEBI}), the resultant expression yields the modified form of the EBID action
\begin{equation}
    \begin{split}
        S_{EBI} &= -\frac{1}{16 \pi G_5}\int \mathrm{d^5}x \sqrt{-g} \biggr[R + f(\phi) b^2 \left(1-\sqrt{1+\frac{F_{M N}F^{M N}}{2 b^2}}\right) \\ &-\frac{1}{2}\partial_{\mu}\phi \partial^{M}\phi -V(\phi)\biggr]\,.
        \label{actionEBImodified}
    \end{split}
\end{equation}
 
As the parameter $b \rightarrow \infty$, the action collapses to the Einstein-Maxwell-dilaton theory presented in Sec.~\ref{EMDspec}. For a nice review of Born-Infeld electrodynamics, we refer the readers to \cite{Alam2021Nov}. In the next Section, we show how to achieve the equivalence between expressions (\ref{actionEBI}) and (\ref{actionEBImodified}).

\subsection{Equivalence between actions}

To show the equivalence between the Born--Infeld terms in Eqs.~\eqref{actionEBI} and \eqref{actionEBImodified}, we
start from the determinant form
\begin{equation}
\mathcal{L}_{BI}
= b^2\left[\sqrt{-g}-\sqrt{-\det\!\left(g_{M N}+ \frac{F_{M N}}{b}\right)}\right]\,,
\label{BIlagrangianappend}
\end{equation}
where $b\equiv(2\pi\alpha')^{-1}$. Using the identity $\det{AB}=\det{A}\det{B}$, we get
\begin{equation}
\det\!\left(g_{M N}+\frac{F_{M N}}{b}\right)
=\det(g_{M P})\;
\det\!\left(\delta^P_{\ N}+\frac{F^P_{\ N}}{b}\right)
=g\;\det\!\left(\delta^P_{\ N}+\frac{F^P_{\ N}}{b}\right)\,,
\label{eq:det_factor}
\end{equation}

Therefore, we reduced the problem to evaluating $\det(\mathbf{1}+X)$ with
\(
X^P_{\ N}\equiv F^P_{\ N}/b
\).

For the background used in this chapter, we have a purely magnetic field with a single nonzero component,
\(
F_{23}=B
\),
so $X^P_{\ N}$ has only one nontrivial antisymmetric $2\times2$ block in the $(2,3)$ subspace and is otherwise
diagonal. Explicitly,
\begin{equation}
    \left.X^P_{\ N}\right|_{(2,3)}
=
\frac{1}{b}
\begin{pmatrix}
0 & F^{2}{}_{3}\\[2pt]
F^{3}{}_{2} & 0
\end{pmatrix}
\implies
\mathbf{1}+X
=
\mathrm{diag}\left(1,1,\ \mathbf{1}_{2}+X_{(2,3)}\right)\,,
\end{equation}

Thus,
\begin{equation}
\det(\mathbf{1}+X)
=\det\left(\mathbf{1}_{2}+X_{(2,3)}\right)
=1-\left(\frac{F^{2}{}_{3}}{b}\right)\left(\frac{F^{3}{}_{2}}{b}\right)
=1+\frac{F_{23}F^{23}}{b^{2}}\,.
\label{eq:det1plusX}
\end{equation}

In this background, the invariant $F^2\equiv F_{\mu\nu}F^{\mu\nu}$ satisfies $F^2=2\,F_{23}F^{23}$, hence \eqref{eq:det1plusX} can be rewritten as
\begin{equation}
\det(\mathbf{1}+X)=1+\frac{F^2}{2b^2}\,,
\label{eq:det_final}
\end{equation}

Combining \eqref{eq:det_factor} and \eqref{eq:det_final} gives
\begin{equation}
\sqrt{-\det\!\left(g_{M N}+\frac{F_{MN}}{b}\right)}
=\sqrt{-g}\,\sqrt{1+\frac{F^2}{2b^2}}\,,
\end{equation}

Substituting this result into \eqref{BIlagrangianappend} yields the desired square-root form
\begin{equation}
\boxed{
\mathcal{L}_{BI}
=b^2\sqrt{-g}\left[1-\sqrt{1+\frac{F_{MN}F^{MN}}{2b^2}}\right]\,.
}
\end{equation}

This establishes the equivalence between the determinant Born--Infeld action \eqref{actionEBI} and the modified form
\eqref{actionEBImodified} for the magnetic background considered in this work.

\subsection{Solution to the EBID equations}\label{sec:EBID_solutions}

In this subsection we derive, in detail, the equations of motion that follow from the Einstein-Born-Infeld-dilaton (EBID) action and then reduce them to ordinary differential equations using the anisotropic magnetic Ans\"atze.

We start from the modified EBID action (Eq.~\ref{actionEBImodified}), written in the convenient Born--Infeld form
\begin{equation}
    \begin{split}
        S_{\rm EBID}
&=-\frac{1}{16\pi G_5}\int d^5x\,\sqrt{-g}\left[
R-\frac12(\partial\phi)^2 -V(\phi)
+ f(\phi)\,b^2\Bigl(1-\mathcal{S}\Bigr)
\right]\,,\\
\mathcal{S}&\equiv\sqrt{1+\frac{F^2}{2b^2}}\,,
\qquad
F^2\equiv F_{MN}F^{MN}\,.\label{eq:EBID_action12}
    \end{split}
\end{equation}
with $b\equiv (2\pi\alpha')^{-1}$. Only the Born--Infeld term depends on $A_N$ (through $F_{MN}$). This implies
\begin{equation}
    \begin{split}
        \delta F_{MN} &= \nabla_M\delta A_N-\nabla_N\delta A_M=2\nabla_{[M}\delta A_{N]}\,,\\
\delta F^2 &=2\,F^{MN}\delta F_{MN}\,,
\qquad
\delta\mathcal{S}
=\frac{1}{2\mathcal{S}}\delta\!\left(\frac{F^2}{2b^2}\right)
=\frac{1}{4b^2\mathcal{S}}\,\delta F^2\,.
    \end{split}
\end{equation}

From this, we obtain
\begin{align}
\delta S_{\rm BI}
&=-\frac{1}{16\pi G_5}\int d^5x\,\sqrt{-g}\;
f(\phi)\,b^2\bigl(-\delta\mathcal{S}\bigr)\nonumber\\
&=-\frac{1}{16\pi G_5}\int d^5x\,\sqrt{-g}\;
f(\phi)\,b^2\left(-\frac{1}{4b^2\mathcal{S}}\right)\delta F^2\\
&=\frac{1}{16\pi G_5}\int d^5x\,\sqrt{-g}\;
\frac{f(\phi)}{4\mathcal{S}}\;2F^{MN}\delta F_{MN}
=\frac{1}{16\pi G_5}\int d^5x\,\sqrt{-g}\;
\frac{f(\phi)}{\mathcal{S}}\,F^{MN}\nabla_M\delta A_N\,.\nonumber
\end{align}

Integrating by parts and dropping the boundary term,
\begin{equation}
    \delta S_{\rm BI}
=-\frac{1}{16\pi G_5}\int d^5x\;\delta A_N\;
\nabla_M\!\left(\sqrt{-g}\,\frac{f(\phi)}{\mathcal{S}}F^{MN}\right)\,,
\end{equation}

Since $\delta A_N$ is arbitrary, we obtain the gauge field equation of motion
\begin{equation}\label{maxwelleom12}
\boxed{
\nabla_M\!\left(\frac{f(\phi)}{\mathcal{S}}\,F^{MN}\right)=0
\Longleftrightarrow
\partial_M\!\left[\sqrt{-g}\,\frac{f(\phi)}{\mathcal{S}}\,F^{MN}\right]=0\,.
}
\end{equation}

Varying \eqref{eq:EBID_action12} with respect to $\phi$ gives
\begin{equation}
    \delta S_{\phi}
=-\frac{1}{16\pi G_5}\int d^5x\,\sqrt{-g}\left[
-\partial_M\phi\,\partial^M(\delta\phi)-V_{,\phi}\,\delta\phi
+ f_{,\phi}\,b^2(1-\mathcal{S})\,\delta\phi
\right]\,,
\end{equation}

Integrating the kinetic variation by parts,
\begin{equation}
    -\partial_M\phi\,\partial^M(\delta\phi)
= -\nabla_M(\partial^M\phi\,\delta\phi) + (\nabla_M\partial^M\phi)\,\delta\phi\,,
\end{equation}
and dropping the surface term yields
\begin{equation}
    \delta S_{\phi}
=-\frac{1}{16\pi G_5}\int d^5x\,\sqrt{-g}\;\delta\phi\left[
\nabla^2\phi -V_{,\phi} + f_{,\phi}\,b^2(1-\mathcal{S})
\right]\,,
\end{equation}

This implies
\begin{equation}
    \nabla^2\phi -\frac{\partial V}{\partial\phi}
+\frac{\partial f}{\partial\phi}\,b^2\left(1-\mathcal{S}\right)=0\,,
\end{equation}

Therefore, the dilaton equation is
\begin{equation}\label{dilatoneom12}
\boxed{
\partial_M\!\left(\sqrt{-g}\,\partial^M\phi\right)
-\sqrt{-g}\Bigl[V_{,\phi}-f_{,\phi}\,b^2(1-\mathcal{S})\Bigr]=0\,.
}
\end{equation}

Varying with respect to $g^{MN}$ gives the standard Einstein equation
\begin{equation}\label{einsteineom12}
\boxed{
R_{MN}-\frac12 g_{MN}R = T_{MN}\,.
}
\end{equation}
where $T_{MN}$ comes from the matter Lagrangian
\begin{equation}
    \mathcal{L}_{\rm m}
=-\frac12(\partial\phi)^2 -V(\phi) + f(\phi)\,b^2(1-\mathcal{S})\,.
\end{equation}

We can use the following relations
\begin{equation}
    \begin{split}
        \delta\sqrt{-g}&=-(1/2)\sqrt{-g}\,g_{MN}\delta g^{MN}\,,\\
        \delta(\partial\phi)^2&=\delta(g^{MN}\partial_M\phi\partial_N\phi)
=\partial_M\phi\,\partial_N\phi\;\delta g^{MN}\,,\\
    \delta F^2
        &=\delta(g^{MP}g^{NQ}F_{MN}F_{PQ})=2F_M{}^{P}F_{NP}\,\delta g^{MN}\,,\\
        \delta\mathcal{S}&=\frac{1}{4b^2\mathcal{S}}\delta F^2\,,
    \end{split}
\end{equation}
to find
\begin{equation}
    \begin{split}
    T_{MN}
&=\frac{1}{2}\left(
\partial_M\phi\,\partial_N\phi
-\frac12\, g_{MN}(\partial\phi)^2
-g_{MN}V(\phi)
\right)\\
&+\frac{f(\phi)}{2}\left[
\frac{F_{MP}F_N{}^{P}}{\mathcal{S}}
+b^2\, g_{MN}\bigl(1-\mathcal{S}\bigr)
\right]\,.
    \end{split}
\end{equation}

We now impose the anisotropic magnetic Ans\"atze 
\begin{equation}
    \begin{split}
        ds^2&=\frac{L^2 e^{2A(z)}}{z^2}\Bigl[-g(z)dt^2+\frac{dz^2}{g(z)}
+dx_1^2+e^{B^2 z^2}(dx_2^2+dx_3^2)\Bigr]\,,\\
\phi&=\phi(z)\,,\qquad
F_{23}=B\,.
\label{ansatze12}
    \end{split}
\end{equation}
so the background magnetic field points along the $x_1$ direction and breaks spatial $SO(3)$ to $SO(2)$ in the $(x_2,x_3)$ plane.

We define the common warp factor $W(z)\equiv L e^{A(z)}/z$ and $C(z)\equiv B^2 z^2$ so that the nonzero metric components are $g_{tt}=-W^2 g$, $g_{zz}=W^2/g$, $g_{11}=W^2$, $g_{22}=g_{33}=W^2 e^{C}$ and 
$\sqrt{-g}=W^5 e^{C}=L^5 e^{5A(z)+B^2 z^2}/z^5$.

The invariant field strength is
\begin{equation}\label{eq:F2mag12}
    \begin{split}
        F^2
=2F_{23}F^{23}
=2B^2\,g^{22}g^{33}
=
\frac{2B^2 z^4}{L^4}\,e^{-4A(z)-2B^2 z^2}\,,\\
\therefore
\mathcal{S}(z)=\sqrt{1+\frac{B^2 z^4}{b^2L^4}\,e^{-4A(z)-2B^2 z^2}}\,.
    \end{split}
\end{equation}

For this background, $F_{23}=B$, the only nonzero $F^{MN}$ components are $(23)$ and $(32)$; moreover all background functions depend only on $z$. Hence each component of $\partial_M\!\left[\sqrt{-g}\,(f/\mathcal{S})F^{MN}\right]$ either vanishes, because $F^{zN}=0$, or reduces to an $x_2/x_3$ derivative of a $z$-dependent quantity, and therefore vanishes identically.

Because the trace term cancels in differences, it is convenient to use
\begin{equation}
    G^M{}_N-G^{M'}{}_{N'}=T^M{}_N-T^{M'}{}_{N'} \Longleftrightarrow
R^M{}_N-R^{M'}{}_{N'}=T^M{}_N-T^{M'}{}_{N'}\,.
\end{equation}

A direct computation from the Christoffel symbols of \eqref{ansatze12} shows that the following combinations reduce
to simple ODEs. Since the magnetic term $F_{MP}F_N{}^{P}$ contributes only when $M,N\in\{2,3\}$, one has
$T^t{}_t=T^{1}{}_{1}$, so $R^t{}_t-R^{1}{}_{1}=0$.
Evaluating $R^t{}_t-R^{1}{}_{1}$ for \eqref{ansatze12} gives
\begin{equation}\label{einstein1}
\boxed{
g''(z)+\left(3A'(z)-\frac{3}{z}+2B^2 z\right)g'(z)=0\,.
}
\end{equation}

Using $R^z{}_z-R^t{}_t=T^z{}_z-T^t{}_t$ and inserting the scalar stress tensor yields an equation involving
$A$ and $\phi$. After simplifying with $C(z)=B^2 z^2$ (so $C'=2B^2 z$, $C''=2B^2$), one obtains
\begin{equation}\label{einstein2}
\boxed{
A''(z)+\frac{2A'(z)}{z}-A'(z)^2+\frac{1}{6}\phi'(z)^2
+\frac{2B^2}{3}\left(1+B^2 z^2\right)=0\,.
}
\end{equation}

Now the magnetic stress does contribute, and
\begin{equation}
    T^{2}{}_{2}-T^{1}{}_{1}
=
f(\phi)\,\frac{F^{23}F_{23}}{\mathcal{S}}
=
f(\phi)\,\frac{B^2 z^4}{L^4}\,\frac{e^{-4A-2B^2 z^2}}{\mathcal{S}}\,.
\end{equation}

Evaluating $R^{2}{}_{2}-R^{1}{}_{1}$ for \eqref{ansatze12} and equating to the above gives a first-order relation that
can be written as
\begin{equation}\label{einstein3}
\boxed{
g'(z)+g(z)\left(-\frac{2}{z}+2B^2 z+3A'(z)\right)
-\frac{z\,e^{-2A(z)-2B^2 z^2}}{2L^2}\,\frac{f(z)}{\mathcal{S}(z)}=0\,.
}
\end{equation}

Equivalently, \eqref{einstein3} can be solved algebraically for $f(z)$, yielding the useful reconstruction formula
\begin{equation}
    f(z)=\frac{2L^2 e^{2A(z)+2B^2 z^2}}{z^2}\,\mathcal{S}(z)\,
\Bigl[z g'(z)+g(z)\bigl\{-2+2B^2 z^2+3zA'(z)\bigr\}\Bigr]\,,
\end{equation}

Inserting \eqref{einstein1}--\eqref{einstein3} to eliminate $g''$ and $f$-derivatives, one may solve algebraically for the potential as a function of $z$:
\begin{equation}\label{einstein4}
\boxed{\,
\begin{aligned}
\displaystyle
V(z)
&=
-\,b^2\bigl(1-\mathcal{S}(z)\bigr)\,f(z)\\
&\quad+\frac{e^{-2A(z)}z^2}{2L^2}\Big[
-2g(z)\,\mathcal{K}(z)+ z\big\{\mathcal{M}(z)\,g'(z)-z g''(z)\big\}
\Big]\,.
\end{aligned}}
\end{equation}

The purely geometric combinations are
\begin{equation}
    \begin{split}
        \mathcal{K}(z)&\equiv 12-10B^2 z^2+4B^4 z^4
+6zA'(z)\bigl(-3+2B^2 z^2\bigr)+9z^2A'(z)^2+3z^2A''(z)\,,\\
\mathcal{M}(z)&\equiv 9-6B^2 z^2-9zA'(z)\,.
    \end{split}
\end{equation}

Finally, inserting \eqref{ansatze12} and \eqref{eq:F2mag12} into the dilaton equation \eqref{dilatoneom12} gives
\begin{equation}\label{dilatoneq}
\boxed{\,
\begin{aligned}
\phi''(z)
&+\left(\frac{g'(z)}{g(z)}+3A'(z)-\frac{3}{z}+2B^2 z\right)\phi'(z)\\
&-\frac{L^2 e^{2A(z)}}{g(z)\,z^2}\left[
V_{,\phi}
- f_{,\phi}\,b^2\bigl\{1-\mathcal{S}(z)\bigr\}
\right]=0\,.
\end{aligned}}
\end{equation}

As in the isotropic case, we impose asymptotically AdS boundary conditions
\begin{equation}\label{boundaryconditions12}
g(0)=1\,,\qquad A(0)=0\,,\qquad \phi(0)=0\,.
\end{equation}
together with horizon regularity $g(z_h)=0$ for the black hole branch.  Equipped with
\eqref{einstein1}--\eqref{dilatoneq} and these boundary conditions, one can implement the potential reconstruction
method to obtain closed-form solutions for $A(z)$, $g(z)$, $\phi(z)$, $f(z)$ and $V(z)$. 

It is instructive to verify that the magnetized EBID background reduces smoothly to the isotropic EMD setup developed
in Sec.~\ref{EMDspec} when the magnetic field is switched off.  Taking $B\to 0$ in the Ans\"atze
\eqref{ansatze12} removes the spatial anisotropy, $e^{B^{2}z^{2}}\!\to\!1$, and sets the background field strength to
zero, so that $F_{23}=B\to 0$ implies $F^{2}\to 0$ and hence the Born--Infeld square-root factor approaches unity,
\(\sqrt{1+F^{2}/(2b^{2})}\to 1\).  In this limit, the Born--Infeld contribution to the stress tensor and to the dilaton
equation vanishes identically because it is proportional to \(1-\sqrt{1+F^{2}/(2b^{2})}\), and the remaining
independent background equations collapse to the $\mu=0$ EMD equations (equivalently $A_t\equiv 0$) derived in
Sec.~\ref{EMDspec}: the blackening-function equation \eqref{einstein1} reduces to the $\mu=0$ EMD equation for $g(z)$,
the warp-factor/dilaton constraint \eqref{einstein2} becomes the EMD relation between $A(z)$ and $\phi(z)$, and the
dilaton equation \eqref{dilatoneq} reduces to the EMD dilaton equation with only $V_{,\phi}$ as a source.  Likewise,
the reconstruction formula for $V(z)$ in \eqref{einstein4} reduces to its EMD counterpart once $B=0$.
Finally, the additional equation that is used at $B\neq 0$ to reconstruct the gauge--dilaton coupling $f(z)$ becomes
redundant when $F_{MN}=0$, as expected: with no background gauge field, $f(\phi)$ drops out of the background
Einstein--dilaton dynamics and cannot be fixed from the geometry alone. 

\subsection{Born-Infeld parameter $b$}

With the magnetized EBID background determined in Sec.~\ref{sec:EBID_solutions}, the remaining input needed to fully specify the model is
the Born--Infeld parameter $b$, which sets the nonlinear scale of the gauge sector. In string-theory language, it is
related to the fundamental string length by
\(
b=(2\pi\alpha')^{-1}
\),
so fixing $b$ amounts to fixing the effective $\alpha'$ scale of the holographic description. As in the EMD analysis,
we determine this parameter by matching a benchmark non-perturbative observable: the static quark--antiquark potential
extracted from the Wilson loop, or equivalently the string tension.

The Wilson loop is the gauge-invariant observable
\(
W(C)=\mathrm{Tr}\,\mathcal{P}\exp\,\bigl(i\oint_C A_\mu\,dx^\mu\bigr)
\),
whose expectation value yields the heavy-quark free energy. In earlier EBID-inspired soft-wall models
\cite{Dudal:2014jfa}, the Wilson loop was not sufficient to constrain $b$ because those backgrounds did not reproduce
a robust linear potential at large separation \cite{Karch2010Dec}. A key advantage of the present self-consistent EBID
construction is that, in the confined (thermal AdS) phase, the Wilson loop exhibits an area law, so the string
tension becomes a reliable matching criterion.

The computation proceeds as usual in holography: the Wilson loop is evaluated by a classical string worldsheet whose
action is the Nambu--Goto functional in the \emph{string-frame} metric. For the magnetized Ans\"atze, \eqref{ansatze12} this
metric is obtained from the Einstein-frame one by
\begin{equation}
    \begin{split}
(g_s)_{MN}&=e^{\sqrt{\frac{2}{3}}\phi}\,g_{MN}\,, \\
ds_{s}^2&= \frac{L^2 e^{2 A_{s}(z)}}{z^2}\biggl[-g(z)\,dt^2 + \frac{dz^2}{g(z)} + dx_{1}^2
+ e^{B^2 z^2}\bigl(dx_{2}^2+dx_{3}^2\bigr) \biggr] \,,
\label{stringmetric}
    \end{split}
\end{equation}
with
\(
A_{s}(z)=A(z)+\sqrt{\frac{1}{6}}\,\phi(z)
\).
The saddle-point prescription gives
\begin{equation}
\langle W(C)\rangle \sim e^{-S_{NG}^{\text{on-shell}}}\,,
\end{equation}
so that the large-separation behavior of the on-shell Nambu--Goto action determines the string tension
\(\sigma\) through \(V_{Q\bar Q}(r)\sim \sigma\,r\).

We fix $b$ by matching the $B=0$ string tension to lattice expectations. Concretely, we use the quenched estimate
\(\sqrt{\sigma}=0.43\pm0.02~\text{GeV}\), obtained by fitting the static potential to a Cornell form with a linear
term $\sigma r$ (see also \cite{Bohra:2019ebj}). The central value \(\sqrt{\sigma}=0.43~\text{GeV}\) corresponds, in
our setup, to
\begin{equation}
b=0.044~\text{GeV}^2\,.
\end{equation}
and to assess the stability of subsequent results we allow a conservative $10\%$ variation around this value.
For completeness, this choice corresponds to \(\ell_s/L=1.92\).

Because the Wilson loop probes a static heavy quark--antiquark pair, it is natural to use it to set the effective
string scale without confronting dynamical string breaking. The resulting ratio $L/\ell_s$ is not large, implying
that $\alpha'$ corrections may not be parametrically suppressed; this is a common limitation in bottom-up holographic
models \cite{Mahapatra:2019uql,Noronha2009Oct,Gubser2008Apr,Andreev2006Apr}, and has been discussed recently in
\cite{Rougemont:2023gfz}. Models engineered to produce larger $L/\ell_s$ typically require additional structure
(e.g.\ extra parameters or special scaling in the dilaton potential), which can obscure predictivity
\cite{Gursoy2009Mar}. While this limitation may affect the quantitative precision of the model, our conclusions are
primarily qualitative and remain stable under moderate variations of $b$ and $a$: numerically, the observables
studied below shift only by a few tens of MeV, within the uncertainty band suggested by lattice data.

For the remainder of this work, we therefore adopt the benchmark value
\(
b=0.044~\text{GeV}^2
\),
which reproduces the $B=0$ string tension at vanishing temperature and magnetic field.

\newpage

\section{Thermodynamics in the magnetized EBID model}\label{Bthermo}

This Section analyzes the thermodynamics of the magnetized EBID setup by comparing the black hole and thermal AdS solutions and assessing their global stability. It presents the Hawking temperature $T(z_h)$ for several magnetic-field values, showing a minimum temperature and the usual stable large-black-hole versus unstable small-black-hole branches, and it uses the free-energy difference $\Delta F$ to identify a first-order Hawking--Page transition. 

\subsection{Temperature and free energy difference}

In the magnetized EBID model under consideration, two distinct gravity solutions arise. The first is a black hole solution with a horizon located at $z = z_h$, while the second corresponds to the thermal AdS configuration. Given the presence of multiple solutions, we will examine their global thermodynamic stability. 

Since both the calculations and results closely resemble those encountered in the previously studied Einstein-Maxwell-dilaton system, for instance, as seen in Sec.~\ref{thermo}, we will present only a concise discussion here.
\begin{figure}[htb!]
\begin{minipage}[b]{0.48\linewidth}
\centering
\includegraphics[width=2.8in,height=2.3in]{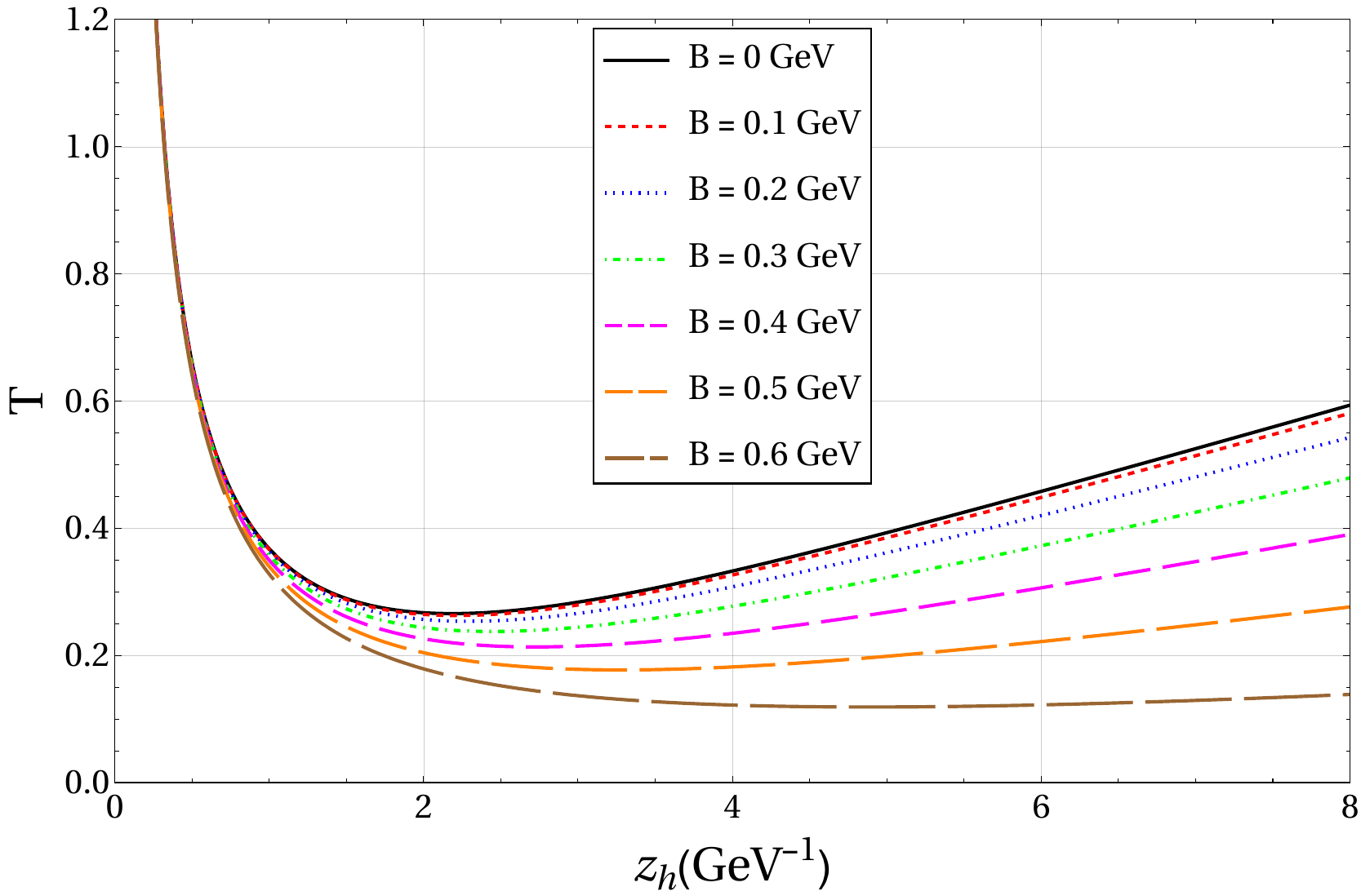}
\caption{\small Temperature $T$ as a function of horizon radius $z_h$ for various values of the magnetic field $B$. In units of GeV.}
\label{TvszhBH}
\end{minipage}
\hspace{0.4cm}
\begin{minipage}[b]{0.48\linewidth}
\centering
\includegraphics[width=2.8in,height=2.3in]{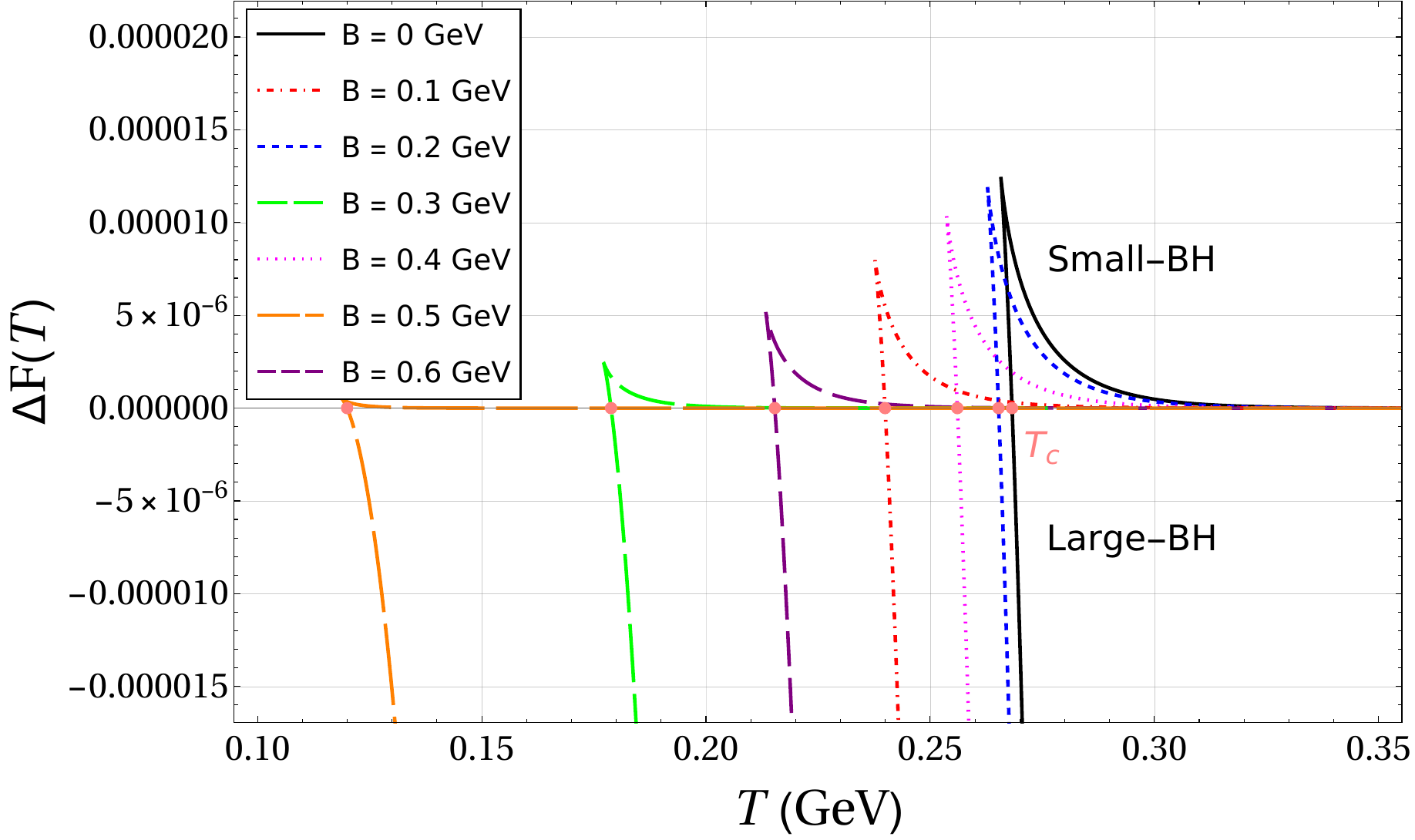}
\caption{\small The free energy difference $\Delta F$ as a function of temperature $T$ for various values of the magnetic field $B$. In units of GeV.}
\label{FvsTBH}
\end{minipage}
\end{figure}
 
In Fig.~\ref{TvszhBH}, we depict the dependence of the Hawking temperature on the horizon radius for various magnetic field strengths. While the thermal AdS phase exists at all temperatures, we identify a minimum temperature below which no black hole solution is present. Above this threshold, the system admits two distinct black hole solutions: a large black hole and a small black hole. 

The large black hole solution (small $z_h$) exhibits a decreasing temperature as the horizon radius increases. It possesses a positive specific heat, making it thermodynamically stable. Conversely, the small black hole solution (large $z_h$) shows an increasing temperature trend with respect to the horizon radius. This solution has a negative specific heat, rendering it thermodynamically unstable.

In Fig.~\ref{FvsTBH}, we present the free energy difference $\Delta F$ between the black hole and thermal AdS phases. It is evident that the dominance of the free energy between the large black hole and thermal AdS phases shifts as the temperature varies. Specifically, at high temperatures, the free energy of the large black hole phase is the lowest, indicating it as the true global minimum in this regime. Conversely, at low temperatures, the free energy of the thermal AdS phase is the lowest, establishing it as the global minimum in the low-temperature regime.

In contrast, the free energy of the small black hole phase remains higher than both the thermal AdS and large black hole phases at all temperatures, making it thermodynamically disfavored throughout. Consequently, a first-order phase transition occurs between the large black hole and thermal AdS phases as the temperature changes. This corresponds to the well-known Hawking-Page phase transition \cite{Hawking1983Dec}. 

The temperature at which the free energy difference vanishes determines the transition temperature $T_c$. At $B=0$, this transition temperature is found to be $270~\text{MeV}$, consistent with the earlier determination.

\subsection{Phase structure}\label{sec:phase_structure_B}

Interestingly, the thermodynamic characteristics described above remain valid even for small but finite magnetic field values ($B \leq 0.6$). The EBID gravity system continues to exhibit both an unstable small black hole phase and a stable large black hole phase, with the thermal AdS phase dominating at low temperatures. 

\begin{figure}[htb!]
	\centering
	\includegraphics[height=7cm,width=11cm]{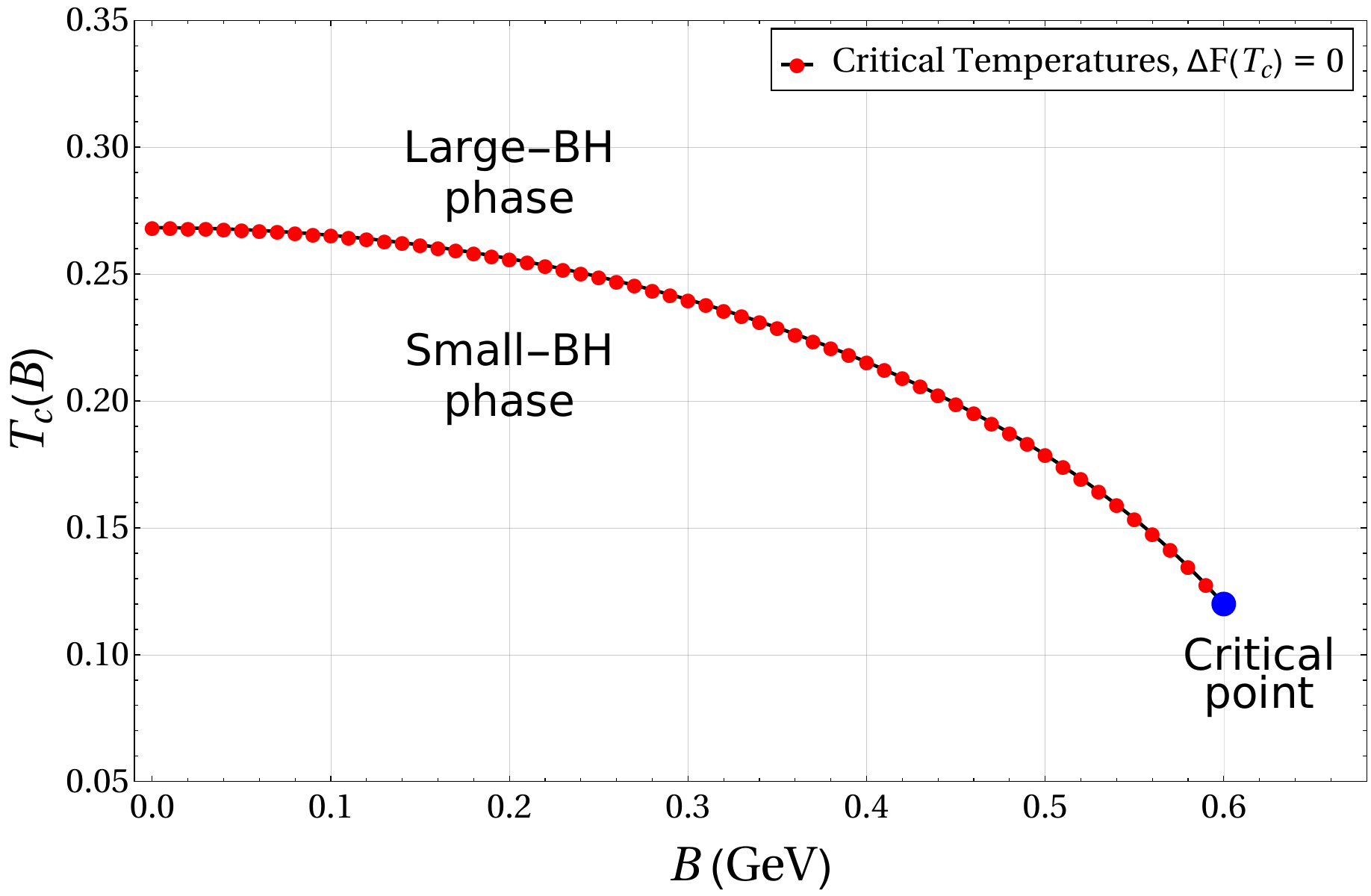}
	\caption{\small The variation of the deconfined transition temperature $T_c$ with magnetic field. In units of GeV.}
	\label{TcvsB}	
\end{figure}

Importantly, the Hawking-Page thermal AdS/black hole phase transition persists at finite magnetic fields. The key distinction lies in the transition temperature, which decreases monotonically as the magnetic field increases. The dependence of $T_c$ on $B$ is illustrated in Fig.~\ref{TcvsB}.

As previously discussed, the thermal AdS and black hole phases correspond to the confined and deconfined phases, respectively, in the boundary theory. Consequently, the thermodynamic results presented above indicate inverse magnetic catalysis in the deconfinement sector. This holographic prediction aligns with lattice QCD findings \cite{Bali:2011qj}. 

Additionally, although explicit results are not shown here for brevity, the constructed EBID gravity model predicts a decrease (increase) in the quark-antiquark string tension with the magnetic field when the pair is aligned parallel (perpendicular) to it. This prediction is once again consistent with recent lattice results \cite{Bonati:2014ksa,Bonati:2016kxj}. 

It is important to emphasize that these holographic QCD features in the presence of $B$ are highly robust, maintaining their qualitative nature for different values of $a$ and $b$, despite variations in the magnitude of $T_c(B)$. 

Since the primary objective is to explore the effects of the magnetic field on quarkonium melting—a process involving the dissociation of quark-antiquark bound states and intrinsically linked to the deconfinement sector—we now proceed with the analysis by considering the deconfined black hole phase.

We note that the thermodynamical discussion here is quite similar to the one done in Sec.~\ref{fimuspec}. Therefore, in a way, $\mu$ and $B$ produce somewhat similar effects in these models: EMD and EBID. 

\newpage

\section{Equations of motion of vector mesons in a magnetic field}\label{sec:mesonfield}

In this Section, we derive the bulk equations of motion for vector meson (gauge field) fluctuations in the magnetized EBID background. 

\subsection{Geometrical quantities}\label{sec:geom_quant}

To compute spectral functions we need the equations of motion for the bulk gauge field fluctuations in the EBID
background. Starting from the Born--Infeld sector written in determinant form in Sec.~\ref{modEBID}, we decompose the
field strength into background plus fluctuations,
\begin{equation}
F_{MN}=\bar F_{MN}+\tilde F_{MN}\,,
\end{equation}
and introduce the background-dependent matrix
\begin{equation}
\mathcal{G}_{MN}\equiv g_{MN}+\frac{\bar F_{MN}}{b}\,,
\end{equation}
so that \(g_{MN}+\frac{F_{MN}}{b}=\mathcal{G}_{MN}+\frac{\tilde F_{MN}}{b}\). Expanding the square root of the
determinant to quadratic order in $\tilde F$ gives 
\begin{equation}
    \begin{split}
        \sqrt{-\det\!\left(\mathcal{G}+\frac{\tilde F}{b}\right)}
\simeq
\sqrt{-\det(\mathcal{G})}\Bigg[
1+\frac{1}{2b}\mathrm{Tr}\,(\mathcal{G}^{-1}\tilde F)
+\frac{1}{8b^2}\big\{\mathrm{Tr}\,(\mathcal{G}^{-1}\tilde F)\big\}^2\\
-\frac{1}{4b^2}\mathrm{Tr}\,\big\{(\mathcal{G}^{-1}\tilde F)^2\big\}
+\cdots
\Bigg]\,.
\label{BIlagrangianexpansion}
    \end{split}
\end{equation}

For the spectral functions, we work at vanishing spatial momentum, so the fluctuations do not carry $(x_2,x_3)$
dependence and in particular $\tilde F_{23}=0$, which makes the linear trace term vanish in the fluctuation sector
considered below.

We now evaluate the geometrical objects entering \eqref{BIlagrangianexpansion} for the magnetic background
\eqref{ansatze12}, where the only nonzero background field-strength component is
\begin{equation}
\bar F_{23}=B\,,\qquad \bar F_{32}=-B\,,
\label{fieldtensoransatze}
\end{equation}
so $\mathcal{G}_{MN}$ differs from $g_{MN}$ only in the $(2,3)$ block. In the coordinate ordering
\((t,z,x_1,x_2,x_3)\) one has
\begin{equation}
\mathcal{G}_{MN}=
\begin{pmatrix}
g_{tt} & 0      & 0      & 0        & 0\\
0      & g_{zz} & 0      & 0        & 0\\
0      & 0      & g_{11} & 0        & 0\\
0      & 0      & 0      & g_{22}   & \frac{B}{b}\\
0      & 0      & 0      & -\frac{B}{b} & g_{33}
\end{pmatrix}\,.
\label{metrictensor}
\end{equation}

Since only the $(2,3)$ sub-block is nondiagonal, the determinant factorizes into the product of the diagonal entries
times the determinant of that $2\times2$ block:
\begin{align}
\det(\mathcal{G})
&=g_{tt}\,g_{zz}\,g_{11}\,
\det\!\begin{pmatrix} g_{22} & \frac{B}{b}\\[2pt] -\frac{B}{b} & g_{33}\end{pmatrix}
=g_{tt}\,g_{zz}\,g_{11}\left(g_{22}g_{33}+\frac{B^2}{b^2}\right)\,,
\label{metricdeterminant}
\end{align}
It is convenient to define
\begin{equation}
X \;\equiv\; g_{22}g_{33}+\frac{B^2}{b^2}\,,
\end{equation}
The inverse of the $(2,3)$ block is
\[
\begin{pmatrix} g_{22} & \frac{B}{b}\\[2pt] -\frac{B}{b} & g_{33}\end{pmatrix}^{-1}
=\frac{1}{X}\begin{pmatrix} g_{33} & -\frac{B}{b}\\[2pt] \frac{B}{b} & g_{22}\end{pmatrix}\,,
\]
so the full inverse matrix reads
\begin{equation}
\mathcal{G}^{MN}=
\begin{pmatrix}
\frac{1}{g_{tt}} & 0 & 0 & 0 & 0\\
0 & \frac{1}{g_{zz}} & 0 & 0 & 0\\
0 & 0 & \frac{1}{g_{11}} & 0 & 0\\
0 & 0 & 0 & \frac{g_{33}}{X} & -\frac{B}{bX}\\
0 & 0 & 0 & \frac{B}{bX} & \frac{g_{22}}{X}
\end{pmatrix}\,.
\label{metricinverse}
\end{equation}

Finally, we decompose $\mathcal{G}^{MN}$ into its symmetric and antisymmetric parts,
\(\mathcal{G}^{MN}=G^{MN}+S^{MN}\) with
\(G^{MN}=\tfrac12(\mathcal{G}^{MN}+\mathcal{G}^{NM})\) and
\(S^{MN}=\tfrac12(\mathcal{G}^{MN}-\mathcal{G}^{NM})\). This yields
\begin{equation}
G^{MN}=
\begin{pmatrix}
\frac{1}{g_{tt}} & 0 & 0 & 0 & 0\\
0 & \frac{1}{g_{zz}} & 0 & 0 & 0\\
0 & 0 & \frac{1}{g_{11}} & 0 & 0\\
0 & 0 & 0 & \frac{g_{33}}{X} & 0\\
0 & 0 & 0 & 0 & \frac{g_{22}}{X}
\end{pmatrix}\,,
\qquad
S^{MN}=
\begin{pmatrix}
0 & 0 & 0 & 0 & 0\\
0 & 0 & 0 & 0 & 0\\
0 & 0 & 0 & 0 & 0\\
0 & 0 & 0 & 0 & -\frac{B}{bX}\\
0 & 0 & 0 & \frac{B}{bX} & 0
\end{pmatrix}\,.
\label{symantisym}
\end{equation}
These expressions are the basic ingredients needed to write the quadratic fluctuation action and, from it, the
equations of motion for the gauge field perturbations in the magnetized EBID background.

\subsection{Fluctuation equations at $q=0$}\label{sec:BI_fluct_EOM}

To compute the spectral functions, we need the linearized equations of motion for the vector fluctuations in the
magnetized EBID background. As in the EMD part of this dissertation, we work with the vector combination
\(V_M\equiv(A_{L\,M}+A_{R\,M})/2\) and impose the radial gauge \(V_z=0\) \cite{Dudal:2015kza}. We take plane-wave
fluctuations
\begin{equation}
V_i(t,x,z)=e^{-i\omega t+i q\cdot x}\,V_i(z)\,,\qquad i=1,2,3\,,
\end{equation}
and, to simplify the numerics and to match the setup used previously for spectral functions, we restrict to
vanishing spatial momentum on the boundary, \(\ q= 0\)\,.

The starting point is the bulk gauge field equation derived in Sec.~\ref{sec:EBID_solutions}, Eq.~(\ref{maxwelleom12}), which we now write
with the background\,+\,fluctuation split \(F_{MN}=\bar F_{MN}+\tilde F_{MN}\),
\begin{equation}\label{fluctuationeom}
\partial_M\!\left[
\sqrt{-g}\,\frac{f(\phi)}{\sqrt{1+\frac{(\bar F+\tilde F)^2}{2b^2}}}\,
\bigl(\bar F^{MN}+\tilde F^{MN}\bigr)
\right]=0\,,
\qquad
\tilde F_{MN}\equiv\partial_M V_N-\partial_N V_M\,.
\end{equation}
For the background of Sec.~\ref{sec:EBID_solutions} one has \(\bar F_{23}=B\) and all other \(\bar F_{MN}=0\)\,. At \(\vec q=0\) the
fluctuations satisfy \(\partial_{2,3}V_M=0\), hence \(\tilde F_{23}=0\)\,. As a consequence the mixed contraction
\(\bar F_{MN}\tilde F^{MN}\propto \bar F_{23}\tilde F^{23}\) vanishes, so the Born--Infeld square-root factor
\(\sqrt{1+F^2/(2b^2)}\) does not receive a linear correction in this fluctuation sector. Therefore, keeping only
terms linear in \(\tilde F\) in \eqref{fluctuationeom} yields a closed linear equation for the perturbations.

An equivalent and convenient way to organize the linearized dynamics is to use the quadratic expansion of the
determinant form of the Born--Infeld action derived in Sec.~\ref{sec:geom_quant}, Eq.~(\ref{BIlagrangianexpansion}). Introducing
\(\mathcal{G}_{MN}\equiv g_{MN}+\bar F_{MN}/b\) and decomposing \(\mathcal{G}^{MN}=G^{MN}+S^{MN}\) as in
Eq.~(\ref{symantisym}), the linear term \(\mathrm{Tr}\,(\mathcal{G}^{-1}\tilde F)\) vanishes here: the contraction of the
symmetric part with the antisymmetric \(\tilde F\) is identically zero, while the antisymmetric part has support
only in the \((2,3)\) block and \(\tilde F_{23}=0\) at \( q=0\)\,. The remaining quadratic term produces an
effective Maxwell-type action in the open-string geometry,
\begin{equation}
S^{(2)}_{\mathrm{BI}}
=
-\frac{1}{16\pi G_5}\int d^5x\,
\frac{f(z)}{4}\,\sqrt{-\mathcal{G}}\;
G^{MP}G^{NQ}\,\tilde F_{MN}\tilde F_{PQ}\,,
\end{equation}
whose Euler--Lagrange equation is the ``modified'' fluctuation equation used in \cite{Dudal:2014jfa},
\begin{equation}\label{fluctuationEOM}
\partial_M\!\left(\sqrt{-\mathcal{G}}\,f(z)\,G^{MP}G^{NQ}\,\tilde F_{PQ}\right)=0\,.
\end{equation}
Equations \eqref{fluctuationeom} and \eqref{fluctuationEOM} are equivalent for the present background and the
\( q=0\) fluctuation sector, but \eqref{fluctuationEOM} makes the anisotropic structure transparent through the
effective metric \(G^{MN}\).

We now reduce \eqref{fluctuationEOM} to an ordinary differential equation. In radial gauge \(V_z=0\) and at
\( q=0\), the only nonzero fluctuation components are
\[
\tilde F_{zi}=\partial_z V_i,\qquad
\tilde F_{ti}=-\partial_t V_i=i\omega V_i,
\]
and all background quantities depend only on \(z\)\,. Taking \(N=i\) in \eqref{fluctuationEOM} gives
\begin{equation}
\partial_z\!\left(\sqrt{-\mathcal{G}}\,f(z)\,G^{zz}G^{ii}\,\partial_z V_i\right)
+\sqrt{-\mathcal{G}}\,f(z)\,G^{tt}G^{ii}\,\omega^2 V_i = 0\,,
\end{equation}
Dividing by \(\sqrt{-\mathcal{G}}\,f(z)\,G^{zz}G^{ii}\) yields
\begin{equation}
\boxed{
\partial_z^2 V_i
+\partial_z\!\left[\ln\!\Big(\sqrt{-\mathcal{G}}\, f(z)\, G^{zz} G^{ii}\Big)\right]\partial_z V_i
-\frac{G^{tt}}{G^{zz}}\,\omega^2 V_i = 0\,.}
\label{parallelfluceq}
\end{equation}
where \(i=1\) corresponds to polarization parallel to the magnetic field (\(x_1\)-direction), while \(i=2\) (or \(3\))
corresponds to transverse polarization. The anisotropy enters through \(G^{11}\neq G^{22}=G^{33}\), as in Sec.~\ref{sec:geom_quant}, and Eq.~(\ref{parallelfluceq}) reduces to the isotropic EMD fluctuation equation when \(B\to0\)\,.

\newpage
\section{Spectral functions at finite magnetic field}\label{sec:spectral_finiteB}

In the previous chapters, we developed a unified holographic framework to access in-medium properties of heavy mesons
through real-time correlators. In the EMD model (Secs.~\ref{thermo}--\ref{metingmu}), we studied how temperature
and baryon density modify the vector spectral function and lead to the progressive broadening and disappearance of
quasi-bound peaks. In Secs.~\ref{modEBID}--\ref{sec:mesonfield} we extended the construction to a magnetized plasma by introducing a constant
background magnetic field through a self-consistent EBID background, establishing its thermodynamics (Sec.~\ref{Bthermo}) and
deriving the linearized fluctuation equations in the anisotropic geometry (Sec.~\ref{sec:mesonfield}). We now use these results to
compute the thermodynamically relevant spectral functions in the deconfined phase at finite $B$.

Heavy quarkonia, such as the $J/\psi$, have long been regarded as sensitive probes of deconfinement because color
screening and in-medium scattering increase the probability of dissociation in the quark--gluon plasma
\cite{Matsui:1986dk}. In phenomenology, it is common to separate suppression mechanisms into hot nuclear matter (HNM)
effects---directly tied to the properties of the QGP---and cold nuclear matter (CNM) effects associated with the
initial nuclear environment \cite{Zhao2023Jun}. The equilibrium holographic setup employed here is designed to model
HNM physics: it captures strong-coupling modifications of real-time correlators in a thermal medium (and, in the EBID
case, in an external magnetic field), while CNM effects such as nuclear PDFs or absorption are outside its scope.
From first principles, lattice studies also indicate that quarkonium need not melt exactly at $T_c$ and may survive
into the deconfined phase up to temperatures of order $1$--$2\,T_c$ in some channels \cite{Asakawa:2003re}, motivating
a detailed analysis of how spectral peaks broaden and disappear as medium parameters change.

A magnetic field introduces an additional qualitative feature: it breaks spatial $SO(3)$ invariance and therefore
splits the vector channel into polarizations \emph{parallel} and \emph{transverse} to $B$. In our EBID model,
this anisotropy is encoded in the open-string background \(\mathcal G_{MN}\) and the effective metric \(G^{MN}\) that
governs the gauge fluctuations (Sec.~\ref{sec:mesonfield}). Consequently, the spectral function becomes polarization dependent,
\(\rho_{\parallel}(\omega)\neq\rho_{\perp}(\omega)\), and the melting pattern can differ along and across the magnetic
field. In what follows, we work in the deconfined black hole phase (for $T>T_c(B)$ determined in Sec.~\ref{Bthermo}), reduce the
general fluctuation equation \eqref{parallelfluceq} to the explicit ODEs for the two polarizations, and compute the
retarded correlators using the same numerical strategy developed previously in Sec.~\ref{procespec}. This allows us to quantify how
$B$ modifies peak positions and widths, and hence how the magnetic field affects the dissociation of heavy quarkonium
in a strongly coupled plasma.

\subsection{Parallel and perpendicular equations of motion}

In order to compute the spectral functions numerically we adhere to the prescription already shown here in this work in Secs.~\ref{procespec}, \ref{meltingmu0} and \ref{metricmu}.

From (\ref{parallelfluceq}), we can see that for the case where the fluctuation is parallel to the background magnetic field, the relevant equation of motion is given by 
\begin{equation}\label{paraeom}
\boxed{\begin{split}
        \partial_z^2 V_1 &+ \partial_z \left[\ln\left(\frac{e^{-A(z)} f(z) g(z)}{L z} \sqrt{e^{2 B^2 z^2+ 4 A(z)} L^4 + 4 B^2 \pi^2 z^4 \alpha{'}^2} \right)\right]\partial_z V_1  \\&+ \frac{\omega^2}{g(z)^2}  V_1 = 0\,, 
    \end{split}}
\end{equation}

By (\ref{parallelfluceq}), if, however, the fluctuation is perpendicular to the magnetic field, we have the expression
\begin{equation}\label{perpeneom}
\boxed{\begin{split}
        \partial_z^2 V_2 &+ \partial_z \left[\ln\left(\frac{L^3 f(z) g(z) e^{3 A(z)+B^2 z^2} \sqrt{L^4 e^{4 A(z)+2 B^2 z^2}+4 \pi ^2 \alpha{'} ^2 B^2 z^4}}{L^4 z e^{4
   A(z)+2 B^2 z^2}+B^2 z}  \right)\right]\partial_z V_2 \\&+ \frac{\omega^2}{g(z)^2}  V_2 = 0 \,.
    \end{split}}
\end{equation}

All the analytical and numerical analysis that will follow is done for these two equations of motion. As we follow the exact same procedure as detailed beforehand in Sec.~\ref{procespec}, we can simply show the relevant quantities for the procedure and the numerical results.

\subsection{Relevant quantities for the numerical scheme}

Up to the second order, see equivalent in the EMD model in (\ref{tortEMD}), the approximation to the tortoise coordinate $r_*$, defined by the relation $\partial_{r_*} = -g(z) \partial_z$, for the EBID model gives us
\begin{equation}
\boxed{\begin{split}
        r_*(z) &\approx  \frac{z_h^2 \left(3 a-B^2\right)+e^{z_h^2 \left(B^2-3 a\right)}-1}{144 z_h^6 \left(B^2-3a\right)^2} \Bigl[z \Bigl(-4 z_h^4 \left(3 a-B^2\right) \left(z^2 \left(3 a-B^2\right)-30\right)\\&+24 z z_h^3
   \left(B^2-3 a\right)+12 z_h^6 \left(B^2-3 a\right)^2+15 z^2-90 z z_h+243 z_h^2\Bigl)\\&-144
   z_h^3 \tanh ^{-1}\left(\frac{z}{z-2 z_h}\right) \Bigl]\,.
    \end{split}}
\end{equation}

The Frobenius analysis, in the EBID case, leads one to the indicial equation
\begin{equation}
	\beta(\beta-4)=0\,,
\end{equation}
yielding $\beta=0$ or $\beta=4$. 

Therefore, as the difference between roots is, also in the EMD case (\ref{frobEMD}), an integer, we obtain the expansion solution in the near-boundary regime as 
\begin{align}\label{seriesebid}
    \psi_1(z)  &= z^4 \sum_{k=0}^{+\infty} a_k z^k \,, \\
    \psi_2(z) &= C \ln(z) \psi_1(z)+ \sum_{k=0}^{+\infty} b_k z^k \,.
\end{align}

In the EMD model, the shown coefficients are (\ref{power1}). For the considered equations of motion, in the EBID model, both parallel (\ref{paraeom}) and perpendicular (\ref{perpeneom}), have their own distinct coefficients when we plug the series solution (\ref{seriesebid}) to their respective differential equation. As an example, we find that the first coefficient of $\psi_1(z)$ can be written as 
\begin{align}
        a_1^{\text{parallel}} &= \frac{1}{12} \left(-16 B^2-\omega ^2\right) \,, \\
        a_1^{\text{perpendicular}} &= \frac{1}{12} \left(96 a^2+96 a B^2-24 a-8 B^2-\omega ^2\right)	\,.
\end{align}  

The other coefficients, of course, can be computed and also are functions of $a$, $B$ and $\omega$.

As we have stated the required quantities, we can proceed to show the numerical results to the spectral functions computed for the EBID model.
\subsection{Numerical results: Parallel case}
\begin{figure}[htb!]
	\centering
 \subfigure[$T=T_c$]{\includegraphics[width=0.45\linewidth]{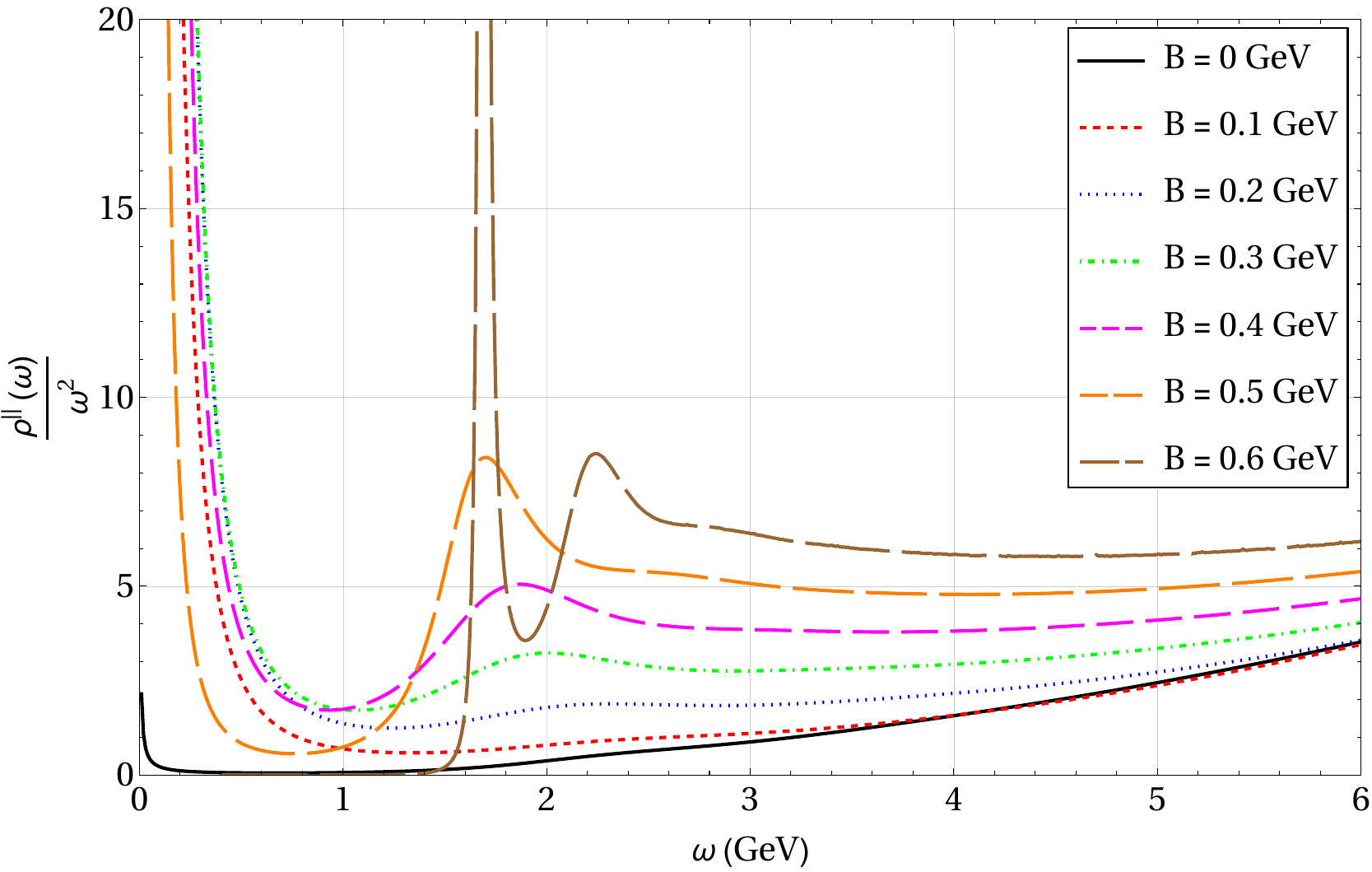}}
  \subfigure[$T=1.5T_c$]{\includegraphics[width=0.45\linewidth]{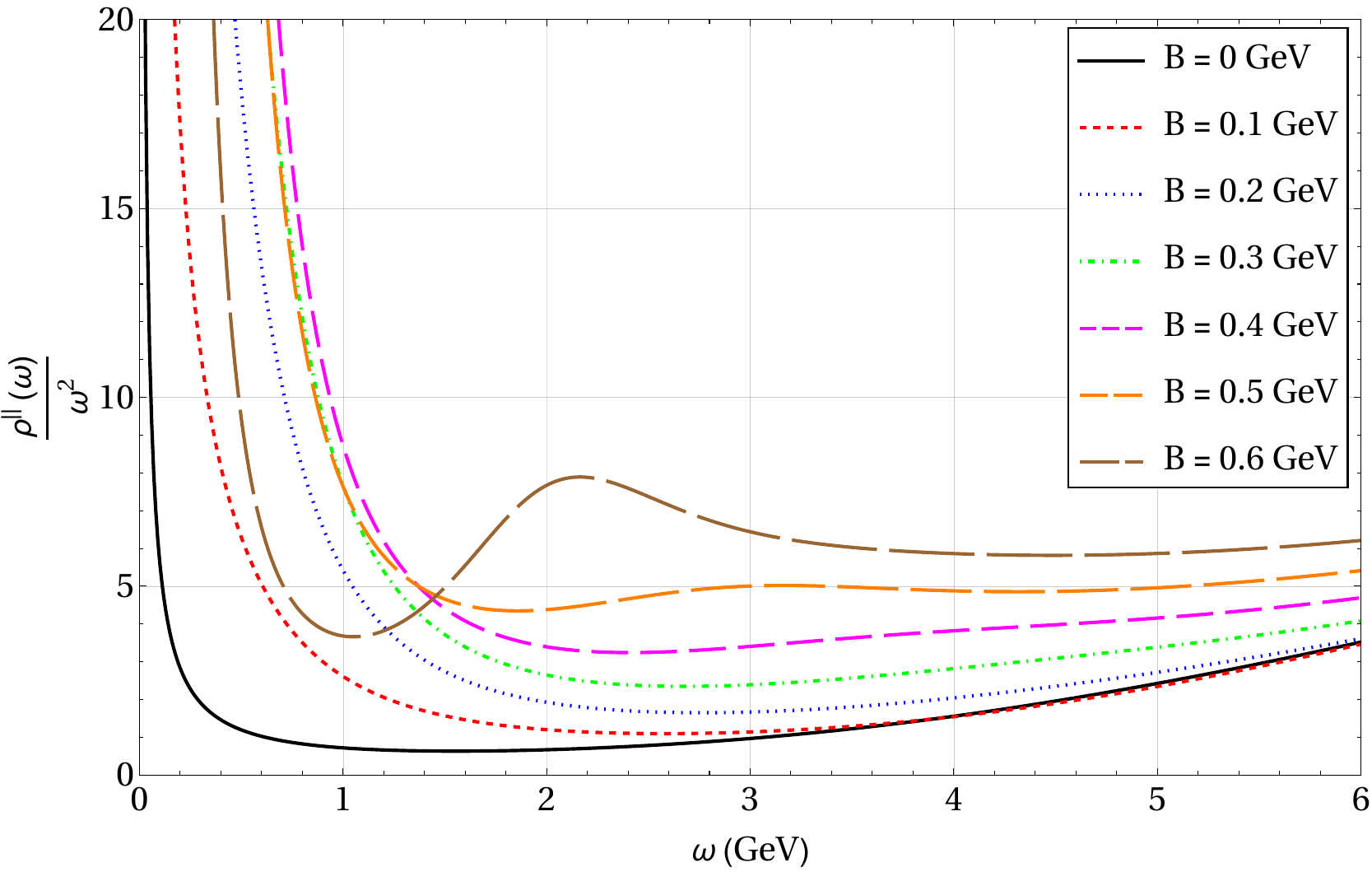}}
	\caption{\small Spectral function as a function of $\omega$ in the parallel direction for different magnetic fields. The temperature is fixed at $T=T_c$ (left) and $T=1.5~T_c$ (right). }
	\label{spf_diffB_parallel}	
\end{figure}

We first examine the case where the polarization is aligned with the background magnetic field. The parametrized spectral function as a function of $\omega$ is presented in Fig.~\ref{spf_diffB_parallel} for different background magnetic field strengths at the deconfinement temperature $T=T_c$ and at $T=1.5~T_c$.\footnote{In Fig.~\ref{spf_diffB_parallel}, $\rho/\omega^2$ (rather than $\rho$) is plotted to facilitate comparison with lattice QCD results \cite{Ding2019Feb,Petreczky2006Jan}. It is important to note that these lattice results do not account for the effects of the background magnetic field.} 

Notably, at lower $B$ values, the spectral peaks exhibit broadening, whereas at higher magnetic fields, multiple spectral peaks emerge. This behavior suggests a reduced impact of the magnetic field on quarkonium states, in line with the findings of \cite{Zhao2023Jun}. This, in turn, implies that the magnetic field suppresses quarkonium dissociation, a phenomenon indicative of magnetic catalysis. Interestingly, lattice simulations \cite{DElia:2021tfb} also indicate an enhanced string tension along the direction of the magnetic field, suggesting that traces of this anisotropic confinement could be reflected in the quarkonium spectra.
\begin{figure}[htb!]
  \centering
  \subfigure[$B=0.0~\text{GeV}$]{\includegraphics[width=0.45\linewidth]{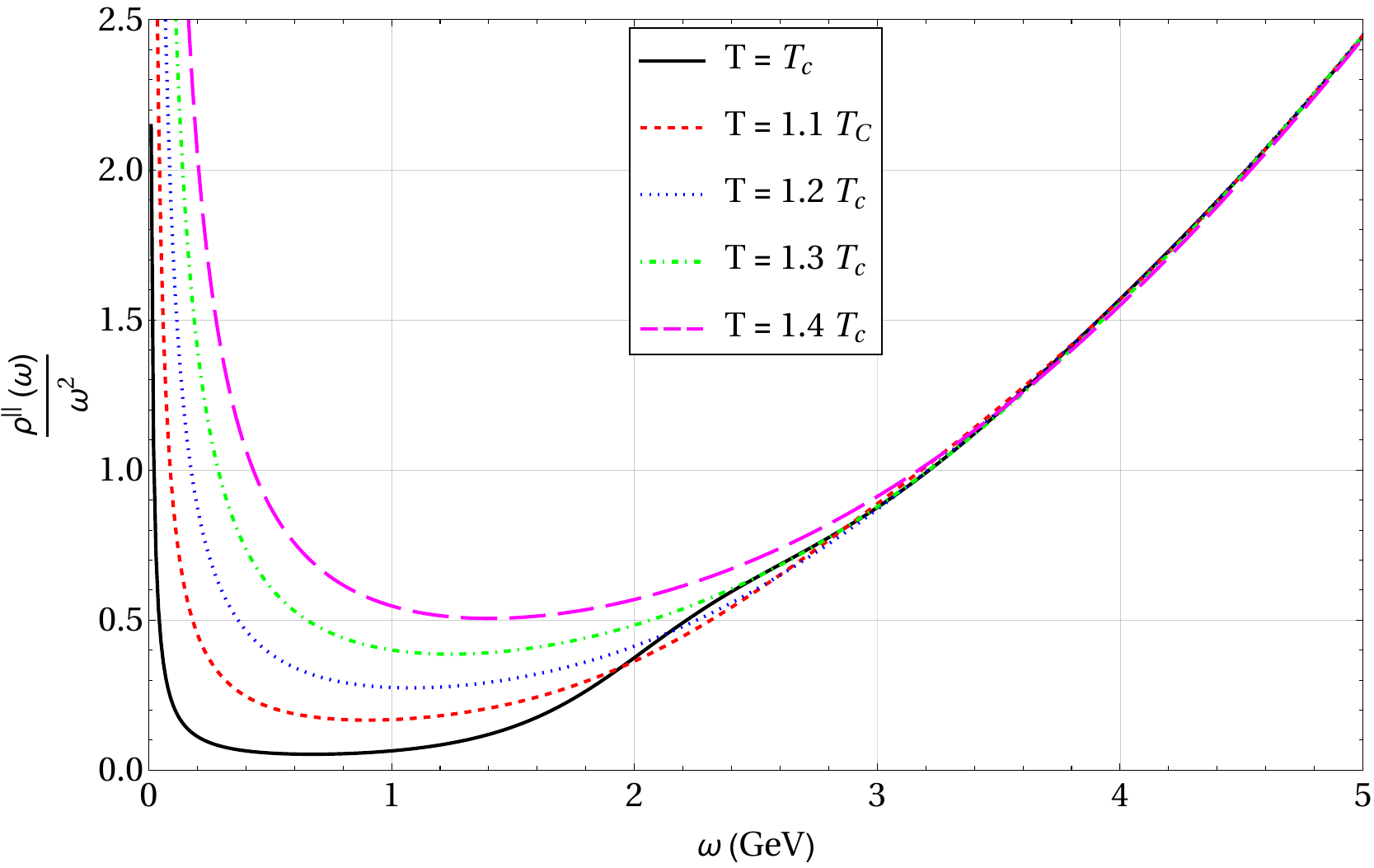}}
  \label{spf_fixedB_parallelB0}
  \subfigure[$B=0.2~\text{GeV}$]{\includegraphics[width=0.45\linewidth]{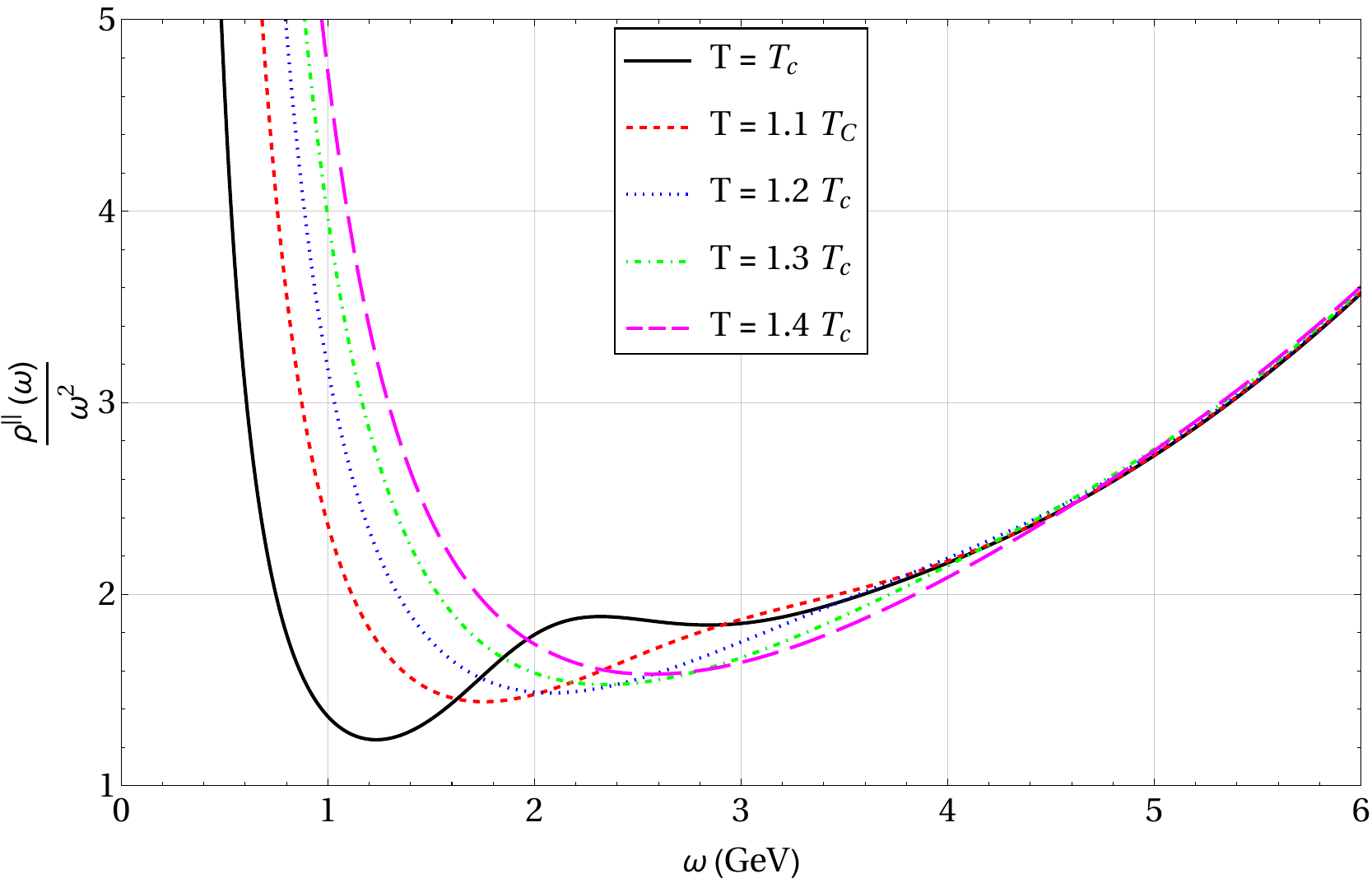}} 
  \label{spf_fixedB_parallelBPt2} \\
  \subfigure[$B=0.4~\text{GeV}$]{\includegraphics[width=0.45\linewidth]{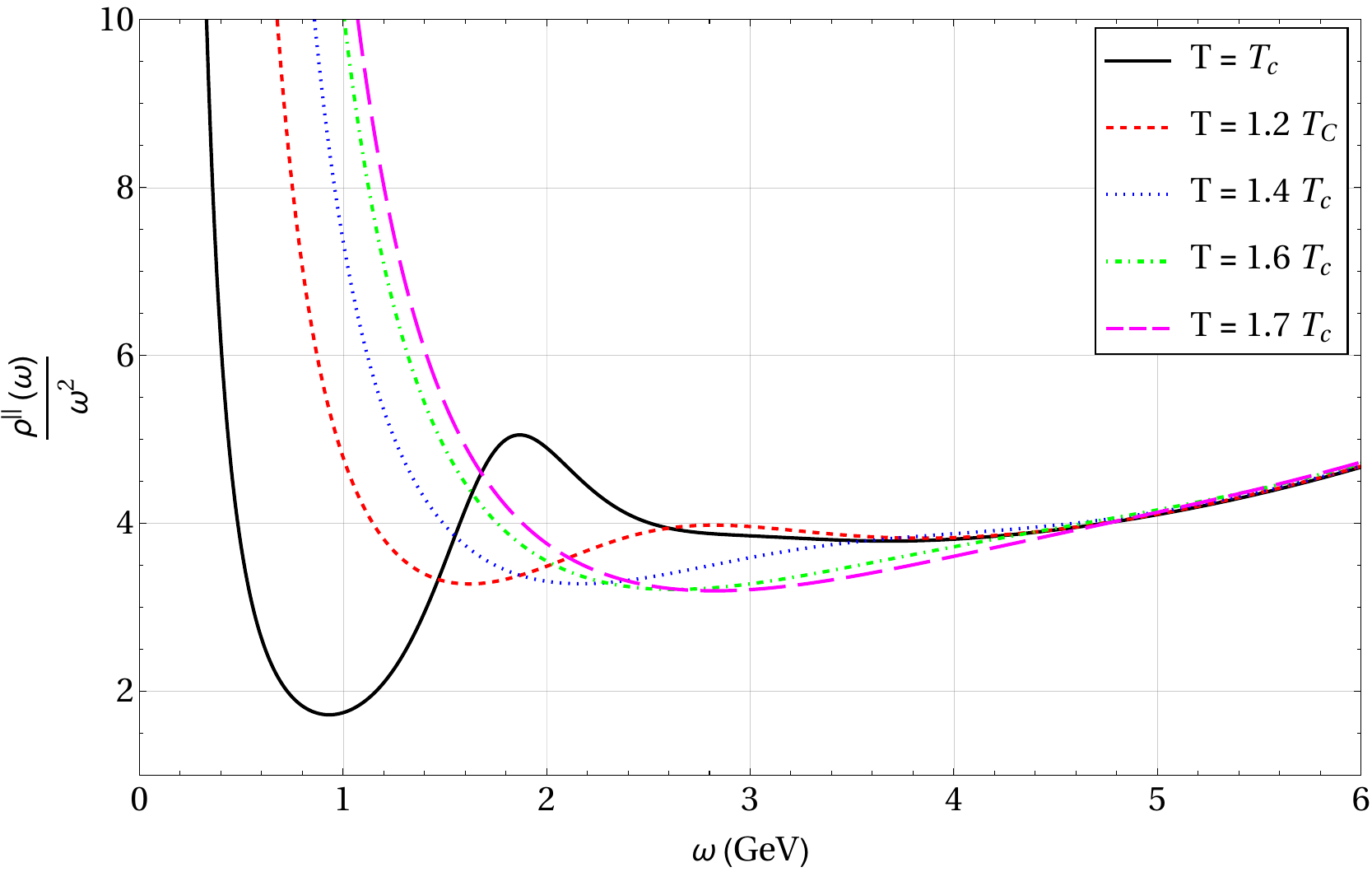}}
  \label{spf_fixedB_parallelBPt4}
  \subfigure[$B=0.6~\text{GeV}$]{\includegraphics[width=0.45\linewidth]{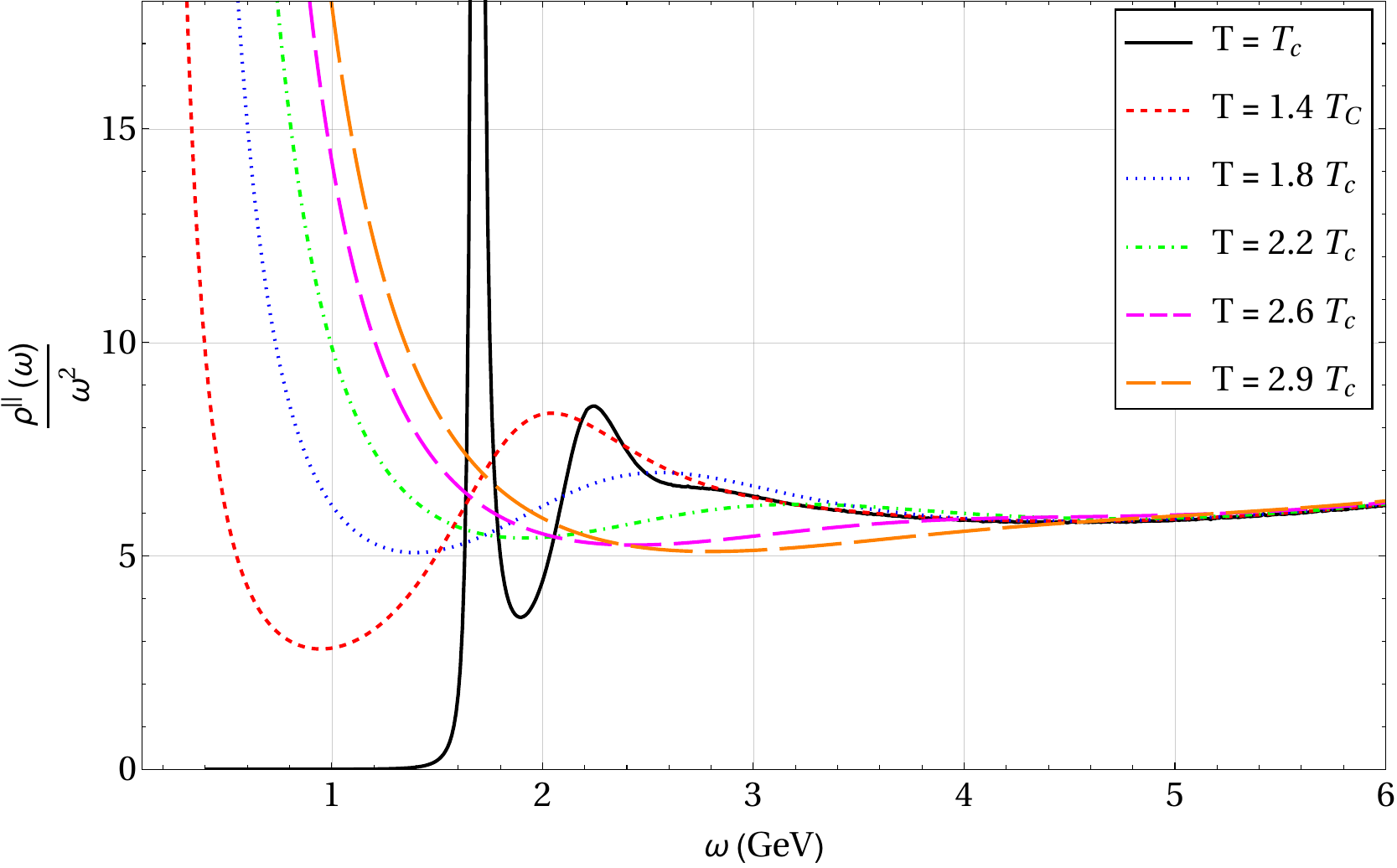}}
  \label{spf_fixedB_parallelBPt6}
  \caption{\small Variation of the spectral function with $\omega$ for varying temperatures with fixed $B$ in the parallel direction.}
  \label{spf_fixedB_parallel}
\end{figure}

It is important to note that $T_c$ depends non-trivially on $B$. At $B=0$, the critical temperature is $268~\text{MeV}$, whereas for $B=0.6$, it decreases significantly to $119~\text{MeV}$. This indicates that higher magnetic field values lead to a substantial reduction in the critical temperature. 

As a result, the broadening of the spectral peaks at large $B$ values may be associated with the lower temperatures at which bound quark states can still persist. To further illustrate this, Fig.~\ref{spf_fixedB_parallel} presents the variation of the spectral function for different temperatures and magnetic field strengths. We observe that for a fixed $B$, spectral peaks appear at low temperatures, gradually widening as the temperature increases, and eventually vanishing altogether. 

This effect is particularly evident in Fig.~\ref{spf_fixedB_parallel}(c), where for $B=0.4$, peaks are visible up to approximately $T\sim 1.28~T_c=275~\text{MeV}$, suggesting the survival of heavy quark bound states up to this temperature, while no peaks are observed beyond this point. Similarly, for $B=0.5$, peaks persist even up to $T\sim 1.6~T_c=286~\text{MeV}$. These findings are consistent with lattice QCD results, which indicate that quarkonia can survive well beyond the deconfinement transition \cite{Asakawa:2003re}.

\begin{figure}[htb!]
	\centering
	\includegraphics[height=6cm,width=9cm]{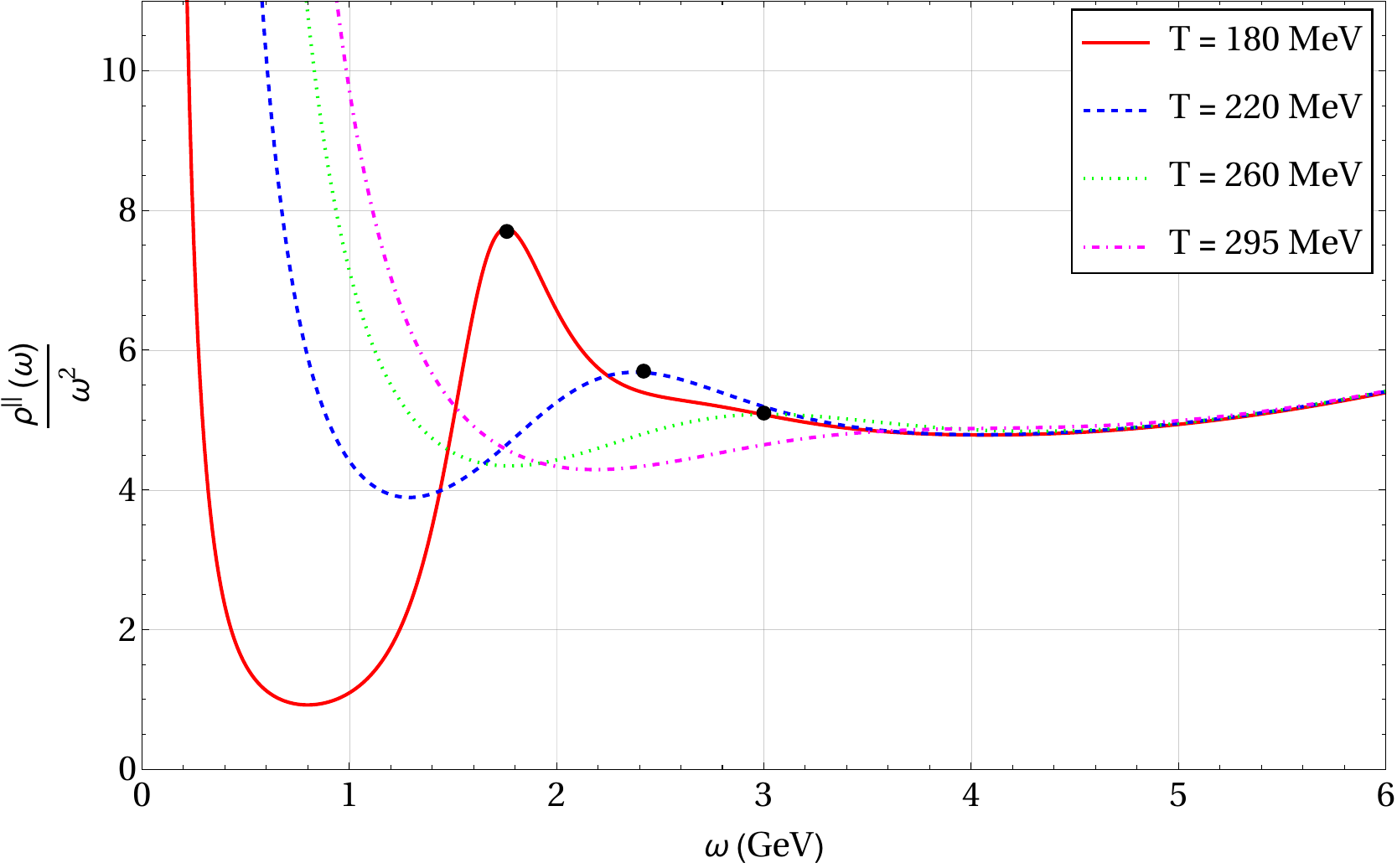}
	\caption{\small The vanishing of local maxima as an increase in temperature. }
	\label{localM}	
\end{figure}

To quantify the disappearance of spectral peaks, we compute the derivative of the spectral function with respect to $\omega$ and identify the local maxima. The temperature at which the local maxima vanish determines the melting temperature, as we can see in Fig.~\ref{localM}.

For $B=0.5~\text{GeV}$ and $b=0.044~\text{GeV}^2$, Fig.~\ref{localM} illustrates the disappearance of the local maxima, marked by a black dot, as the temperature increases to approximately $T\sim 295~\text{MeV}$. This temperature is then interpreted as the melting temperature for this state.

\begin{figure}[htb!]
	\centering
	\includegraphics[height=6cm,width=9cm]{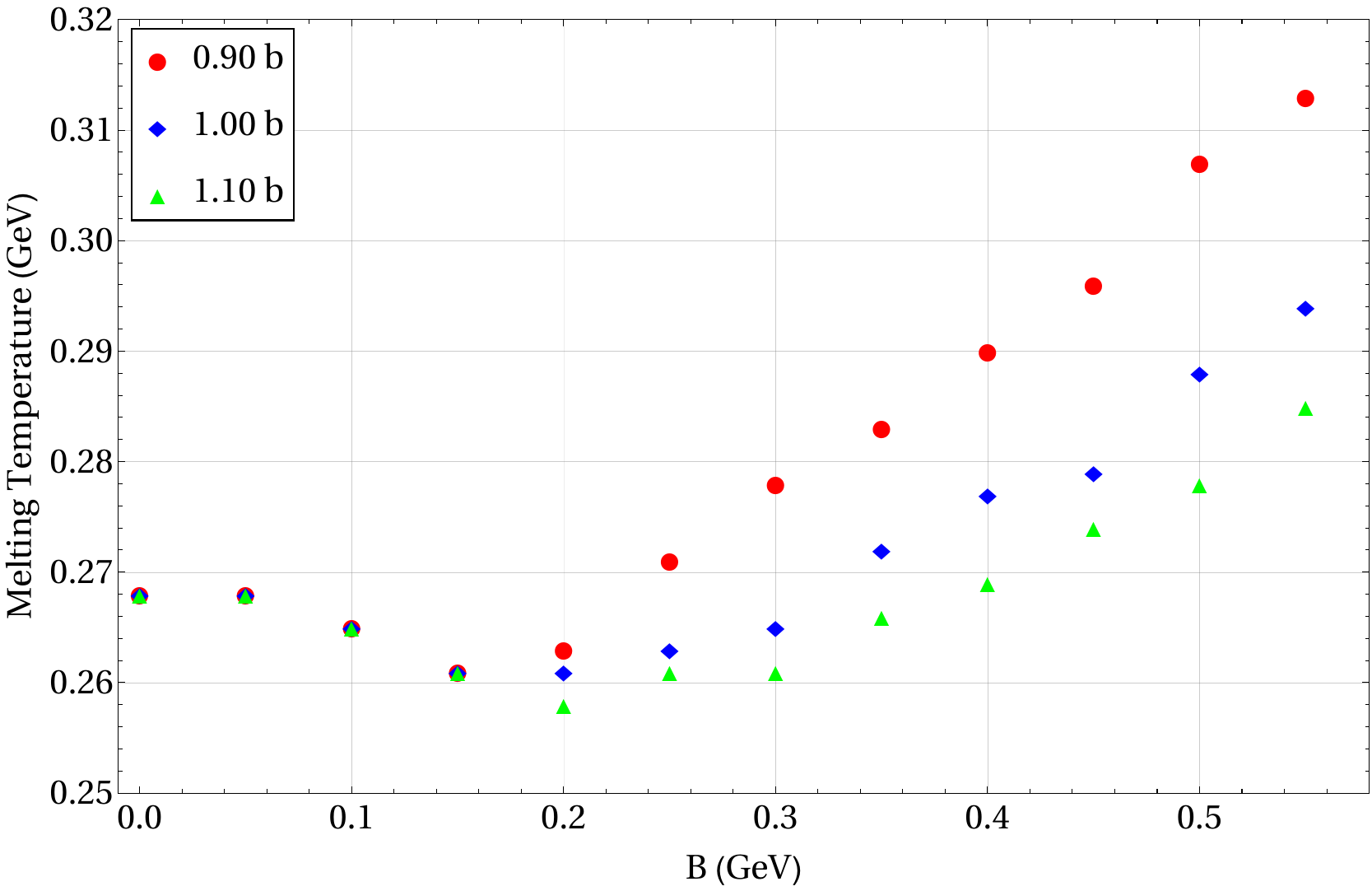}
	\caption{\small The variation of the melting temperature with magnetic field. }
	\label{meltingtempvsB}	
\end{figure}

We also examine how the melting temperature varies with the magnetic field, as shown in Fig.~\ref{meltingtempvsB}. The analysis reveals that the effect of the magnetic field on the melting temperature is non-trivial and intricate. Specifically, the melting temperature initially decreases with increasing magnetic field before subsequently rising.

In Fig.~\ref{meltingtempvsB}, we also present results incorporating a $10\%$ variation in $b$, corresponding to a $10\%$ variation in the string tension. This accounts for the uncertainty reported in the lattice QCD determination of the string tension, cf.~the previously mentioned $\sqrt{\sigma} =0.43\pm0.02~\text{GeV}$ from \cite{Bali:2000gf}. The findings here indicate that the transition from inverse magnetic catalysis to magnetic catalysis in the melting temperature as the magnetic field increases is a robust feature, holding for all values within $b=0.044\pm10\%$. 

At present, no lattice QCD results are available for direct comparison with this intriguing behavior. However, the observed non-monotonic dependence of the melting temperature may be linked to the non-monotonicity reported in the parallel/perpendicular lattice string tensions at higher magnetic field values, as seen in \cite[Fig.~8]{DElia:2021tfb}. Since stronger (weaker) confinement between heavy quarks is expected to correspond to a higher (lower) melting temperature of their bound states, such a connection is plausible. Nevertheless, it is important to note that the polarization directions of charmonia are not necessarily correlated with the orientations of their constituent quark-antiquark pairs relative to the magnetic field. Thus, this reasoning should be considered a heuristic rather than a definitive explanation.

Before concluding this section, we clarify the terminology used in this chapter. In the literature, ``inverse magnetic
catalysis'' (IMC) and ``magnetic catalysis'' (MC) often refer to the \emph{thermodynamic} deconfinement line
$T_c\,(B)$. In our EBID thermodynamic analysis (Sec.~\ref{sec:phase_structure_B}), the deconfinement transition temperature decreases with $B$
over the range considered, i.e.\ it exhibits IMC behavior. When we refer here to an ``IMC $\rightarrow$ MC'' trend, we
do \emph{not} mean that $T_c\,(B)$ itself turns around. Rather, the statement concerns the \emph{real-time spectral
observables} and the melting criterion extracted from them. Concretely, we define the melting temperature from the
spectral-function curves as the temperature at which the last local maximum (the last identifiable resonance-like
peak) disappears, as illustrated in Fig.~\ref{localM}. The resulting melting temperature $T_m(B)$ is summarized in Fig.~\ref{meltingtempvsB}: as
the magnetic field increases, $T_m(B)$ first decreases (IMC-like behavior) up to an intermediate field and then
increases again (MC-like behavior). Fig.~\ref{meltingtempvsB} also shows that this non-monotonic pattern is robust under a
$\pm 10\%$ variation of the Born--Infeld parameter $b$ (equivalently, of the fitted string tension), indicating that
the observed ``IMC $\rightarrow$ MC'' language in this chapter refers to the magnetic-field dependence of the
\emph{spectral-function peak evolution} (and the corresponding $T_m$ extracted from it), not to a reversal of the
thermodynamic deconfinement transition.

\subsection{Numerical results: Perpendicular case}

\begin{figure}[htb!]
	\centering
 \subfigure[$T=T_c$]{\includegraphics[width=0.45\linewidth]{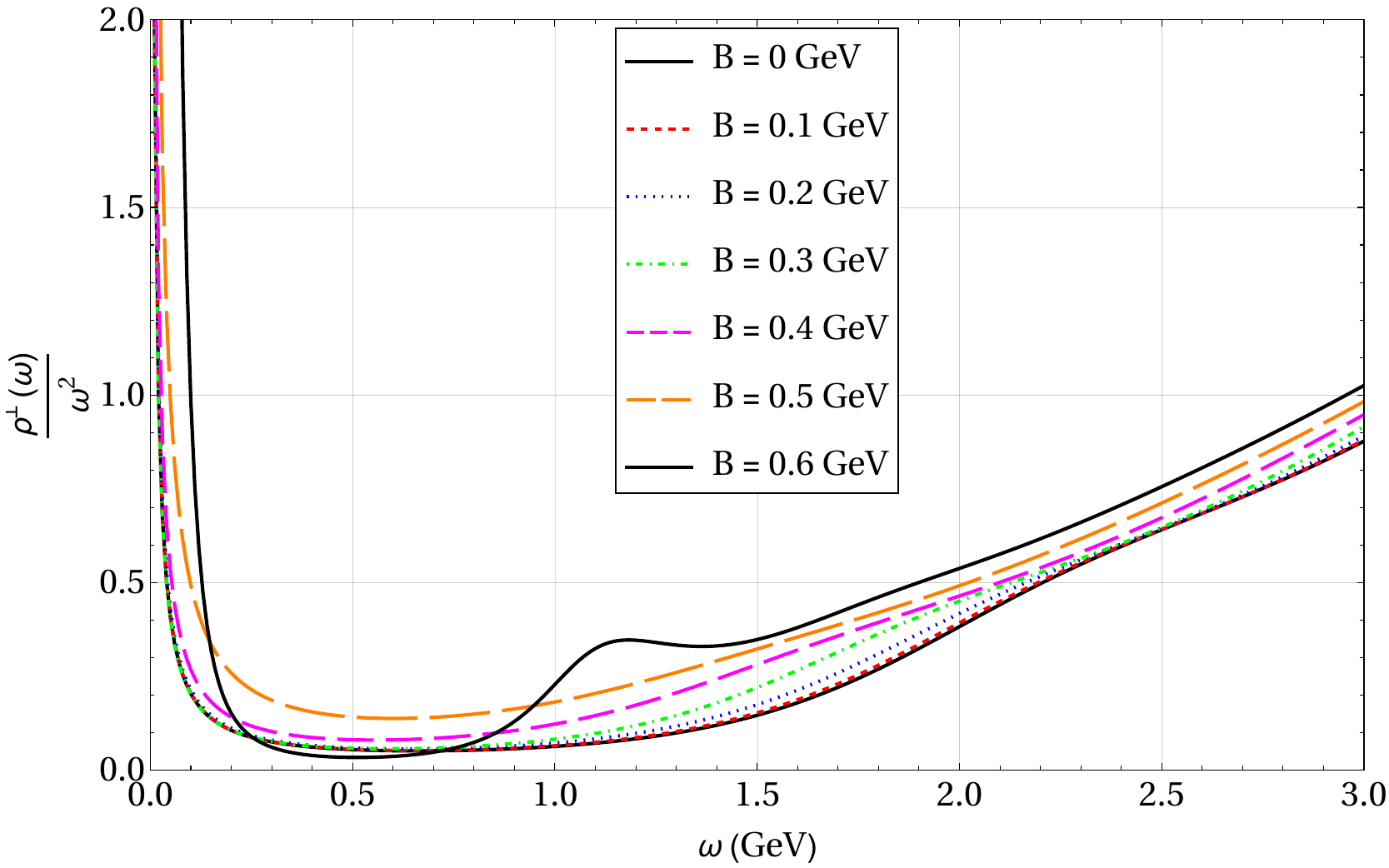}}
  \subfigure[$T=1.5~T_c$]{\includegraphics[width=0.45\linewidth]{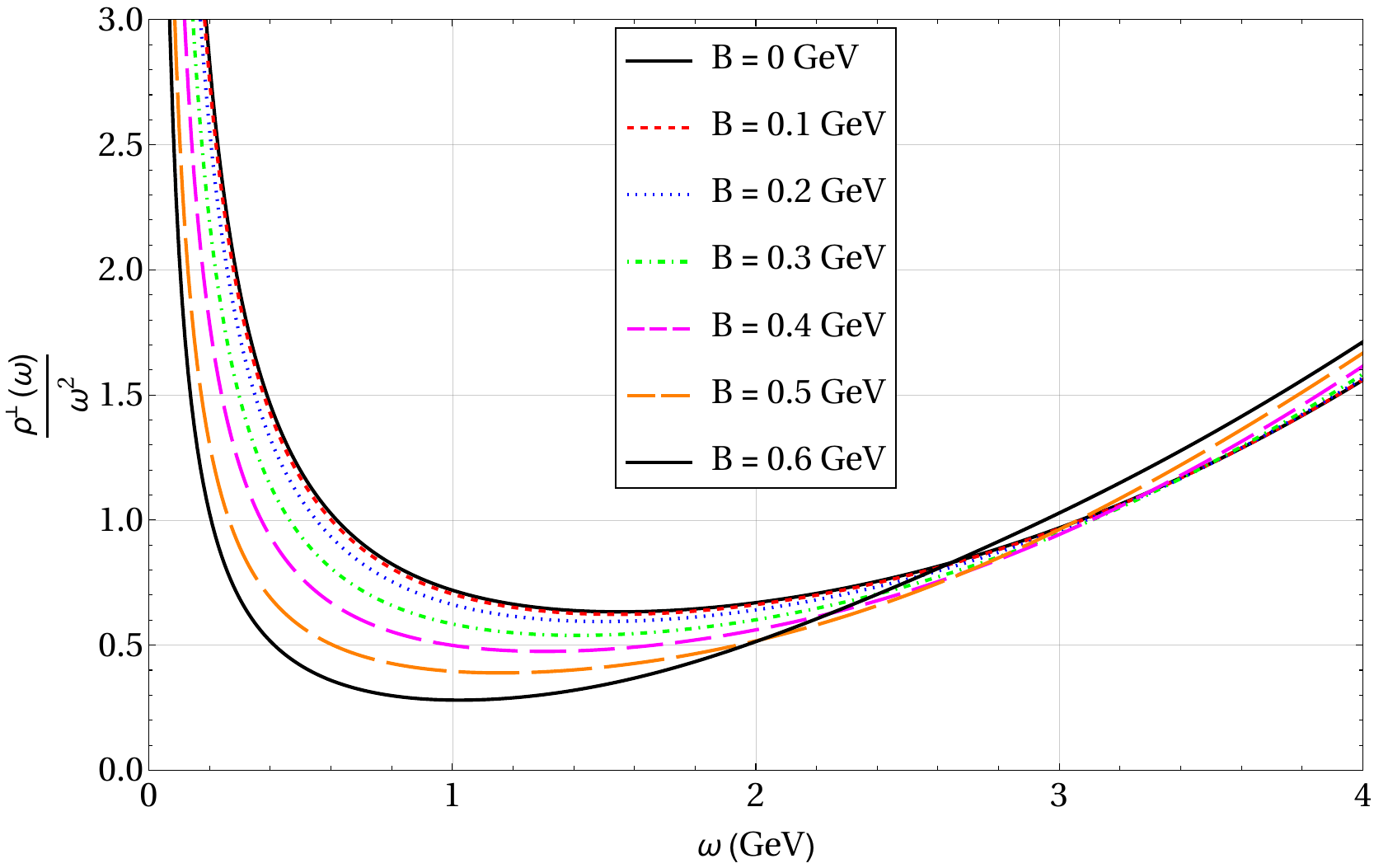}}
	\caption{\small Spectral function as a function of $\omega$ in the perpendicular direction for different magnetic fields. The temperature is fixed at $T=T_c$ (left) and $T=1.5~T_c$ (right). }
	\label{spf_diffB_perp}	
\end{figure}

\begin{figure}[htb!]
  \centering
  \subfigure[$B=0.0~\text{GeV}$]{\includegraphics[width=0.45\linewidth]{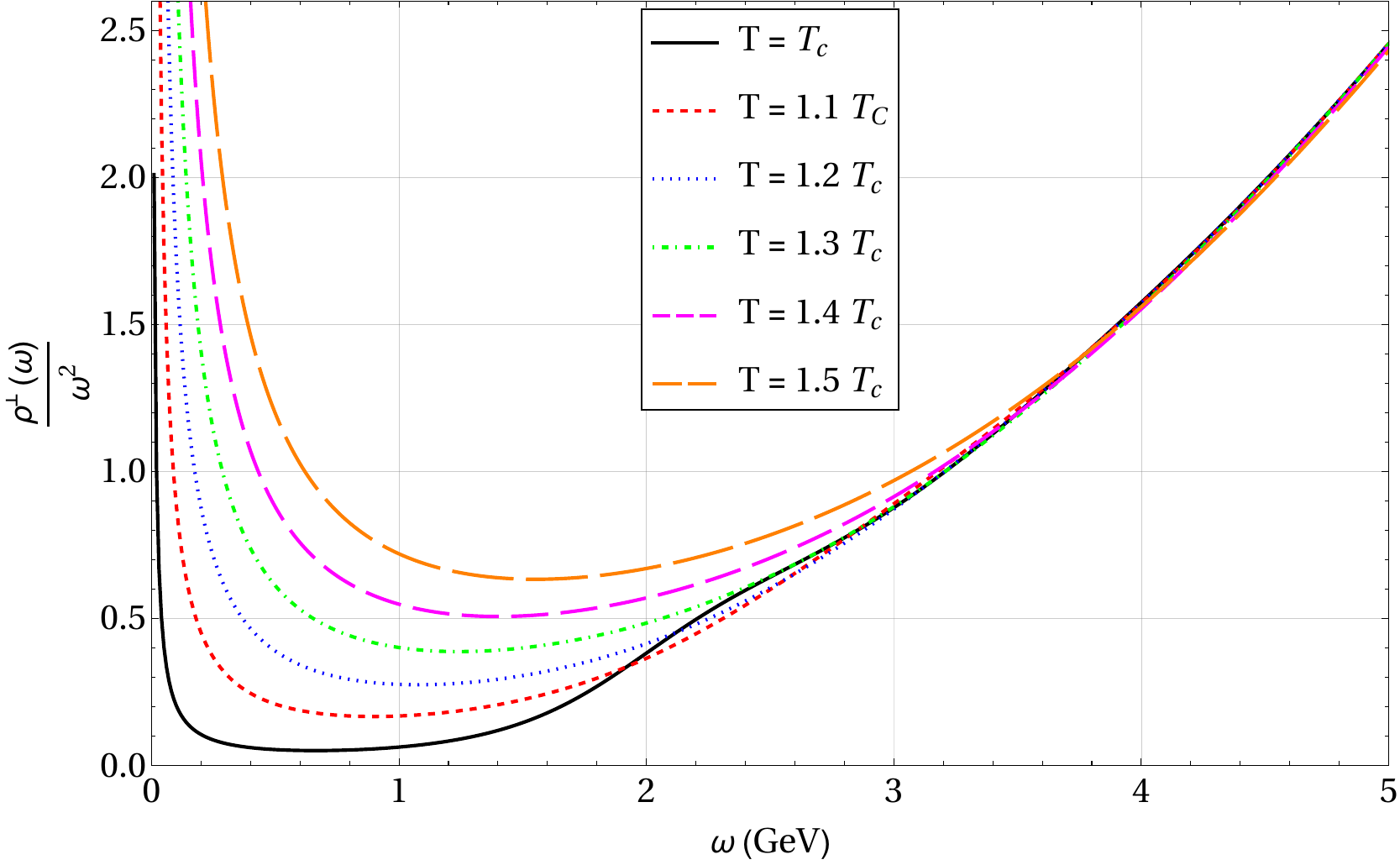}}
  \subfigure[$B=0.2~\text{GeV}$]{\includegraphics[width=0.45\linewidth]{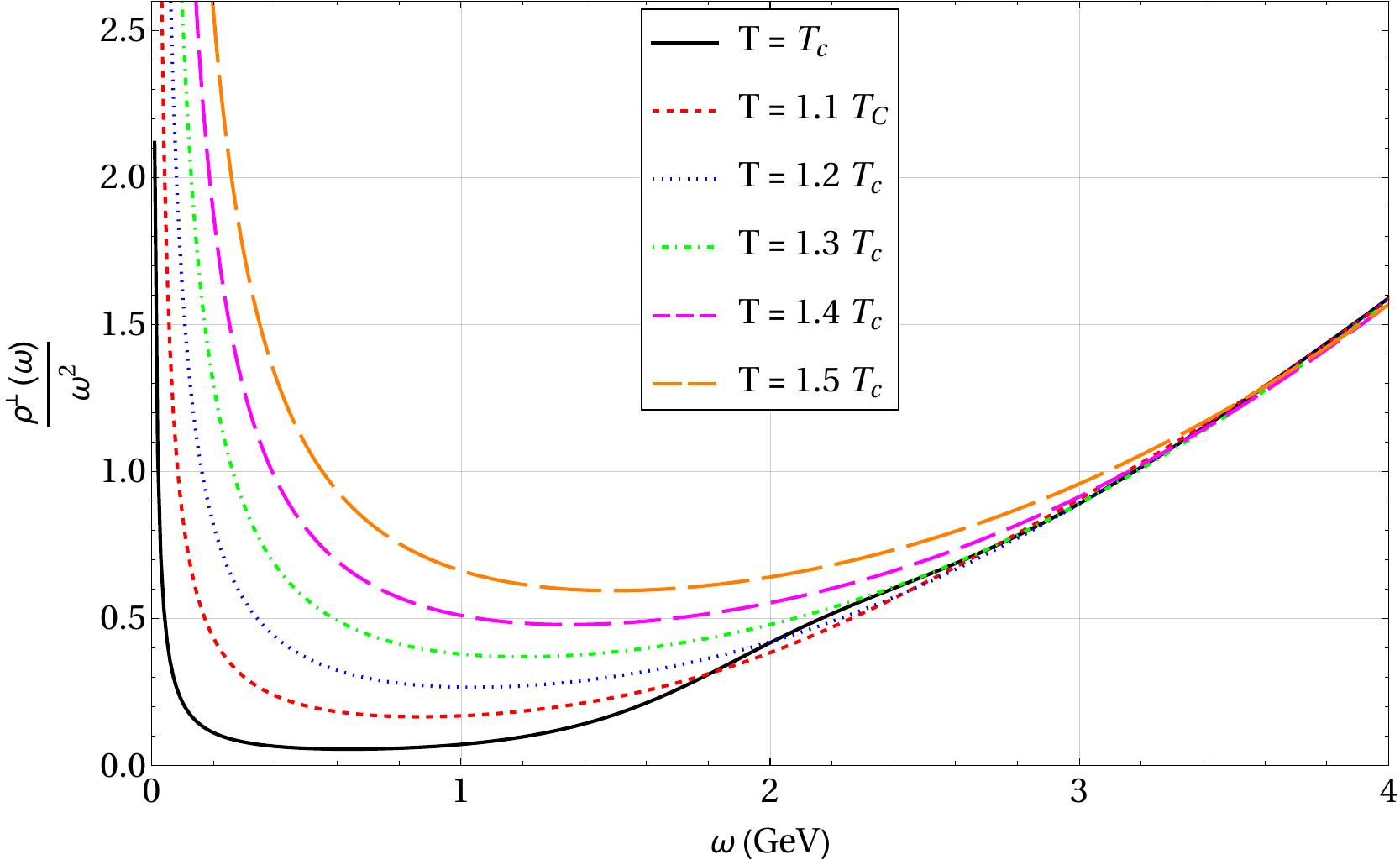}} \\
  \subfigure[$B=0.4~\text{GeV}$]{\includegraphics[width=0.45\linewidth]{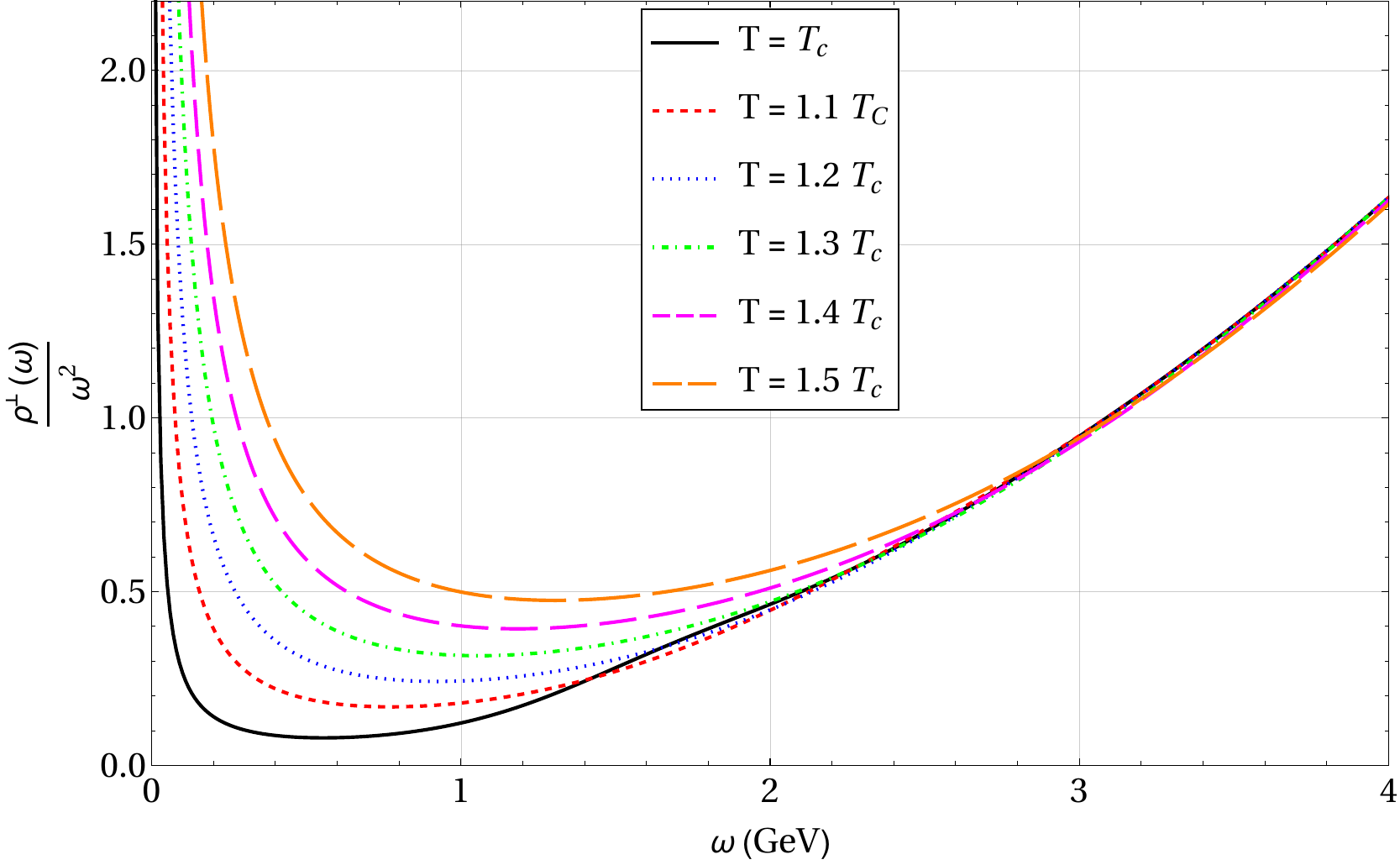}}
  \subfigure[$B=0.6~\text{GeV}$]{\includegraphics[width=0.45\linewidth]{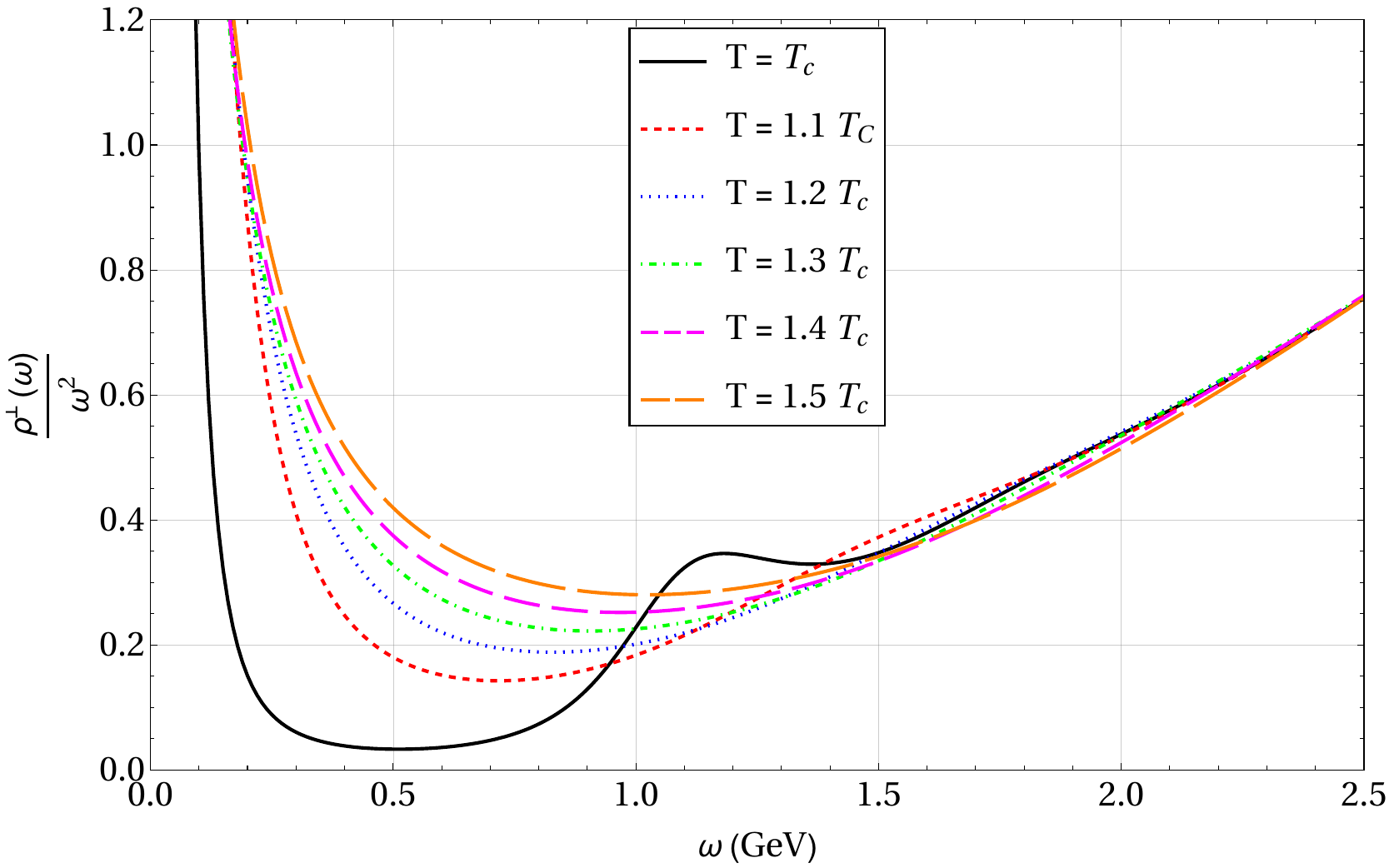}}
  \caption{\small Variation of the spectral function with $\omega$ for fixed values of $B$ with varying temperatures in the perpendicular direction.}
  \label{spf_fixedB_perpendicular}
\end{figure}

We now examine the case where the polarization is perpendicular to the magnetic field. The behavior of the spectral function for different temperature and magnetic field values is shown in Fig.~\ref{spf_diffB_perp} and Fig.~\ref{spf_fixedB_perpendicular}. 

The thermal evolution of the spectral peak remains similar to that observed in the parallel case as the temperature varies. In particular, quarkonium melting progresses with increasing temperature, indicating that, as in the parallel case, higher temperatures serve as a catalyst for the melting process.

However, notable differences arise when compared to the parallel case. Specifically, no spectral peaks are observed even at $T=T_c$ for magnetic field values up to $B \sim 0.5$, indicating that quarkonia melt before the onset of deconfinement. This stands in contrast to the parallel magnetic field case, where peaks persisted up to $T=1.6~T_c$. The same trend holds for other magnetic field values as well, implying that quarkonium bound states can survive at relatively higher temperatures when the magnetic field is applied in the parallel direction compared to the perpendicular direction. 

At $T=1.1~T_c$, spectral peaks are absent for all magnetic field strengths, suggesting that quarkonium has completely melted regardless of the presence or magnitude of the magnetic field. These findings highlight the dissociation of quarkonia at elevated temperatures. 

Thus, the overall holographic analysis suggests that, similar to the behavior of string tension, the magnetic field induces anisotropic effects in quarkonium melting, leading to stronger suppression in the perpendicular direction. Importantly, Fig.~\ref{spf_fixedB_perpendicular} (a) is identical to Fig.~\ref{spf_fixedB_parallel} (a) because, when the magnetic field is turned off, the spatial anisotropy disappears.

\newpage

\section{The membrane approach to compute the spectral functions}\label{memb}

In this Section, we recompute the heavy-quarkonium spectral functions using the membrane-paradigm (radial flow)
approach \cite{Iqbal:2008by,BibEntry2024Jan}. The idea is to rewrite the linearized fluctuation problem as a first-order
flow equation for a complex conductivity $\sigma(\omega,z)$, which can be integrated numerically from the horizon to
the boundary. Horizon regularity supplies the initial condition, and the boundary value determines the retarded
correlator and spectral function. The resulting $\rho(\omega)$ must agree with the real-time prescription used in
Secs.~\ref{procespec}, \ref{meltingmu0} and \ref{metricmu}, providing a nontrivial consistency check. 

\subsection{The dual current $j^\mu$}

We start from the quadratic fluctuation dynamics of the Born--Infeld sector derived in Sec.~\ref{sec:geom_quant}.
Up to an overall normalization (which does not affect the flow equation), the relevant bulk action can be written as
\begin{equation}
S^{(2)}_{\rm BI}
\propto
-\int d^5x\;
\frac{f(\phi)}{4}\,\sqrt{-\mathcal{G}}\;
G^{MP}G^{NQ}\,\tilde F_{MN}\tilde F_{PQ}\,,
\label{fluctuationaction}
\end{equation}
where $\mathcal{G}_{MN}=g_{MN}+\bar F_{MN}/b$ and $G^{MN}$ is the symmetric (open-string) inverse metric introduced in
Sec.~\ref{sec:geom_quant}. The corresponding equation of motion is Eq.~(\ref{fluctuationEOM}),
\begin{equation}
\partial_M\!\left(\sqrt{-\mathcal{G}}\,f(\phi)\,G^{MP}G^{NQ}\,\tilde F_{PQ}\right)=0\,.
\label{fluctuationEOMA}
\end{equation}

Following \cite{Iqbal:2008by}, we define the radial canonical momentum (dual current) on constant-$z$ slices by
\begin{equation}
j^\mu
\equiv
-\frac{\delta \mathcal{L}^{(2)}}{\delta(\partial_z V_\mu)}
=
\sqrt{-\mathcal{G}}\,f(\phi)\,G^{zz}G^{\mu\nu}\,\tilde F_{z\nu}\,.
\label{mpcurrent}
\end{equation}
where we work in radial gauge $V_z=0$ and raise indices with the same effective metric that appears in
\eqref{fluctuationEOMA}. In the channels considered below, only one component $\nu=\mu$ is active, so
$j^\mu=\sqrt{-\mathcal{G}}f\,G^{zz}G^{\mu\mu}\tilde F_{z\mu}$.

Because the magnetic background breaks spatial $SO(3)$, we treat separately the longitudinal (parallel) channel
\((t,x_1)\) and the transverse channel \((t,x_3)\).

\subsection{Membrane: Parallel case}

We take fluctuations with momentum along $x_1$ during the derivation and set $k\to0$ at the end, in analogy with the
standard membrane-paradigm construction. In Fourier space, we write
\(
V_\mu(t,x_1,z)=e^{-i\omega t+i k x_1}V_\mu(z)
\),
and define the gauge-invariant longitudinal electric field
\begin{equation}
E_L(z)\;\equiv\;\tilde F_{x_1 t}(z)
=\partial_{x_1}V_t-\partial_t V_{x_1}
=i k\,V_t+i\omega\,V_{x_1}\,,
\end{equation}

Taking $N=t$ and $N=x_1$ in \eqref{fluctuationEOMA} and using that background functions depend only on $z$ gives
\begin{align}
\partial_z j^t + i k\,\sqrt{-\mathcal{G}}\,f(\phi)\,G^{tt}G^{x_1x_1}\,E_L &=0\,, \label{eq:mp_t}\\
\partial_z j^{x_1} + i \omega\,\sqrt{-\mathcal{G}}\,f(\phi)\,G^{tt}G^{x_1x_1}\,E_L &=0\,, \label{eq:mp_x}\\
-i\omega\,j^t + i k\,j^{x_1} &=0\,, \label{eq:mp_cons}
\end{align}
where the last line is the $(N=z)$ component (current conservation on the slice).

Next, the Bianchi identity
\(
\partial_z\tilde F_{x_1 t}+\partial_{x_1}\tilde F_{t z}+\partial_t\tilde F_{z x_1}=0
\)
becomes, in Fourier space and radial gauge,
\begin{equation}
\partial_z E_L
- i k\,\tilde F_{z t}
+ i\omega\,\tilde F_{z x_1}=0\,.
\label{bianchiidentity}
\end{equation}
Using \eqref{mpcurrent} to express \(\tilde F_{z t}\) and \(\tilde F_{z x_1}\) in terms of \(j^t\) and \(j^{x_1}\),
\[
\tilde F_{z t}=\frac{j^t}{\sqrt{-\mathcal{G}}\,f(\phi)\,G^{zz}G^{tt}}\,,\qquad
\tilde F_{z x_1}=\frac{j^{x_1}}{\sqrt{-\mathcal{G}}\,f(\phi)\,G^{zz}G^{x_1x_1}}\,,
\]
Eq.~\eqref{bianchiidentity} becomes
\begin{equation}
\partial_z E_L
-\frac{i k}{\sqrt{-\mathcal{G}}\,f(\phi)}\frac{j^t}{G^{zz}G^{tt}}
+\frac{i \omega}{\sqrt{-\mathcal{G}}\,f(\phi)}\frac{j^{x_1}}{G^{zz}G^{x_1x_1}}
=0\,.
\end{equation}

We now define the longitudinal conductivity as the ratio of the radial current to the electric field,
\begin{equation}
\sigma_L(\omega,k;z)\equiv \frac{j^{x_1}(\omega,k;z)}{E_L(\omega,k;z)}\,,
\label{conductivity}
\end{equation}
Differentiating and using \eqref{eq:mp_x} together with the Bianchi relation yields a closed first-order equation for
\(\sigma_L\). In the zero-momentum limit \(k\to0\), one obtains the standard Riccati flow equation
\begin{equation}
\partial_z\sigma_L
=i\omega\,\sqrt{-\frac{\mathcal{G}_{zz}}{\mathcal{G}_{tt}}}
\left[\frac{\sigma_L^2}{\zeta_L(z)}-\zeta_L(z)\right],
\label{PDFsigmaL}
\end{equation}
where we introduced
\begin{equation}
\zeta_L(z)\equiv
f(\phi)\,
\sqrt{-\frac{\mathcal{G}}{\mathcal{G}_{zz}\mathcal{G}_{tt}}}\,
G^{x_1x_1}\,,
\end{equation}
Regularity of the physical solution at the horizon fixes the initial condition
\begin{equation}
\sigma_L(\omega,z_h)=\zeta_L(z_h)\,,
\end{equation}
so \eqref{PDFsigmaL} becomes an initial-value problem that can be integrated from \(z=z_h\) to the boundary \(z=0\),
yielding \(\sigma_L(\omega,0)\).

The spectral function in the parallel channel then follows from the same convention used throughout this work,
\begin{equation}
\boxed{\rho_\parallel(\omega)=\omega\,\Re\big[\sigma_L(\omega,0)\big]\,.}
\end{equation}
together with the Kubo relation
\begin{equation}
\sigma_L(\omega)=-\frac{G_R^\parallel(\omega)}{i\omega}\,,
\qquad
\rho_\parallel(\omega)=-\Im\,G_R^\parallel(\omega)\,.
\end{equation}

Within numerical precision, the resulting \(\rho_\parallel(\omega)\) agrees with the real-time computation in
Sec.~\ref{sec:spectral_finiteB}, as expected.

\subsection{Membrane: Perpendicular case}

For the transverse polarization, we take fluctuations along \(x_3\), with momentum again chosen along \(x_1\) during
the derivation. Defining \(E_T(z)\equiv \tilde F_{x_3 t}(z)\) and
\begin{equation}
\sigma_T(\omega,k;z)\equiv \frac{j^{x_3}(\omega,k;z)}{E_T(\omega,k;z)}\,,
\label{conductivitytransverse}
\end{equation}
one finds, in complete analogy with the longitudinal case, the flow equation at \(k\to0\),
\begin{equation}
\partial_z\sigma_T
=i\omega\,\sqrt{-\frac{\mathcal{G}_{zz}}{\mathcal{G}_{tt}}}
\left[\frac{\sigma_T^2}{\zeta_T(z)}-\zeta_T(z)\right]\,,
\label{PDEsigmaT}
\end{equation}
with
\begin{equation}
\zeta_T(z)\equiv
f(\phi)\,
\sqrt{-\frac{\mathcal{G}}{\mathcal{G}_{zz}\mathcal{G}_{tt}}}\,
G^{x_3x_3}\,,
\end{equation}
Imposing horizon regularity again fixes \(\sigma_T(\omega,z_h)=\zeta_T(z_h)\), and integrating to the boundary gives
\(\sigma_T(\omega,0)\). The transverse spectral function is then
\begin{equation}
\boxed{\rho_\perp(\omega)=\omega\,\Re\big[\sigma_T(\omega,0)\big]\,.}
\end{equation}
As in the parallel channel, the membrane result agrees with the real-time spectral functions of
Sec.~\ref{sec:spectral_finiteB} within numerical accuracy.

\newpage

\section{Conclusions and outlook}\label{conc}

In Secs.~\ref{qcdover} and \ref{adsover}, we reviewed the conceptual chain that motivates holographic approaches to
strongly coupled QCD-like matter. Starting from zero and finite temperature QFT, we discussed real-time correlators
and spectral functions, the role of black hole thermodynamics in gauge/gravity duality, and the logic of Maldacena’s
conjecture. We then introduced the holographic dictionary and the real-time prescriptions that later allow us to
compute retarded correlators and spectral densities from bulk fluctuations.

In Sec.~\ref{spectra}, we revisited the prototypical holographic models for hadron spectroscopy, namely the hard-wall
and soft-wall constructions. These models illustrate how breaking conformal invariance in AdS produces discrete,
Regge-like spectra from normalizable bulk modes, but they also highlight well-known shortcomings. In particular,
standard soft-wall models introduce the dilaton profile by hand and do not arise as solutions of coupled Einstein
equations, which limits their thermodynamic consistency.

The main results of this dissertation begin with the construction of two \emph{self-consistent} bottom-up holographic
setups designed to study in-medium mesonic physics through real-time observables. First, we built an
Einstein--Maxwell--dilaton (EMD) model to describe a QCD-like plasma at finite temperature and chemical
potential. Second, we constructed an Einstein-Born-Infeld-dilaton (EBID) model to incorporate an external magnetic
field in an anisotropic but thermodynamically controlled background. In both cases, the defining feature is
mathematical self-consistency: the background geometries are obtained by solving the coupled bulk equations of motion,
so thermodynamics and fluctuations are computed within the same gravitational saddle.

In Secs.~\ref{EMDspec}--\ref{metingmu}, we solved the coupled EMD system using a potential reconstruction strategy and
constructed a dynamical holographic QCD model that exhibits confinement/deconfinement via a Hawking--Page-type
competition between thermal AdS and black hole saddles. A key phenomenological ingredient is the non-quadratic
dilaton profile, which enables non-linear Regge trajectories and improves the description of heavy (and selected
exotic) meson spectra relative to purely quadratic soft-wall profiles. The resulting mass spectra obtained from the
Schr\"odinger-like eigenvalue problem are in good agreement with experimental systematics within the scope of the
model.

At finite temperature, we analyzed the deformation of the effective Schr\"odinger potential and computed vector
spectral functions using the real-time holographic prescription developed earlier. The progressive broadening and
eventual disappearance of spectral peaks provides a real-time characterization of sequential melting, consistent with
the intuitive picture that bound states become destabilized as thermal scales grow. The same conclusions are visible
both in the effective-potential analysis and in the spectral densities, reinforcing the interpretation of the peaks as
quasi-bound excitations in the deconfined medium.

We then extended the analysis to finite baryon chemical potential $\mu$, where the charged black hole branch governs
the deconfined phase. Increasing $\mu$ was found to enhance in-medium dissociation: spectral peaks broaden more
rapidly and disappear at lower temperatures than in the $\mu=0$ case, indicating accelerated melting in a dense
environment. This qualitative trend is relevant for strongly interacting matter at finite baryon density, including
the intermediate-energy regime of heavy-ion collisions and the dense-matter conditions relevant for compact-object
astrophysics.

An additional feature of the finite density thermodynamics is the existence of small and large black hole branches
above a minimum temperature, with distinct stability properties. Deep within each branch, spectral functions display
markedly different behavior: sharp peaks persist in the regime associated with the small-black hole branch, while the
large-black hole branch exhibits substantially broader structures and eventual loss of identifiable resonances. We
emphasize that, in our conventions, spectral functions evolve continuously across the thermodynamic transition: the
transition reorganizes the dominant saddle, while the real-time correlators change smoothly as functions of $(T,\mu)$.
In this sense, spectral functions provide a complementary probe of the plasma, sensitive to the gradual dissolution of
quasi-bound excitations rather than an abrupt order-parameter jump.

In Secs.~\ref{modEBID}--\ref{memb}, we turned to magnetized plasmas and developed a single-flavor EBID model that
incorporates a background magnetic field while retaining a controlled thermodynamic structure, including a
magnetic-field-dependent Hawking--Page transition. In contrast to purely Maxwell-based constructions, the Born--Infeld
sector provides an effective way to encode nonlinear electromagnetic response and a more direct coupling between the
external field and charged constituents in the vector channel. In practice, this coupling becomes crucial once one
seeks to describe how an external field can influence the internal dynamics of a neutral bound state through its
charged constituents.

A central thermodynamic outcome of the EBID model is that the deconfinement transition temperature decreases with
increasing magnetic field over the accessible range, consistent with inverse magnetic catalysis trends reported in
lattice studies. Building on the fluctuation analysis developed in Sec.~\ref{sec:mesonfield}, we computed anisotropic spectral
functions for vector modes polarized parallel and transverse to the magnetic field. The magnetic field breaks spatial
$SO(3)$ invariance and splits the vector channel, leading to distinct melting patterns in the two polarizations. Our
results indicate that thermal effects remain the dominant driver of dissociation, but that the magnetic field can
produce sizable quantitative shifts and an anisotropic survival pattern. In particular, bound-state remnants can
persist to comparatively higher temperatures in the longitudinal channel than in the transverse one, and the
characteristic melting behavior displays a non-monotonic dependence on $B$ within the model’s parameter range.

Finally, in Sec.~\ref{memb} we recomputed the same spectral functions using the membrane-paradigm (radial flow)
approach \cite{Iqbal:2008by}, formulating the problem as a first-order conductivity flow equation integrated from the
horizon to the boundary. The membrane method agrees with the direct real-time computation within numerical precision,
providing a strong consistency check of both the fluctuation equations and our spectral-function extraction
(conventionally, throughout this work, $\rho(\omega)=-\Im G^{R}(\omega)$).

Both the EMD and EBID models are bottom-up constructions and should be interpreted as effective holographic
descriptions rather than exact duals of QCD. In particular, the extracted ratio $L/\ell_s$ in the EBID fit suggests
that higher-derivative $\alpha'$ corrections may not be parametrically suppressed, so quantitative predictions may be
sensitive to stringy corrections. Nevertheless, the qualitative conclusions reported here---sequential melting with
temperature, enhanced dissociation at finite density, inverse magnetic-catalysis behavior for $T_c(B)$, and
polarization-dependent melting in a magnetic background---are robust under reasonable parameter variations within the
model and under the two independent real-time extraction methods employed.

Several extensions are natural. On the EMD side, the same real-time machinery can be used to compute transport
coefficients in the hydrodynamic limit, and to investigate how a non-quadratic Regge trajectory impacts heavy-flavor
diffusion and relaxation. On the EBID side, it would be particularly interesting to compute magnetic-field-dependent
susceptibilities and diffusion coefficients, including their anisotropy, which are directly relevant for
magnetohydrodynamic response and elliptic-flow phenomenology. More broadly, recent progress on including quantum and
higher-curvature corrections in holographic models \cite{daRocha:2021xwq,daRocha:2023waq,Ferreira-Martins:2019svk,daRocha:2024lev}
suggests a systematic route to improving quantitative control over thermodynamics and transport. In addition, the
membrane-paradigm analysis can be generalized to incorporate horizon ``soft hair'' effects along the lines discussed
in \cite{Ferreira-Martins:2021cga}. These directions go beyond the scope of the present dissertation, but they provide
clear paths for refining self-consistent holographic QCD models and confronting increasingly precise experimental and
lattice constraints.

\newpage

\bibliography{mybib}
\bibliographystyle{unsrt}
\end{document}